\documentclass[12pt,english]{oxarticle}
\pdfoutput=1

\usepackage{graphicx, float, array, xspace, amscd, amsmath, amsthm, amssymb, latexsym, bbm, longtable, tikz-cd, mathtools,caption,textcomp}
\usepackage[all,pdf]{xy}\SelectTips {cm}{}
\usepackage[sort&compress, comma, square, numbers]{natbib}
\usepackage{html}
\usepackage{navigator}

\let\SS=\S 

\renewcommand{\a}{\alpha}
\renewcommand{\b}{\beta}
\newcommand{\g}{\gamma}\newcommand{\G}{\Gamma}

\newcommand{\z}{\zeta}

\renewcommand{\l}{\lambda}

\renewcommand{\P}{\Pi}

\renewcommand{\S}{\sum}

\newcommand{\ch}{\chi}
\newcommand{\ps}{\psi}

\DeclareFontFamily{OT1}{pzc}{}
\DeclareFontShape{OT1}{pzc}{m}{it}{<-> s * [1.200] pzcmi7t}{}
\DeclareMathAlphabet{\mathpzc}{OT1}{pzc}{m}{it}

\newcommand{\cA}{\mathcal{A}}

\newcommand{\cK}{\mathcal{K}}

\newcommand{\cN}{\mathcal{N}}
\newcommand{\cO}{\mathcal{O}}

\newcommand{\ccS}{\mathpzc S}

\newcommand{\IC}{\mathbb{C}}

\newcommand{\IP}{\mathbb{P}}
\newcommand{\IQ}{\mathbb{Q}}

\newcommand{\IZ}{\mathbb{Z}}

\font\eightrm=cmr8 at 8pt
\font\csc=cmcsc10

\DeclareFontFamily{U}{wncy}{}
\DeclareFontShape{U}{wncy}{m}{n}{<->wncyr10}{}
\DeclareSymbolFont{mcy}{U}{wncy}{m}{n}
\DeclareMathSymbol{\sha}{\mathord}{mcy}{"58}

\newcommand*{\myalign}[2]{\multicolumn{1}{#1}{#2}}
\newcommand{\varstr}[2]{\vrule height #1 depth #2 width0pt}

\newcommand{\place}[3]{\vbox to0pt{\kern-\parskip\kern-7pt
                             \kern-#2truein\hbox{\kern#1truein #3}
                             \vss}\nointerlineskip}
                             
\newcommand{\capt}[3]{\parbox{#1}{\renewcommand{\baselinestretch}{1.0}
                                                           \caption{\label{#2}\small\it #3}}}

\newcommand{\cy}{Calabi--Yau\xspace}

\newcommand{\cys}{Calabi--Yau manifolds\xspace}
\newcommand{\K}{K\"{a}hler\xspace}

\newcommand{\beq}{\begin{equation}}
\newcommand{\eeq}{\end{equation}}
\newcommand{\beqnn}{\begin{equation*}}
\newcommand{\eeqnn}{\end{equation*}}

\newcommand{\fref}[1]{Figure~\ref{#1}}
\newcommand{\tref}[1]{Table~\ref{#1}}
\newcommand{\sref}[1]{\SS\ref{#1}}

\newcommand{\hodgenos}{(h^{1,1},\,h^{2,1})}

\newcommand{\ii}{\text{i}} 

\newcommand{\+}{\hphantom{-}}

\newcommand{\cicy}[2]{\begin{matrix} #1\end{matrix}\!\left[\begin{matrix}#2 \end{matrix}\right]}

\newcommand{\smallcicy}[2]{\footnotesize{\begin{matrix} #1\end{matrix}\!\left[\begin{matrix}#2 \end{matrix}\right]}}

\newcommand{\displaycicy}[5]{\parbox[c]{#1}{\footnotesize{$#2$\hfill\raisebox{#3}{\includegraphics[width=#4]{#5}}}\hspace*{0.1in}}}

\newcommand{\lrarr}{\longrightarrow }    

\newcommand{\quotient}[1]{_{\hskip-2pt\lower1pt\hbox{$/$}\lower2pt\hbox{\hskip-1pt$#1$}}}

\renewcommand{\baselinestretch}{1.1}

\numberwithin{equation}{section}
\proofmodefalse
\begin{document}
\pagestyle{empty}
\begin{center}
\null\vskip0.3in
{\LARGE \textsc{Hodge Numbers for CICYs with Symmetries\\[1ex]
 of Order Divisible by 4}\\[0.59in]}
{\csc Philip Candelas$^1$, Andrei Constantin$^2$  \\
and \\
Challenger Mishra$^3$\\[1.3cm]}
{\it $^1$Mathematical Institute\hphantom{$^1$}\\
University of Oxford\\
Radcliffe Observatory Quarter\\ 
Woodstock Road,
Oxford OX2 6GG, UK\\[4ex]
$^2$Department of Physics and Astronomy\hphantom{$^2$}\\
Uppsala University, \\ 
SE-751 20, Uppsala, Sweden\\[4ex]
$^3$Rudolf Peierls Centre for Theoretical Physics\hphantom{$^3$}\\
University of Oxford\\
1 Keble Road, 
Oxford OX1 3NP, UK\\
}

\vfill
{\bf Abstract\\[2ex]}
\parbox{6.0in}{\setlength{\baselineskip}{14pt}
We compute the Hodge numbers for the quotients of complete intersection Calabi-Yau three-folds by groups of orders divisible by $4$. We make use of the polynomial deformation method and the counting of invariant K\"ahler classes. The quotients studied here have been obtained in the automated classification of V.~Braun. Although the computer search found the freely acting groups, the Hodge numbers of the quotients were not calculated. The freely acting groups, $G$, that arise in the classification are either $\IZ_2$ or contain $\IZ_4$, $\IZ_2{\times}\IZ_2$, $\IZ_3$ or $\IZ_5$ as a subgroup. The Hodge numbers for the quotients for which the group $G$ contains $\IZ_3$ or $\IZ_5$ have been computed previously. This paper deals with the remaining cases, for which $G\supseteq \IZ_4$ or $G\supseteq \IZ_2{\times}\IZ_2$. We also compute the Hodge numbers for 99 of the 166 CICY's which have $\IZ_2$ quotients.
}

\end{center}
\begingroup
\baselineskip=14pt
\tableofcontents
\endgroup
\newpage
\setcounter{page}{1}
\pagestyle{plain}
\section{Introduction}
Compactifications of the heterotic string are based on smooth Calabi-Yau three-folds \cite{Candelas:1985en}. This approach remains a promising avenue from string theory to realistic particle physics phenomenology \cite{Braun:2005ux, Braun:2005bw, Anderson:2009mh, Bouchard:2005ag, Braun:2011ni, Anderson:2011ns, Anderson:2012yf, Anderson:2013xka}. We are led to seek \cys with small Hodge numbers by a desire to find realistic models constructed with a minimum of complexity. While not universally the case, since the very special class of Gross-Popescu manifolds
\cite{Gross:2001as} yield a small number of manifolds that have small Hodge numbers and are nevertheless simply connected, the great majority of known \cys with small Hodge numbers are realised as quotients of simply connected manifolds by a freely acting group. The process of taking a quotient by a freely acting group does double duty, by first reducing the Hodge numbers, but also by creating a multiply connected manifold. Flux lines around the irreducible paths of the manifold then allow the breaking of the gauge group to the Standard Model group.

A large number of examples of this type resulted from the work initiated in~\cite{Candelas:2008wb} and completed through the automated scan carried out by Volker Braun in~\cite{Braun:2010vc}. Braun's scan led to a complete classification of all free linear actions of finite groups on complete intersection Calabi-Yau (CICY) manifolds embedded in products of projective spaces.\footnote{The embeddings used in this classification were those of the CICY list Ref.~\cite{Candelas:1987kf}. Though the existence of a certain symmetry of a CICY does not depend on the embedding, the linearity of the action does. As such, one expects that Braun's classification is not complete, if different equivalent embeddings are considered.} 

For model building, the properties of the quotient manifolds $X/G$ are of prime importance. In particular, the Hodge numbers of the quotients $h^{1,1}(X/G)$ and $h^{2,1}(X/G)$ play a central role. While Braun's scan gave a complete listing of the freely acting symmetries, the individual Hodge numbers of the quotients were not calculated, though of course the difference
$$
2\Big( h^{1,1}(X/G) - h^{2,1}(X/G) \Big)~=~\frac{\chi(X)}{|G|}
$$
follows immediately from the fact that the Euler number divides by the order of the group.

Braun found that there are 166 manifolds for which $G$ is precisely $\IZ_2$, and for all cases, where $|G|>2$, that $G$ is either $\IZ_3,~\IZ_4,~\IZ_2{\times}\IZ_2$, or $\IZ_5$, or $G$ contains at least one of these groups as a subgroup. The computation of the Hodge numbers, for the case $G\supseteq\IZ_5$ was given in~\cite{Candelas:2008wb}, for the case $G\supseteq\IZ_3$ the majority of the cases were studied in~\cite{Candelas:2008wb} while the remaining cases were studied in~\cite{Candelas:2010ve}. The remaining cases, for which $G\supseteq \IZ_4$ or 
$G\supseteq \IZ_2{\times}\IZ_2$, are the subject of the present work. This completes the calculation of Hodge numbers for the CICY quotients, except for $\IZ_2$-quotients. Though for many of the $\IZ_2-$quotients we are able to compute Hodge numbers. These cases are discussed in the appendix. 
\begin{figure}[!t]
\begin{center}
\includegraphics[width=6.4in]{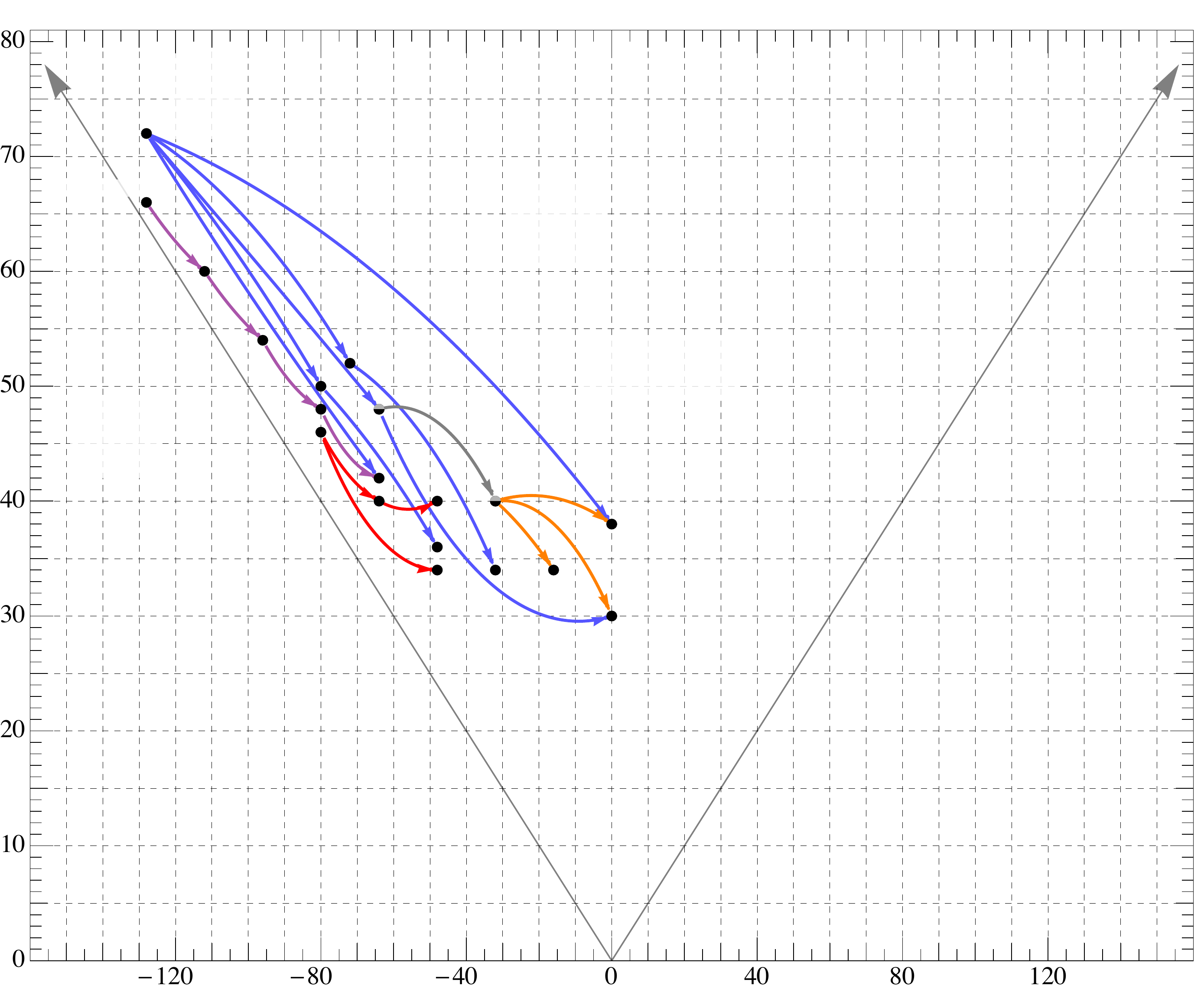}
\vskip0pt 
\place{1.0}{5.15}{\renewcommand{\arraystretch}{0.8}\scalebox{0.7}
{$\boldsymbol{\cicy{P^1\\ P^1\\ P^1\\ P^1}
{2\hspace*{-10pt}\\2\hspace*{-10pt}& \\2\hspace*{-10pt} & \\2\hspace*{-10pt}}}$}}
\place{0.25}{4.48}{\scalebox{0.7}{$\boldsymbol{\cicy{P^7\,}{2,2,2,2\\}}$}}
\place{0.4}{3.25}{\renewcommand{\arraystretch}{0.8}\scalebox{0.7}{
$\boldsymbol{\cicy{P^1\\ P^1 \\P^3}{2\hspace*{-6pt} & 0\\ 0\hspace*{-6pt} & 2\\ 2\hspace*{-6pt} & 2\\}}$}}
\place{0.9}{3.45}{\includegraphics[width=0.8in]{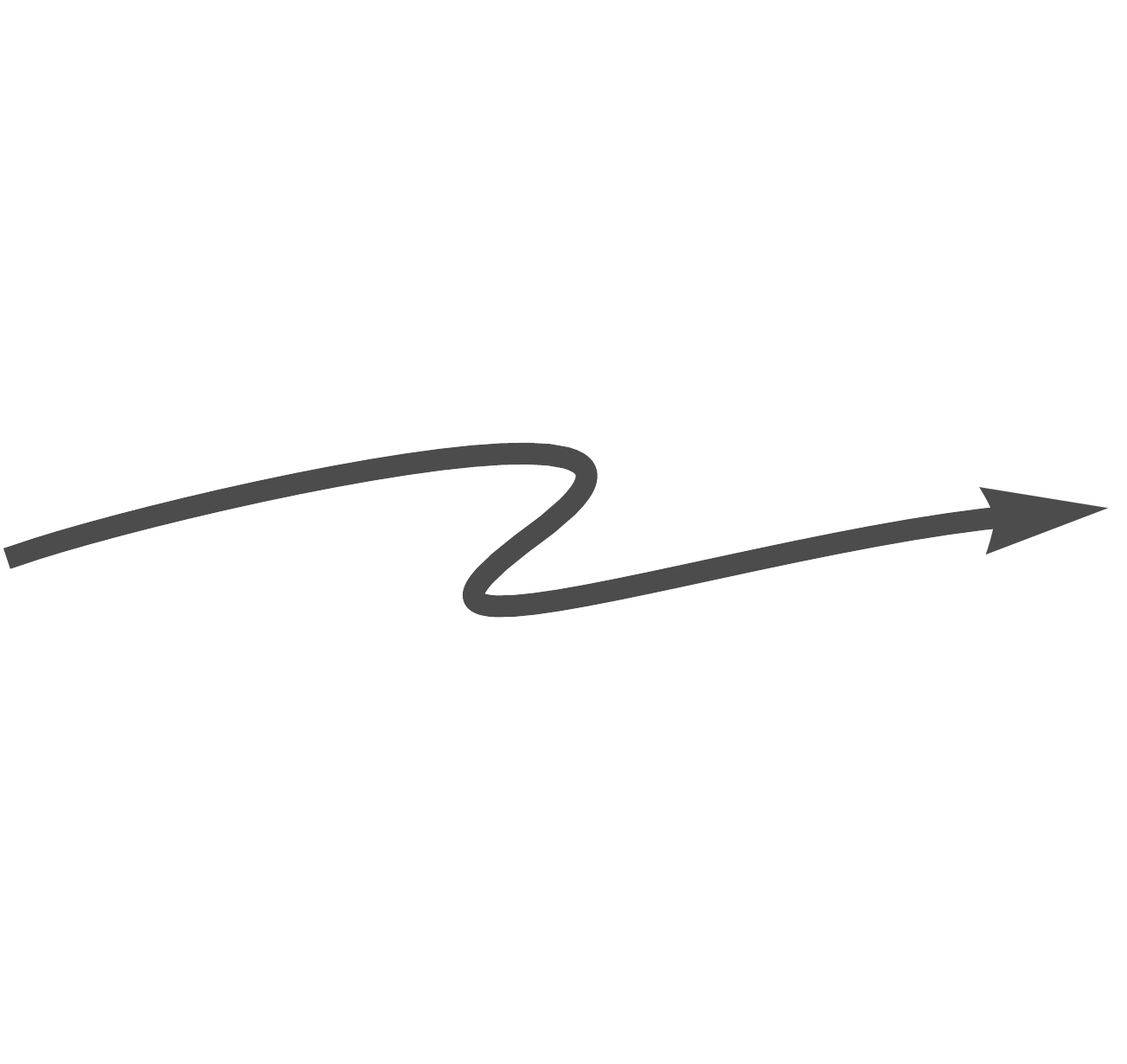}}
\place{3.25}{3.6}{\renewcommand{\arraystretch}{0.8}\scalebox{0.7}
{$\boldsymbol{\cicy{P^4\\ P^4}{2~2~1~0~0\\ 0~0~1~2~2\\}}$}}
\place{2.7}{3.45}{\includegraphics[width=0.5in]{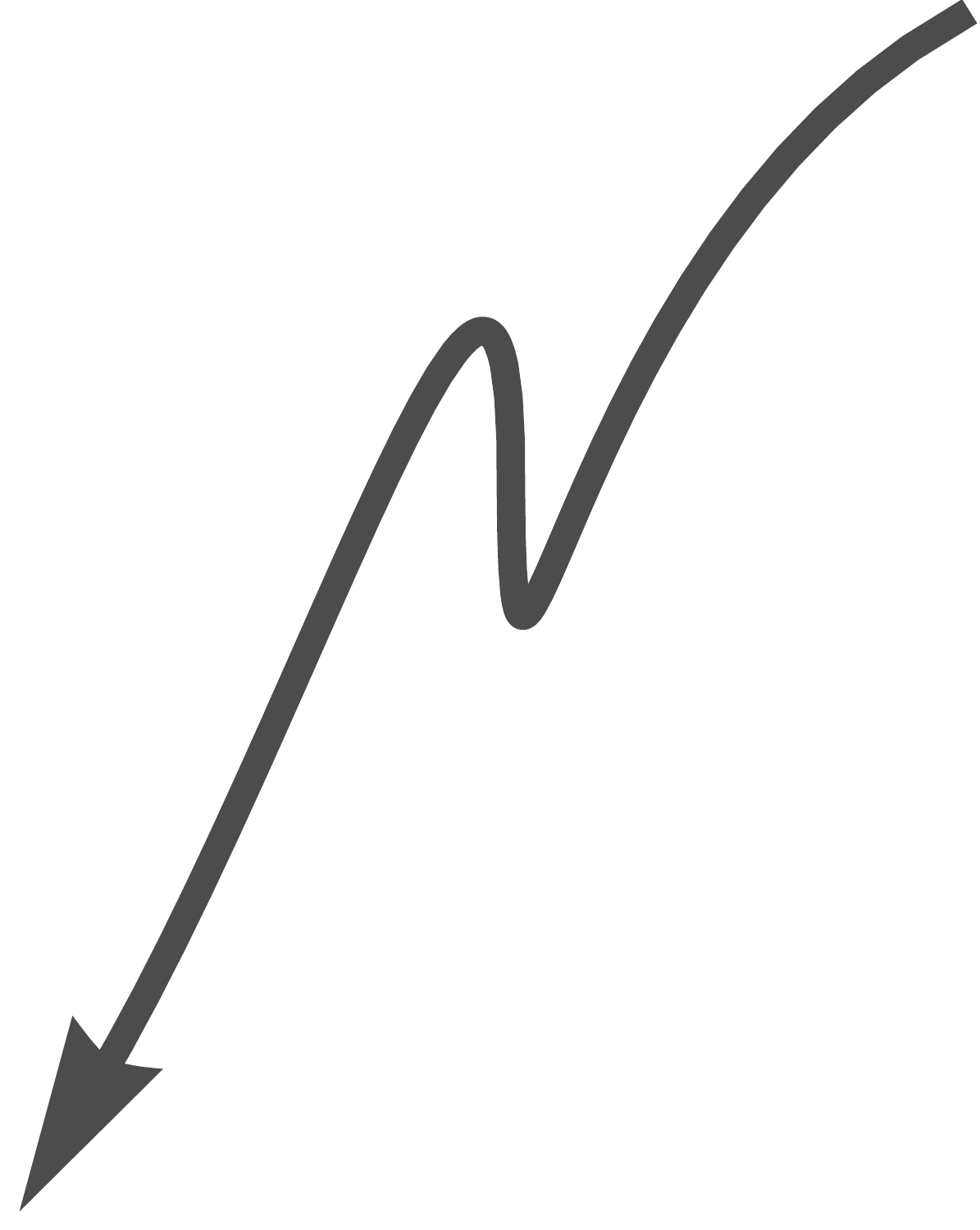}}
\place{2.75}{4.4}{\renewcommand{\arraystretch}{0.8}\scalebox{0.7}{$\boldsymbol{
\cicy{P^1\\ P^1 \\ P^4}{ 
2\hspace*{-6pt} & 0 \hspace*{-6pt} & 0\\ 
2\hspace*{-6pt} & 0 \hspace*{-6pt} & 0\\ 
1\hspace*{-6pt} & 2 \hspace*{-6pt} & 2\\}}$}}
\place{2.1}{4.1}{\includegraphics[width=0.6in]{curlyarrow.pdf}}
\capt{6in}{FourGroupWeb}{The web of CICY manifolds that admit free automorphisms by $\IZ_4$ or 
$\IZ_2{\times}\IZ_2$. The Euler number, $\ch = \frac12(h^{1,1}{-}h^{2,1})$, is plotted horizontally while the height, $h^{1,1}{+}h^{2,1}$, is plotted vertically. The oblique axes correspond to the Hodge numbers $h^{1,1}$ and $h^{2,1}$.}
\end{center}
\end{figure}
\begin{figure}[!t]
\begin{center}
\includegraphics[width=6.4in]{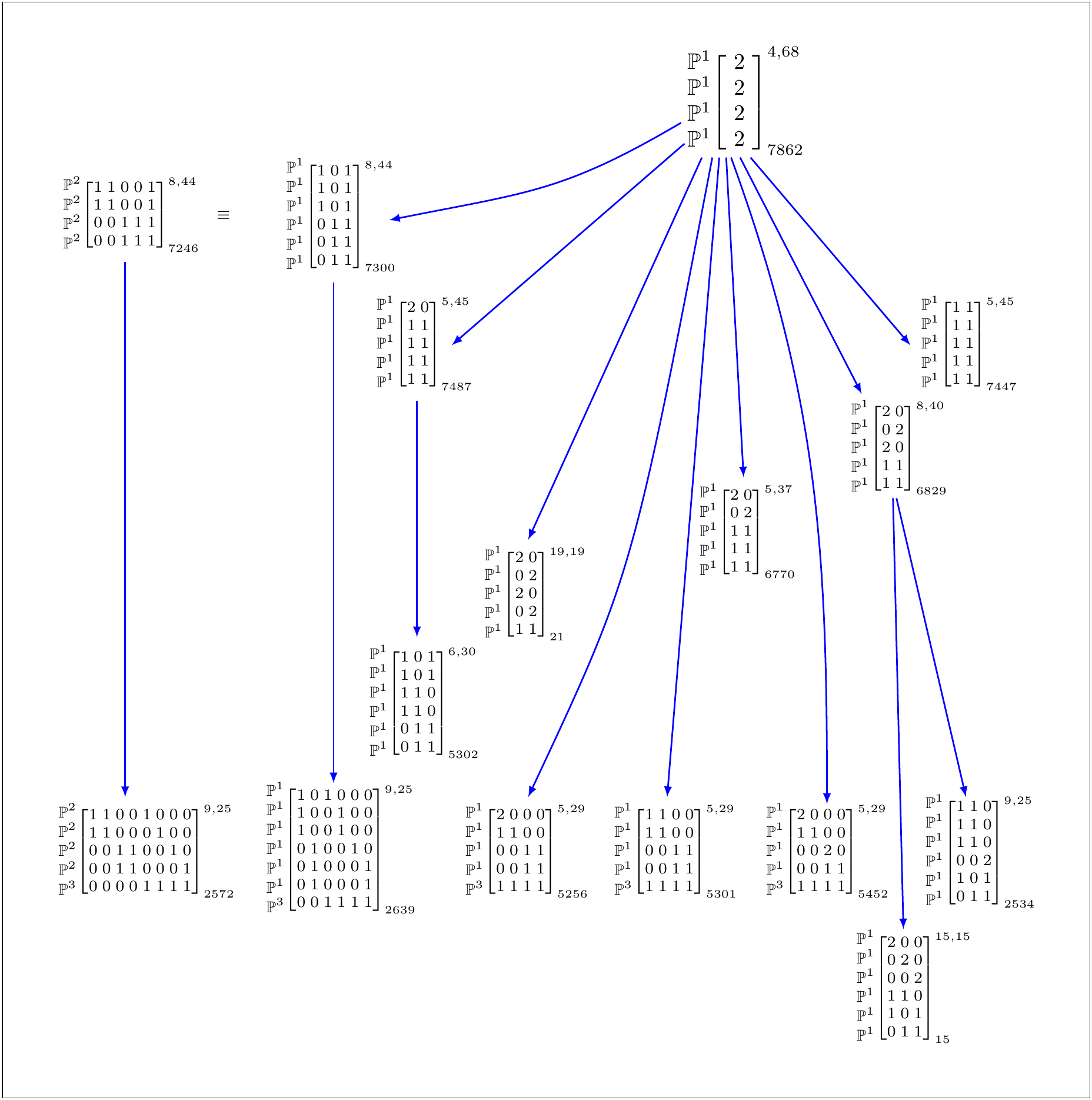}
\vskip3pt
\capt{5.8in}{splits1}{The CICY Web with the tetraquadric as the source manifold. Configuration matrices are decorated with superscripts, indicating the Hodge numbers and subscripts, indicating the position in the CICY list, which is available at \cite{cicylist2}.}
\end{center}
\end{figure}
\subsection{Webs of CICY quotients}
At first sight, the CICY list appears to be a menagerie of manifolds.\footnote{The CICY list can be downloaded online, see Ref.~\cite{cicylist2}.}  Though true to some extent, this impression is mitigated by the fact that, among these manifolds, there are some very interesting spaces. The manifolds that admit symmetry groups of large order are particularly interesting. A second point is that the class has some structure owing to the process of splitting and contraction (a detailed explanation may be found in~\cite{Candelas:2008wb} or~\cite{Hubsch:1992nu}). Indeed the process of repeatedly splitting the matrices that could not be further contracted was the process used to generate the list. Since this process relates the manifolds of the list it is natural that it should also relate the quotients \cite{Candelas:2010ve}.

There are 45 CICY matrices for which $G{\,\supseteq\,}\IZ_4$ or $G{\,\supseteq\,}\IZ_2{\times}\IZ_2$. This somewhat overstates the number of manifolds since a number of the matrices seem to correspond to manifolds that are the same.  The 45 matrices correspond to 19 distinct pairs of Hodge numbers~$\hodgenos$. We present the corresponding web in~\fref{FourGroupWeb}. From this web we obtain others. Those corresponding to 
$G=\IZ_4$ are shown in \fref{Z4Web}, and those corresponding to $G=\IZ_2{\times}\IZ_2$ are shown 
in~\fref{Z2Z2Web}.

Although the number of manifolds is overstated there are nevertheless many cases to consider since a given manifold can admit several and, in some cases many, distinct group actions. An example is provided~by
\beq
X_{7861} ~=~\IP^7[2,2,2,2]~.
\notag\eeq
For this configuration, \tref{tab:SymmetryGroups} lists 21 distinct groups and these give rise to 45 distinct group~actions.
\begin{figure}[p]
\begin{center}
\includegraphics[width=5.75in]{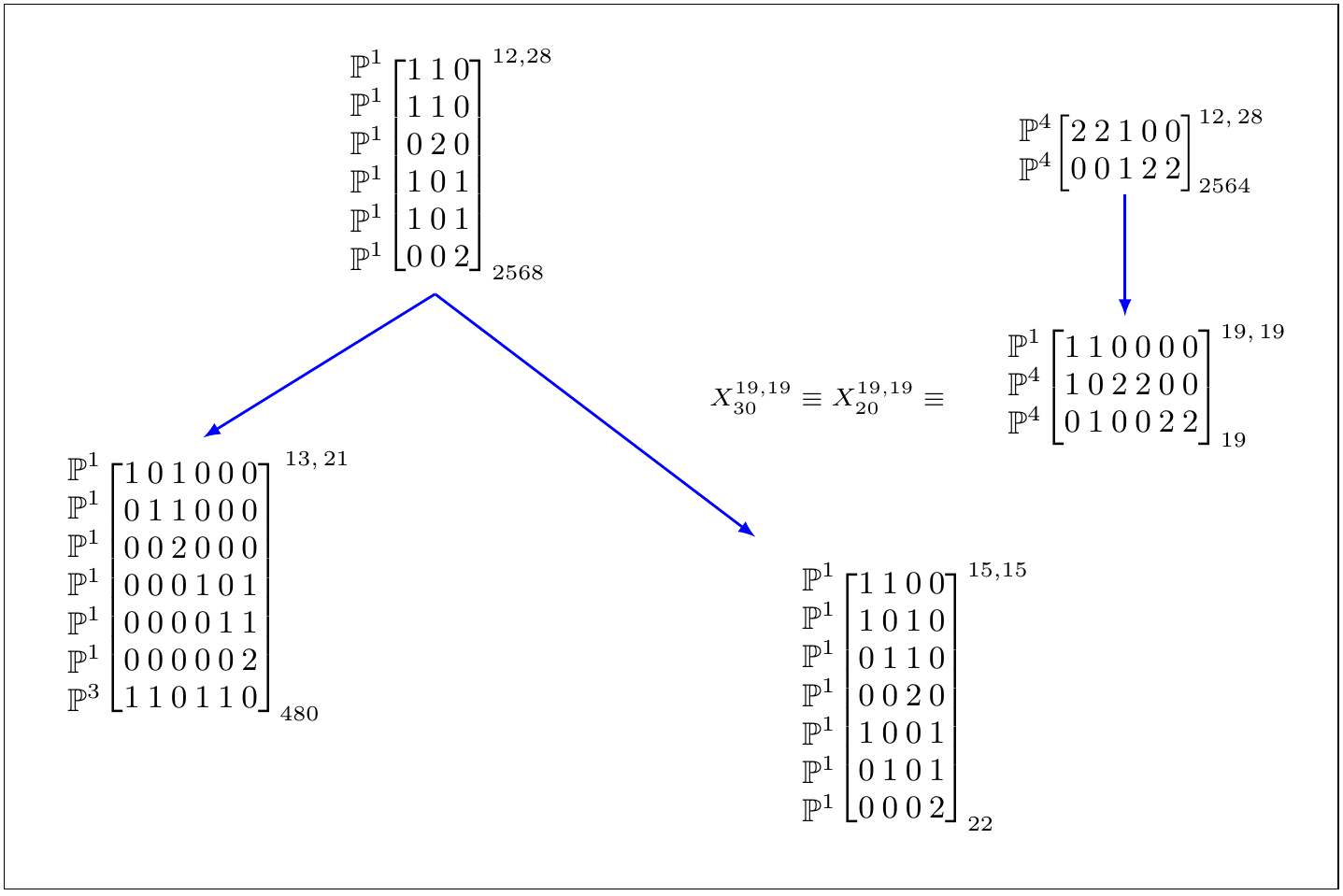}
\vskip5pt
\includegraphics[width=5.75in]{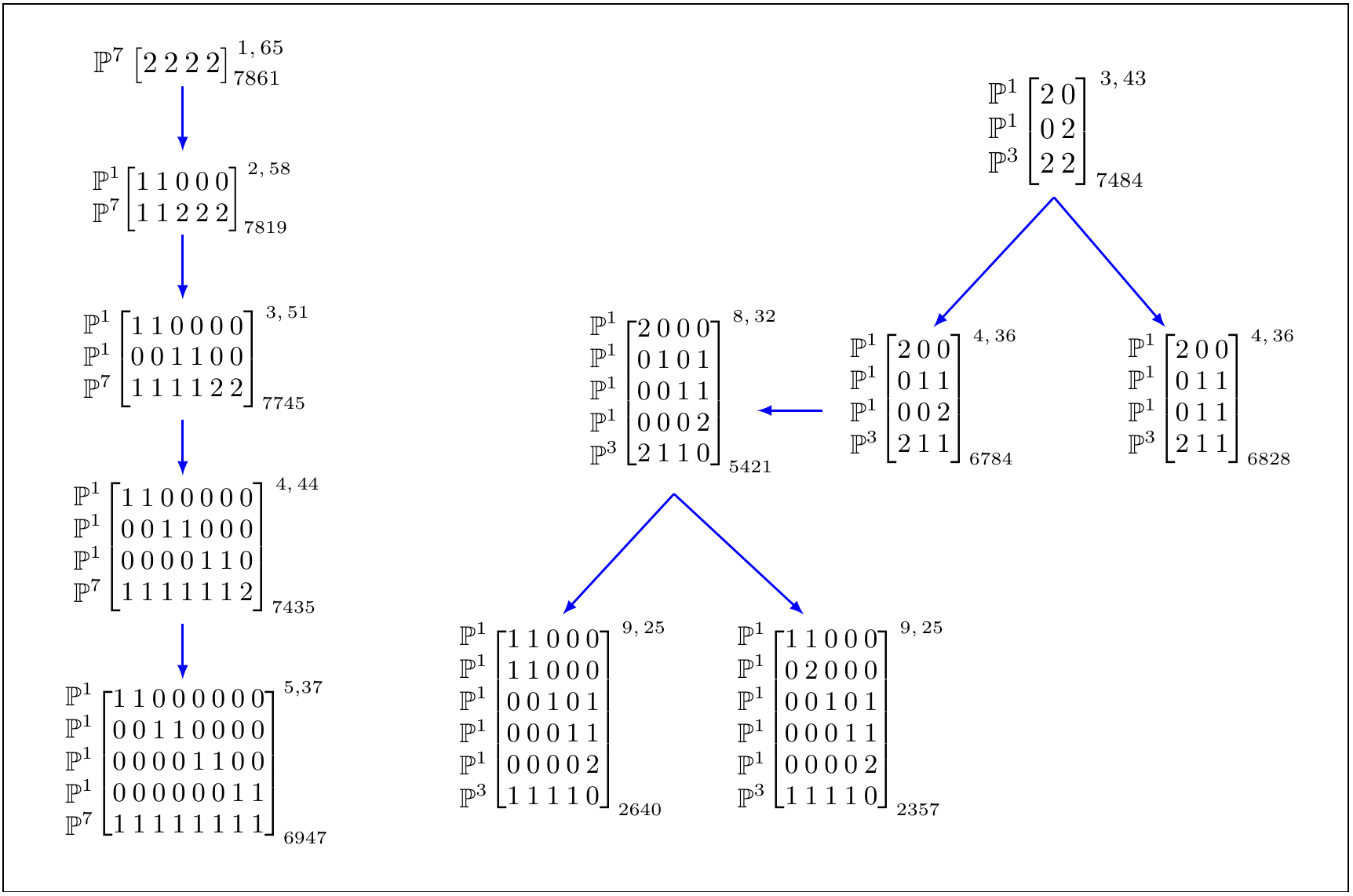}
\vskip10pt
\capt{5.75in}{splits23}{The top diagram shows a CICY Web with parent manifolds $X^{12,28}_{2564}$ and  $X^{12,28}_{2568}$. The diagram below shows CICY Webs with parent manifolds $X^{1,65}_{7861}$ and $X^{3,43}_{7484}$. The conifold transitions with $\IP^7 [2\ 2\ 2\ 2]$ as the parent are described in detail in \fref{FlowerWeb}.}
\end{center}
\end{figure}
\begin{figure}[!t]
\begin{center}
\includegraphics[width=6.4in]{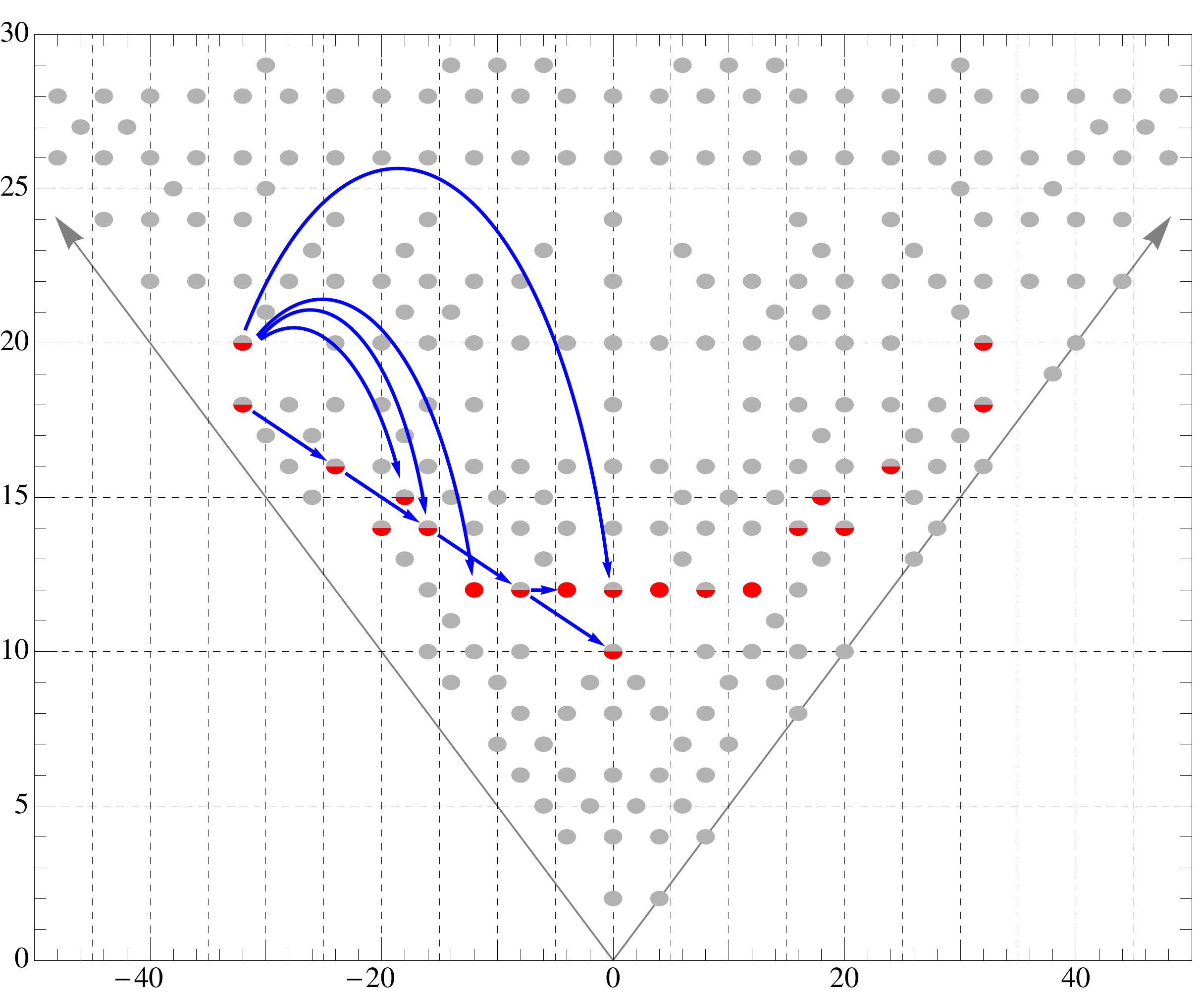}
\vskip3pt
\capt{6.0in}{Z4Web}{The web of $\IZ_4$ quotients of CICY manifolds and their mirrors. The red points indicate $\IZ_4$ quotients whose Hodge numbers fall onto sites previously unoccupied, while the bicoloured points correspond to previously occupied sites.}
\end{center}
\end{figure}
\begin{figure}[!t]
\begin{center}
\includegraphics[width=6.4in]{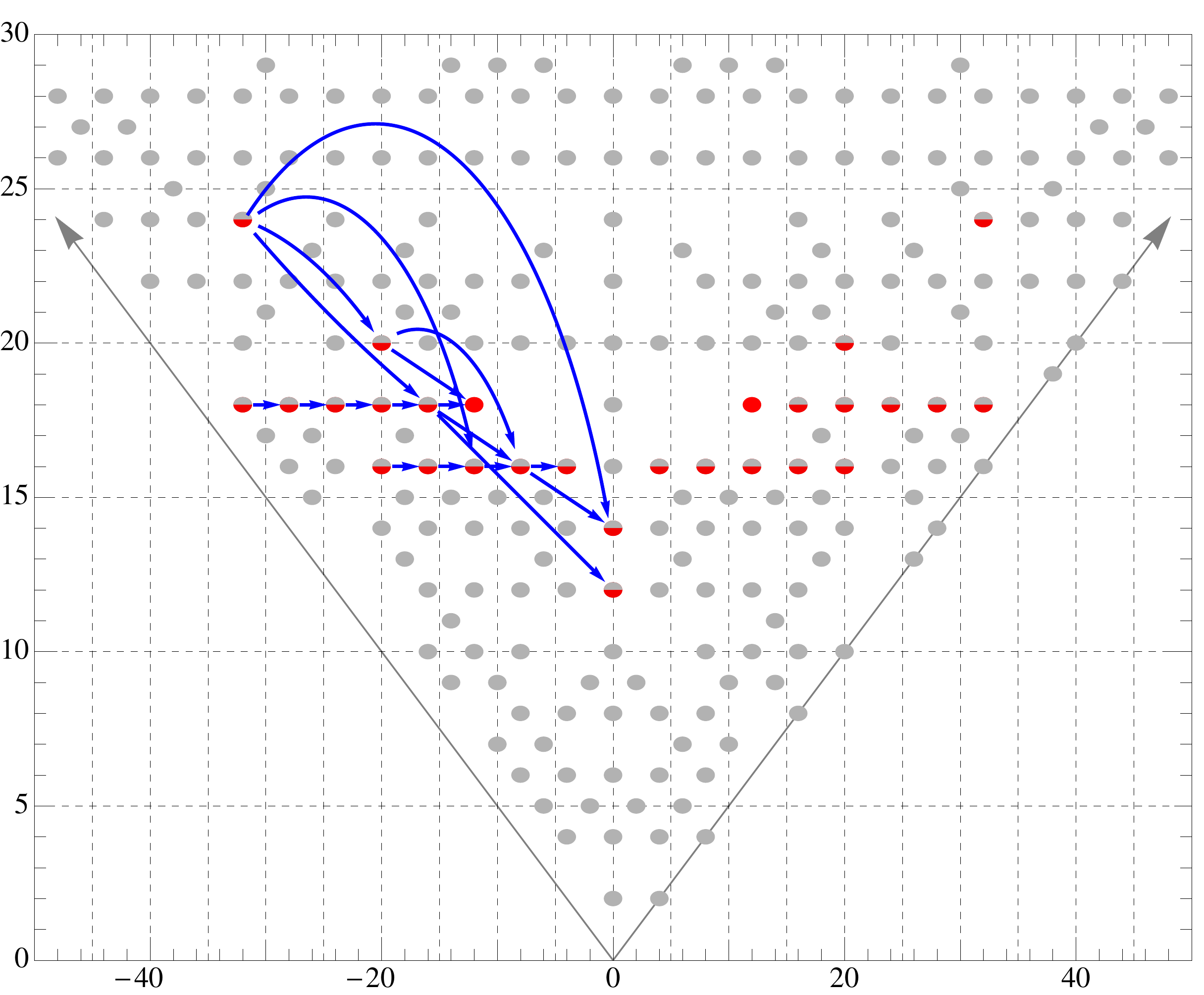}
\vskip3pt
\capt{6.35in}{Z2Z2Web}{The web of $\IZ_2{\times} \IZ_2$ quotients of CICY manifolds and their mirrors. The red points indicate $\IZ_2{\times}\IZ_2$ quotients whose Hodge numbers fall onto sites previously unoccupied, while the bicoloured points correspond to previously occupied sites.}
\end{center}
\end{figure}
\begin{figure}[!t]
\begin{center}
\includegraphics[width=6.3in]{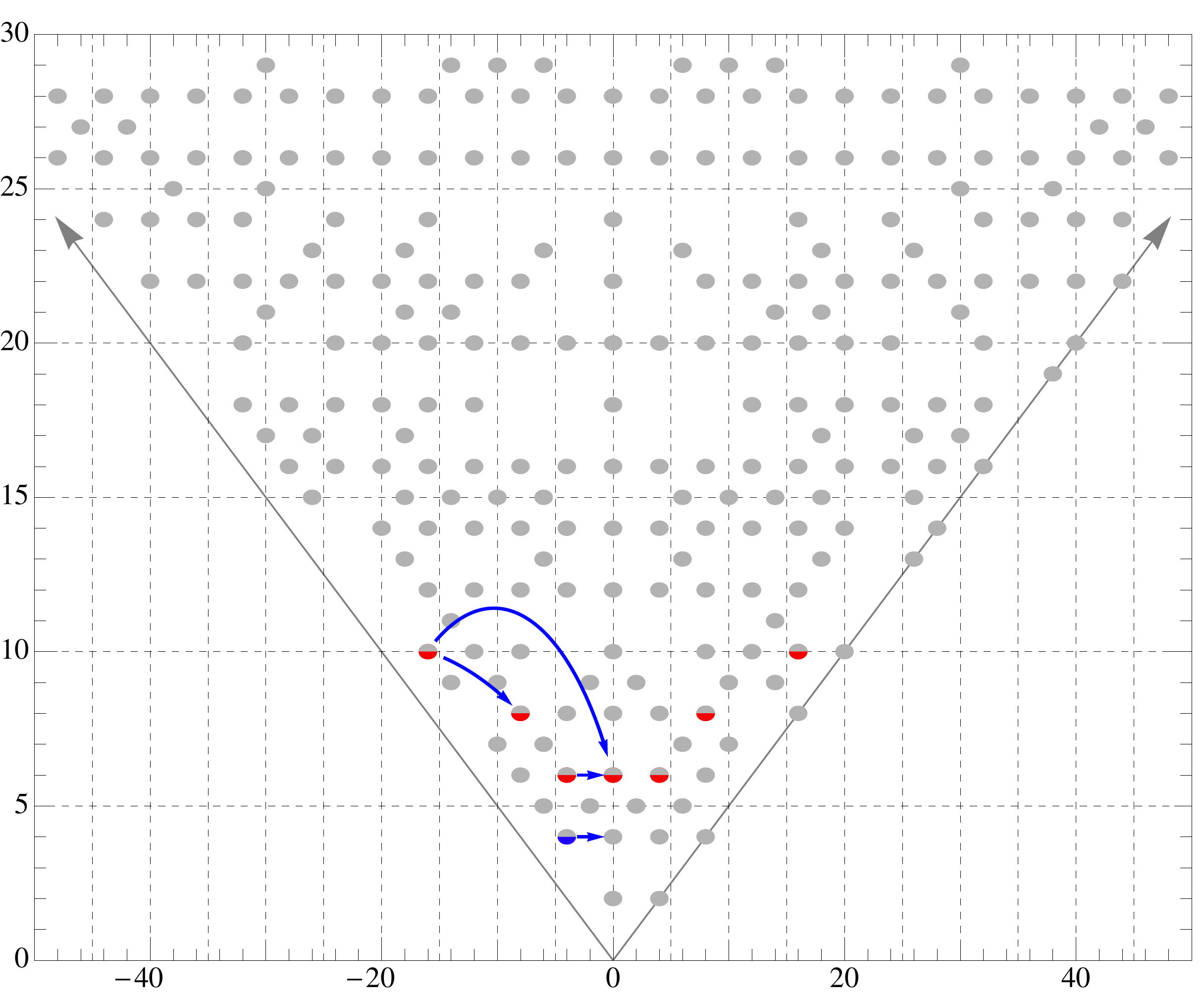}
\vskip3pt
\capt{6in}{Z8Web}{The overlapping web of $\IZ_8$ and $\IQ_8$--quotients of CICY manifolds and their mirrors. The grey and red points indicate quotients whose Hodge numbers fall onto sites previously occupied. The grey and blue point is not part of the web and indicates the manifold $\IP^7 [2\  2\ 2\ 2]/G$ with $|G|=32$. The conifold transition originating there, is to a Gross-Popescu manifold with Hodge numbers $(2,2)$ (see \SS 5 of Ref.~\cite{Candelas:2008wb}).}
\end{center}
\end{figure}

\begin{figure}[!t]
\begin{center}
\includegraphics[width=6.4in]{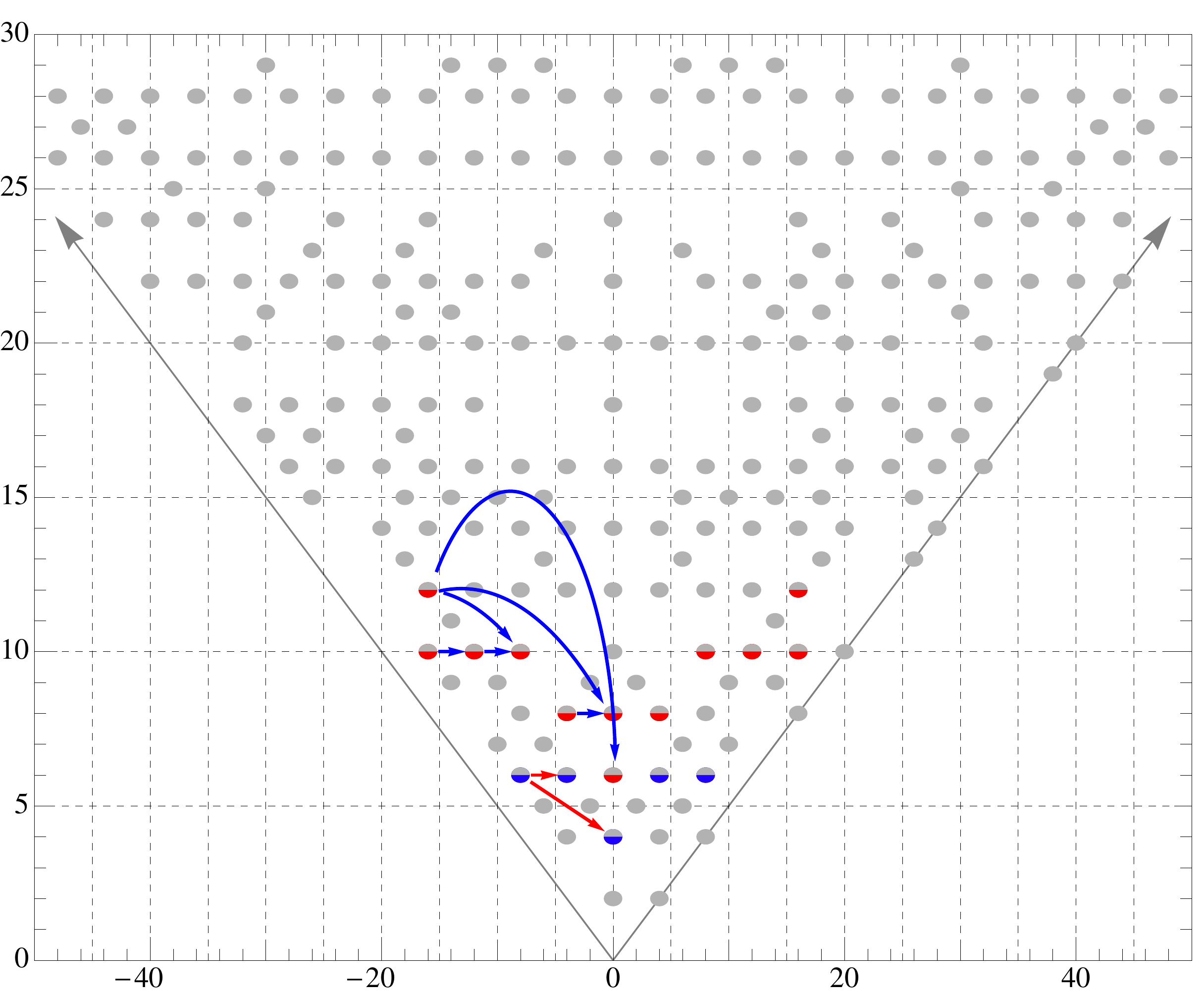}
\vskip3pt
\capt{6.2in}{Z2Z4Web}{The webs of $\IZ_2 {\times} \IZ_4$, $\IZ_4 {\rtimes} \IZ_4$ and $\IZ_2 {\times} \IZ_8$--quotients of CICY manifolds and their mirrors. The webs of $\IZ_4 {\rtimes} \IZ_4$ and $\IZ_2 {\times} \IZ_8$--quotients overlap and contain the blue and grey points connected by red arrows. The $\IZ_2 {\times} \IZ_4$ web corresponds to the red and gray points connected by blue arrows. }
\end{center}
\end{figure}
\subsection{Redundancy of the representations}\label{redun}
While it is far from being the case that the manifolds of the list are classified by their Hodge numbers\footnote{There are 7921 matrices in the CICY list, of which at least 2590 are known to be distinct as classical manifolds, but there are only 266 distinct pairs $\hodgenos$ of Hodge numbers~\cite{He:1990pg}.}, nevertheless there is some redundancy in that some of the configurations with the same Hodge numbers correspond to the same manifold. The process of finding freely acting symmetries will tend to pick out identical manifolds, since if two configurations correspond to identical manifolds and one is symmetric, then so is the other. Where we have found suspected identities between manifolds we have not attempted to prove, in all cases, that some of these manifolds are in fact the same, though in some cases we do. In any event, different representations of the same manifold are often useful. Symmetries, for example, may be more evident in one representation than another.
Braun's classification, as well as previous work, has identified linearly represented symmetries. It is possible, and we will identify examples in the following, for one representation of a manifold to admit a linear representation of a symmetry while another does not.

Consider a first example of the redundancy. The CICY's
\beq
X_{7246}~=~\smallcicy{\IP^2\\ \IP^2\\ \IP^2\\ \IP^2\\}{1&1&0&0&1\\ 1&1&0&0&1\\ 0&0&1&1&1\\ 0&0&1&1&1\\}
\hskip30pt\text{and}\hskip30pt
X_{7300}~=~
\smallcicy{\IP^1\\ \IP^1\\ \IP^1\\ \IP^1\\ \IP^1\\ \IP^1\\}{1&0&1\\ 1&0&1\\ 1&0&1\\ 0&1&1\\ 0&1&1\\ 0&1&1\\}
\label{Order12Groups}\eeq
both have $(h^{1,1},\,h^{2,1})=(8,44)$ and both admit freely acting symmetries of order 12, corresponding to the groups $\IZ_{12}$ and $\text{Dic}_3$. We observe also that the del Pezzo surface\footnote{Our convention here is that $\text{dP}_n$ denotes the del Pezzo surface of degree $n$. Thus the del Pezzo surface corresponding to $\IP_2$ blown up in $k$ generic points is, with this convention, $\text{dP}_{9-k}$.} 
$\text{dP}_6$ can be represented both as
\beqnn
\text{dP}_6\,=~\smallcicy{\IP^2\\ \IP^2}{1&1\\ 1&1\\} \hskip20pt\text{or}\hskip20pt
\normalsize\text{dP}_6\,=~\smallcicy{\IP^1\\ \IP^1\\ \IP^1\\}{1\\1\\1\\}
\eeqnn
so each of the two CICY's above corresponds to a hypersurface in $\text{dP}_6{\times}\text{dP}_6$.
\goodbreak
In this case the manifolds are the same\footnote{We are grateful to Rhys Davies for pointing out the following easy argument.}, 
since in each case, we have a linear system of anti-canonical hypersurfaces in
$\text{dP}_6{\times}\text{dP}_6$, so the only question is whether we cover the same part of the moduli space.  But both constructions give the entire 44-parameter family, so they must be the same.

Given this identity, should there not also be a `hybrid' matrix
\beqnn
\smallcicy{\IP^1\\ \IP^1\\ \IP^1\\ \IP^2\\ \IP^2}{1&0&0&1\\ 1&0&0&1\\ 1&0&0&1\\ 0&1&1&1\\ 0&1&1&1\\}_{\raisebox{5pt}{~.}}
\eeqnn
Indeed there is, this is matrix 7206 of the list, it also has $(h^{1,1},\,h^{2,1})=(8,44)$ and is assumed to be identical to the previous two representations. In Braun's classification, the hybrid appears with a maximal group  $\IZ_6$. Thus, this manifold also admits the symmetry groups of order 12, but not all the elements of the groups are represented linearly.

Another interesting case derives from the above. This is the split of $X_{7246}$:
\beq\footnotesize
\cicy{\IP^1\\ \IP^2\\ \IP^2\\ \IP^2\\ \IP^2\\}{0&0&0&0&1&1\\ 1&1&0&0&1&0\\ 1&1&0&0&1&0\\ 
0&0&1&1&0&1\\ 0&0&1&1&0&1\\}_{\raisebox{5pt}{~.}}
\label{SplitOf7246}\eeq
A quick calculation reveals the Euler number as $\chi{\,=\,}0$ and the practised reader will see that by contracting, say, the second and also the last $\IP^2$ we arrive at the split bicubic
\beq\footnotesize
\cicy{\IP^2\\ \IP^2\\ \IP^2}{1&1\\ 3&0\\ 0&3}^{19,19}_{\chi=0}
\label{SplitBicubic}\eeq
which also has $\chi{\,=\,}0$. The significance of this is that, under a conifold transition, the Euler number changes by twice the number of nodes. In this case, since the Euler number does not change, there are no nodes. Thus the last two configurations above correspond to the same manifold. Now the configuration \eqref{SplitOf7246} is not in the CICY list owing to the fact that, in constructing the list, extended matrices of the type we have just seen, that are related to matrices of the list by redundant splits, were suppressed. The interesting point is that
\eqref{SplitOf7246} inherits the linear group actions of $\IZ_{12}$ and $\text{Dic}_3$ from~$X_{7246}$. Thus the split bicubic must also admit these as freely acting symmetries, a fact that was not otherwise known. However these symmetries are not linearly realised on the configuration~\eqref{SplitBicubic}. A lesson is that there are very probably nonlinearly realised symmetries of the CICY manifolds, of which we are not aware. Some of these may correspond to linear actions on extended CICY matrices.

Returning to the matter of redundancy among the matrices. We have discussed the redundancy due to the two ways of representing $\text{dP}_6$. It turns out that $\text{dP}_4$ (for us, this is $\IP^2$ blown up in five points) also appears in our matrices in two ways. The first is the well known presentation $\IP^4[2,2]$. Another is
\beq
\text{dP}_4~=~\footnotesize\cicy{\IP^1\\ \IP^1 \\ \IP^1\\}{1\\ 1\\ 2\\}_{\raisebox{5pt}{~.}}\normalsize
\label{newdP4}\eeq
To see this note that, by taking coordinates $x_i$, $y_i$, $z_i$, in the three $\IP^1$'s, we can realise the space on the right as the zero locus of an equation
\beq
A(y,z)\,x_1 + B(y,z)\,x_2~=~0
\label{ABeq}\eeq
where $A$ and $B$ are polynomials of bidegree $\left(1, 2\right)$ in their arguments. For generic $(y,z)\in\IP^1{\times}\IP^1$, this yields a unique solution for $(x_1,x_2)$, as a point of $\IP^1$. However there will be four points, for sufficiently general $A$ and $B$, such that
$A(y,z)=B(y,z)=0$,
and for these points the solutions to~\eqref{ABeq} yield a complete $\IP^1$'s worth of $x$'s. In this way we see that we can think of the space on the right of \eqref{newdP4} as $\IP^1{\times}\IP^1$ blown up in four points, and this is equivalent to $\IP^2$ blown up in five points. It remains to check that these points ``are in general position'', that is, no three on a line. This is most simply done in the context of an example and we do this in~\sref{subsec:delPezzo}.

The relevance of the remarks above is that we find among our matrices presumed identities such as
\beqnn
X_{2564}~=~\smallcicy{\IP^4\\ \IP^4}{1&2&2&0&0\\ 1&0&0&2&2}_{-32}^{12,28} ~~~~\text{and}~~~~~
X_{2568}~=~
\smallcicy{\IP^1 \\ \IP^1\\ \IP^1\\ \IP^1\\ \IP^1\\ \IP^1}
{ ~1 &1 & 0  ~\\
  ~1 &1 & 0  ~\\
  ~0 &2 & 0 ~\\
  ~1 &0 & 1 ~\\
  ~1 &0 & 1 ~\\
    ~0 &0 & 2~\\}_{-32}^{12,28}
\eeqnn
as well as the corresponding splits and hybrids of these.

Now, as we have observed, the surface $\text{dP}_4$ can be obtained by blowing up $\IP^2$ in five points. It follows that $h^{1,1}(\text{dP}_4)= 6$. So by recognising that the configurations above are hypersurfaces in $\text{dP}_4$ we have explained the fact that $h^{1,1}=12$ for these configurations, a fact that was not immediately apparent from the configurations themselves. We make extensive use of this method to calculate $h^{1,1}$ for these `difficult' cases. By working out how the groups act on the exceptional lines of the $\text{dP}_4$'s we are also able to compute $h^{1,1}$ for the quotients.
\subsection{Unexpected symmetries}
The webs of symmetric CICY's, that we discuss here, have groups $G\supseteq\IZ_4$ or $G\supseteq\IZ_2{\times}\IZ_2$. The groups that arise, by the process of splitting and contraction are, for the most part, two-groups\footnote{Two-groups are groups such that the order of the group, and so the order of each of its subgroups, is a power of 2.}, as is easily appreciated from a glance at \tref{tab:SymmetryGroups} at the end of this introduction. For example, the tetraquadric has groups
\beqnn
\smallcicy{\IP^1\\ \IP^1\\ \IP^1\\ \IP^1\\}{2\\ 2\\ 2\\ 2\\} :\hskip20pt\normalsize
\begin{minipage}[c][65pt][c]{3.2in}
\begin{gather*}
\IZ_{2}\,,~\IZ_{4}\,,~\IZ_{2}{\times} \IZ_{2}\,,~\IZ_{8}\,,~\IZ_{4}{\times} \IZ_{2}\,,~\IQ_8\,, \\ 
\IZ_{4}{\times} \IZ_{4}\,,~\IZ_{4}{\rtimes} \IZ_{4}\,,~ \IZ_{8}{\times} \IZ_{2}\,,~ \IZ_{8}{\rtimes} \IZ_{2}\,,~ \IZ_{2}{\times} \IQ_8~\\
\end{gather*}
\end{minipage}
\eeqnn
`Unexpected' groups arise occasionally, however. The particularly symmetric split
\beqnn
X_{7447}~=~\smallcicy{\IP^1\\ \IP^1\\ \IP^1\\ \IP^1\\ \IP^1\\}{1&1\\ 1&1\\ 1&1\\ 1&1\\ 1&1\\}\hskip50pt
\eeqnn
can admit a freely acting $\IZ_5$ that corresponds to the cyclic permutation of the five $\IP^1$ spaces, and admits in fact a freely acting $\IZ_5{\times}\IZ_2{\times}\IZ_2$. A second example of this phenomenon could be the manifold $X_{7300}$
of \eqref{Order12Groups}. This is a (double) split of the tetraquadric and admits the freely acting groups  
$\IZ_{12}=\IZ_3{\times}\IZ_4$ and 
$\text{Dic}_3=\IZ_3{\rtimes}\IZ_4$.
\subsection{Layout of the paper}
In many cases, we use the polynomial deformation method to calculate the Hodge numbers. That is we compute the number of free parameters for the most general symmetric polynomials and we find $h^{2,1}$ from this. The number of invariant \K classes follows, since, as already remarked, the Euler number divides by the order of the group. The method and the circumstances for which it gives a reliable count are described in~\sref{subsec:prelims}. We discuss here also the alternate approach of calculating the action of the group on the \K-forms. The cases where the manifold is a hypersurface in products of $\text{dP}_6$'s and $\text{dP}_4$'s leads to some very classical and beautiful algebraic geometry.

We turn next to the actual calculations for the 45 CICY's. As has been remarked, there is a partial ordering imposed by splitting. We follow~\fref{FourGroupWeb}. There are five `sources' in this diagram, corresponding to the labelled CICY's. We begin, in~\sref{sec:tetraquadricweb}, with the successive splits of the tetraquadric. For each symmetric split we record the groups and the Hodge numbers corresponding to the quotients.

In \sref{sec:fourquadrics} we consider the split manifolds that descend from $\IP^7[2,2,2,2]$. These are the manifolds of~\fref{FlowerWeb}. It is interesting to remark that there is a `sink' in~\fref{FourGroupWeb}, corresponding to a manifold that descends from both $\IP^7[2,2,2,2]$ and the tetraquadric:
\beqnn
\displaycicy{5.75in}{
X_{6947}~=~~
\cicy{\IP^1 \\ \IP^1\\ \IP^1\\ \IP^1\\ \IP^7}
{ ~1 &1&0&0 & 0 & 0 & 0 & 0 ~\\
  ~0 &0&1&1 & 0 & 0 & 0 & 0~\\
  ~0 &0&0 &0& 1 & 1 & 0 & 0~\\
  ~0 &0&0 &0& 0 & 0 &1 & 1 ~\\
  ~1 &1&1 &1& 1 & 1 & 1& 1~\\}_{-64}^{5,37}}{-1.6cm}{1.4in}{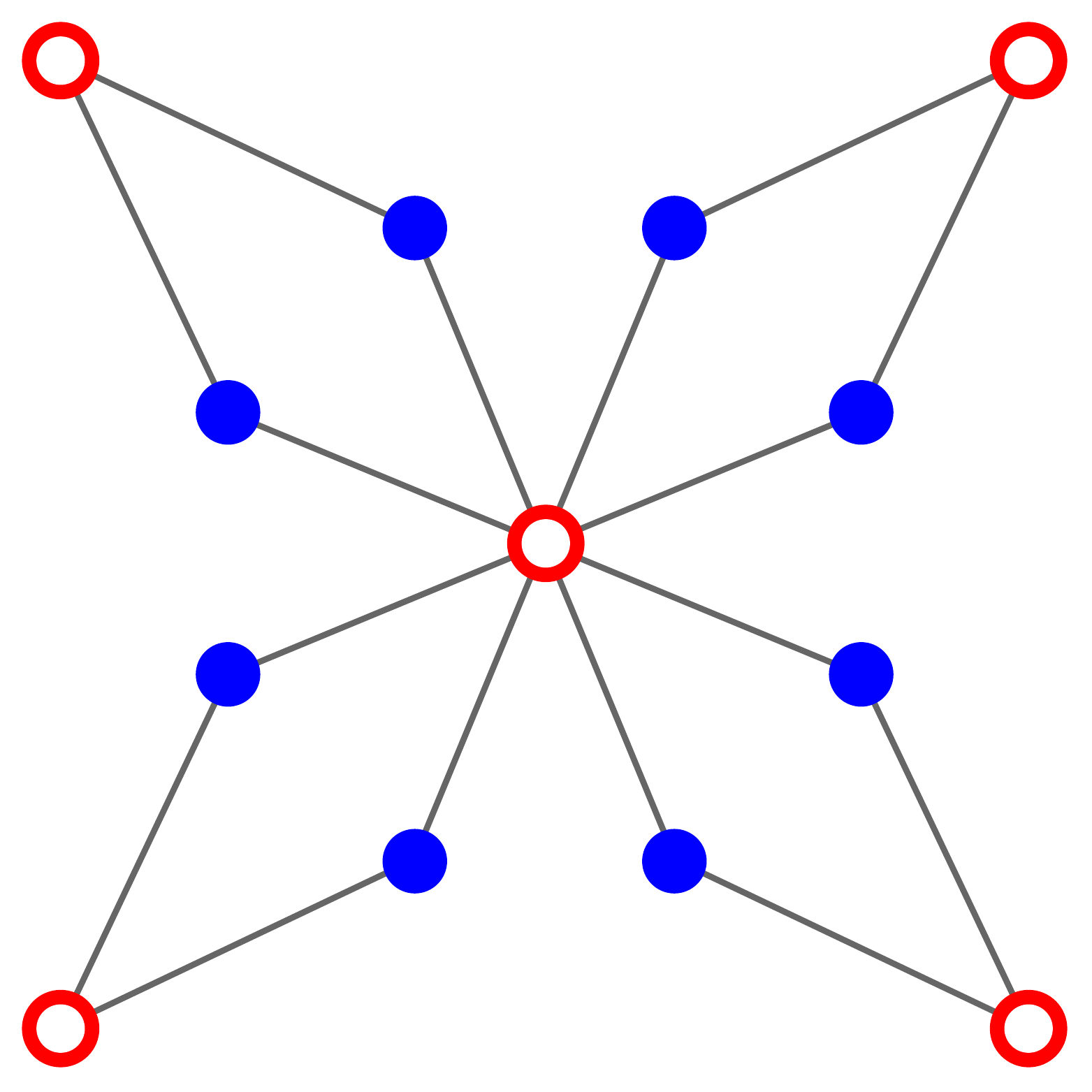}
\eeqnn
It is elementary, but interesting, to observe that, if we contract the $\IP^1$'s we return to $\IP^7[2,2,2,2]$. While, if we contract the $\IP^7$, we return to the tetraquadric.

In~\sref{sec:RemainingSources} we follow the splits of the two remaining sources
\beqnn
X_{7484}~=~~
\smallcicy{\IP^1\\\IP^1 \\\IP^3}
{ ~2& 0~ \\
  ~0& 2~ \\
  ~2& 2~ \\}_{-80}^{\,3,\,43}
~~~\text{and}~~~
X_{2564}~=~~
\smallcicy{\IP^4\\\IP^4}
{~1&2&2& 0& 0~ \\
  ~1&0&0& 2& 2~ \\}_{-32}^{12,\,28}
\eeqnn
There are more matrices here than is apparent from~\fref{FourGroupWeb}, owing to the occurrence of multiple representations, as noted previously.

Of the 166 manifolds that admit a freely acting $\IZ_2$ symmetry, we have presented Hodge numbers of quotients of 45 in the body of the paper. In Appendix~\ref{app:Z2quotients}, we list a further 53 manifolds and the Hodge numbers of their $\IZ_2$--quotients.
\subsection{How to navigate this paper}
This paper can be read in a linear fashion. However, the reader might also wish to look up the properties of a particular CICY. The following methods are efficient:
\begin{itemize}
\item If you know the CICY number: for example, you are looking up the CICY $X_{6947}$. The quickest way is to look up the manifold in \tref{tab:SymmetryGroups}, in the following subsection, this directs to the discussion of each of the manifolds. Alternatively, use the PDF search facility, on the electronic version, for 6947. This leads rapidly to all the ocurrencies of this space.\smallskip
\item If you have an explicit matrix, but do not know the CICY number, then draw the diagram for the configuration and then contract the matrix as far as possible. The contraction will lead to one (or more) of the `source' manifolds of \fref{FourGroupWeb}. From there proceed to \sref{sec:tetraquadricweb}, \sref{sec:fourquadrics} or \sref{sec:RemainingSources}, as appropriate. It is then easy to identify the relevant configuration from the diagram.
\end{itemize}
\newpage
\subsection{Table of freely acting symmetries}\label{subsec:symms}
We now present a table of the freely acting symmetries of those CICYs that possess symmetries of order divisible by 4. There are a total of 45 such manifolds, many of which are equivalent to others in this list.
\hfill
\newlength{\myht}
\newlength{\mydp}
\newlength{\mywd}
\newsavebox{\mybox}
%
\newcommand{\entry}[2]{\settowidth{\mywd}{\footnotesize${}\quotient{#2}$}%
\hspace{\mywd}%
\sbox{\mybox}{\footnotesize$#1\quotient{#2}$}%
\settoheight{\myht}{\usebox{\mybox}}\addtolength{\myht}{6pt}%
\settodepth{\mydp}{\usebox{\mybox}}\addtolength{\mydp}{5pt}%
\vrule height\myht width0pt depth\mydp\usebox{\mybox}}
\newcommand{\simpentry}[1]{%
\sbox{\mybox}{\footnotesize$#1$}%
\settoheight{\myht}{\usebox{\mybox}}\addtolength{\myht}{12pt}%
\settodepth{\mydp}{\usebox{\mybox}}\addtolength{\mydp}{10pt}%
\vrule height\myht width0pt depth\mydp\usebox{\mybox}}
\def\str{\vrule height14pt depth8pt width0pt}
\setlength{\LTcapwidth}{5in}
\begin{center}
\begin{longtable}{|c|>{\quad}l<{\quad}|l|}
\caption{\label{tab:SymmetryGroups}\mbox{Symmetry groups of CICYs with 
$G\supseteq \IZ_4$ or $G\supseteq \IZ_2{\times}\IZ_2$.}}\\
\hline 
\str\textbf{CICY \#} &
\textbf{\hfil Symmetry Groups} &
\hfil\textbf{References} \\ \hline 
\endfirsthead
\hline 
\str\textbf{CICY \#} &
\textbf{\hfil Symmetry Groups} &
\hfil\textbf{References} \\ \hline 
\endhead
\hline\hline \multicolumn{3}{|r|}{{\str Continued on next page}} \\ \hline
\endfoot
\hline\hline\multicolumn{3}{|c|}{\str}\\ \hline
\endlastfoot
\hline\hline
\varstr{13pt}{8pt}
15 & $\IZ_{2}$, $\IZ_{2}{\times} \IZ_{2}$  & \SS\ref{TQ_FurSpl_X8,40}, \tref{TQSplit14quotients}   \tabularnewline
\hline \varstr{13pt}{8pt}
19 & $ \IZ_2$, $ \IZ_4$, $ \IZ_2 {\times} \IZ_2$, $ \IZ_8$, $ \IZ_4 {\times} \IZ_2$, $ \IQ_8 $ & \SS\ref{REM_X12,28_2}, \tref{M9Split1quotients}   \tabularnewline
\hline \varstr{13pt}{8pt}
20 & $ \IZ_2$, $ \IZ_4$, $ \IZ_2 {\times} \IZ_2$ & \SS\ref{REM_OTHER_X8,40}, \tref{M8Split1quotients}  \tabularnewline
\hline 21 & 
\begin{minipage}[c][43pt][c]{3in}
\begin{gather*}
\IZ_2, \ \IZ_4, \ \IZ_2 {\times} \IZ_2, \ \IZ_8, \ \IZ_4 {\times} \IZ_2, \ \IQ_8,  \ \IZ_4 {\times} \IZ_4, \\
\hskip -47pt \IZ_4 {\rtimes} \IZ_4, \ \IZ_8 {\times} \IZ_2,  \ \IZ_8 {\rtimes} \IZ_2, \ \IZ_2 {\times} \IQ_8  \\
\end{gather*}
\end{minipage} &  \SS\ref{TQ_2_2}, \tref{TQSplit4quotients} \tabularnewline
\hline \varstr{13pt}{8pt}  
22 & $ \IZ_2$, $ \IZ_2 {\times} \IZ_2$ &  \SS\ref{REM_X12,28_1}, \tref{TQSplit18quotients} \tabularnewline
\hline \varstr{13pt}{8pt}  
30 & $ \IZ_2$, $ \IZ_4$ &  \SS\ref{REM_OTHER_X19,19} \tabularnewline
\hline \varstr{13pt}{8pt}  
480 & $ \IZ_2$, $ \IZ_4$, $ \IZ_2 {\times} \IZ_2$ & \SS\ref{REM_X3,43_3}, \tref{dP4_480}  \tabularnewline
\hline \varstr{13pt}{8pt}  
2357 & $ \IZ_2$, $ \IZ_2 {\times} \IZ_2$ & \SS\ref{REM_X3,43_3}, \tref{dP4_2357} \tabularnewline
\hline \varstr{13pt}{8pt}  
2534 & $ \IZ_2$, $ \IZ_2 {\times} \IZ_2$ & \SS\ref{TQ_FurSpl_X8,40}, \tref{dP4_2534} \tabularnewline
\hline \varstr{13pt}{8pt}  
2564 & $ \IZ_2$, $ \IZ_4$, $ \IZ_2 {\times} \IZ_2$, $ \IZ_8$, $ \IZ_4 {\times} \IZ_2$, $ \IQ_8 $ & \SS\ref{REM_X12,28_2}, \tref{M9quotients} \tabularnewline
\hline \varstr{13pt}{8pt}  
2566 & $ \IZ_2$, $ \IZ_2 {\times} \IZ_2$ &  \SS\ref{sec:12_28_2566}, \tref{dP4_2566} \tabularnewline
\hline \varstr{13pt}{8pt}  
2568 & $ \IZ_2$, $ \IZ_4$, $ \IZ_2 {\times} \IZ_2$, $ \IZ_4 {\times} \IZ_2$ & \SS\ref{REM_X12,28_1}, \tref{dP4_2568} \tabularnewline
\hline \varstr{13pt}{8pt}  
2572 & $ \IZ_2$, $ \IZ_4$ & \SS\ref{TQ_X8,44}, \tref{TQSplit19quotients} \tabularnewline
\hline \varstr{13pt}{8pt}  
2639 & $ \IZ_2$, $ \IZ_4$ & \SS\ref{TQ_X8,44}, \tref{TQSplit19quotients} \tabularnewline
\hline \varstr{13pt}{8pt}  
2640 & $ \IZ_2$, $ \IZ_2 {\times} \IZ_2$ & \SS\ref{REM_X3,43_3}, \tref{dP4_2357} \tabularnewline
\hline \varstr{13pt}{8pt}  
5256 & $ \IZ_2$, $ \IZ_2 {\times} \IZ_2$ & \SS\ref{TQOtherSplitsX5,29}, \tref{TQSplit11quotients} \tabularnewline
\hline \varstr{13pt}{8pt}  
5301 & $ \IZ_2$,  $ \IZ_4$, $ \IZ_2 {\times} \IZ_2$ & \SS\ref{TQOtherSplitsX5,29}, \tref{TQSplit10quotients} \tabularnewline
\hline \varstr{13pt}{8pt}  
5302 & $ \IZ_2$, $ \IZ_2 {\times} \IZ_2$ & \SS\ref{TQFurtherSplits1}, \tref{TQSplit13quotients} \tabularnewline
\hline \varstr{13pt}{8pt}  
5421 & $ \IZ_2$, $ \IZ_2 {\times} \IZ_2$ &\SS\ref{REM_X3,43_3}, \tref{dP4_5421} \tabularnewline
\hline \varstr{13pt}{8pt}  
5452 & $ \IZ_2$,  $ \IZ_4$, $ \IZ_2 {\times} \IZ_2$ & \SS\ref{TQOtherSplitsX5,29}, \tref{TQSplit10quotients} \tabularnewline
\hline \varstr{13pt}{8pt}  
6715 & $ \IZ_2$, $ \IZ_2 {\times} \IZ_2$ &  \SS\ref{TransposeTQSeqOfSplitsX5,37}, \tref{TQSplit6quotients} \tabularnewline
\hline \varstr{13pt}{8pt}  
6784 & $ \IZ_2$, $ \IZ_2 {\times} \IZ_2$ & \SS\ref{REM_X3,43_2}, \tref{M3Split1quotients}  \tabularnewline
\hline \varstr{13pt}{8pt}  
6788 & $ \IZ_2$, $ \IZ_2 {\times} \IZ_2$ &   \SS\ref{TransposeTQSeqOfSplitsX5,37}, \tref{TQSplit6quotients} \tabularnewline
\hline \varstr{13pt}{8pt}  
6826 & $\IZ_{2}$, $\IZ_4$, $\IZ_{2}{\times} \IZ_{2}$ &  \SS\ref{REM_OTHER_X8,40}, \tref{M8quotients} \tabularnewline
\hline \varstr{13pt}{8pt}  
6828 & $\IZ_{2}$, $\IZ_{2}{\times} \IZ_{2}$ & \SS\ref{REM_X3,43_2}, \tref{M3Split1quotients} \tabularnewline
\hline \varstr{13pt}{8pt}  
6829 & $\IZ_{2}$, $\IZ_{2}{\times} \IZ_{2}$ & \SS\ref{TQ_2_2}, \tref{dP4_6829} \tabularnewline
\hline \varstr{13pt}{8pt}  
6836 & $ \IZ_2$, $ \IZ_4$, $ \IZ_2 {\times} \IZ_2$, $ \IZ_8$, $ \IZ_4 {\times} \IZ_2$, $ \IQ_8$, $ \IZ_4 {\rtimes} \IZ_4$, $ \IZ_8 {\times} \IZ_2 $ & \SS\ref{TransposeTQSeqOfSplitsX5,37}, \tref{TQSplit5quotients} \tabularnewline
\hline \varstr{13pt}{8pt}  
6927 & $ \IZ_2$, $ \IZ_4$, $ \IZ_2 {\times} \IZ_2$, $ \IZ_4 {\times} \IZ_2$ &   \SS\ref{TransposeTQSeqOfSplitsX5,37}, \tref{TQSplit7quotients}  \tabularnewline
\hline \varstr{13pt}{8pt}  
6947 & $ \IZ_2$, $ \IZ_4$, $ \IZ_2 {\times} \IZ_2$, $ \IZ_8$, $ \IZ_4 {\times} \IZ_2$, $ \IQ_8$, $ \IZ_4 {\rtimes} \IZ_4$, $ \IZ_8 {\times} \IZ_2 $ & \SS\ref{TransposeTQSeqOfSplitsX5,37}, \tref{TQSplit5quotients} \tabularnewline
\hline \varstr{13pt}{8pt}  
7246 & $\IZ_2$, $ \IZ_3$, $ \IZ_4$, $ \IZ_6$, $ \IZ_3 {\rtimes} \IZ_4$, $ \IZ_{12}$ &  \SS\ref{TQ_X8,44}, \tref{TQSplit17quotients} \tabularnewline
\hline \varstr{13pt}{8pt}  
7300 & $\IZ_2$, $ \IZ_3$, $ \IZ_4$, $ \IZ_6$, $ \IZ_3 {\rtimes} \IZ_4$, $ \IZ_{12}$ &  \SS\ref{TQ_X8,44}, \tref{TQSplit17quotients} \tabularnewline
\hline \varstr{13pt}{8pt}  
7435 & $\IZ_{2}$, $\IZ_{2}{\times} \IZ_{2}$ & \SS\ref{TransposeTQSeqOfSplitsX4,44}, \tref{TTQSplit3quotients} \tabularnewline
\hline \varstr{13pt}{8pt}  
7447 & $ \IZ_2$, $ \IZ_2 {\times} \IZ_2$, $ \IZ_5$, $ \IZ_{10}$, $ \IZ_{10} {\times} \IZ_2 $ & \SS\ref{TQ_X5,45}, \tref{TQSplit1quotients} \tabularnewline
\hline \varstr{13pt}{8pt}  
7462 & $\IZ_{2}$, $\IZ_{2}{\times} \IZ_{2}$ &  \SS\ref{TransposeTQSeqOfSplitsX4,44}, \tref{TTQSplit3quotients} \tabularnewline
\hline \varstr{13pt}{8pt}  
7484 & $\IZ_{2}$, $\IZ_4$, $\IZ_{2}{\times} \IZ_{2}$  &  \SS\ref{REM_X3,43_1}, \tref{M3quotients} \tabularnewline
\hline \varstr{13pt}{8pt}  
7487 & $\IZ_{2}$, $\IZ_{2}{\times} \IZ_{2}$ & \SS\ref{TQ_X5,45}, \tref{TQSplit2quotients} \tabularnewline
\hline \varstr{13pt}{8pt}  
7491 & $\IZ_{2}$, $\IZ_{2}{\times} \IZ_{2}$ & \SS\ref{TransposeTQSeqOfSplitsX4,44}, \tref{TTQSplit3quotients}  \tabularnewline
\hline \varstr{13pt}{8pt}  
7522 & $\IZ_{2}$, $\IZ_{2}{\times} \IZ_{2}$ & \SS\ref{TransposeTQSeqOfSplitsX4,44}, \tref{TTQSplit3quotients} \tabularnewline
\hline \varstr{13pt}{8pt}  
7714 & $\IZ_{2}$, $\IZ_{2}{\times} \IZ_{2}$ & \SS\ref{TransposeTQSeqOfSplitsX3,51}, \tref{M4Split1quotients} \tabularnewline
\hline \varstr{13pt}{8pt}  
7735 & $ \IZ_2$, $\IZ_4$, $\IZ_2 {\times} \IZ_2$, $\IZ_4 {\times} \IZ_2$ & \SS\ref{TransposeTQSeqOfSplitsX3,51}, \tref{TTQSplit2quotients} \tabularnewline
\hline \varstr{13pt}{8pt}  
7745 & $ \IZ_2$, $\IZ_4$, $\IZ_2 {\times} \IZ_2$, $\IZ_4 {\times} \IZ_2$ &\SS\ref{TransposeTQSeqOfSplitsX3,51}, \tref{TTQSplit2quotients}  \tabularnewline
\hline \varstr{13pt}{8pt}  
7819 & $\IZ_{2}$, $\IZ_{2}{\times} \IZ_{2}$ &  \SS\ref{TransposeTQSeqOfSplitsX2,58}, \tref{TTQSplit1quotients}  \tabularnewline
\hline \varstr{13pt}{8pt}  
7823 & $\IZ_{2}$, $\IZ_{2}{\times} \IZ_{2}$ &  \SS\ref{TransposeTQSeqOfSplitsX2,58}, \tref{TTQSplit1quotients}  \tabularnewline
\hline 7861 & 
\begin{minipage}[c][85pt][c]{3.51in}
\begin{gather*}
\IZ_2, \IZ_4, \IZ_2 {\times} \IZ_2, \IZ_8, \IZ_4 {\times} \IZ_2, \IQ_8, \IZ_2 {\times} \IZ_2 {\times} \IZ_2, \IZ_4 {\times} \IZ_4,\\
\IZ_4 {\rtimes} \IZ_4, \IZ_8 {\times} \IZ_2, \IZ_4 {\times} \IZ_2 {\times} \IZ_2, \IZ_2 {\times} \IQ_8,(\IZ_4 {\times} \IZ_2) {\rtimes} \IZ_4,\\
\hskip -59pt \IZ_8 {\times} \IZ_4, \IZ_8 {\rtimes} \IZ_4, \  (\IZ_8 {\times} \IZ_2) {\rtimes} \IZ_2, \IZ_8 {\rtimes} \IZ_4, \\
\hskip -20pt \IZ_4 {\times} \IZ_4 {\times} \IZ_2,\IZ_2 {\times} (\IZ_4 {\rtimes} \IZ_4), \IZ_4 {\rtimes} \IQ_8, \IZ_2 {\times} \IZ_2 {\times} \IQ_8  \\
\end{gather*}
\end{minipage} & \SS\ref{sec:TTQ}, \tref{TTQquotients}\tabularnewline
\hline 7862 & 
\begin{minipage}[c][45pt][c]{3in}
\begin{gather*}
\hskip -45pt \IZ_{2},~\IZ_{4},~\IZ_{2}{\times} \IZ_{2},~\IZ_{8},~\IZ_{4}{\times} \IZ_{2},~\IQ_8, \\ 
\IZ_{4}{\times} \IZ_{4},~\IZ_{4}{\rtimes} \IZ_{4},~ \IZ_{8}{\times} \IZ_{2},~ \IZ_{8}{\rtimes} \IZ_{2},~ \IZ_{2}{\times} \IQ_8~\\
\end{gather*}
\end{minipage} & \SS\ref{sec:TQ}, \tref{TQquotients}
\end{longtable}
\end{center}
\begin{figure}[H]
\begin{center}
\includegraphics[width=6.4in]{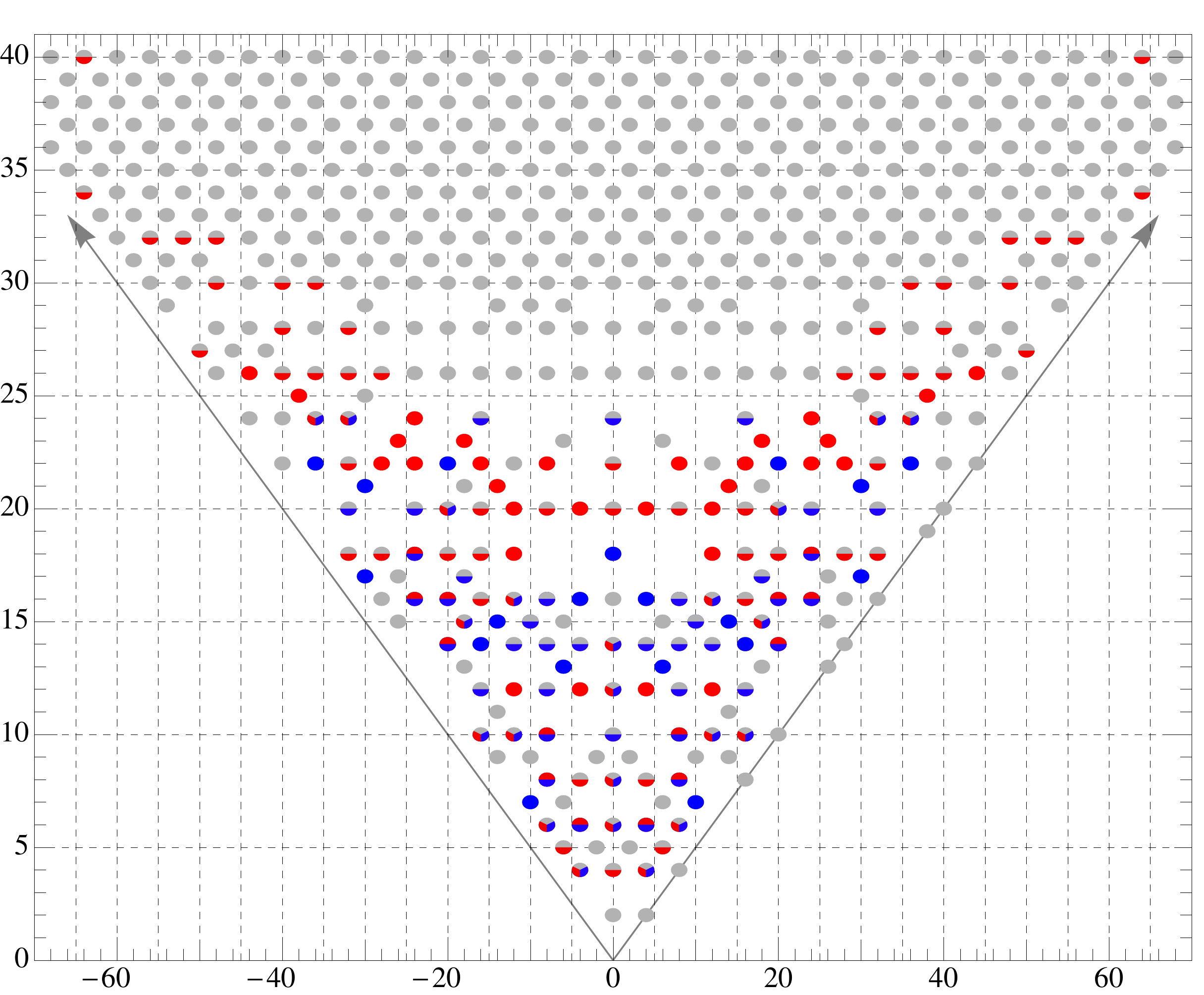}
\vskip3pt
\capt{5.8in}{TipHodgePlot}{The tip of the Hodge number plot for all the Calabi-Yau three-folds that we know. The grey points are the manifolds of the Kreuzer-Skarke list, CICYs, generalized CICYs, toric CICYs, resolutions of toric conifolds, Gross-Popescu manifolds, the manifold of V.\ Braun with Hodge numbers (1,1), manifolds obtained through hyperconifold transitions and other manifolds studied in Refs.~\cite{Yau:1986gu, Green:1987cr, Rodland:1998pm, Kreuzer:2000xy, Gross:2001as, Kreuzer:2003xx, Klemm:2004km, Tonoli:2004ps, Kreuzer:2007ni, Candelas:2007ac, Hua:2007fq, Kapustka:2007pc, Bouchard:2007mf, Batyrev:2008rp, Braun:2009qy, Kapustka:2010pr, Garbagnati:2010um, Stapledon:2010mo, Davies:2011fr, Davies:2011is, Braun:2011hd, Freitag:2011st, Filippini:2011rf, Borisov:2012xx, bini2012, Anderson:2015iia}, as well as the mirrors of the foregoing. The blue points correspond to the CICY quotients previously studied in \cite{Candelas:2008wb, Candelas:2010ve}. The red points correspond to CICY quotients studied in the present paper together with their mirrors. Monochrome points indicate quotients whose Hodge numbers fall onto sites previously unoccupied, while the multicoloured points correspond to multiply occupied sites.}
\end{center}
\end{figure}
\newpage
\section{Preliminaries}\label{subsec:prelims}
\vskip-10pt
\subsection{The polynomial deformation method}
The polynomial deformation method, proposed in \cite{Witten:1985xc}, has been used in the literature to compute the number of harmonic $(2,1)$-forms on Calabi-Yau manifolds defined as complete intersections of hypersurfaces given by homogeneous polynomials in products of projective spaces. The method relies on the observation that in many such cases the coefficients of the defining polynomials give a complete and non-redundant parametrisation of the complex structure moduli space. In the present section we will describe the polynomial deformation method in detail, first for CICY manifolds, and then for smooth quotients thereof. 

Let $X\,{\subset}\, {\cal A}$ denote a CICY manifold embedded in a product of $m$ projective spaces, with ${\cal A}={\IP}^{n_1}{\times}{\ldots}{\times}\, \IP^{n_m}$. The manifold $X$ is defined as the common zero locus of homogeneous polynomials $p_1,\ldots, p_K$. The manifold is smooth if the hypersurfaces meet transversally, that is 
${\rm d}p_1\wedge\ldots\wedge {\rm d}p_K\neq0$ on $X$. The coefficients specifying the defining polynomials can be changed, thus altering the complex structure of~$X$. The deformation class of $X$ is then specified by the configuration matrix, which collects the multi-degrees of the defining polynomials: 
\begin{equation}\label{eq:confmatr}
X~=~~
\cicy{\IP^{\,n_1} \\[4pt] \vdots\\[4pt] \IP^{\,n_m}}
{ ~q^1_1 & &\ldots && q^1_K \\[7pt]
  ~\vdots & &\ddots && \vdots \\[4pt]
  ~q^m_1 & &\ldots && q^m_K}_{\chi(X)}^{h^{1,1}(X),\ h^{2,1}(X)}
\end{equation}
where $\chi(X)$ stands for the Euler number of $X$. The complex dimension of $X$ is given by $ \sum_r n_r {-} K$, while the vanishing of the first Chern class of $X$ (the Calabi-Yau condition), corresponds to $\sum_a q^r_a = n_r+1$, for each $r\in\{1,\ldots,m\}$. One can associate a diagram to the configuration matrix \eqref{eq:confmatr} by drawing a blue disk for each polynomial, a red annulus for each projective space and connecting the $r$-th projective space with the $a$-th polynomial by $q^r_a$ lines. There will be many examples of these diagrams in the following.

The polynomial deformation method for computing the number of complex structure deformations of a CICY $X$ proceeds by the following three steps: 
\begin{itemize}
\item[$1.$] Compute the number of coefficients in the defining polynomials:
\beq
N_{\rm coeffs}~=~\sum_{a=1}^K~ \prod_{r=1}^m~ \frac{(q^r_a+n_r)!}{(q^r_a)!\,n_r!}
\notag\eeq
\newpage
\item[$2.$] Subtract the number of parameters corresponding to the freedom to redefine the homogeneous coordinates of the $m$ projective spaces:
\beq
N_{\rm c.r.}~=~\sum_{r=1}^m \left( n_r^2-1\right)
\notag\eeq 
\item[$3.$] Subtract the number of parameters corresponding to the freedom to redefine the defining polynomials, $N_{\rm p.r.}$. This step is not always straightforward. In the simplest case, in which there are no relations between the defining polynomials, the only available polynomial redefinition is an overall re-scaling of each polynomial, which gives a total number of $K$ to be subtracted from $N_{\rm coeffs}$. The computation for the generic case is best illustrated through a few examples, which we pursue below.
\end{itemize}
\subsection{Examples}\label{sec:examples_intro}
We choose our examples from the class of manifolds discussed in the following sections. A~simple illustration of the polynomial deformation method is afforded by the tetraquadric, a manifold defined as a hypersurface of multi-degree $(2,2,2,2)$ in a product of four $\IP^1$ spaces:
\begin{equation}\label{eq:TQintro}
\displaycicy{5.25in}{
X_{7862}~=~~
\cicy{\IP^1 \\ \IP^1\\ \IP^1\\ \IP^1}
{ ~2 \!\!\!\!\\
  ~2\!\!\!\! & \\
  ~2\!\!\!\! & \\
  ~2\!\!\!\!}_{-128}^{4,68}}{-1.1cm}{0.95in}{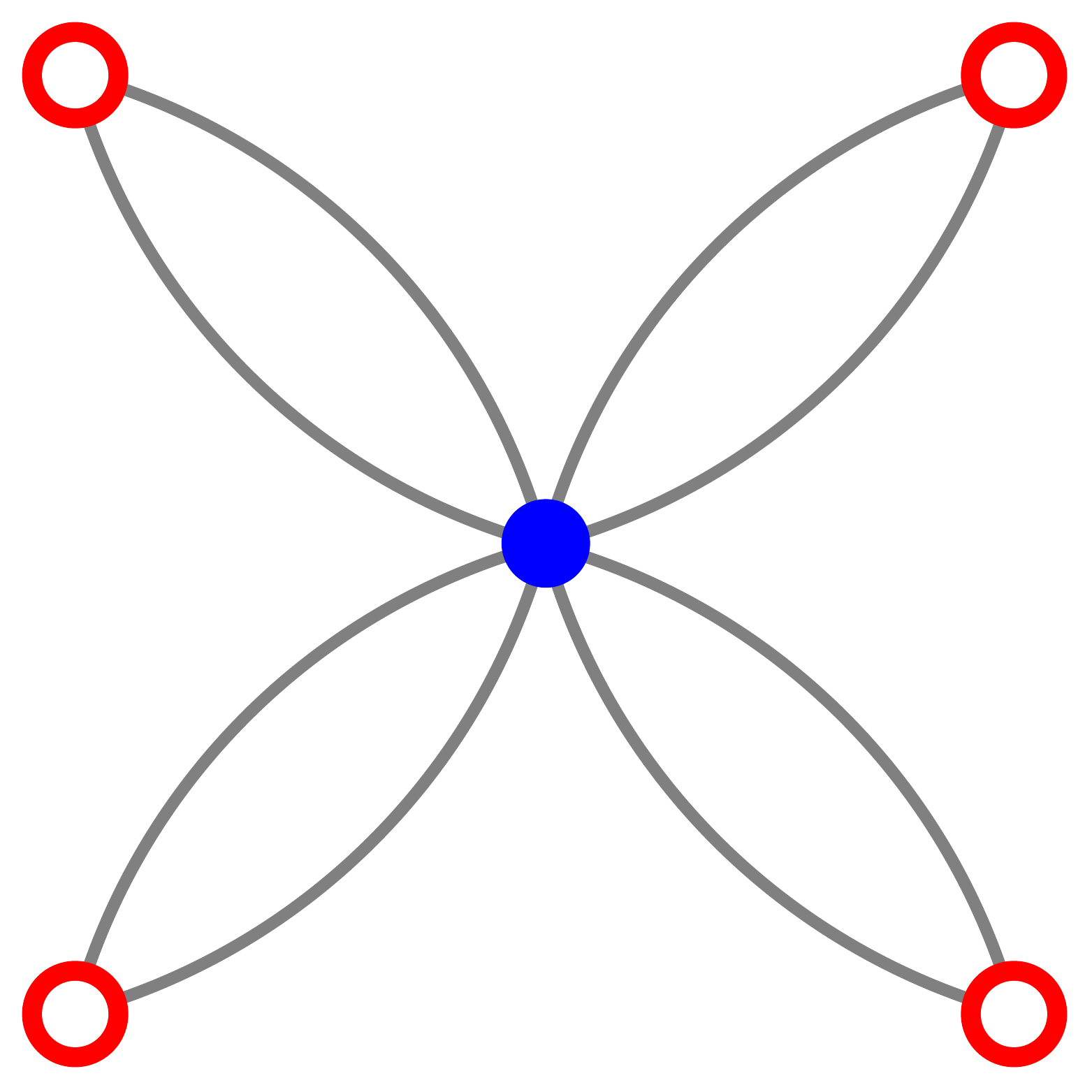}
\end{equation}
For this manifold, we have: 
\beq
h^{2,1}(X)~=~\left(\frac{3!}{2!}\right)^4 - 4\cdot(2^2-1)-1 ~=~ 68~.
\notag\eeq

The second example involves polynomials which have the same (multi)-degrees, these polynomials, $p^\a$, can be redefined by linear transformations $p^\a\to M^\a{}_\b\,p^\b$.
\vspace{-5pt}
\begin{equation}\label{eq:TTQintro}
\displaycicy{5.25in}{
X_{7861}~=~~
\cicy{\IP^7\,\,}{ ~2& \! 2 &\!2 &\!2~ \\}_{-128}^{\,1,\,65}}
{-1.1cm}{0.95in}{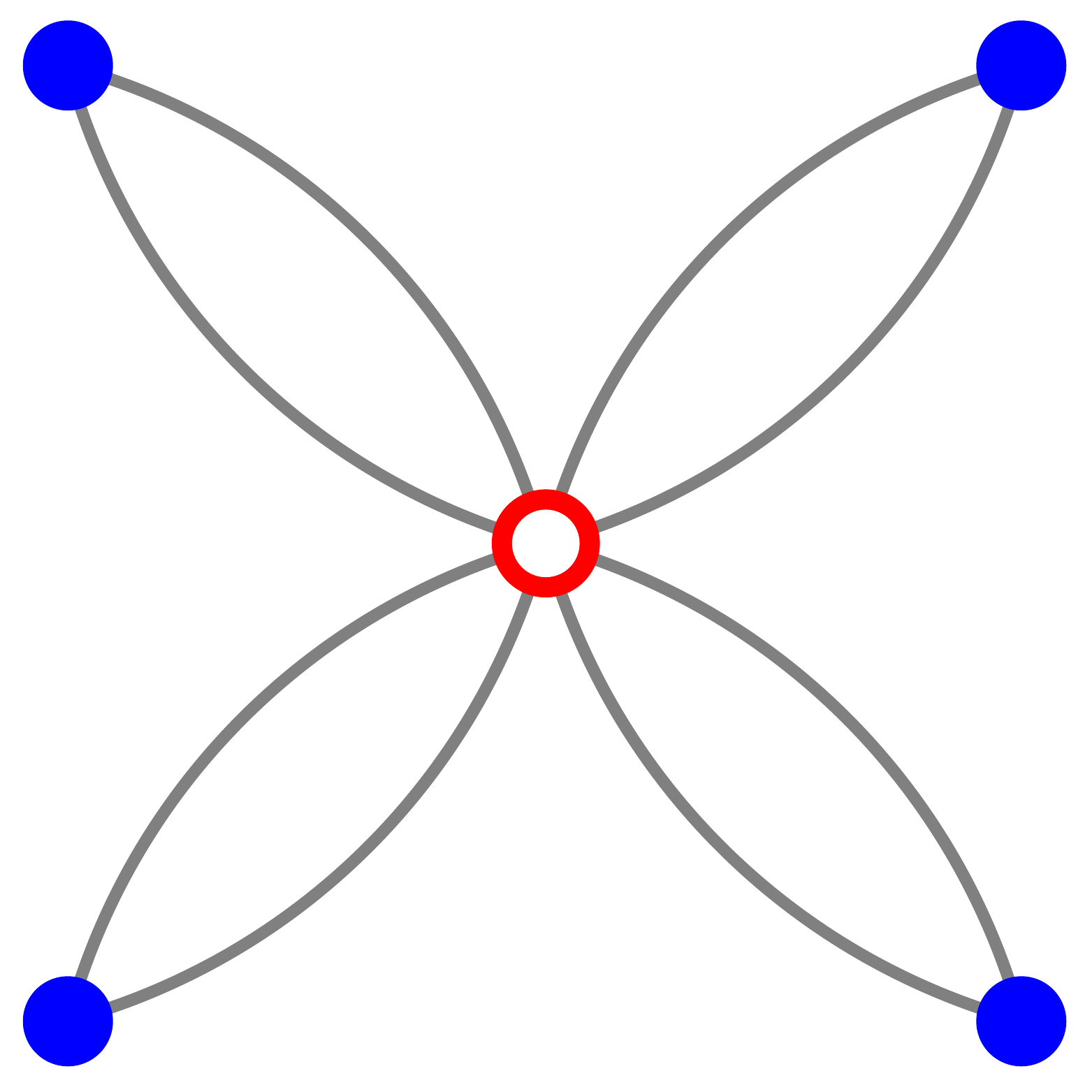}
\end{equation}
In this case, we have $N_{\rm p.r.}=16$, hence: 
\beq
h^{2,1}(X)~=~4\cdot\frac{9!}{7!\cdot 2!} - (8^2-1)-16 ~=~ 65~.
\notag\eeq 

In the third example, the computation of $N_{\rm p.r.}$ is somewhat more involved. 
\begin{equation}\label{eq:TQSplit17intro}
\displaycicy{5.25in}{
X_{7246}~=~~
\cicy{\IP^2 \\ \IP^2\\ \IP^2\\ \IP^2}
{ ~1 &1 & 0& 0  & 1~\\
  ~1 & 1&0& 0 & 1~\\
  ~0 & 0&1& 1 & 1~\\
  ~0&0 &1& 1& 1  ~\\}_{-72}^{8,44}}
{-1.15cm}{1.8in}{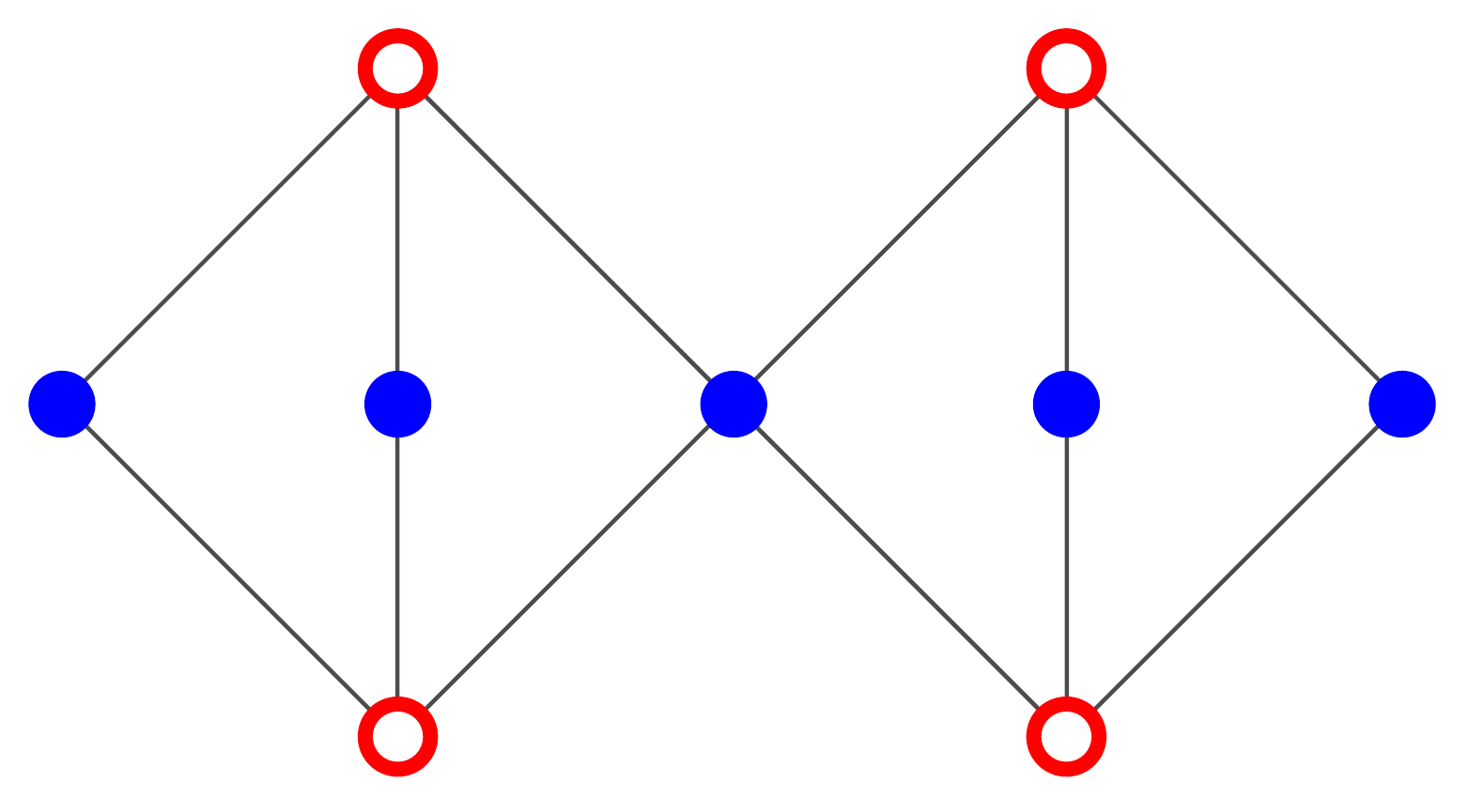}
\end{equation}
Let us denote the polynomials corresponding to the columns of the configuration, taken in order, by $p^\a$, 
$\a=1,\ldots,5$. In this case, there is a $4$ parameter freedom to redefine $p^1$ and $p^2$ and another $4$ parameter freedom to redefine $p^3$ and $p^4$. We may also redefine $p^5$ by
\beq
p^5 ~\to~ \l\,p^5 + \sum_{k=1}^4\widetilde{p}_k\, p^k 
\notag\eeq
where $\l$ is a scale, $\widetilde{p}_1$ and $\widetilde{p}_2$ have multidegree $(0,0,1,1)$ and $\widetilde{p}_3$ and 
$\widetilde{p}_4$ have multidegree $(1,1,0,0)$. Each $\widetilde{p}_k$ has 9 degrees of freedom, so would give a total of $4\cdot 9+1=37$ parameters. However, there is an over-counting since the 4 products 
$p^1\,p^3,\,p^2\,p^3,\,p^1\,p^4, p^2\,p^4$ are counted twice. This leaves $37-4=33$ parameters corresponding to redefinitions of $p_5$. 
Thus we have: 
\beq
h^{2,1}(X)~=~4\cdot\left(\frac{3!}{2!\cdot1!}\right)^2 + \left(\frac{3!}{2!\cdot1!}\right)^4 - 4\cdot(3^2-1) - (2\cdot 4 + 33)~= ~44~.
\notag\eeq 

The fourth example corresponds to a manifold that can be obtained from the manifold~\eqref{eq:TTQintro} through a conifold transition:
\vspace{-20pt}
\begin{equation}\label{eq:TTQSplit1intro}
\displaycicy{5.25in}{\vrule height0pt depth50pt width0pt
X_{7819}~=~~
\cicy{\IP^1\\ \IP^7}
{ ~ 1 & 1 & 0 & 0 & 0~\\
~1 & 1 & 2& 2 &2 ~ \\}_{-112}^{\,2,\,58}}
{-0.6cm}{1.0in}{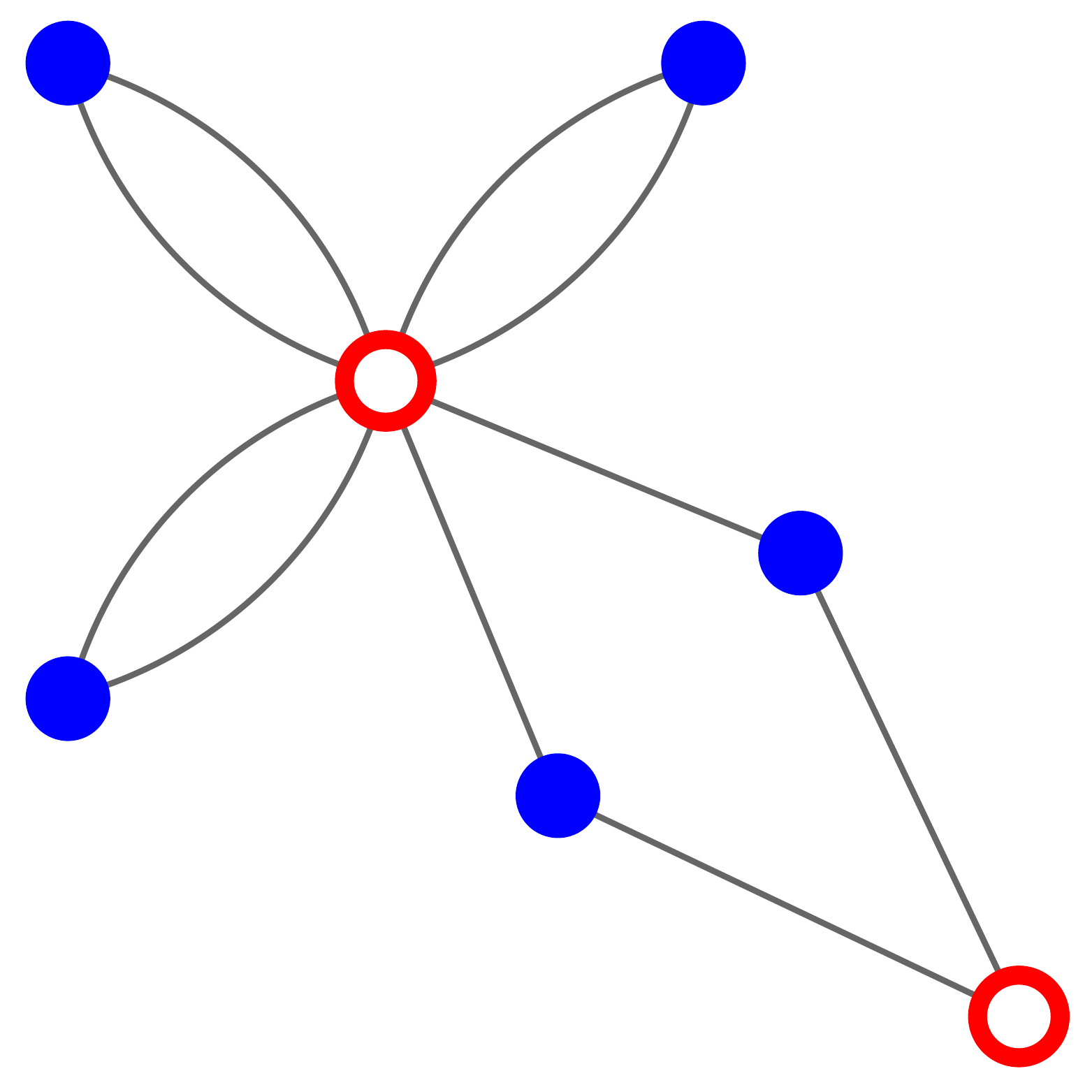}
\end{equation}
\vskip-40pt
In this case, the counting of polynomial redefinitions goes as follows. There is a $2{\times}2$ matrix, so 4 parameters, of redefinitions associated with the first two polynomials and $3{\times}3$ matrix, so 9 parameters, associated with the last three polynomials. Furthermore, the first two polynomials, $p_1$ and $p_2$ are linear in the homogeneous coordinates 
$\{x_0,x_1\}$ of $\IP^1$:
\beq
\begin{aligned}
p_1 = A_1(y)\,x_0 + B_1(y)\,x_1~,\\[3pt]
p_2 = A_2(y)\,x_0 + B_2(y)\,x_1~.
\end{aligned}
\notag\eeq
Since the coordinates $x_0,x_1$ cannot both vanish, we have the determinant
\beq
A_1(y)\,B_2(y) - A_2(y)\,B_1(y)~=~0~,
\notag\eeq
which corresponds to the vanishing of a polynomial of degree $2$ in the coordinates of $\IP^7$. Multiples of this determinant may be added to any of the last three polynomials, giving $3$ extra parameters. Thus $N_{\rm p.r.}=13+3$ and we obtain in this way 
\beq
h^{2,1}(X)~=~2\cdot\frac{8!}{7!\cdot1!}\cdot \frac{2!}{1!\cdot1!} + 3\cdot\frac{9!}{7!\cdot2!}  - \Big( (2^2-1)+(8^2-1) \Big)-16 ~=~ 58~.
\notag\eeq
 
The final example corresponds to a manifold that will not appear in the following sections, owing to the fact that it does not admit a symmetry of order 4, however we discuss it here since the computation of $N_{\rm p.r.}$ involves an extra element. This manifold admits a smooth quotient by $\IZ_3$ and was discussed in \cite{Candelas:2008wb}. The manifold corresponds to the following configuration matrix:
\begin{equation}\label{eq:M11}
\displaycicy{5.25in}{
X_{7664}~=~~
\cicy{\IP^1\\\IP^1\\\IP^1\\ \IP^5}
{ ~ 1 & 0 & 0 & 1 & 0~\\
 ~ 0 & 1 & 0 & 1 & 0~\\
  ~ 0 & 0 & 1 & 1 & 0~\\
~1 & 1 & 1& 0 & 3 ~ \\}_{-90}^{\,5,\,50}}
{-1.1cm}{1.5in}{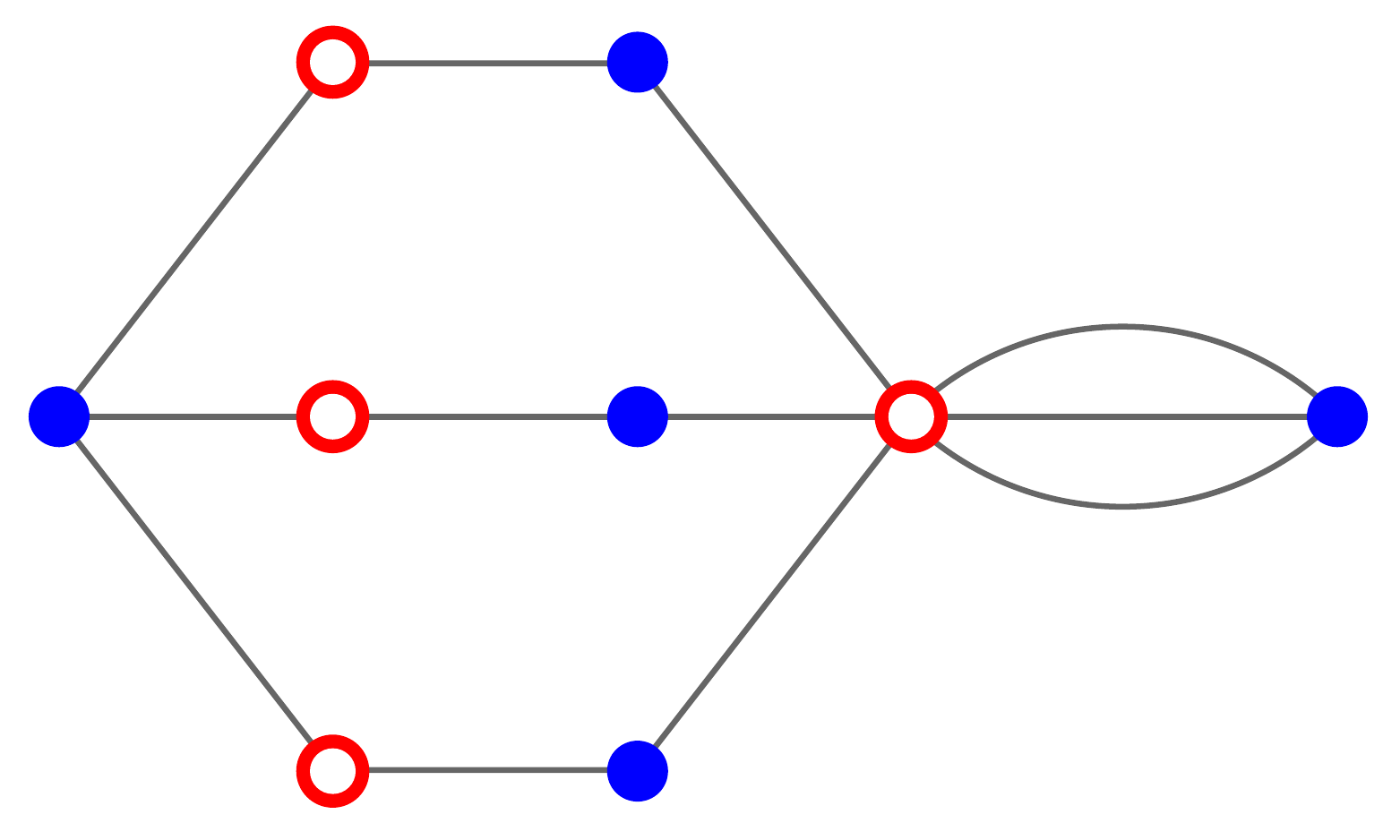}
\end{equation}
In this case, apart from the $5$ overall re-scalings of the defining polynomials, there exists an additional redefinition, which can be understood as follows. For fixed coordinates in $\IP^5$, the equation $p_1=0$ gives a unique solution for $x_1{:}\,x_0$, and hence, the locus $p_1=0$ and, further, the manifold $X$ intersect the first~$\IP^1$ in a single point. Similarly, there are unique points in the second and the third $\IP^1$ space that correspond to the intersection with $X$, for a given set of coordinates in $\IP^5$. As such, the fourth polynomial can be regarded as a polynomial of degree $3$ in the coordinates of $\IP^5$, and it can be used in order to redefine $p_5$. Consequently, we arrive to the following counting of parameters:
\beq
h^{2,1}(X)~=~3\cdot\frac{2!}{1!\cdot1!}\cdot \frac{6!}{5!\cdot1!} + \left(\frac{2!}{1!\cdot 1!}\right)^3 +\frac{8!}{5!\cdot 3!}  - \Big(3\cdot(2^2-1)+(6^2-1)\Big)-6 ~=~ 50~.
\notag\eeq 
\subsection{Understanding the polynomial deformation method}\label{sec:sequences1}
The validity of the polynomial deformation method, as outlined above, was studied in~\cite{Green:1987rw}. The procedure provides a complete and non-redundant parametrisation of the complex structure moduli space if (i) the parameter counting agrees with $h^{2,1}(X)$ and (ii) the associated diagram is not one-leg-decomposable. (A connected diagram is called one-leg-decomposable if the complement of a single leg is disconnected.) 

Without repeating the full discussion of~\cite{Green:1987rw}, we would like to review the cohomology computations that underlie the polynomial deformation method. To this end, we need to relate the cohomology of $X$ to bundle-valued cohomologies. Through Dolbeault's theorem, we have the identification:
\beq\label{eq:cohrel1}
H^{p,q}\left(X\right)~ \cong~ H^q\left(X,\,\wedge^p TX^*\right)~.
\eeq
Moreover, on a Calabi-Yau three-fold, the multiplication with the projectively unique and nowhere vanishing holomorphic 3-form, induces the isomorphism:
\beq\label{eq:cohrel2}
H^q\left(X,TX\right)~\cong~H^q\left( X,\,\wedge^2 TX^*\right)~.
\eeq
Together, \eqref{eq:cohrel1} and \eqref{eq:cohrel2} imply $H^{2,1}(X)~\cong~H^1(X,TX)$. If $X$ is a CICY manifold, we can compute $H^1(X,TX)$ using the normal bundle sequence and the Euler sequence, as discussed below. 

Let $X\,{\subset}\,\cA$ be a Calabi-Yau three-fold, with ${\cal A}=\mathbb{P}^{n_1}{\times}{\ldots}{\times}\, \IP^{n_m}$, defined as the common zero locus of the $K$ polynomials, $p_1,\ldots, p_K$. We will assume that $X$ is not a direct product, or equivalently, that the configuration matrix is in-decomposable (i.e.~it cannot be written in block diagonal form with more than one block), or equivalently, that the associated diagram is connected.  Let $\cN\rightarrow \cA$ denote the line bundle sum whose sections are the defining polynomials, and $N=\cN|_X$, the restriction of $\cN$ to $X$, denote the normal bundle of $X$. Explicitly, the normal bundle $\cN$ can be obtained from the configuration matrix~\eqref{eq:confmatr} as the sum of line bundles: 
\beq
\cN~=~ \bigoplus_{a=1}^K \cO_{\cA}({\bf q}_a)~.
\notag\eeq 

  Then the normal bundle sequence reads: 
\begin{equation*}\label{eq:nbsequence}
0\rightarrow TX\rightarrow T{\cal A}|_X\rightarrow N\rightarrow 0~.
\end{equation*}
This short exact sequence induces the long exact sequence in cohomology: 
\beq
\begin{array}{lllllllll}
0 & \lrarr &H^0(X,TX)&\lrarr&H^0(X,T\cA|_X)&\lrarr&H^0(X,N)&\lrarr&\\[3pt]
& \lrarr &H^1(X,TX)&\lrarr&H^1(X,T\cA|_X)&\lrarr&H^1(X,N)&\lrarr&\\[3pt]
& \lrarr &H^2(X,TX)&\lrarr&H^2(X,T\cA|_X)&\lrarr&H^2(X,N)&\lrarr&\\[3pt]
& \lrarr &H^3(X,TX)&\lrarr&H^3(X,T\cA|_X)&\lrarr&H^3(X,N)&\lrarr &0
\end{array}
\notag\eeq 
which implies: 
\beq
\begin{aligned}
H^1(X,TX) \cong \left( H^0(X,N) / H^0(X,T\cA|_X)\right) \oplus {\rm Ker}\left( H^1(X,T\cA|_X)\rightarrow H^1(X,N)\right)~,
\end{aligned}
\notag\eeq 
where we have used the fact that, for a Calabi-Yau three-fold that is not a direct product, we have $H^{0}(X,TX)\cong H^{3,1}(X)= 0$. 
Since $N$ is a sum of line bundles, its cohomology can be relatively easily computed from line bundle cohomology on $\cA$, using the Koszul spectral sequence. The cohomology of $T\cA|_X$ can be obtained from the Euler sequence, restricted to~$X$:
\begin{equation*}\label{eq:Eulersequence}
0\rightarrow {\cal O}_X^{\oplus m}\rightarrow S\rightarrow T{\cal A}|_X\rightarrow 0~,
\end{equation*}
where $\cO_X$ denotes the trivial line bundle on $X$ (or, rather, the sheaf of holomorphic functions on $X$, also called the structure sheaf of $X$), $S=\bigoplus_{r=1}^m{\cal O}_X({\bf e}_r)^{\oplus (n_r+1)}$, and ${\bf e}_r$ are the standard unit vectors in $m$ dimensions. The associated long exact sequence in cohomology,
\beq
\begin{array}{lllllllll}
0 & \lrarr &H^0(X,{\cal O}_X^{\oplus m})&\lrarr&H^0(X,S)&\lrarr&H^0(X,T{\cal A}|_X)&\lrarr&\\[3pt]
& \lrarr &H^1(X,{\cal O}_X^{\oplus m})&\lrarr&H^1(X,S)&\lrarr&H^1(X,T{\cal A}|_X)&\lrarr&\\[3pt]
& \lrarr &H^2(X,{\cal O}_X^{\oplus m})&\lrarr&H^2(X,S)&\lrarr&H^2(X,T{\cal A}|_X)&\lrarr&\\[3pt]
& \lrarr &H^3(X,{\cal O}_X^{\oplus m})&\lrarr&H^3(X,S)&\lrarr&H^3(X,T{\cal A}|_X)&\lrarr &0
\end{array}
\notag\eeq 
leads to the following identifications: 
\begin{align*}
H^0(X,T\cA|_X) &\cong H^0(X,S)/ \IC^m\\
H^1(X,T\cA|_X) &\cong H^{1}(X,S)
\end{align*}
where we have used the fact that, for spaces defined by in-decomposable configuration matrices, $H^0(X,{\cal O}_X)\cong H^3(X,{\cal O}_X)\cong \IC$ and the cohomology groups $H^1(X,{\cal O}_X)$ and $H^2(X,{\cal O}_X)$ are trivial, results which follow from the application of Bott's formula for line bundles on projective spaces, the K\"unneth formula for cohomology of bundles over a direct product of spaces and the Koszul resolution of the restriction $\cO_{\cA}\rightarrow \cO_X$. 
With the above identifications, we have: 
\beq\label{eq:H21}
\begin{aligned}
H^{2,1}(X) ~\cong ~& \frac{ H^0(X,N)}{H^0(X,S)/\IC^m} \oplus {\rm Ker}\left( H^1(X,S)\rightarrow H^1(X,N)\right)~.
\end{aligned}
\eeq

The expression \eqref{eq:H21} is completely general, subject only to the assumption that the configuration matrix is in-decomposable. All the cohomologies involved are line bundle cohomologies, which can be computed using the Koszul spectral sequence (see e.g.~\cite{Hubsch:1992nu,Anderson:2013qca}). To relate it with the polynomial deformation method, we note the following: 
\begin{itemize}
\item[1.] The dimension of the cohomology group $H^0(X,N)$ corresponds to the difference: 
\beq
N_{\rm coeffs} - N_{\rm p.r.}
\notag\eeq 
\item[2.] The term $H^0(X,S)/ \IC^m$ corresponds to the (projective) coordinate redefinitions of the ambient space, $N_{\rm c.r.}$. 
\end{itemize}

For the case that
\beq
\begin{aligned}
{\rm Ker}\left( H^1(X,S)\rightarrow H^1(X,N)\right)\cong~ 0
\end{aligned}
\notag\eeq 
we obtain the simple relation: 
\beq\label{eq:h21formula}
h^{2,1}(X) = h^0(X,N)- \left(h^0 (X,S) - m\right)~.
\eeq

When it fails, the polynomial deformation method gives a wrong result for ${\rm dim}\!\left( H^0(X,N)\right)$. This has to be so, since in general the computation of $H^0(X,N)$ involves contributions from higher iterations in the Koszul spectral sequence. Some of these contributions can be accounted for by using the artifices described above in Section~\ref{sec:examples_intro}. 

In the following sections we will apply formula~\eqref{eq:h21formula} in order to compute the Hodge numbers for quotient manifolds for which the polynomial deformation method, as well as the counting of K\"ahler parameters described below are not applicable. Given a group action $G\times X\rightarrow X$, and an equivariant structure of the normal bundle $N$ specified by the action on the defining polynomials, the cohomology groups involved in Eq.~\eqref{eq:h21formula} split into representations of $G$. The part which corresponds to the trivial representation descends to the quotient manifold. Thus by counting the multiplicity of the trivial representation in a given cohomology group on~$X$ one can obtain the dimension of the cohomology group in question on $X/G$. 

\subsection{The polynomial deformation method for quotient manifolds}
Let $G\times X \rightarrow X$ be a free group action of the finite group $G$ on the CICY three-fold~$X$ defined as above, as the common zero locus of $K$ polynomials, $p_1,\ldots, p_K$, in ${\cal A}=\mathbb{P}^{n_1}{\times}{\ldots}{\times}\, \IP^{n_m}$. We can pick $g_1,\ldots,g_k$ to be a set of generators of $G$. The action can be specified by two sets of matrices, $\{\gamma(g_1),\ldots,\gamma(g_k)\}$ and $\{\rho(g_1),\ldots,\rho(g_k)\}$, representing the action on the coordinates and the action on the polynomials. Note that the matrices $\{\gamma(g)\}$ and $\{\rho(g)\}$ do not form a representation in the usual sense; rather, they are obtained by multiplying a linear representation of $G$ with a permutation representation, corresponding to the part of the action of $G$ that permutes the embedding projective spaces (see \cite{Braun:2010vc} for more details). 

For the smooth quotient $X/G$, the counting of complex structure parameters using the polynomial deformation method can be systematised in the following steps: 
\begin{itemize}
\item[1.] Find a basis of invariant polynomial vectors $\{v_i=(p^i_1,\ldots,p^i_K)\}$. The dimension of this basis corresponds to the number of coefficients in the specialised polynomials that define manifolds with the given symmetry. For the manifolds discussed in the present paper, a basis of invariant polynomial vectors was given in each case in \cite{Braun:2010vc}.
\item[2.] Count the number of coordinate redefinitions consistent with the given group action. Let $C_x$ be a block-diagonal matrix with the $i$-th block representing a general linear transformation of the homogeneous coordinates of the $i$-th projective space. Then $C_x$ is consistent with the given group action if 
\beq\label{eq:Cx}
[C_x, \gamma(g_i)] = 0~,
\eeq
for all group generators $g_i$. The number of free parameters in the matrix $C_x$ that remain after solving \eqref{eq:Cx} minus the number of projective re-scalings consistent with the given symmetry corresponds to $N_{\rm c.r.}$ for the quotient manifold. In order to find the number of allowed projective re-scalings, one starts with a matrix $\widetilde C_x$ of the same size and block-diagonal structure as $C_x$, in which each block is a multiple of the identity matrix with an arbitrary multiplication factor. Solving equations analogous to \eqref{eq:Cx}, leaves a certain number of free parameters, which correspond to the allowed projective re-scalings. 
\item[3.] Similarly, in order to find the number of polynomial redefinitions $N_{\rm p.r.}$, one starts with a matrix $C_p$ representing the most general redefinition by linear transformations and then counts the number of free parameters that are left after solving the equations $[C_p,\rho(g_i)]=0$, analogous to \eqref{eq:Cx}. The cases which contain ``embeddings" and redefinitions with determinants are more subtle. Below, we outline the computation for a case in which determinants are present in the computation of $N_{\rm p.r.}$.
\end{itemize}

\subsection{An example: detailed computation of $h^{2,1}$ by parameter counting}

We start with the manifold given by the following configuration:
\begin{equation}\label{eq:TTQSplit2_intro}
\displaycicy{5.25in}{
X_{7745}~=~~
\cicy{\IP^1\\ \IP^1\\ \IP^7}
{ ~ 1 & 1 & 0 & 0& 0 & 0 ~\\
  ~ 0 & 0 & 1 & 1&0 & 0 ~\\
  ~ 1 & 1 & 1 & 1& 2& 2~ \\}_{-96}^{\,3,\,51}}
{-1.4cm}{0.9in}{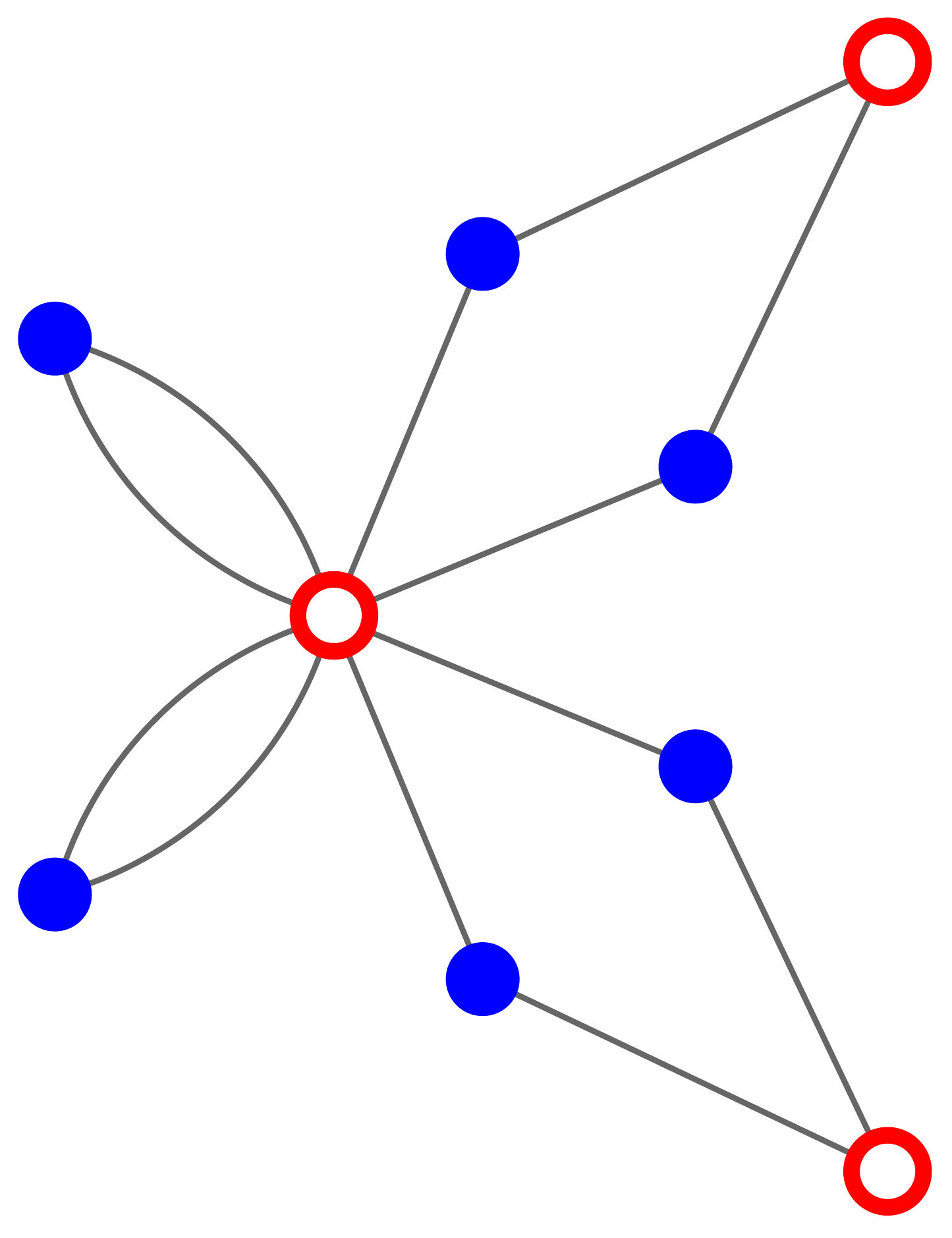}
\end{equation}
The counting of polynomial redefinitions is similar to the counting done for the manifold~\eqref{eq:TTQSplit1intro} above. There are $12$ redefinitions associated with linear transformations of polynomials of the same multi-degrees. Furthermore, there is a determinant condition associated with the first and the second pairs of polynomials, which give two polynomials of multi-degree~$(0,0,2)$ which can be added to each of the last two polynomials. This gives $4$ additional redefinitions. Hence, we obtain: 
\beq
h^{2,1}(X)~=~4\cdot\frac{8!}{7!\cdot1!}\cdot \frac{2!}{1!\cdot1!} + 2\cdot\frac{9!}{7!\cdot 2!}  - 
\Big(2\cdot(2^2-1)+(8^2-1)\Big)-16 ~=~51~.
\notag\eeq

The manifold $X$ admits a free action of $\IZ_{4}$, which can be specified by two matrices $\gamma(g)$ and $\rho(g)$, where $g$ is a generator of $\IZ_4$. The matrix $\gamma(g)$ corresponds to the action on the twelve ambient space coordinates, while $\rho(g)$ gives the action on the six defining polynomials:
\begin{align}
\gamma(g)~&=~\footnotesize
\left[\begin{array}{rrrr}
0 & \+0 & \+1 & \+0\\
0 & 0 & 0 & -1\\
1 & 0 & 0 & 0 \\
0 & 1 & 0 & 0 
\end{array}\right] \oplus
\left[\begin{array}{rrrrrrrr}
1 & \+0 & \+0 & \+0 & \+0 & \+0 & \+0 & \+0\\
0 & \ii & 0 & 0 & 0 & 0 & 0 & 0\\
0 & 0 & -1 & 0 & 0 & 0 & 0 & 0\\
0 & 0 & 0 & -\ii & 0 & 0 & 0 & 0\\
0 & 0 & 0 & 0 & 1 & 0 & 0 & 0\\
0 & 0 & 0 & 0 & 0 & \ii & 0 & 0\\
0 & 0 & 0 & 0 & 0 & 0 & -1 & 0\\
0 & 0 & 0 & 0 & 0 & 0 & 0 & -\ii
\end{array}\right]
\label{eq:gamma}\\[25pt]
\rho(g)~&=~\footnotesize
\left[\begin{array}{rrrr}
0 & \+0 & \+1 & \+0\\
0 & 0 & 0 & 1\\
1 & 0 & 0 & 0 \\
0 & -1 & 0 & 0 
\end{array}\right] \oplus
\left[\begin{array}{rr}
1& \+0\\
0& -1
\end{array}
\right]
\label{eq:rho}\end{align}

More explicitly, denoting the coordinates of the $\IP^{1}$ spaces by $x_{\b}$ and $y_{\b}$
respectively, with $\b\in\IZ_2$ and those of the $\IP^{7}$ space by $z_{m}$, the polynomial
equations that define the manifold can then be written as:

\beq\label{eq:polyns}
\begin{aligned}
p^{\alpha}\left(x_{\beta},z_{m}\right)  ~=~  \sum\limits _{m,\beta}P_{\b,m}^{\,\alpha}\,x_{\beta}\,z_{m}~,~~~ & ~~~~q^{\alpha}\left(y_{\beta},z_{m}\right)  ~=~  \sum\limits _{m,\beta}Q_{\b,m}^{\,\alpha}\,y_{\beta}\,z_{m}~,\\
r^{\alpha}\left(z_{m}\right) ~ =& ~  \sum\limits _{m,n}R_{m,n}^{\,\alpha}\,z_{m}\,z_{n}~.
\end{aligned}
\eeq
where $p^{\alpha}$, $q^{\alpha}$ and $r^{\alpha}$ denote the three pairs of polynomials, with $\a\in\IZ_2$ and $m\in\IZ_{8}$. With these notations, the action \eqref{eq:gamma}, \eqref{eq:rho} can be recast in the following form:
\beq\label{eq:action}
\begin{array}{lllllllll}
x_{\beta}  &\rightarrow&  (-1)^{\beta}\,y_{\beta}~,&~~~~y_{\beta} &\rightarrow&  x_{\beta}~,&~~~~z_{m}& \rightarrow& \ii^{m}z_{m}~,\\[3pt]
p^{\alpha} & \rightarrow& q^{\alpha}~,&~~~~q^{\alpha}&  \rightarrow& (-1)^{\alpha}\,p^{\alpha}~,&~~~~r^{\alpha}& \rightarrow&  (-1)^{\alpha}\,r^{\alpha}~.
\end{array}
\eeq

The generic form of the polynomials \eqref{eq:polyns} is not consistent with this action. To achieve that, we need to impose certain relations between the coefficients $P_{\beta,m}^{\,\alpha}$,
$Q_{\beta, m}^{\,\alpha}$ and $R_{m,n}^{\,\alpha}$, namely:
\beq
Q_{\beta, m}^{\,\alpha}  ~=~  (-1)^{\beta}\,\ii^{m}P_{\beta, m}^{\,\alpha}  ~=~  (-1)^{\alpha}\,\ii^{-m}P_{\beta, m}^{\,\alpha}~,~~~~
\ii^{m+n}R_{m,n}^{\,\alpha}  ~=~  (-1)^{\alpha}\,R_{m,n}^{\,\alpha}~.
\notag\eeq 

This implies that the coefficients $Q_{\beta, m}^{\,\alpha}$ are completely obtainable from the coefficients $P_{\beta, m}^{\,\alpha}$, subject to the condition
$(-1)^{\beta}\,\ii^{m}=(-1)^{\alpha}\,\ii^{-m}$, which, in turn, implies that $\alpha+\beta-m$ must be an even integer.
If $\alpha+\beta$ is odd (of which there are two cases), $m\in\{1,3,5,7\}$, while if $\alpha+\beta$ is even (of which also, there are two cases), $m\in\{0,2,4,6\}$, leading to a total of 16 valid tuples $\{\alpha,\beta,m\}$. This implies that there are 16 non-zero terms in the $p^{\alpha}$
polynomial, consistent with the given $\IZ_{4}$ action.
The third condition involving the $r^{\alpha}$ polynomials reduces to the condition $m+n\equiv0\ \text{mod\ 4}$, for $\alpha=0$, and $m+n\equiv2\ \text{mod\ 4}$, for $\alpha=1$. Together, these yield a total of $20$ distinct cases, implying that there are $20$ terms in the restricted $r^{\alpha}$~polynomials. Thus the total number of free coefficients in the restricted polynomials, consistent with the given $\IZ_{4}$ action is $16+20=36$, which agrees with the number the number of polynomial invariants obtained in \cite{Braun:2010vc}.

In order to find the number of coordinate redefinitions, consistent with the $\IZ_4$ action \eqref{eq:action}, we introduce a matrix $C_x$, as described in the previous section. Solving the equation $[C_x,\gamma(g)]=0$ leaves $18$ free parameters in $C_x$, of the original $72$ parameters. Out of these, we need to subtract the $2$ independent projective re-scalings of the coordinates consistent with the action \eqref{eq:action}. Thus $N_{\rm c.r.}(X/\IZ_4)=16$. Similarly, we introduce the matrix $C_p$ which corresponds to linear transformations among the defining polynomials of equal multi-degrees. Solving the equation $[C_p,\rho(g)]=0$ leaves $4$ free parameters in $C_p$. 

To account for the polynomial determinants that can be added to the last two polynomials, we note that these can be written as:
\beq
\Delta^{1}=\text{Det}\left[\begin{array}{cr}
\sum\limits _{m}P_{0,m}^{\,0}\,z_{m} & ~\sum\limits _{m}P_{1,m}^{\,0}\,z_{m}\\[12pt]
\sum\limits _{m}P_{0,m}^{\,1}\,z_{m} & ~\sum\limits _{m}P_{1,m}^{\,1}\,z_{m}
\end{array}\right]~,~~
\Delta^{2}=\text{Det}\left[\begin{array}{cc}
\sum\limits _{m}Q_{0,m}^{\,0}\,z_{m} & ~\sum\limits _{m}Q_{1,m}^{\,0}\,z_{m}\\[12pt]
\sum\limits _{m}Q_{0,m}^{\,1}\,z_{m} & ~\sum\limits _{m}Q_{1,m}^{\,1}\,z_{m}
\end{array}\right]~.
\notag\eeq 

For the manifold $X$, these lead to $4$ polynomial redefinitions of the form:
\beq
r^{\alpha}(z)\rightarrow r^{\alpha}(z)+\sum\limits _{\beta}\lambda_{\beta}\,\Delta^{\beta}~.
\notag\eeq 
However, under the action \eqref{eq:action}, $\Delta^{1}\leftrightarrow-\Delta^{2}$. It follows that there is actually only one invariant quantity constructed out of the two determinants, $\Delta^{1}-\Delta^{2}$, which can be added to the two polynomials $r^{\alpha}$. Thus, we have only
two polynomial redefinitions coming from the determinant constraints, as opposed
to four for the covering manifold.

Finally, we obtain: 
\beq
h^{2,1}(X/\IZ_4) = 36 - 16 - 6 = 14~.
\notag\eeq 
This implies $h^{1,1}(X/\IZ_{4})=2$, since the Euler characteristic of the quotient manifold is $-96/4=-24$. This is consistent with the counting of K\"ahler parameters. 

\subsection{Invariant K\"ahler forms}\label{sec:KahlerParameters}
We turn now to a consideration of the counting of the \K parameters. The K\"ahler parameters are most easily counted if the embedding of the CICY manifold is favourable, that is, for CICY manifolds for which the entire second cohomology descends from the ambient space, thought of as a product of projective spaces. A precise statement is given in \cite{Anderson:2013xka}, which we review below. An extension that we discuss in the following concerns the case where the $h^{1,1}$ of the manifold can be explained by virtue of it being embedded in a product of del Pezzo and projective spaces. In these cases we count separately the homology invariants of the del Pezzo part.

We begin with the case of a favourable embedding. For this case we seek to gain an explicit understanding of the cohomology group $H^{1,1}(X)$. An application of Dolbeault's theorem leads to the identification:
\beq
H^{1,1}(X)~\cong~ H^1(X,TX^*)~.
\notag\eeq 
For a Calabi-Yau three-fold, Serre duality formula gives
\beq
H^q(X,TX^*)~\cong~ H^{3-q}(X,TX)^*~.
\notag\eeq 
Combining these two formulas, we obtain $H^{1,1}(X)~\cong~H^2(X,TX)^*$. The bundle-valued cohomology $H^2(X,TX)$ can be computed using the tangent bundle and the Euler sequences, as done in Section~\ref{sec:sequences1}. Assuming that the configuration matrix defining $X$ is in-decomposable, we obtain:
\beq
\begin{aligned}
H^2(X,TX)~&\cong~{\rm Coker}\left(H^1(X,S)\rightarrow H^1(X,N)\right)\oplus{\rm Ker}\left(H^2(X,T{\cal A}|_X)\rightarrow H^2(X,N)\right)~, \\[4pt] 
H^2(X,T{\cal A}|_X)~&\cong~ H^2(X,S)\oplus {\rm Ker}\left(\mathbb{C}^m\rightarrow H^3(X,S)\right)~.
\end{aligned}
\notag\eeq 

The part of $H^{2}(X,TX)$ which descends from the second cohomology of the ambient space corresponds to the $\mathbb{C}^m$ term in the second equation, where $m$ is the number of projective spaces in the ambient space $\cA$. Thus, a CICY three-fold $X$ is favourable if the following conditions are satisfied: 
\begin{equation}\label{eq:favour_conds}
  {\rm Coker}\left(H^1(X,S)\rightarrow H^1(X,N)\right)=0\;,\quad  H^2(X,S)=0\; . 
\end{equation} 
A necessary, but not sufficient condition for \eqref{eq:favour_conds} to hold is that $h^{1,1}(X)=m$. A sufficient, however slightly too strong, condition for $X$ to be favourable is
\begin{equation}
 h^1(X,N)=h^2(X,S)=0\;~.\label{favourprac}
\end{equation} 

It follows that for a favourable CICY three-fold $X$, $h^{1,1}(X)$ is given by the number of projective spaces in the ambient space. In this case, a basis of the second cohomology of $X$ is given by the pullbacks of the hyperplane classes of the embedding projective spaces.

If $X$ admits a linearly represented free action of the finite group $G$, and $X$ is favourably embedded in a product of projective spaces $\cA$, one can easily count the number of K\"ahler parameters, $h^{1,1}(X/G)$. The action of $G$ on $\cA$ consists of two parts, one which permutes projective spaces of the same dimension, and one which acts internally on the coordinates of each projective space. As such, $h^{1,1}(X/G)$ equals the number of orbits of the permutation part of the action. 

To give an example, consider the manifold \eqref{eq:TTQSplit2_intro} discussed in the previous section, for which 
$\cA {\,=\,}\IP^1{\times}\IP^1{\times}\IP^7$, and its $\IZ_4$--quotient. The manifold is embedded favourably, hence $h^{1,1}(X) = 3$. The $\IZ_4$ action permutes the two $\IP^1$ spaces; consequently, $h^{1,1}(X/\IZ_4) = 2$, in agreement with the results obtained above. 
\subsection{Manifolds embedded in products of del Pezzo surfaces}\label{subsec:delPezzo}
We note, on several occasions throughout this paper, that some of our CICY manifolds can be thought of as submanifolds of products of del Pezzo surfaces. The del Pezzo surfaces are smooth algebraic surfaces with ample anticanonical bundle $-K$. These are $\IP^1{\times}\IP^1$, $\IP^2$ and $\IP^2$ blown up in $k$, $1\leq k\leq8$ points. 
We here denote by $\text{dP}_n$ the surface obtained by blowing up $\IP^2$ in $9-n$ points, in general position. This is then the del Pezzo surface of degree $n$. Where the degree of the surface is the self intersection number, $K^2$, of its anti-canonical~class.

del Pezzo surfaces contain special lines and the configuration of these lines has been the subject of much study in classical algebraic geometry (for a recent text see \cite{zbMATH06053083}). One way to study how the symmetries of the manifold act on the \K classes is to study how the symmetries permute the special lines, since there is a correspondence between the lines and cohomology classes. This study was carried out for the three-generation manifold in \SS2 of~\cite{Braun:2009qy} realised as a hypersurface in a product of two del Pezzo surfaces of degree 6. In the following we will work this through for manifolds that are embedded in products of del Pezzo surfaces of degree $4$, which play an important role in the class of manifolds we consider here. The fact that the class of hypersurfaces in products of del Pezzo surfaces, and products of del Pezzo surfaces with projective spaces is a fruitful place to look for \cys with symmetries was appreciated by Bini and Favale~\cite{bini2012}.

\subsubsection{del Pezzo surfaces of degree 4\label{subsubsec:dP4}} 
\vskip-10pt
For a thorough discussion of del Pezzo surfaces of degree 4 see also \cite{reid1972complete}. They are surfaces isomorphic to $\IP^2$ blown up in $5$ points, in general position, which in this case amounts to no $3$ points on a line. Each blow-up introduces an exceptional divisor, isomorphic to $\IP^1$, with self-intersection $-1$. We denote these curves by $E_i$, with $i\in\{1,\ldots,5\}$. Together with the pullback of the hyperplane class from $\IP^2$, denoted by $H$, these curves span $H_2(\text{dP}_4,\IZ)$. 
Since the $E_i$ curves are disjoint, and $H$ can always be chosen such that it misses the blown-up points, we have the  intersection numbers $H\cdot H {\,=\,} 1$, $H\cdot E_i {\,=\,} 0$ and $E_i\cdot E_j {\,=} -\delta_{ij}$.
In terms of these classes, the anti-canonical class is given by: 
\begin{equation*}
-K ~=~ 3H - \sum\nolimits_i E_i~,
\end{equation*}
and we see that $K^2 = 4$ which is the degree of the surface.
In addition, $\text{dP}_4$ contains $11$ other curves with self-intersection $-1$. For each pair $(i,j)$, with $i<j$, there is an exceptional line $L_{ij}$ that intersects both $E_i$ and $E_j$, so $10$ such lines. The $L_{ij}$ line corresponds to the line in $\IP^2$ that passes through the points blown-up to $E_i$ and~$E_j$. Since any two hyperplanes in $\IP^2$ intersect at a point, we have $H\cdot L_{ij} = 1$. In homology, these intersection relations identify the $L_{ij}$ as: 
\begin{equation}\label{eq:Lij}
L_{ij}  ~=~ H - E_i - E_j~.
\end{equation}
\begin{figure}[!t]
\vskip-10pt
\begin{center}
\includegraphics[width=2.8in]{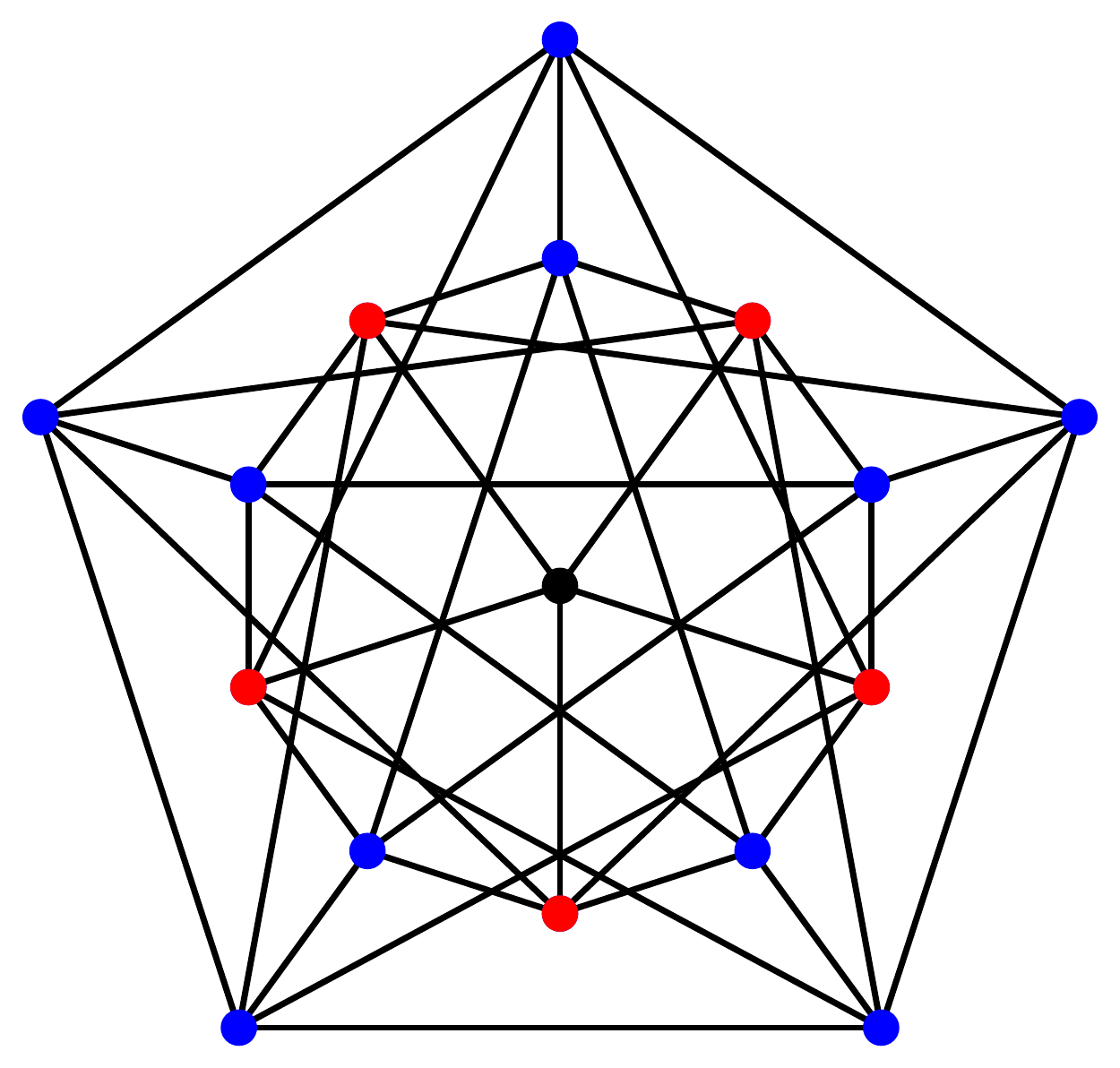}
\vskip4pt
\capt{5.7in}{clebsch_graph}{The Clebsch Graph showing the intersections of the -1 curves of a del Pezzo surface of degree 4. The $16$ vertices in the graph correspond to the curves. Two such curves intersect if the corresponding vertices are connected. From the graph, one sees that each curve meets exactly $5$ other curves. The vertices in red and blue correspond to the lines  $E_i$ and $L_{ij}$ respectively. The black vertex at the centre corresponds to $Q$.}
\end{center}
\vskip-15pt
\end{figure}
Finally, there is a unique quadric $Q$ passing through the $5$ blown-up up points of $\IP^2$, that satisfies 
$Q\cdot E_i=1$ (a smooth quadric, $\IP^2[2]$, is a $\IP^1$).
This curve intersects a generic hyperplane $H$ in two points. These intersections define the class $Q$ as: 
\begin{equation*}
Q  ~=~ 2H - \sum\nolimits_i E_i~.
\end{equation*}
We summarise the intersection numbers of the lines by a table
\begin{align}
E_i . E_j ~&= -\delta_{ij}~, & H.H ~&=\+ 1~, & Q.Q ~&=-1~, & L_{ij}. L_{ij}~&= -1~,\notag \\[3pt]
H. Q ~&=\+2~,  & H. E_i~&=\+0~,  & H. L_{ij}~&=\+1~, & L_{ij}. L_{kl} ~&=\+ 0~;\ \text{if}~\ \{i,j\}\cap\{k,l\}=\emptyset~,\notag\\[3pt]
E_i . L_{ij}~&=\+1~, & E_i . Q ~&=\+1~, & L_{ij} . Q ~&=\+0~,  & E_i . L_{jk}~&=\+ 0~;
\ \text{if}~\  i\notin\{j,k\}~. \label{eq:IntersectionNumbers}
\end{align}
\vskip-10pt
The combinatorics of the intersection of the lines is summarised by the Clebsch graph of \fref{clebsch_graph}.

It is a classical fact also that a $\text{dP}_4$ surface can be represented as a transverse intersection of two quadrics in 
$\IP^4$, so $\IP^4[2, 2]$.  At the end of \sref{redun}, we came across another representation of $\text{dP}_4$ embedded in a product of three $\IP^1$s:
\beq
\text{dP}_4~=~\smallcicy{\IP^1\\ \IP^1 \\ \IP^1\\}{1\\ 1\\ 2\\}_{\raisebox{5pt}{~.}}\normalsize
\label{dP4eq}\eeq
In the following we see how to identify the 16 exceptional lines for each of these two representations. 

Our first task is to show that this surface \eqref{dP4eq} is indeed $\text{dP}_4$. Since $\text{dP}_4$ is $\IP^2$ blown up in 5 points, we note:
\begin{equation*}
c_2(\text{dP}_4)~=~c_2(\IP^2)+5~=~8~,
\end{equation*}
since each blow-up replaces a point with a $\IP^1$. 
It is easy to check that the numerical invariants for $c_2$ and $c_1 ^2$, computed from the representation \eqref{dP4eq} match those of $\text{dP}_4$. We present further strong evidence of the equality in \eqref{dP4eq} by showing that the surface contains sixteen exceptional lines with the correct properties.  
\begin{figure}[!t]
\begin{center}
\vskip5pt
\includegraphics[width=3.02in]{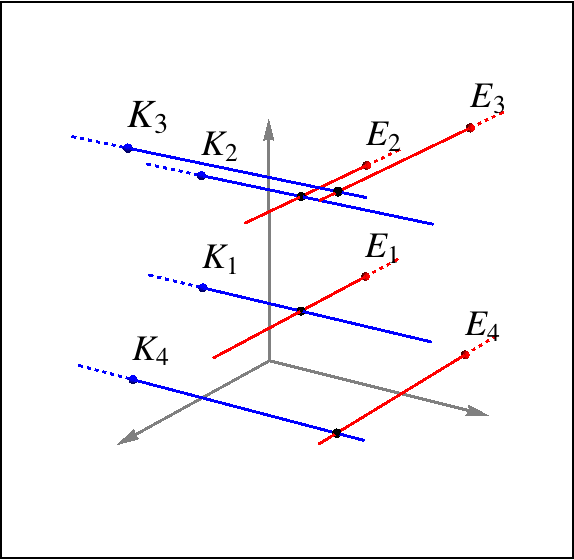}\hskip20pt
\raisebox{0.0in}{\includegraphics[width=3.1in]{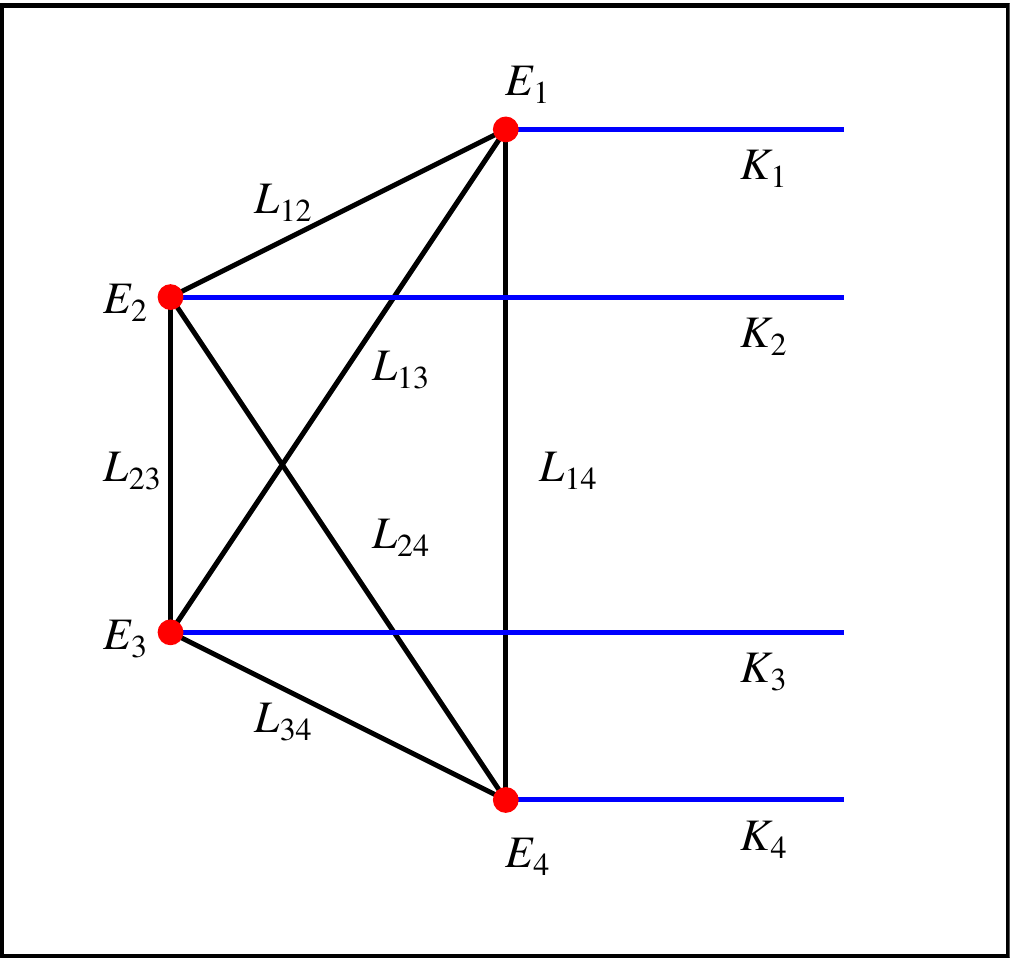}}
\vskip15pt
\raisebox{-0.025in}{\includegraphics[width=3.1in]{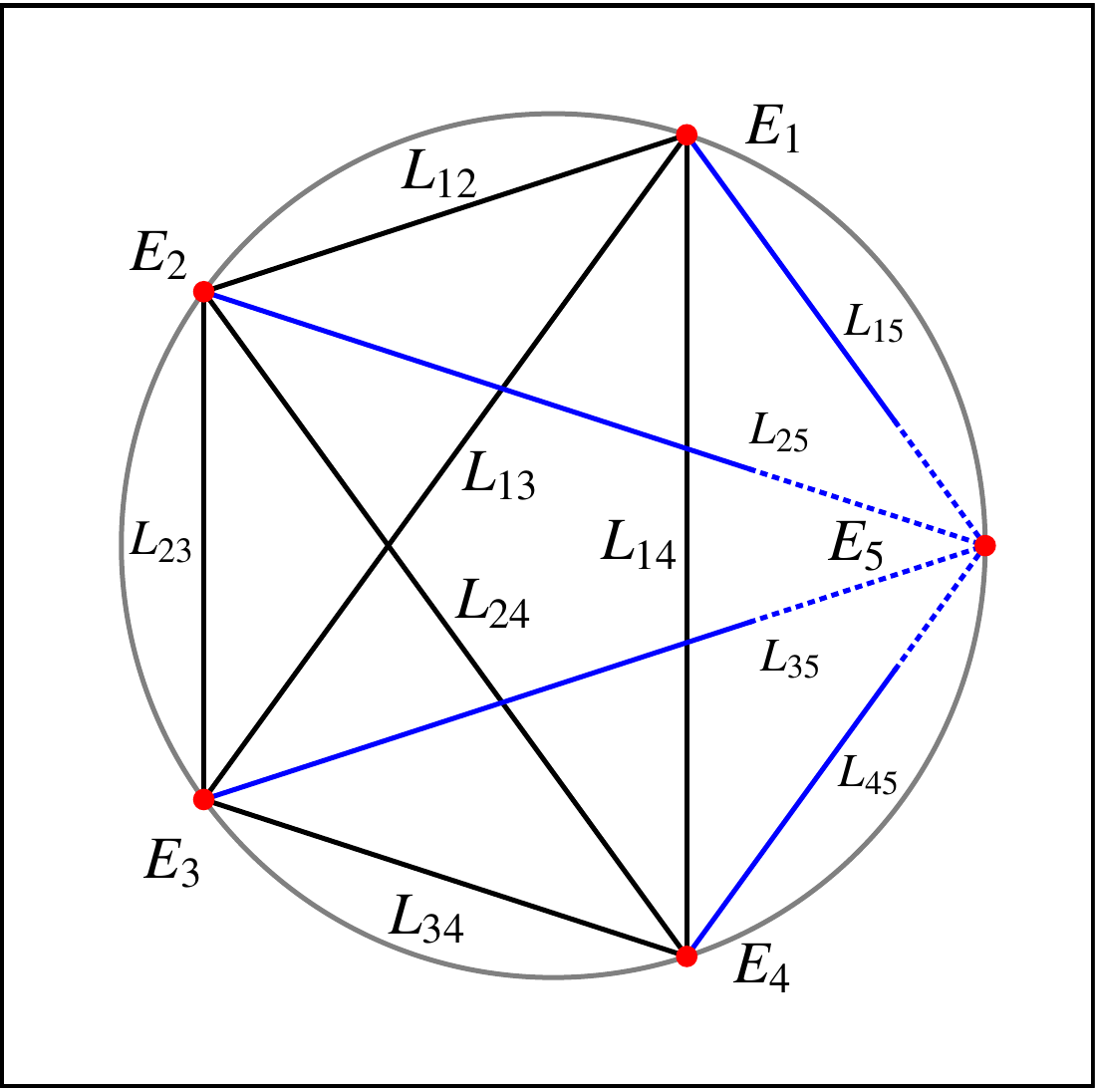}}
\vskip0pt 
\place{0.2}{3.75}{$\boxed{\boldsymbol{A}}$}
\place{3.5}{3.75}{$\boxed{\boldsymbol{B}}$}
\place{1.84}{0.43}{$\boxed{\boldsymbol{C}}$}
\place{0.7}{3.95}{\Large$x$}
\place{2.5}{4.05}{\Large$y$}
\place{1.4}{6.0}{\Large$z$}
\vskip5pt
\capt{6.2in}{dP4_special_points}{Steps in obtaining the -1 lines of $\text{dP}_4$: (A) eight of the sixteen -1 lines of $\text{dP}_4$ in~$\IC^3$. (B) Fourteen of the sixteen -1 lines projected onto the $yz$-plane. (C) The sixteen -1 lines of $\text{dP}_4$ in the $X_3\neq0$ patch of $\IP^2$. The lines $\cK_i$ $(=L_{i5})$ meet in the limit 
$y \rightarrow \infty$. The circle represents the unique quadric $Q$ passing through the $5$ blown~up points of 
$\IP^2$.}
\end{center}
\end{figure}
To begin seeing these lines, in this representation, consider the equation implicit in \eqref{dP4eq} written below:
\beq
A(y,z)\,x_1 + B(y,z)\,x_2~=~0
\label{dP4AB}\eeq
where $x,\ y,\ z$ are coordinates of the three $\IP^1$'s taken in order and $A$ and $B$ are polynomials of bidegree 
$\left(1, 2\right)$ in their arguments. For general $(y,z)\in \IP^1{\times}\IP^1$, the equation above yields a unique $(x_1,x_2)\in\IP^1$. However, there will be 4 points $(y_*,z_*)$ such that $A(y_*,z_*){=}B(y_*,z_*){=}0$. When this happens, the solution to \eqref{dP4AB} will be an entire $\IP^1$. These in fact correspond to the exceptional curves 
$E_i$, $i\in\{1,2,3,4\}$. In this way we see that the equation~\eqref{dP4AB} yields a $\IP^1{\times}\IP^1$ blown up in 4 general points. It is another classical fact that blowing up $\IP^1{\times}\IP^1$ in 4 points is equivalent to blowing up $\IP^2$ in 5 points. This will be subsumed in the following explicit construction. 

In order to refer this calculation to $\IP^2$ and find the remaining exceptional lines, we first choose affine coordinates 
$x{\,=\,}x_2/x_1, y{\,=\,}y_2/y_1, z{\,=\,}z_2/z_1$. A simple choice of a specific equation, that will also be useful later, is
\beq
(1+y z^2)+x(y+z^2)~=~0~.
\label{dP4explicit}\eeq
Four lines are then associated with the simultaneous roots of the equations
\beq
1+y z^2~=~0~~~\text{and}~~~y+z^2~=~0~.
\label{eq:FourLines}\eeq
The roots correspond to the intersections of the four lines with the $(y,z)$ plane, see \fref{dP4_special_points}A.
One sees that \eqref{dP4explicit} is symmetric under the interchange $x\leftrightarrow y$ so we obtain another set of four $-1$ lines by swapping $x$ and $y$ in the above argument. So now we have a total of eight $-1$ lines. These lines, shown in \fref{dP4_special_points}A are parallel to either the $x$ or $y$ axis and are referred to as $E_i$ and $\cK_i$ respectively, $i \in \{1,2,3,4\}$. 

Let $L_{ij}$ denote the exceptional line that intersects $E_i$ and $E_j$.  In total, there are $6$ such lines. Similarly, there are $6$ lines that meet the lines $\cK_i$ and $\cK_j$ with $i\ne j$, which we denote by~$\cK_{ij}$. The set of $L_{ij}$ lines is the same as the set of $\cK_{ij}$ lines. In fact 
$L_{ij}=\cK_{mn}$ if $\{i,j\} \cap \{m,n\}=\emptyset$. The four intersection points, denoted as solid black points in \fref{dP4_special_points}A can be easily obtained by solving for  $E_i\cap\cK _j$ for $i\ne j$. If we now project to the $yz$-plane, the $-1$ lines, are as shown in \fref{dP4_special_points}(B). We then map the lines to~$\IP^2$ 
\beq\label{eq:maptoP2}\begin{split}
\IP^1 \times \IP^1 &\dashrightarrow \IP^2 \\
(1,y) \times (1,z) &\longmapsto (y,\,z,\,1)
\end{split}\eeq
By considering the limit $y\to\infty$, we see that the four $\cK_i$ lines meet at $(1,0,0)$, thereby revealing the location of the fifth blow up point $E_5$; see \fref{dP4_special_points}C. Thus the 5 blown up points in $\IP^2$ in our example are given by: 
$\{(-\z^2, \z , 1)\,|\, \z ^4{=}1\} \cup (1,0,0)$. We can now see that $\cK_i$ is in fact $L_{i5}$. The sixteenth line is the quadratic $Q$ that intersects the five $E_i$. In our case this is given by the second of Eqs.~\eqref{eq:FourLines}. In homogeneous coordinates $(X_1, X_2, X_3)$ this is the quadric $X_1 X_3 + X_2^2{\,=\,}0$. The first of 
Eqs.~\eqref{eq:FourLines} is the cubic $X_3^3 +X_1X_2^2{\,=\,}0$. This intersects $Q$ in the five $E_i$ with a double contact at $E_5$.
In summary the sixteen lines are given by the five $E_i$, the ten $L_{ij}$ and $Q$.

Many of the intersection relations of \eqref{eq:IntersectionNumbers} between the exceptional lines are intuitively clear from \fref{dP4_special_points}. The point of intersection of a curve with the exceptional lines $E_i$ is determined by the tangent direction to the curve at~$E_i$ in \fref{dP4_special_points}. This makes it clear that $Q.L_{ij}{\,=\,}0$, for example. 

Let us turn now to the representation of $\text{dP}_4$ as the surface $\IP^4[2,2]$. In this representation it is also easy to see the lines but harder to see the map to $\IP^2$. Although this is well-known, we pause to indicate the map.

Note first that $\IP^4[2,2]$ has, by a process that is by now very familiar, $15{+}15{-}3{-}25\,{=}\,2$ complex structure parameters. It is convenient, in order to be able to write the lines explicitly, to make a simple choice of quadrics
\begin{equation}\label{dP4num2}
\begin{split}
w_1^2+w_2^2+w_3^2+w_4^2+w_5^2~&=~0\\[3pt]
w_1^2+2 w_2^2+3 w_3^2+4 w_4^2+5 w_5^2~&=~0~.\\
\end{split}
\end{equation}
This choice will also be useful to us in \sref{sec:X2566}, since it manifests a $\IZ_2 {\times }\IZ_2$ symmetry. 

Choose now 5 points, $e_k$, $k{=}1, .. ,5$, in general position on $\IP^2$, that is such that no three lie on a line. We may choose coordinates $(x,\,y,\,z)$ such that the first four points are $e_1=(1,0,0)$, $e_2=(0,1,0)$, $e_3=(0,0,1)$ and $e_4=(1,1,1)$. Having exhausted the coordinate freedom we write for the fifth point 
$e_5=(\alpha,\beta,\gamma)$. This freedom to choose the fifth point corresponds to the two degrees of freedom computed above. Note that since $e_5$ does not lie on any of the lines that join the other four points, none of the coordinates $(\alpha,\beta,\gamma)$ vanish and no two are equal. Consider now the cubics $\psi(x)$ that pass through the five $e_k$. A general cubic in the coordinates $(x, y, z)$ has 10 coefficients. Requiring that the cubic passes through the five $e_k$ imposes five conditions so there are 5 linearly independent cubics that satisfy the condition. A specific choice is 
\beq\begin{split}
\ps_1~&=~\b(\a-\g)\, y(x-z) (\g x - \a z) \\[2pt]
\ps_2~&=~\g\, z \Big( \b(\a -\g)\,x^2 + (\b\g - \a^2)\, xy + \a(\a - \b)\, yz \Big)\\[2pt]
\ps_3~&=~\b\, y \Big( \a(\b-\g)\, z^2 + (\g^2 - \a\b)\, zx + \g(\a-\g)\, xy \Big)\\[2pt]
\ps_4~&=~\b\g\, yz \Big( (\g -\b) x + (\a - \g) y + (\b -\a) z \Big)\\[2pt]
\ps_5~&=~\g\,z \Big( \g(\b-\a)\, xy + \a(\g - \b)\, yz + \b(\a-\g)\, zx \Big)
\end{split}\notag\eeq
We think of the $\ps_k$ as being proportional to homogenous coordinates on $\IP^4$. It is straightforward to check that the $\ps_k$ satisfy the two quadratic relations
\beq\begin{split}
\g (\a - \b)\,\ps_2\ps_3 + \a(\b - \g)\, \ps_2\ps_4 + \b (\a - \g)\, \ps_1\ps_5 + (\b\g - \a^2)\,\ps_3\ps_5 
+ \a(\a - \b)\,\ps_4\ps_5~&=~0~,\\[5pt]
\b(\g - \a)\,(\ps_2\ps_4 - \ps_3\ps_5) + \g(\a - \b)\,\ps_1\ps_4 + \b(\g - \b) \, \ps_1\ps_5 ~&=~0~.
\end{split}\notag\eeq
\vskip-20pt
We could now proceed to study the $-1$ lines of $\IP^4[2,2]$ and how these relate to curves of \fref{dP4_special_points}. 
Rather than do this in general, let us, with a certain prescience, choose the surface with $(\a,\b,\g){\,=\,}(-3,-2,1)$. With this choice we can make a linear change of basis $w_i=A_{ij}\ps_j$ with
\beq
A~=~\footnotesize\left(
\begin{array}{rrrrr}
                            1 & \+ -1     & -\frac{3}{2}           & \frac{3}{2}               & -3 \\[4pt]
       \frac{3 }{2} \ii & 0           & -2 \ii                       & 0                   	   & 4 \ii \\[4pt]
 \sqrt{\frac{3}{2}} & \sqrt{6} & -\sqrt{\frac{3}{2}} & -\sqrt{\frac{27}{2}} & -\sqrt{6} \\[4pt]
        \frac{1}{2}\ii & 4 \ii        & 0                             & 6 \ii                          & 0 \\[4pt]
                           0 & -3          & \frac{1}{2}             & -\frac{9}{2}             & -1 \\[4pt]
\end{array}
\right)
\notag\eeq
such that the two quadrics above become equivalent to the quadrics \eqref{dP4num2}. To find the lines for these quadrics we seek lines $\IP^1\hookrightarrow\IP^4$ that lie in the surface defined by the quadrics. Thus we set
$w_i=a_i u + b_i v$ with $(u,v)$ coordinates on the $\IP^1$. By choice of $u$ and $v$ we may take 
\beq
w~=~(u,\, v,\, a_3\, u + b_3\, v,\,  a_4\, u + b_4\, v,\,  a_5\, u + b_5\, v)
\notag\eeq
and choose the coefficients so that $w$ satisfies \eqref{dP4num2} identically in $(u,v)$. This requires
\beq
a_5^2~=~9~,~~~b_3^2~=-6~,~~~b_4^2~=~9~,~~~ b_5^2~=-4~,~~~a_4~=~\frac29\, a_5 b_4 b_5~,~~~
a_3~= -\frac16 a_5 b_3 b_5~.
\notag\eeq
There is a choice of sign associated with solving for each of $a_5,\, b_3,\,b_4,\,b_5$, so there are 16 solutions. The corresponding lines are shown in \tref{dP4Lines} where the lines are also identified.

The birational map $\IP^2\dashrightarrow \text{dP}_4$ given by $w=A\ps$ is well defined apart from the five 
points~$e_k$. The inverse map is defined everywhere and is given by the relations
\beq
\frac{x}{z}~=~  \frac{\ii \sqrt{6} (w_1 + w_4) - 3 w_3}{\sqrt{6} (w_1 + \ii w_2 ) + w_3}~,~~~
\frac{y}{z}~= -\frac{2 \left(\sqrt{6} (w_1 - \ii w_2 )+ w_3\right)}{\sqrt{6}( w_1 + w_5) - 2 w_3}~.
\notag\eeq
It is a pleasure to check that the lines of \tref{dP4Lines} map to $\IP^2$ as they should. 

Although we have described the configuration of lines in some detail, in the following it will suffice to consider the action of the groups on the cohomology basis $\{H,E_k\}$. It is quite often the case that a group element will map an $E$-line to an $L$-line, say, and then the relation \eqref{eq:Lij} is used to re-express $L_{ij}$ as $H{\,-\,}E_i {\,-\,}E_j$.
\begin{center}
\begin{longtable}{|c|l|}
\captionsetup{width=0.6\textwidth}
\caption{\it Special lines for the $\text{dP}_4$ defined by \eqref{dP4num2}. }\label{dP4Lines} \\
\hline \multicolumn{1}{|c|}{\str\textbf{Line}}&  \multicolumn{1}{|c|}{\str\textbf{~Representation~}} \\ \hline 
\endfirsthead
\hline 
\textbf{~Line~} & \textbf{~Representation~}\\ \hline 
\endhead

\hline\hline \multicolumn{2}{|r|}{{\str Continued on next page}} \\ \hline
\endfoot

\endlastfoot

\hline\hline
\varstr{14pt}{7pt}  $Q$ &  $\left(u,v,-\sqrt{6} u-\ii \sqrt{6} v,-3 v+4 \ii  u,\+3 u+2 \ii v\right)$\\
\varstr{14pt}{7pt}  $E_1$ &  $\left(u,v,-\sqrt{6} u-\ii \sqrt{6} v,\+3 v-4 \ii  u,\+3 u+2 \ii v\right)$\\
\varstr{14pt}{7pt}  $E_2$ & $\left(u,v,-\sqrt{6} u-\ii \sqrt{6} v,-3 v+4 \ii  u,-3 u-2 \ii v\right)$\\
\varstr{14pt}{7pt}  $E_3$ & $\left(u,v,-\sqrt{6} u+\ii \sqrt{6} v,\+3 v+4 \ii  u,\+3 u-2 \ii v\right)$\\
\varstr{14pt}{7pt}  $E_4$ & $\left(u,v,\+\sqrt{6} u-\ii \sqrt{6} v,-3 v-4 \ii  u,-3 u+2 \ii v\right)$\\
\varstr{14pt}{7pt}  $E_5$ & $\left(u,v,\+\sqrt{6} u+\ii \sqrt{6} v,-3 v+4 \ii  u,\+3 u+2 \ii v\right)$\\
 \varstr{14pt}{7pt}  $L_{1,2}$ & $\left(u,v,-\sqrt{6} u-\ii \sqrt{6} v,\+3 v-4 \ii  u,-3 u-2 \ii v\right)$\\
  \varstr{14pt}{7pt}  $L_{1,3}$ & $\left(u,v,-\sqrt{6} u+\ii \sqrt{6} v,-3 v-4 \ii  u,\+3 u-2 \ii v\right)$\\
 \varstr{14pt}{7pt}  $L_{1,4}$ & $\left(u,v,\+\sqrt{6} u-\ii \sqrt{6} v,\+3 v+4 \ii  u,-3 u+2 \ii v\right)$\\
 \varstr{14pt}{7pt}  $L_{1,5}$ & $\left(u,v,\+\sqrt{6} u+\ii \sqrt{6} v,\+3 v-4 \ii  u,\+3 u+2 \ii v\right)$\\
 \varstr{14pt}{7pt}  $L_{2,3}$ & $\left(u,v,-\sqrt{6} u+\ii \sqrt{6} v,\+3 v+4 \ii  u,-3 u+2 \ii v\right)$\\
 \varstr{14pt}{7pt}  $L_{2,4}$ & $\left(u,v,\+\sqrt{6} u-\ii \sqrt{6} v,-3 v-4 \ii  u,\+3 u-2 \ii v\right)$\\
 \varstr{14pt}{7pt}  $L_{2,5}$ & $\left(u,v,\+\sqrt{6} u+\ii \sqrt{6} v,-3 v+4 \ii  u,-3 u-2 \ii v\right)$\\
 \varstr{14pt}{7pt}  $L_{3,4}$ & $\left(u,v,\+\sqrt{6} u+\ii \sqrt{6} v,\+3 v-4 \ii  u,-3 u-2 \ii v\right)$\\
 \varstr{14pt}{7pt}  $L_{3,5}$ & $\left(u,v,\+\sqrt{6} u-\ii \sqrt{6} v,\+3 v+4 \ii  u,\+3 u-2 \ii v\right)$\\
 \varstr{14pt}{7pt}  $L_{4,5}$ & $\left(u,v,-\sqrt{6} u+\ii \sqrt{6} v,-3 v-4 \ii  u,-3 u+2 \ii v\right)$\\
\hline 
\end{longtable}
\vskip-20pt
\end{center}
\subsubsection{Quotients of $X_{2568} \subset$ $\text{dP}_4\times \text{dP}_4$}\label{sec:X2568}
In this subsection, we compute the Hodge numbers for the quotients of the manifold $X_{2568}$ by studying the group actions on the $-1$ lines of $\text{dP}_4$, and in particular on the homology basis $\{H,E_1,E_2,E_3,E_4,E_5\}$. The number of linear invariant homology classes under this action will give the $h^{1,1}$ of the quotient. We begin with the following representation of $X_{2568}$:
\vskip-12pt
\begin{equation}\label{eq:dP4_2568}
\displaycicy{5.25in}{
X_{2568}~=~~
\cicy{\IP^1 \\ \IP^1\\ \IP^1\\ \IP^1\\ \IP^1\\ \IP^1}
{ ~1 &1 & 0  ~\\
  ~1 &1 & 0  ~\\
  ~0 &2 & 0 ~\\
  ~1 &0 & 1 ~\\
  ~1 &0 & 1 ~\\
    ~0 &0 & 2~\\}_{-32}^{12,28}}{-1cm}{2.4in}{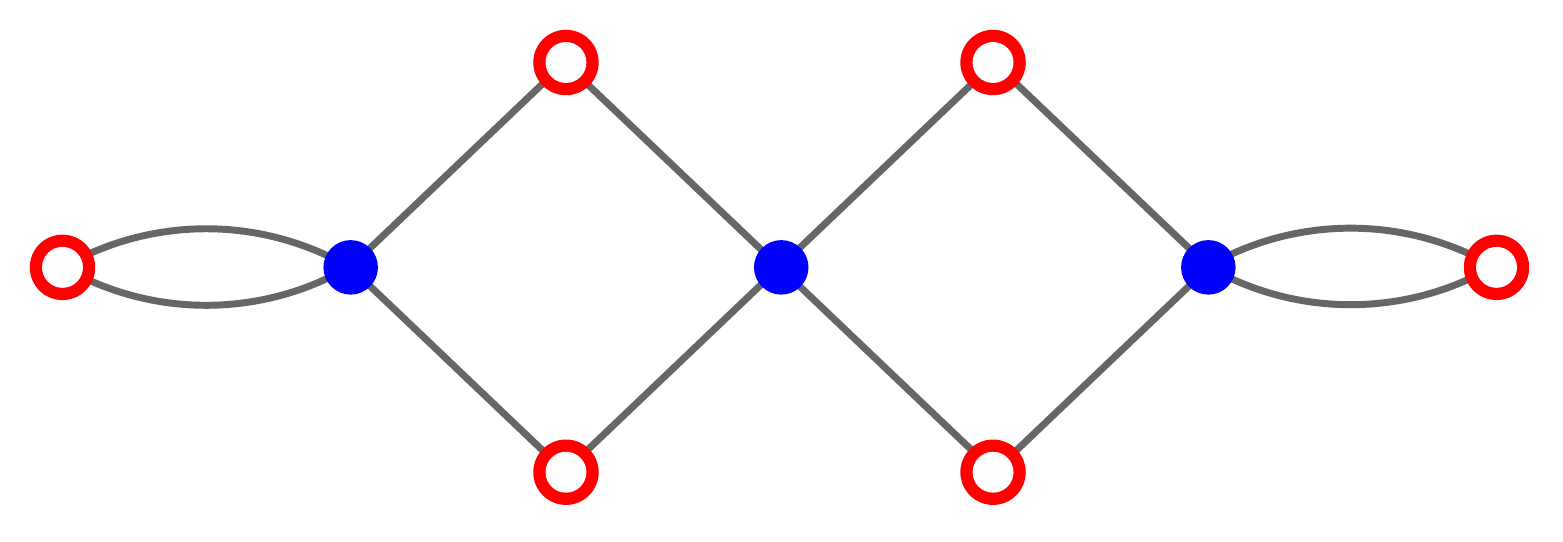}
\end{equation}
The largest symmetry group acting on this manifold is $\IZ_2{\times}\IZ_4$. For the covering manifold, the number of complex structure parameters, seen by the polynomial deformation method is $19$, which undercounts the true number. Thus the polynomial deformation method is inapplicable in this case. However, since $ X_{2568}\subset\text{dP}_4{\times}\text{dP}_4$, we can apply the previous discussion of lines on $\text{dP}_4$ to compute $h^{1,1}$ for the quotients. In view of the fact that we know the Euler number for the quotient, this allows us to compute $h^{2,1}$ also. 

The cohomology group $H^2(X_{2568})$ descends from that of the product of the two $\text{dP}_4$ spaces, that is from $H^2(\text{dP}_4{\times}\text{dP}_4)$. We are thus interested in the group action on the combined cohomology basis of the two $\text{dP}_4$ factors, $\{H,\,E_i,\,\widetilde{H},\,\widetilde{E}_i\}$, with $1\leq i \leq5$. Although Braun lists 42 (freely acting) symmetry actions on the combined set of coordinates and the polynomials for \eqref{eq:dP4_2568}, only 14 of them act distinctly on the co-ordinates. For our purpose, it suffices to consider only the distinct actions on the coordinates. \tref{dP4_2568_Symm_Action} lists these actions. 

We apply our discussion of the last subsection to obtain the group action on the cohomology for each symmetry representation. All cohomology actions are linear and can be expressed in terms of the set of matrices defined in \eqref{eq:CohMatDefs2}. We will repeat this calculation for other manifolds, in the following, and these definitions of the matrices $P_i$ and $Q_i$, $i{=}1,2,3$, will hold in these cases also. For each symmetry action listed in \tref{dP4_2568_Symm_Action}, the corresponding group action on the cohomology basis is shown in \tref{dP4_2568_Symm_Action_Coh}. The table also lists the cohomology invariants corresponding to each symmetry. Since the equations describing the exceptional curves depend on the choice of coefficients in the defining polynomials of a CICY quotient, different choices will yield different cohomology actions and hence different invariants in principle. However, for all coefficient choices that yield a smooth manifold and sixteen distinct special lines, the number of invariants is fixed. This is the $h^{1,1}$ of the quotient.

Consider the first $\IZ_2$ action listed in \tref{dP4_2568_Symm_Action}. With a generic choice of coefficients in the defining polynomials of this quotient of \eqref{eq:dP4_2568}, the lines transform in the following way: 
\beq\begin{split}
&H \rightarrow H~, ~~E_1\leftrightarrow E_2~,~~E_3\leftrightarrow E_4~,~~E_5 \rightarrow E_5~,\\
&\widetilde{H}\rightarrow \widetilde{H}~,~~\widetilde{E}_1\leftrightarrow \widetilde{E}_2~,~~
\widetilde{E}_3\leftrightarrow \widetilde{E}_4~,~~\widetilde{E}_5\rightarrow \widetilde{E}_5~.
\end{split}\notag
\eeq
The transformation properties of the lines $E_i$ and $\widetilde{E}_i$, $1\leq i\leq4$, follow from the transformation properties of the roots of the equations $A\,=\,0$ and $B\,=\,0$ that derive from \eqref{dP4AB}. The transformation properties of $H$, $\widetilde{H}$, $E_5$ and $\widetilde{E}_5$ follow from the invariance of the canonical classes $K$ and $\widetilde{K}$ and the preservation of the intersection numbers \eqref{eq:IntersectionNumbers}.
This yields the following 8 cohomology invariants: 
\beq
H,~ E_1+E_2,~E_3+E_4,~E_5;~~\widetilde{H},~ \widetilde{E}_1+\widetilde{E}_2, 
~\widetilde{E}_3+\widetilde{E}_4,~\widetilde{E}_5~. 
\eeq
So $h^{1,1}=8$ for this quotient of $X_{2568}$.
The computation of $h^{1,1}$ for the remaining symmetry actions proceed in a similar way and together with $h^{2,1}$ and the Euler characteristic, they are listed in \tref{dP4_2568}.
The $P$ and $Q$ matrices, of \tref{dP4_2568_Symm_Action_Coh} above, are given by\vskip-10pt
\begin{eqnarray*}
\begin{split}\footnotesize
P_1=\left[
\begin{array}{cccccc}
 1 & 0 & 0 & 0 & 0 & 0 \\
 0 & 0 & 1 & 0 & 0 & 0 \\
 0 & 1 & 0 & 0 & 0 & 0 \\
 0 & 0 & 0 & 0 & 1 & 0 \\
 0 & 0 & 0 & 1 & 0 & 0 \\
 0 & 0 & 0 & 0 & 0 & 1 \\
\end{array}\right], ~
P_2=\left[
\begin{array}{cccccc}
 1 & 0 & 0 & 0 & 0 & 0 \\
 0 & 0 & 0 & 0 & 1 & 0 \\
 0 & 0 & 0 & 1 & 0 & 0 \\
 0 & 0 & 1 & 0 & 0 & 0 \\
 0 & 1 & 0 & 0 & 0 & 0 \\
 0 & 0 & 0 & 0 & 0 & 1 \\
\end{array}
\right],~P_3=\left[
\begin{array}{cccccc}
 1 & 0 & 0 & 0 & 0 & 0 \\
 0 & 0 & 0 & 1 & 0 & 0 \\
 0 & 0 & 0 & 0 & 1 & 0 \\
 0 & 1 & 0 & 0 & 0 & 0 \\
 0 & 0 & 1 & 0 & 0 & 0 \\
 0 & 0 & 0 & 0 & 0 & 1 \\
\end{array}\right];
\end{split}
\end{eqnarray*}
\vspace*{5pt}
\beq\label{eq:CohMatDefs2}
\begin{split}
Q_1&=\footnotesize\left[
\begin{array}{rrrrrr}
 3 & -1 & -1 & -1 & -1 & -2 \\
 1 & 0 & -1 & 0 & 0 & -1 \\
 1 & -1 & 0 & 0 & 0 & -1 \\
 1 & 0 & 0 & 0 & -1 & -1 \\
 1 & 0 & 0 & -1 & 0 & -1 \\
 2 & -1 & -1 & -1 & -1 & -1 \\
\end{array}\right], ~~
Q_2=\footnotesize\left[
\begin{array}{rrrrrr}
 3 & -1 & -1 & -1 & -1 & -2 \\
 1 & 0 & 0 & 0 & -1 & -1 \\
 1 & 0 & 0 & -1 & 0 & -1 \\
 1 & 0 & -1 & 0 & 0 & -1 \\
 1 & -1 & 0 & 0 & 0 & -1 \\
 2 & -1 & -1 & -1 & -1 & -1 \\
\end{array}\right];\\ \\[-10pt] 
Q_3&=\footnotesize\left[
\begin{array}{rrrrrr}
 3 & -1 & -1 & -1 & -1 & -2 \\
 1 & 0 & 0 & -1 & 0 & -1 \\
 1 & 0 & 0 & 0 & -1 & -1 \\
 1 & -1 & 0 & 0 & 0 & -1 \\
 1 & 0 & -1 & 0 & 0 & -1 \\
 2 & -1 & -1 & -1 & -1 & -1 \\
\end{array}\right], ~~
Q_4=\footnotesize\left[
\begin{array}{rrrrrr}
 3 & -1 & -1 & -1 & -1 & -2 \\
 1 & -1 & 0 & 0 & 0 & -1 \\
 1 & 0 & -1 & 0 & 0 & -1 \\
 1 & 0 & 0 & -1 & 0 & -1 \\
 1 & 0 & 0 & 0 & -1 & -1 \\
 2 & -1 & -1 & -1 & -1 & -1 \\
\end{array}\right].
\end{split}
\eeq
\begin{table}[H]
\vspace{15pt}
\begin{center}
\begin{tabular}{| c || c | c | c | c |}
\hline
\myalign{| c||}{\varstr{16pt}{10pt}$~~~~~~~  \Gamma ~~~~~~~$ } &
\myalign{m{1.31cm}|}{$~~~ \IZ_2 $} &
\myalign{m{1.9cm}|}{ $~~~\IZ_2\times\IZ_2$ }  &
\myalign{m{1.31cm}|}{ $~~~\IZ_4 $ }  &
\myalign{m{1.9cm}|}{ $~~~\IZ_2\times\IZ_4 \ \ \ $ }  
\\ \hline\hline
\varstr{14pt}{8pt} $h^{1,1}(X/\Gamma)$ & 8 & 6 & 4 & 3\\
 \hline
\varstr{14pt}{8pt} $h^{2,1}(X/\Gamma)$ & 16 & 10 & 8 & 5 \\
 \hline
\varstr{14pt}{8pt} $\chi(X/\Gamma)$ & $\!\!\!\!-16$ & $\!\!\!\!-8$ & $\!\!\!\!-8$ & $\!\!\!\!-4$\\
 \hline
 \end{tabular}
\vskip 0.3cm
\capt{4.5in}{dP4_2568}{Hodge numbers for the quotients of the manifold~\eqref{eq:dP4_2568}.}
 \end{center}
 \end{table}
\begin{center}
\begin{longtable}{|c|c|c|c|}
\captionsetup{width=0.9\textwidth}
\caption{\it Various symmetry actions on the ambient space of the manifold~\eqref{eq:dP4_2568}. The coordinate patch of the two $\text{dP}_4$'s are chosen to be $(1,x)\times(1,y)\times(1,z)$ and $(1,\widetilde{x}){\times}(1,\widetilde{y}){\times}(1,\widetilde{z})$ respectively.} \label{dP4_2568_Symm_Action} \\

\hline \multicolumn{1}{|c|}{\str\textbf{Index}} &  \multicolumn{1}{|c|}{\str\textbf{~Group~}} & \multicolumn{1}{|c|}{$\mathbf{(\+x,\+y,\+z)}$} &  \multicolumn{1}{|c|}{$\mathbf{(\+\widetilde{x},\+\widetilde{y},\+\widetilde{z})}$} \\ \hline 
\endfirsthead

\hline 
\textbf{Index} &
\textbf{~Group~} &
$\mathbf{(\+x,\+y,\+z)}$ &
$\mathbf{(\+\widetilde{x},\+\widetilde{y},\+\widetilde{z})}$ \\ \hline 
\endhead

\hline\hline \multicolumn{4}{|r|}{{\str Continued on next page}} \\ \hline
\endfoot

\endlastfoot

\hline\hline

\varstr{15pt}{8pt} 1 & $\IZ_2$ & $(-x,-y,-z)$ & $(-\widetilde{x},-\widetilde{y},-\widetilde{z})$ \\
 \hline
\varstr{15pt}{8pt} 2 & $\IZ_2$ & $(\+y,\+x,-z)$ & $(-\widetilde{x},-\widetilde{y},-\widetilde{z})$ \\
 \hline
\varstr{15pt}{8pt} 3 & $\IZ_2$ & $(\+y,\+x,-z)$ & $(\+\widetilde{y},\+\widetilde{x},-\widetilde{z})$ \\
 \hline
\varstr{15pt}{8pt} 4 & $\IZ_4$ &$(\+\widetilde{x},\+\widetilde{y},-\widetilde{z})$ & $(\+y,\+x,\+z)$ \\
 \hline
5 & $\IZ_2{\times}\IZ_2$ &
  \begin{minipage}[c][43pt][c]{1.3in}
  \begin{gather*} 
 (-x,-y,-z) \\ 
 ({x^{-1}},{y^{-1}},{z^{-1}}) \\
  \end{gather*}
  \end{minipage} 
& 
 \begin{minipage}[c][43pt][c]{1.3in}
  \begin{gather*}
  (-\widetilde{x},-\widetilde{y},-\widetilde{z}) \\ 
  ({\widetilde{x}^{-1}},{\widetilde{y}^{-1}},{\widetilde{z}^{-1}})\\
  \end{gather*}
  \end{minipage} 
\\
 \hline
6 & $\IZ_2{\times}\IZ_2$ &
  \begin{minipage}[c][43pt][c]{1.3in}
  \begin{gather*} 
 (-x,-y,-z) \\ 
 ({\+y},{\+x},{z^{-1}}) \\
  \end{gather*}
  \end{minipage} 
& 
 \begin{minipage}[c][43pt][c]{1.3in}
  \begin{gather*}
  (-\widetilde{x},-\widetilde{y},-\widetilde{z}) \\ 
  ({\widetilde{x}^{-1}},{\widetilde{y}^{-1}},{\widetilde{z}^{-1}})\\
  \end{gather*}
  \end{minipage} 
\\
 \hline
 \varstr{16pt}{10pt} 7 & $\IZ_2{\times}\IZ_2$ &
  \begin{minipage}[c][43pt][c]{1.3in}
  \begin{gather*} 
 (\+y,\+x,-z) \\ 
 (-y,-x,{z^{-1}}) \\
  \end{gather*}
  \end{minipage} 
& 
 \begin{minipage}[c][43pt][c]{1.3in}
  \begin{gather*}
  (-\widetilde{x},-\widetilde{y},-\widetilde{z}) \\ 
  ({\widetilde{x}^{-1}},{\widetilde{y}^{-1}},{\widetilde{z}^{-1}})\\
  \end{gather*}
  \end{minipage} 
\\
 \hline
\varstr{16pt}{10pt} 8 & $\IZ_2{\times}\IZ_2$ &
  \begin{minipage}[c][43pt][c]{1.3in}
  \begin{gather*} 
 (\+y,\+x,-z) \\ 
 (-x,-y,{z^{-1}}) \\
  \end{gather*}
  \end{minipage} 
& 
 \begin{minipage}[c][43pt][c]{1.3in}
  \begin{gather*}
  (-\widetilde{x},-\widetilde{y},-\widetilde{z}) \\ 
  ({\widetilde{x}^{-1}},{\widetilde{y}^{-1}},{\widetilde{z}^{-1}})\\
  \end{gather*}
  \end{minipage} 
\\
 \hline
\varstr{16pt}{10pt} 9 & $\IZ_2{\times}\IZ_2$ &
  \begin{minipage}[c][43pt][c]{1.3in}
  \begin{gather*} 
 (\+y,\+x,-z) \\ 
 (-y,-x,z^{-1}) \\ 
  \end{gather*}
  \end{minipage} 
& 
 \begin{minipage}[c][43pt][c]{1.3in}
  \begin{gather*}
  ({\+\widetilde{y}},{\+\widetilde{x}},{-\widetilde{z}})\\
  ({-\widetilde{y}},{-\widetilde{x}},{\widetilde{z}^{-1}})\\
  \end{gather*}
  \end{minipage} 
 \\
 \hline
\varstr{16pt}{10pt} 10 & $\IZ_2{\times}\IZ_2$ &
  \begin{minipage}[c][43pt][c]{1.3in}
  \begin{gather*} 
 (-x,-y,-z) \\ 
 (\+{y},\+{x},{z}^{-1}) \\
  \end{gather*}
  \end{minipage} 
& 
 \begin{minipage}[c][43pt][c]{1.3in}
  \begin{gather*}
  (-\widetilde{x},-\widetilde{y},-\widetilde{z}) \\ 
  ({\+\widetilde{y}},{\+\widetilde{x}},{\widetilde{z}^{-1}})\\
  \end{gather*}
  \end{minipage} 
\\
 \hline
\varstr{16pt}{10pt} 11 & $\IZ_2{\times}\IZ_2$ &
  \begin{minipage}[c][43pt][c]{1.3in}
  \begin{gather*} 
 (\+y,\+x,{-z}) \\
 (-x,-y,{z}^{-1}) \\
  \end{gather*}
  \end{minipage} 
& 
 \begin{minipage}[c][43pt][c]{1.3in}
  \begin{gather*}
  (\+\widetilde{y},\+\widetilde{x},-\widetilde{z}) \\ 
  ({-\widetilde{x}},{-\widetilde{y}},{\widetilde{z}^{-1}})\\
  \end{gather*}
  \end{minipage}  
\\
 \hline
\varstr{16pt}{10pt} 12 & $\IZ_2{\times}\IZ_2$ &
  \begin{minipage}[c][43pt][c]{1.3in}
  \begin{gather*} 
 (\+y,\+x,-z) \\ 
 (-y,-x,z^{-1}) \\ 
  \end{gather*}
  \end{minipage} 
& 
 \begin{minipage}[c][43pt][c]{1.3in}
  \begin{gather*}
  ({\+\widetilde{y}},{\+\widetilde{x}},{-\widetilde{z}})\\
  ({-\widetilde{x}},{-\widetilde{y}},{\widetilde{z}^{-1}})\\
  \end{gather*}
  \end{minipage} 
 \\
 \hline
\varstr{16pt}{10pt} 13 & $\IZ_2{\times}\IZ_4$ &
  \begin{minipage}[c][43pt][c]{1.3in}
  \begin{gather*} 
  (\+y,\+x,{z}^{-1}) \\ 
  (-\widetilde{y},-\widetilde{x},-\widetilde{z}^{-1}) \\ 
  \end{gather*}
  \end{minipage} 
& 
 \begin{minipage}[c][43pt][c]{1.3in}
  \begin{gather*}
  ({\+\widetilde{y}},{\+\widetilde{x}},{\widetilde{z}}^{-1})\\
  (\+x,\+y,\+z)\\
  \end{gather*}
  \end{minipage} \\
\hline
\varstr{16pt}{10pt} 14 & $\IZ_2{\times}\IZ_4$ &
  \begin{minipage}[c][43pt][c]{1.3in}
  \begin{gather*} 
  (-x,-y,-z) \\ 
  (\+\widetilde{y},\+\widetilde{x},\+\widetilde{z}^{-1}) \\ 
  \end{gather*}
  \end{minipage} 
& 
 \begin{minipage}[c][43pt][c]{1.3in}
  \begin{gather*}
  ({-\widetilde{x}},{-\widetilde{y}},{-\widetilde{z}})\\
  (\+x,\+y,\+z)\\
  \end{gather*}
  \end{minipage} \\
\hline
\end{longtable}
\end{center}
\footnotesize
\begin{center}
\begin{longtable}{|c|c|c|c|}
\captionsetup{width=0.9\textwidth}
\caption{\it Symmetry actions on the cohomology basis and the corresponding invariants for the manifold~\eqref{eq:dP4_2568}. The matrices $P_i$ and $Q_i$ are defined in \eqref{eq:CohMatDefs2}.}\label{dP4_2568_Symm_Action_Coh} \\

\hline \multicolumn{1}{|c|}{\str\textbf{Index}}&  \multicolumn{1}{|c|}{\str\textbf{~Group~}} & \multicolumn{1}{|c|}{\str\textbf{Action on Coh Basis}} &  \multicolumn{1}{|c|}{\str\textbf{~~Coh Invariants~~}} \\ \hline 
\endfirsthead

\hline 
\textbf{~Index~} & \textbf{~Group~} & \textbf{Action on Coh Basis} & \textbf{~~Coh Invariants~~} \\ \hline 
\endhead

\hline\hline \multicolumn{4}{|r|}{{\str Continued on next page}} \\ \hline
\endfoot

\endlastfoot

\hline\hline
1 & $\IZ_2$ &
$\left[\begin{array}{cc}
P_1 & 0 \\
0 & P_1
\end{array}\right]$
& \begin{minipage}[c][33pt][c]{2.8in}
\begin{gather*} 
H, E_1+E_2,E_3+E_4,E_5,\\
\widetilde{H}, \widetilde{E}_1+\widetilde{E}_2, \widetilde{E}_3+\widetilde{E}_4,\widetilde{E}_5 \\
\end{gather*}
\end{minipage} \\
 \hline
 
2 & $\IZ_2$ &
$\left[\begin{array}{cc}
Q_1 & 0 \\
0 & P_1
\end{array}\right]$
& \begin{minipage}[c][33pt][c]{2.4in}
\begin{gather*} 
H-E_5, E_1+E_4-E_5,E_2+E_4-E_5,\\
E_3-E_4,~\widetilde{H}, \widetilde{E}_1+\widetilde{E}_2,\widetilde{E}_3+\widetilde{E}_4,\widetilde{E}_5 \\
\end{gather*}
\end{minipage} \\
 \hline
 
3 & $\IZ_2$ &
$\left[\begin{array}{cc}
Q_1 & 0 \\
0 & Q_1
\end{array}\right]$
&\begin{minipage}[c][50pt][c]{2.4in}
\begin{gather*} 
H-E_5, E_1+E_4-E_5,E_2+E_4-E_5,\\
E_3-E_4,\widetilde{H}-\widetilde{E}_5, \widetilde{E}_1+\widetilde{E}_4-\widetilde{E}_5,\\
\widetilde{E}_2+\widetilde{E}_4-\widetilde{E}_5,\widetilde{E}_3-\widetilde{E}_4 \\
\end{gather*}
\end{minipage} \\ \hline

4 & $\IZ_4$ &
$\left[\begin{array}{cc}
0 & Q_4 \\
P_1 & 0
\end{array}\right]$
& \begin{minipage}[c][54pt][c]{2.8in}
\begin{gather*} 
H-E_5+\widetilde{H}-\widetilde{E}_5,
E_3-E_4-(\widetilde{E}_3-\widetilde{E}_4), \\
E_1+E_4-E_5+\widetilde{E}_2+\widetilde{E}_3-\widetilde{E}_5,\\
E_2+E_4-E_5+\widetilde{E}_1+\widetilde{E}_3-\widetilde{E}_5 \\
\end{gather*}
\end{minipage} \\
 \hline
 
5 & $\IZ_2{\times}\IZ_2$ &
$\left[\begin{array}{cc}
P_2 & 0 \\
0 & P_1
\end{array}\right]$, $\left[\begin{array}{cc}
P_1 & 0 \\
0 & P_2
\end{array}\right]$
& \begin{minipage}[c][33pt][c]{2.8in}
\begin{gather*} 
H,E_1+E_2+E_3+E_4,E_5,\\
\widetilde{H},\widetilde{E}_1+\widetilde{E}_2+\widetilde{E}_3+\widetilde{E}_4, \widetilde{E}_5 \\
\end{gather*}
\end{minipage} \\
 \hline
 
6 & $\IZ_2{\times}\IZ_2$ &
$\left[\begin{array}{cc}
P_1 & 0 \\
0 & P_1
\end{array}\right]$, $\left[\begin{array}{cc}
Q_3 & 0 \\
0 & P_2
\end{array}\right]$
& \begin{minipage}[c][33pt][c]{2.8in}
\begin{gather*} 
H-E_5, E_1+E_2-E_5, E_3+E_4-E_5,\\
~\widetilde{H}, \widetilde{E}_1+\widetilde{E}_2+\widetilde{E}_3+\widetilde{E}_4,\widetilde{E}_5 \\
\end{gather*}
\end{minipage} \\
 \hline
 
7 & $\IZ_2{\times}\IZ_2$ &
$\left[\begin{array}{cc}
Q_1 & 0 \\
0 & P_1
\end{array}\right]$, $\left[\begin{array}{cc}
Q_2 & 0 \\
0 & P_2
\end{array}\right]$& \begin{minipage}[c][33pt][c]{2.8in}
\begin{gather*} 
H-E_5, E_1+E_3-E_5, E_2+E_4-E_5,\\
 \widetilde{H}, \widetilde{E}_1+\widetilde{E}_2+\widetilde{E}_3+\widetilde{E}_4,\widetilde{E}_5 \\
\end{gather*}
\end{minipage} \\
 \hline
8 & $\IZ_2{\times}\IZ_2$ &
$\left[\begin{array}{cc}
Q_1 & 0 \\
0 & P_1
\end{array}\right]$, $\left[\begin{array}{cc}
P_3 & 0 \\
0 & P_2
\end{array}\right]$
& \begin{minipage}[c][33pt][c]{2.8in}
\begin{gather*} 
H-E_5, E_1+E_3-E_5, E_2+E_4-E_5,\\
 \widetilde{H}, \widetilde{E}_1+\widetilde{E}_2+\widetilde{E}_3+\widetilde{E}_4,\widetilde{E}_5 \\
\end{gather*}
\end{minipage} \\
 \hline
 9 & $\IZ_2{\times}\IZ_2$ &
$\left[\begin{array}{cc}
Q_1 & 0 \\
0 & Q_1
\end{array}\right]$, $\left[\begin{array}{cc}
Q_2 & 0 \\
0 & Q_3
\end{array}\right]$
& \begin{minipage}[c][33pt][c]{2.8in}
\begin{gather*} 
H-E_5, E_1+E_3-E_5, E_2+E_4-E_5,\\
~ \widetilde{H}-\widetilde{E}_5, \widetilde{E}_1+\widetilde{E}_4-\widetilde{E}_5,\widetilde{E}_2+\widetilde{E}_3-\widetilde{E}_5 \\\end{gather*}
\end{minipage} \\
 \hline
10 & $\IZ_2{\times}\IZ_2$ &
$\left[\begin{array}{cc}
P_2 & 0 \\
0 & P_2
\end{array}\right]$, $\left[\begin{array}{cc}
Q_3 & 0 \\
0 & Q_1
\end{array}\right]$
& \begin{minipage}[c][33pt][c]{2.8in}
\begin{gather*} 
H-E_5, E_1+E_4-E_5, E_2+E_3-E_5,\\
~ \widetilde{H}-\widetilde{E}_5, \widetilde{E}_1+\widetilde{E}_4-\widetilde{E}_5,\widetilde{E}_2+\widetilde{E}_3-\widetilde{E}_5\\\end{gather*}
\end{minipage} \\
 \hline
 11 & $\IZ_2{\times}\IZ_2$ &
$\left[\begin{array}{cc}
Q_1 & 0 \\
0 & Q_1
\end{array}\right]$, $\left[\begin{array}{cc}
P_2 & 0 \\
0 & P_2
\end{array}\right]$
& \begin{minipage}[c][33pt][c]{2.8in}
\begin{gather*} 
H-E_5, E_1+E_4-E_5, E_2+E_3-E_5,\\
~ \widetilde{H}-\widetilde{E}_5, \widetilde{E}_1+\widetilde{E}_4-\widetilde{E}_5,\widetilde{E}_2+\widetilde{E}_3-\widetilde{E}_5 \\\end{gather*}
\end{minipage} \\
 \hline
12 & $\IZ_2{\times}\IZ_2$ &
$\left[\begin{array}{cc}
Q_1 & 0 \\
0 & Q_1
\end{array}\right]$, $\left[\begin{array}{cc}
Q_3 & 0 \\
0 & P_2
\end{array}\right]$
& \begin{minipage}[c][33pt][c]{2.8in}
\begin{gather*} 
H-E_5, E_1+E_4-E_5, E_2+E_3-E_5,\\
~ \widetilde{H}-\widetilde{E}_5, \widetilde{E}_1+\widetilde{E}_4-\widetilde{E}_5,\widetilde{E}_2+\widetilde{E}_3-\widetilde{E}_5 \\\end{gather*}
\end{minipage} \\
 \hline
13 & $\IZ_2{\times}\IZ_4$ &
$\left[\begin{array}{cc}
Q_2 & 0 \\
0 & Q_2
\end{array}\right]$, $\left[\begin{array}{cc}
0 & Q_3 \\
\mathbb{I}_6 & 0
\end{array}\right]$
& \begin{minipage}[c][56pt][c]{2.8in}
\begin{gather*} 
H-E_5+\widetilde{H}-\widetilde{E}_5,\\
E_1+E_2-E_5+\widetilde{E}_1+\widetilde{E}_2-\widetilde{E}_5,\\
E_3+E_4-E_5+\widetilde{E}_3+\widetilde{E}_4-\widetilde{E}_5 \\
\end{gather*}
\end{minipage} \\
 \hline 
 14 & $\IZ_2{\times}\IZ_4$ &
$\left[\begin{array}{cc}
P_1 & 0 \\
0 & P_1
\end{array}\right]$, $\left[\begin{array}{cc}
0 & Q_2 \\
\mathbb{I}_6 & 0
\end{array}\right]$
& \begin{minipage}[c][56pt][c]{2.8in}
\begin{gather*} 
H-E_5+\widetilde{H}-\widetilde{E}_5,\\
E_1+E_2-E_5+\widetilde{E}_1+\widetilde{E}_2-\widetilde{E}_5,\\
E_3+E_4-E_5+\widetilde{E}_3+\widetilde{E}_4-\widetilde{E}_5 \\
\end{gather*}
\end{minipage} \\
 \hline 
\end{longtable}
\end{center}
\normalsize
\subsubsection{Quotients of $X_{2566} \subset$ $\text{dP}_4\times \text{dP}_4$}\label{sec:X2566}\vskip-10pt
The following configuration contains a product of two $\text{dP}_4$'s in two distinct representations. 
\vskip-15pt
\begin{equation}\label{eq:M8Split2ver1}
\displaycicy{5.25in}{
X_{2566}~=~~
\cicy{\IP^1\\\IP^1\\\IP^1\\\IP^4}
{ ~2&0& 0& 0~ \\
  ~1&1& 0& 0~ \\
  ~1&1& 0& 0~ \\
  ~0&1& 2& 2~ \\}_{-32}^{12,\,28}}{-1.1cm}{1.7in}{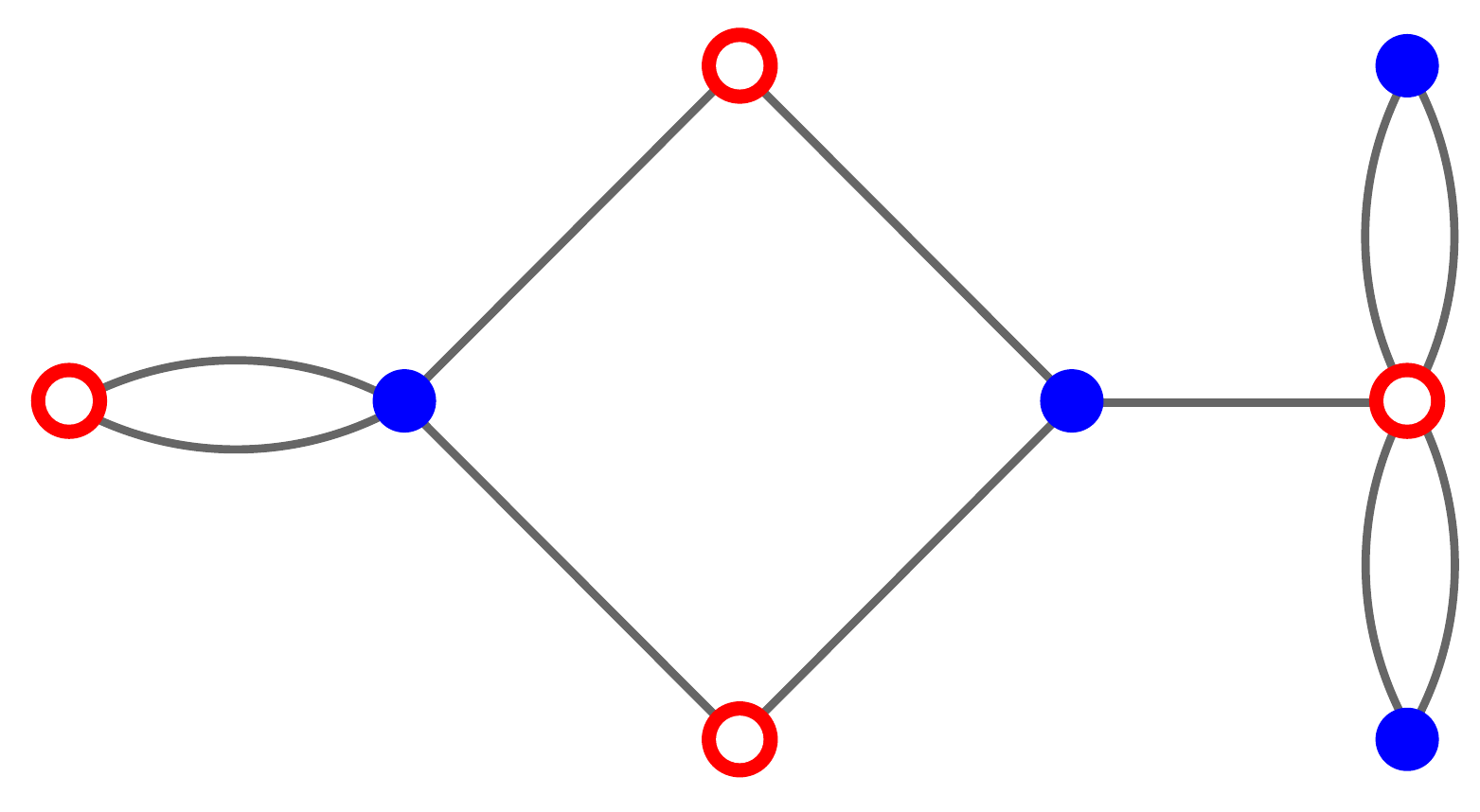}
\end{equation}
This manifold admits quotients by free $\IZ_2{\times}\IZ_2$ actions, one of which is given by the following generators. The computations of the Hodge numbers for the other actions proceed in a similar way. 
\beqnn
R(g)=\footnotesize\left[\begin{array}{ccc}
A & 0 & 0 \\
0 & B & 0\\
0 & 0 & C\\
\end{array}\right] \normalsize
~~~\text{and}~~~ 
R(h)=\footnotesize\left[\begin{array}{ccc}
A' & 0 & 0 \\
0 & B' & 0\\
0 & 0 & C'\\
\end{array}\right]\normalsize
\eeqnn
\vskip-10pt
where
\vskip-15pt 
\beqnn
A=\footnotesize\left[\begin{array}{cc} 1&\+0\\0&-1\\ \end{array}\right]\normalsize;~~ 
A'=\footnotesize\left[\begin{array}{cc} 0&1\\1&0\\ \end{array}\right]\normalsize;~~ 
B=\footnotesize\left[\begin{array}{rrrr} 0&0&1&0\\0&0&0&1\\1&0&0&0\\0&1&0&0 \end{array}\right]\normalsize;~~ 
B'=\footnotesize\left[\begin{array}{rrrr} \+0&\+0&\+\ii&\+0\\0&0&0&-\ii\\-\ii&0&0&0\\0&\ii&0&0 \end{array}\right]
\normalsize;
\eeqnn
and $C$ and $C'$ are diagonal matrices with entries $\{1,-1,-1,1,-1\}$ and $\{1,-1,-1,-1,1\}$ respectively. 
Over $\IP^4$, the special lines take the form shown in \tref{dP4Lines}.

The symmetry action on the cohomology is listed in \tref{dP4SymmAction} below. This action can also be written in terms of matrices acting on the cohomology basis. The advantage in doing so is that one can easily compute the dimension of the eigenspace with eigenvalue 1. This gives the number of invariants under the group action and is related to the $h^{1,1}$ of the quotient manifold under this group action. 
\begin{table}[H]
\setlength{\tabcolsep}{4pt}
\renewcommand{\arraystretch}{1.5}
\vspace{5pt}
\begin{center}
\begin{tabular}{| c || p{0.58cm}| p{0.58cm}| p{0.58cm}| p{0.58cm}| p{0.58cm}| p{0.58cm}| p{0.58cm}| p{0.58cm}| p{0.58cm}| p{0.58cm}| p{0.58cm}| p{0.58cm}| p{0.58cm}| p{0.58cm}| p{0.58cm}| p{0.58cm}|}
\hline
\myalign{| c||}{\varstr{8pt}{6pt}} &
$Q$ & $E_1$ & $E_2$ & $E_3$ & $E_4$ & $E_5$ & $L_{12}$ & $L_{13}$ & $L_{14}$ & $L_{15}$ & $L_{23}$ & $L_{24}$ & $L_{25}$ & $L_{34}$ & $L_{35}$ & $L_{45}$
\\ \hline\hline
\varstr{7pt}{4pt} $g$ & $L_{13}$ & $E_{3}$ & $L_{45}$ & $E_{1}$ & $L_{25}$ & $L_{24}$ & $L_{23}$ & $~Q$ & $L_{34}$ & $L_{35}$ & $L_{12}$ & $E_{5}$ & $E_{4}$ & $L_{14}$ & $L_{15}$ & $E_{2}$\\
 \hline
\varstr{7pt}{4pt} $h$ & $L_{14}$ & $E_{4}$ & $L_{35}$ & $L_{25}$ & $E_{1}$ & $L_{23}$ & $L_{24}$ & $L_{34}$ & $~Q$ & $L_{45}$ & $E_{5}$ &  $L_{12}$ & $E_{3}$ & $L_{13}$ & $E_{2}$ & $L_{15}$   \\
 \hline
\varstr{7pt}{4pt} $g h$ & ${L_{34}}$ & ${L_{25}}$ & ${L_{15}}$ & ${E_{4}}$ & ${E_{3}}$ & ${L_{12}}$ & ${E_{5}}$ & ${L_{14}}$ & ${L_{13}}$ & $E_2$ & ${L_{24}}$ & ${L_{23}}$ & ${E_1}$ & ${~Q}$ & ${L_{45}}$ & ${L_{35}}$\\
 \hline
 \end{tabular}
\vskip 0.3cm
\capt{5.7in}{dP4SymmAction}{Group Action on the special lines of $\text{dP}_4$, for the surface defined by 
\eqref{dP4num2}}
 \end{center}
 \vspace{-24pt}
 \end{table}
\begin{table}[H]
\setlength{\tabcolsep}{4pt}
\renewcommand{\arraystretch}{1.5}
\vspace{0pt}
\begin{center}
\begin{tabular}{| c || l|}
\hline
~Group Generators~ & \myalign{|l|}{\hfil Cohomology Invariants}\\
\hline\hline
\varstr{5pt}{5pt} $g$          & ~~$K$,\quad  $H-E_5$,\quad $E_2-E_5$,\quad $E_4-E_5$~~\\
 \hline
\varstr{5pt}{5pt} $h$          & ~~$K$,\quad  $H-E_5$,\quad  $E_2-E_5$,\quad  $E_3-E_5$~~\\
 \hline
 \varstr{5pt}{5pt} $gh$     & ~~$K$,\quad  $H-E_5$,\quad  $E_1-E_5$,\quad  $E_2-E_5$~~\\
 \hline
\varstr{5pt}{5pt} $g ,\, h$ & ~~$K$,\quad  $H-E_5$,\quad  $E_2-E_5$~~\\
 \hline
 \end{tabular}
\vskip 0.3cm
\capt{4.7in}{coh_invs_2566}{Cohomology Invariants of the group action on the $\text{dP}_4$ defined in \eqref{dP4num2}. In the last row we give the invariants under both $g$ and $h$.}
 \end{center}
 \vspace{-30pt}
 \end{table}
\begin{table}[H]
\begin{center}
\begin{tabular}{| c || c | c |}
\hline
\myalign{| c||}{\varstr{16pt}{10pt}$~~~~~~~  \Gamma ~~~~~~~$ } &
\myalign{m{1.31cm}|}{$\hfil \IZ_2 $} &
\myalign{m{1.9cm}|}{ $\hfil \IZ_2\times\IZ_2$ }
\\ \hline\hline
\varstr{14pt}{8pt} $h^{1,1}(X/\Gamma)$ & 8 & 6\\
 \hline
\varstr{14pt}{8pt} $h^{2,1}(X/\Gamma)$ & 16 & 10 \\
 \hline
\varstr{14pt}{8pt} $\chi(X/\Gamma)$ & $\!\!\!\!-16$ & $\!\!\!\!-8$\\
 \hline
 \end{tabular}
\vskip 0.3cm
\capt{4.5in}{dP4_2566}{Hodge numbers for the quotients of the manifold~\eqref{eq:M8Split2ver1}.}
 \end{center}
 \vspace{-20pt}
 \end{table}
In order to compute the $h^{1,1}$ for the quotients of \eqref{eq:M8Split2ver1} though, we need to know the number of cohomology invariants of the two $\text{dP}_4$'s. We have listed the invariants for the second $\text{dP}_4$ in the representation $\IP^4[2,2]$ in \tref{coh_invs_2566}. Using the methods in the previous subsection, the number of cohomology invariants of the first $\text{dP}_4$ was computed to be 4 for the $\IZ_2$ quotients and 3 for the $\IZ_2\times\IZ_2$ quotients. The corresponding group actions on the coordinates were taken from \cite{Braun:2010vc}.
Finally, we produce the value of $h^{1,1}$ for all smooth quotients of \eqref{eq:M8Split2ver1} in \tref{dP4_2566}. This required in summary, a knowledge of the group action on the cohomologies of two distinct representations of del Pezzo surfaces of degree 4.   
\newpage
\section{The tetraquadric and its splits}\label{sec:tetraquadricweb}
\vskip-10pt
\subsection{The tetraquadric $X^{4,68}$ and its smooth quotients}\label{sec:TQ}
We start with the class of manifolds defined by the following configuration matrix:
\begin{equation}\label{eq:TQ}
\displaycicy{5.25in}{
X_{7862}~=~~
\cicy{\IP^1 \\ \IP^1\\ \IP^1\\ \IP^1}
{ ~2 \!\!\!\!\\
  ~2\!\!\!\! & \\
  ~2\!\!\!\! & \\
  ~2\!\!\!\!}_{-128}^{4,68}}{-1.1cm}{0.95in}{TQ.pdf}
\end{equation}
This manifold has recently been studied in Refs.~\cite{Buchbinder:2013dna, Buchbinder:2014qda, Buchbinder:2014sya, Constantin:2015bea} leading to heterotic models which come very close to the desired properties of the Standard Model. The tetraquadric manifold admits $11$ different smooth quotients \cite{Braun:2010vc}. The finite groups in question, as well as the Hodge numbers for the corresponding quotients are listed in Table~\ref{TQquotients}. The polynomial deformation method correctly reproduces the number of harmonic $(1,2)$--forms on this manifold. Indeed, the defining polynomial 
contains $81$ coefficients, of which $13$ are redundant, corresponding to $12$ coordinate redefinitions and an overall rescaling of the polynomial. Moreover, since the embedding \eqref{eq:TQ} is favourable, we are able to count the number of linearly independent K\"ahler forms for each  quotient. The two methods agree with the independent computation of the Euler number.
\begin{table}[H]
\vspace{12pt}
\begin{center}
\begin{tabular}{l}
\begin{tabular}{| c || c | c | c | c | c | c | }
\hline
\myalign{| c||}{\varstr{16pt}{10pt}$~~~~~~~  \Gamma ~~~~~~~$ } &
\myalign{m{1.31cm}|}{$~~~ \IZ_2 $}&
\myalign{m{1.31cm}|}{$~~~\IZ_4 $}& 
\myalign{m{1.9cm}|}{ $~~~\IZ_2\times\IZ_2 \ \ \ $ }&
\myalign{m{1.31cm}|}{$~~~\IZ_8 $}&
\myalign{m{1.9cm}|}{ $~~~\IZ_2\times\IZ_4 \ \ \ $ }&
\myalign{m{1.31cm}|}{$~~~\IQ_8 $} 
\\ \hline\hline
\varstr{14pt}{8pt} $h^{1,1}(X/\Gamma)$ & 4 & 2 & 4 & 1 & 2 &1\\
 \hline
\varstr{14pt}{8pt} $h^{2,1}(X/\Gamma)$ & 36 & 18 & 20 & 9 & 10 & 9 \\
 \hline
\varstr{14pt}{8pt} $\chi(X/\Gamma)$ & $\!\!\!\!-64$ & $\!\!\!\!-32$ & $\!\!\!\!-32$ & $\!\!\!\!-16$ & $\!\!\!\!-16$ & $\!\!\!\!-16$ \\
 \hline
 \end{tabular}
\\[2cm]
\begin{tabular}{| c || c | c | c | c | c |}
\hline
\myalign{| c||}{\varstr{16pt}{10pt}$~~~~~~~  \Gamma ~~~~~~~$ } &
\myalign{m{1.9cm}|}{ $~~~\IZ_4\times\IZ_4 \ \ \ $ }&
\myalign{m{1.9cm}|}{ $~~~\IZ_4\rtimes\IZ_4 \ \ \ $ } &
\myalign{m{1.9cm}|}{ $~~~\IZ_8\times\IZ_2 \ \ \ $ }&
\myalign{m{1.9cm}|}{ $~~~\IZ_8\rtimes\IZ_2 \ \ \ $ }&
\myalign{m{1.9cm}|}{ $~~~\IZ_2\times \IQ_8 \ \ \ $ }
\\ \hline\hline
\varstr{14pt}{8pt} $h^{1,1}(X/\Gamma)$ & 1 & 1 & 1 & 1 & 1\\
 \hline
\varstr{14pt}{8pt} $h^{2,1}(X/\Gamma)$ & 5 & 5 & 5 & 5 & 5\\
 \hline
\varstr{14pt}{8pt} $\chi(X/\Gamma)$ & $\!\!\!\!-8$ & $\!\!\!\!-8$ & $\!\!\!\!-8$ & $\!\!\!\!-8$ & $\!\!\!\!-8$ \\
 \hline
 \end{tabular}
 \end{tabular}
 \vskip 0.3cm
\capt{5in}{TQquotients}{Hodge numbers for the quotients of the tetraquadric.}
 \end{center}
 \vspace{-12pt}
 \end{table}
\goodbreak
\subsection{Splits of the tetraquadric with a $\IP^1$}
\vskip-10pt
\subsubsection{Two favourable splits with Hodge numbers $(5,45)$}\label{TQ_X5,45}\vskip-8pt
The first split corresponds to the following configuration:
\vspace{4pt}
\begin{equation}\label{eq:TQSplit1}
\displaycicy{5.25in}{
X_{7447}~=~~
\cicy{\IP^1 \\ \IP^1\\ \IP^1\\ \IP^1\\ \IP^1}
{ ~1 &1 ~\\
  ~1 &1 ~\\
  ~1 &1 ~\\
  ~1 &1 ~\\
  ~1 &1 ~\\}_{-80}^{5,45}}{-1.5cm}{0.9in}{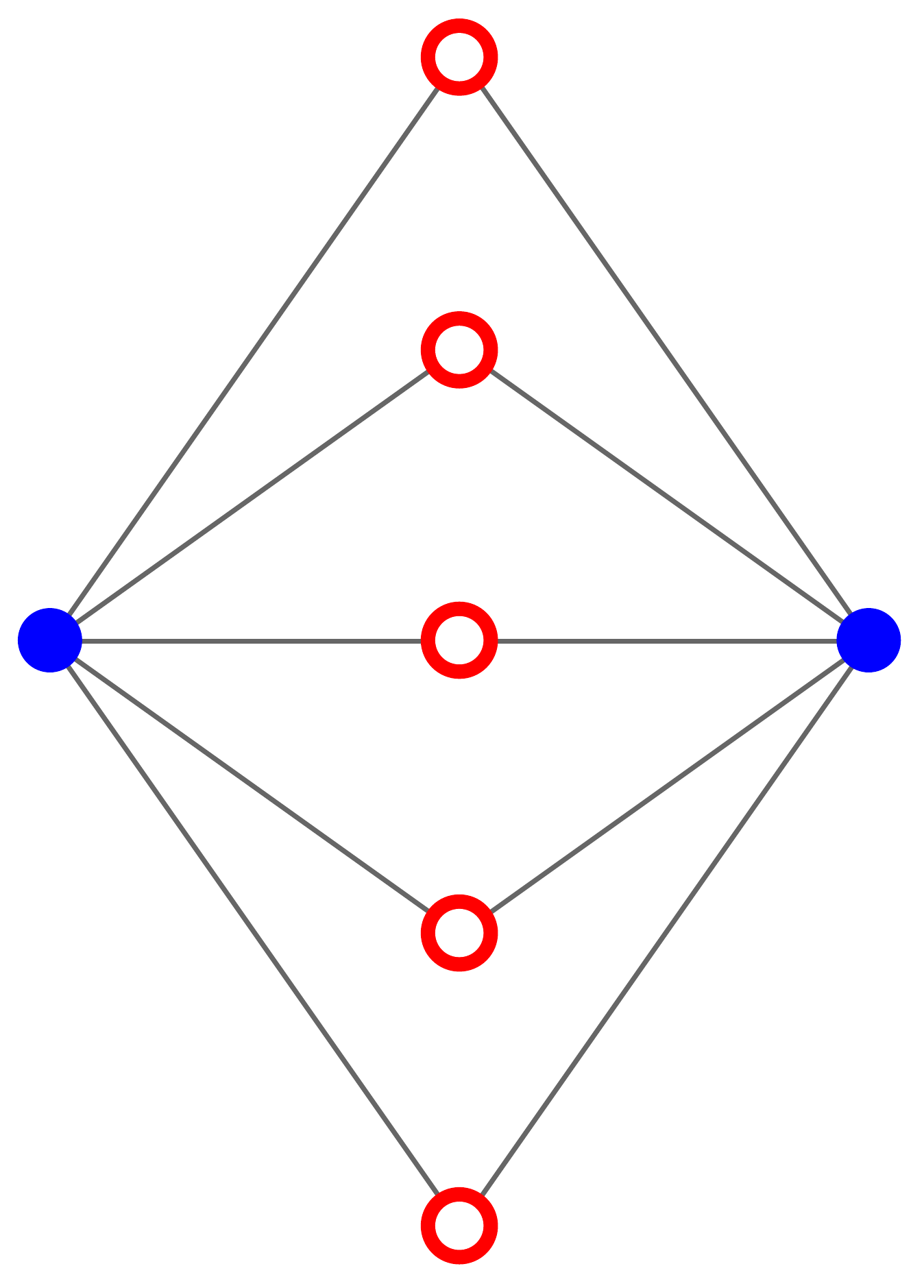}
  \end{equation}
This manifold admits $5$ different smooth quotients, whose Hodge numbers are presented in 
Table~\ref{TQSplit1quotients}. The polynomial deformation method correctly reproduces the number of complex structure parameters. Indeed, the defining polynomials 
contain $65$ coefficients, of which $15$ can be removed by coordinate redefinitions and $4$ by redefinitions of the defining polynomials. Furthermore, the configuration \eqref{eq:TQSplit1} is favourable, so are able to count the number of linearly independent K\"ahler forms for each  quotient. The two methods agree with the independent computation of the Euler number.
\begin{table}[!ht]
\vspace{12pt}
\begin{center}
\begin{tabular}{| c || c | c | c | c | c |}
\hline
\myalign{| c||}{\varstr{16pt}{10pt}$~~~~~~~  \Gamma ~~~~~~~$ } &
\myalign{m{1.31cm}|}{$~~~ \IZ_2 $} &
\myalign{m{1.9cm}|}{ $~~~\IZ_2\times\IZ_2 \ \ \ $ } &
\myalign{m{1.31cm}|}{$~~~ \IZ_5 $} &
\myalign{m{1.31cm}|}{$~~~ \IZ_{10} $} &
\myalign{m{1.9cm}|}{ $~~\IZ_2\times \IZ_{10} \ \ $ }
\\ \hline\hline
\varstr{14pt}{8pt} $h^{1,1}(X/\Gamma)$ & 5 & 5 & 1 & 1 & 1\\
 \hline
\varstr{14pt}{8pt} $h^{2,1}(X/\Gamma)$ & 25 & 15 & 9 & 5 & 3\\
 \hline
\varstr{14pt}{8pt} $\chi(X/\Gamma)$ & $\!\!\!\!-40$ & $\!\!\!\!-20$ & $\!\!\!\!-16$ & $\!\!\!\!-8$ & $\!\!\!\!-4$ \\
 \hline
 \end{tabular}
 \vskip 0.3cm
\capt{4in}{TQSplit1quotients}{Hodge numbers for the quotients of $X^{5,45}$.}
 \end{center}
 \vspace{-12pt}
 \end{table}
 
The second split corresponds to the following configuration:
\begin{equation}\label{eq:TQSplit2}
\displaycicy{5.25in}{
X_{7487}~=~~
\cicy{\IP^1 \\ \IP^1\\ \IP^1\\ \IP^1\\ \IP^1}
{ ~2 &0 ~\\
  ~1 &1 ~\\
  ~1 &1 ~\\
  ~1 &1 ~\\
  ~1 &1 ~\\}_{-80}^{5,45}}{-1.3cm}{1.6in}{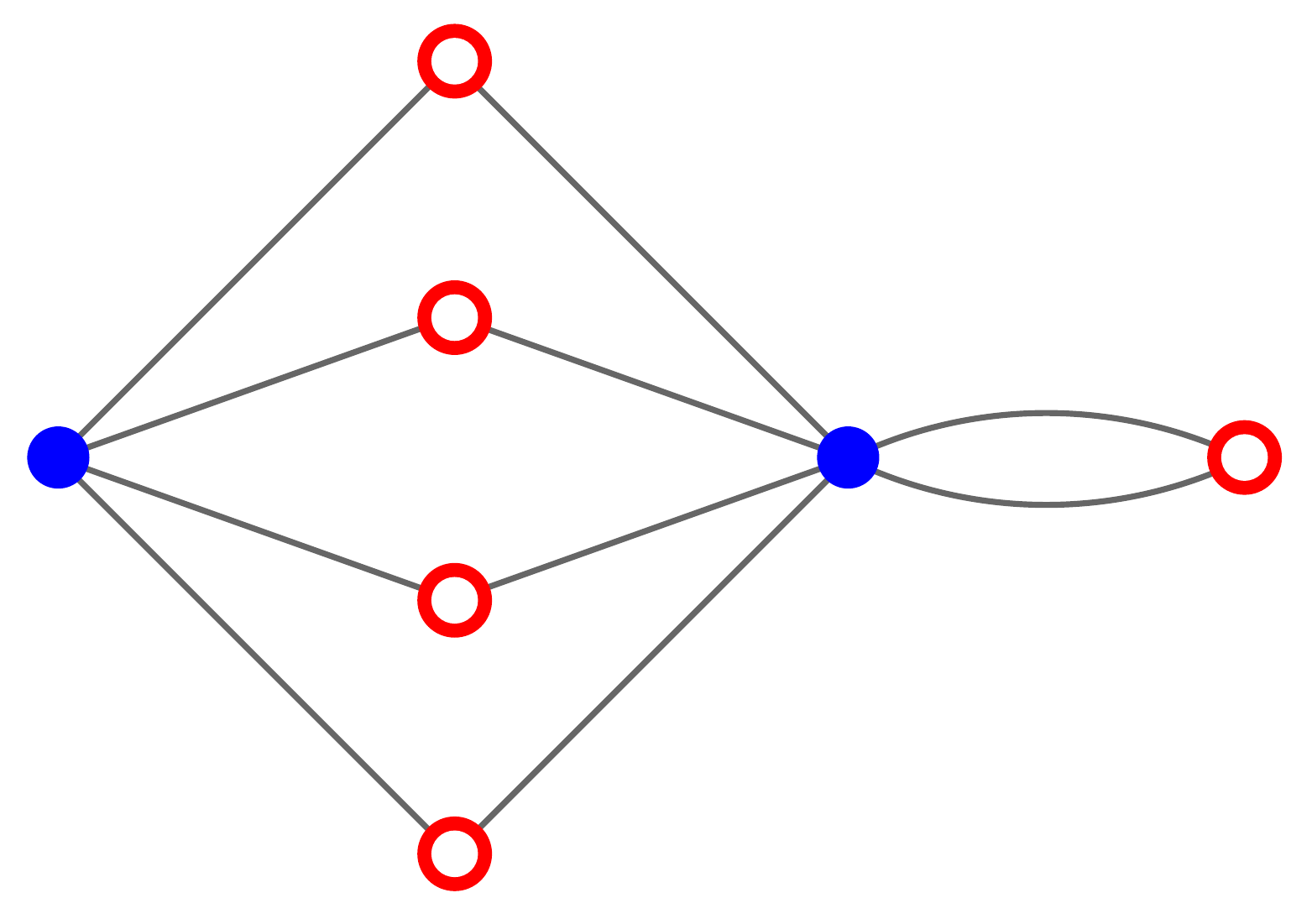}
\end{equation}
It is interesting that this manifold was used in Ref.~\cite{Buchbinder:2014qca} in the construction of a heterotic model with a QCD axion. It admits $6$ different free group actions, which result in two pairs of Hodge numbers, as listed in \tref{TQSplit2quotients}.
\begin{table}[!ht]
\vspace{12pt}
\begin{center}
\begin{tabular}{| c || c | c |}
\hline
\myalign{| c||}{\varstr{16pt}{10pt}$~~~~~~~  \Gamma ~~~~~~~$ } &
\myalign{m{1.31cm}|}{$~~~ \IZ_2 $} &
\myalign{m{1.9cm}|}{ $~~~\IZ_2\times\IZ_2 \ \ \ $ }  
\\ \hline\hline
\varstr{14pt}{8pt} $h^{1,1}(X/\Gamma)$ & 5 & 5 \\
 \hline
\varstr{14pt}{8pt} $h^{2,1}(X/\Gamma)$ & 25 & 15 \\
 \hline
\varstr{14pt}{8pt} $\chi(X/\Gamma)$ & $\!\!\!\!-40$ & $\!\!\!\!-20$ \\
 \hline
 \end{tabular}
 \vskip 0.3cm
\capt{5.8in}{TQSplit2quotients}{Hodge numbers for the quotients of the second split of the tetraquadric, $X^{5,45}$.}
 \end{center}
 \end{table}
\vspace*{-40pt}
\subsubsection{Two non-favourable splits of the tetraquadric: $X^{8,40}$ and $X^{19,19}$ }\label{TQ_2_2}
The manifold $X^{8,40}$ corresponds to the following configuration:
 \begin{equation}\label{eq:TQSplit3}
\displaycicy{5.25in}{
X_{6829}~=~~
\cicy{\IP^1 \\ \IP^1\\ \IP^1\\ \IP^1\\ \IP^1}
{ ~2 &0 ~\\
  ~0 &2 ~\\
  ~2 &0 ~\\
  ~1 &1 ~\\
  ~1 &1 ~\\}_{-64}^{8,40}}{-0.9cm}{1.6in}{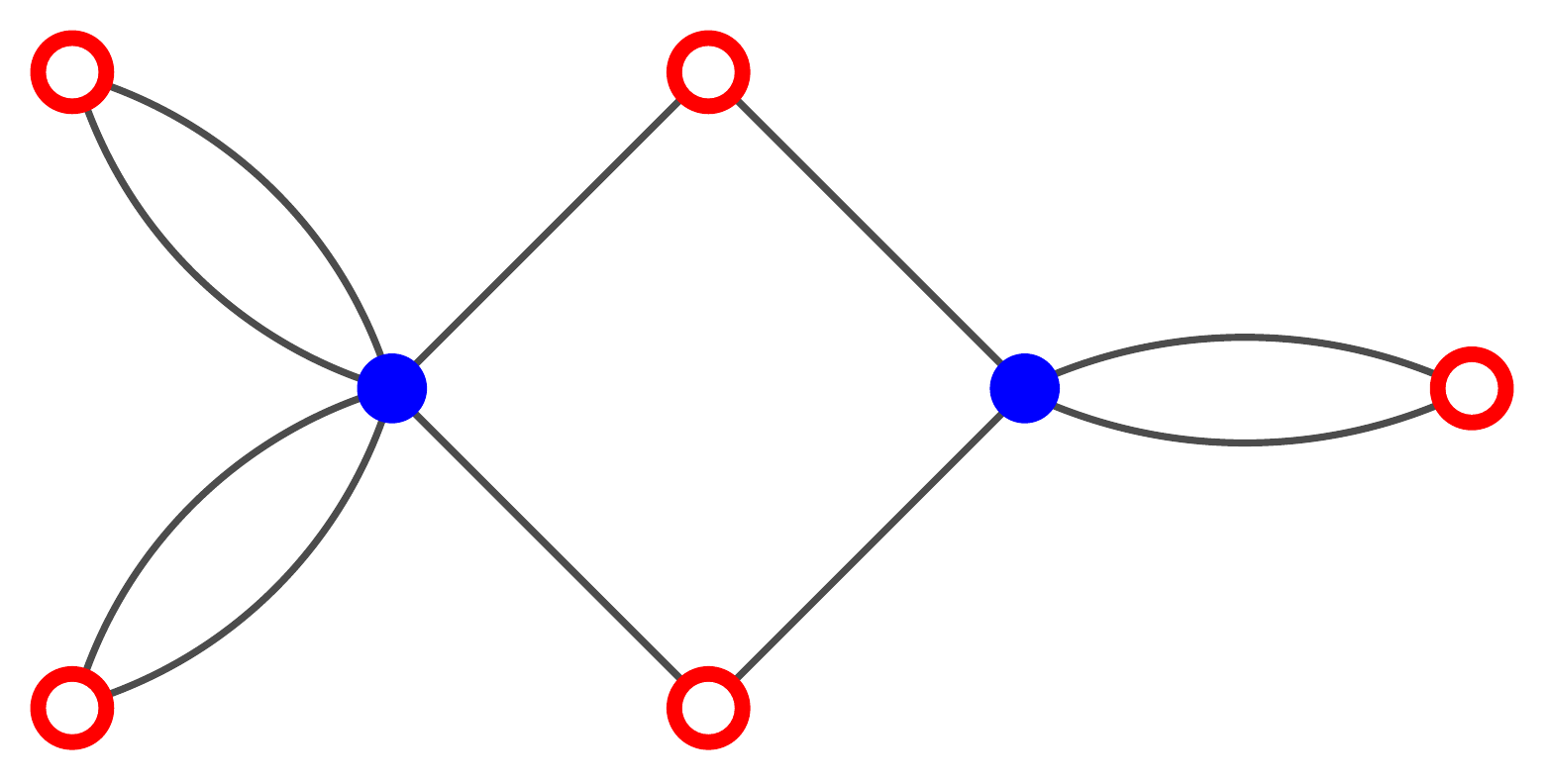}
\end{equation}
The embedding \eqref{eq:TQSplit3} is not favourable. Moreover, the application of the polynomial deformation method does not reproduce the number of complex structure parameters. For these reasons, in order to compute the Hodge numbers for the quotients of this manifold, we make use of the fact that $X_{6829}$ can be regarded as a hypersurface in the product $\text{dP}_4\times\IP^1\times\IP^1$, where the $\text{dP}_4$ surface is defined by the second polynomial. This embedding explains the number of K\"ahler parameters for this manifold since $h^{1,1}(\text{dP}_4)=6$. We begin our computation by listing the distinct group actions on the coordinates in \tref{dP4_6829_Symm_Action}.  The group action on the cohomology basis of the $\text{dP}_4$ and the corresponding invariants are listed in \tref{dP4_6829_Symm_Action_Coh}. Finally, the $h^{1,1}$ for each quotient equals the number of cohomology invariants plus two, since the symmetries listed in \tref{dP4_6829_Symm_Action} do not mix the first and the third $\IP^1$ spaces of \eqref{eq:TQSplit3}.
\begin{table}[!ht]
\vspace{12pt}
\begin{center}
\begin{tabular}{| c || c | c |}
\hline
\myalign{| c||}{\varstr{16pt}{10pt}$~~~~~~~  \Gamma ~~~~~~~$ } &
\myalign{m{1.31cm}|}{$\hfil \IZ_2 $} &
\myalign{m{1.9cm}|}{ $\hfil \IZ_2\times\IZ_2$ }
\\ \hline\hline
\varstr{14pt}{8pt} $h^{1,1}(X/\Gamma)$ & 6 & 5\\
 \hline
\varstr{14pt}{8pt} $h^{2,1}(X/\Gamma)$ & 22 & 13 \\
 \hline
\varstr{14pt}{8pt} $\chi(X/\Gamma)$ & $\!\!\!\!-32$ & $\!\!\!\!-16$\\
 \hline
 \end{tabular}
 \vskip 0.3cm
\capt{1\textwidth}{dP4_6829}{Hodge numbers for the quotients of the manifolds ~\eqref{eq:TQSplit3}.}
 \end{center}
 \vspace{-12pt}
 \end{table}
\begin{center}
\begin{longtable}{|c|c|c|c|c|}
\captionsetup{width=0.9\textwidth}
\caption{\it Various symmetry actions on the ambient space of the manifold~\eqref{eq:TQSplit3}. The coordinate patch of the $\text{dP}_4$ is chosen to be $(1,x){\times}(1,y){\times}(1,z)$. $(p,q)$ and $(r,s)$ are taken to be coordinates of the first and third $\IP^1$ spaces.} \label{dP4_6829_Symm_Action}\\
\hline \multicolumn{1}{|c|}{\str\textbf{Index}} &  \multicolumn{1}{|c|}{\str\textbf{~Group~}} & \multicolumn{1}{|c|}{$\mathbf{(\+x,\+y,\+z)}$} &  \multicolumn{1}{|c|}{$\mathbf{(\+p,\+q)}$} &  \multicolumn{1}{|c|}{$\mathbf{(\+r,\+s)}$} \\ \hline 
\endfirsthead

\hline 
\textbf{Index} &
\textbf{~Group~} &
$\mathbf{(\+x,\+y,\+z)}$ &
$\mathbf{(\+p,\+q)}$ &
$\mathbf{(\+r,\+s)}$ \\ \hline 
\endhead

\hline\hline \multicolumn{5}{|r|}{{\str Continued on next page}} \\ \hline
\endfoot

\endlastfoot

\hline\hline
\varstr{14pt}{8pt} 1 & $\IZ_2$ & $(-x,-y,-z)$ & $(-p,\+q)$ & $(-r,\+s)$ \\
 \hline
\varstr{14pt}{8pt} 2 & $\IZ_2$ & $(\+y,\+x,-z)$ &  $(-p,\+q)$ & $(-r,\+s)$ \\
 \hline
\varstr{16pt}{10pt} 3 & $\IZ_2{\times}\IZ_2$ &
  \begin{minipage}[c][40pt][c]{1.3in}
  \begin{gather*} 
 (-x,-y,-z) \\ 
 ({x^{-1}},{y^{-1}},{z^{-1}}) \\
  \end{gather*}
  \end{minipage} 
& 
 \begin{minipage}[c][40pt][c]{0.8in}
  \begin{gather*}
(\+p,-q) \\ 
(\+q,\+p)\\
  \end{gather*}
  \end{minipage} 
  & 
 \begin{minipage}[c][40pt][c]{0.8in}
  \begin{gather*}
(\+r,-s) \\ 
(\+s,\+r)\\
  \end{gather*}
  \end{minipage} 
  \\
 \hline
 4 & $\IZ_2{\times}\IZ_2$ &
  \begin{minipage}[c][40pt][c]{1.3in}
  \begin{gather*} 
 (-x,-y,-z) \\ 
 ({\+y},{\+x},{z^{-1}}) \\
  \end{gather*}
  \end{minipage} 
& 
 \begin{minipage}[c][40pt][c]{0.8in}
  \begin{gather*}
(\+p,-q) \\ 
(\+q,\+p)\\
  \end{gather*}
  \end{minipage} 
  & 
 \begin{minipage}[c][40pt][c]{0.8in}
  \begin{gather*}
(\+r,-s) \\ 
(\+s,\+r)\\
  \end{gather*}
  \end{minipage} 
\\ \hline
  5 & $\IZ_2{\times}\IZ_2$ &
  \begin{minipage}[c][40pt][c]{1.3in}
  \begin{gather*} 
 (\+y,\+x,-z) \\ 
 ({-x},{-y},{z^{-1}}) \\
  \end{gather*}
  \end{minipage} 
& 
 \begin{minipage}[c][40pt][c]{0.8in}
  \begin{gather*}
(\+p,-q) \\ 
(\+q,\+p)\\
  \end{gather*}
  \end{minipage} 
& 
 \begin{minipage}[c][40pt][c]{0.8in}
  \begin{gather*}
(\+r,-s) \\ 
(\+s,\+r)\\
  \end{gather*}
  \end{minipage} 
 \\ \hline
6 & $\IZ_2{\times}\IZ_2$ &
  \begin{minipage}[c][40pt][c]{1.3in}
  \begin{gather*} 
 (\+y,\+x,-z) \\ 
 ({-y},{-x},{z^{-1}}) \\
  \end{gather*}
  \end{minipage} 
& 
 \begin{minipage}[c][40pt][c]{0.8in}
  \begin{gather*}
 (\+p,-q) \\ 
(\+q,\+p)\\
 \end{gather*}
  \end{minipage} 
  & 
 \begin{minipage}[c][40pt][c]{0.8in}
  \begin{gather*}
 (\+r,-s) \\ 
(\+s,\+r)\\
 \end{gather*}
  \end{minipage}
  \\ \hline
\end{longtable}
\end{center}
\small
\begin{center}
\begin{longtable}{|c|c|c|c|}
\captionsetup{width=0.9\textwidth}
\caption{\it Symmetry actions on the cohomology basis and the corresponding invariants for the manifold~\eqref{eq:TQSplit3}. The matrices $P_i$ and $Q_i$ are defined in \eqref{eq:CohMatDefs2}.} \label{dP4_6829_Symm_Action_Coh} \\

\hline \multicolumn{1}{|c|}{\str\textbf{Index}}&  \multicolumn{1}{|c|}{\str\textbf{~Group~}} & \multicolumn{1}{|c|}{\str\textbf{\begin{minipage}[c][35pt][c]{0.85in}
Action on\\ 
Coh Basis
  \end{minipage}}} &  \multicolumn{1}{|c|}{\str\textbf{~Coh Invariants~}} \\ \hline 
\endfirsthead

\hline 
\textbf{~Index~} & \textbf{~Group~} & \textbf{\begin{minipage}[c][35pt][c]{0.85in}
Action on\\ 
Coh Basis
 \end{minipage}} & \textbf{~Coh Invariants~} \\ \hline 
\endhead

\hline\hline \multicolumn{4}{|r|}{{\str Continued on next page}} \\ \hline
\endfoot

\endlastfoot

\hline\hline
 1 & $\IZ_2$ & $P_1$ & \begin{minipage}[c][22pt][c]{2.1in}
\begin{gather*} 
H, E_1+E_2,E_3+E_4,E_5\\
\end{gather*}
\end{minipage} \\
 \hline
 
2 & $\IZ_2$ & $Q_1$
& \begin{minipage}[c][22pt][c]{3.2in}
\begin{gather*} 
H-E_5, E_1+E_4-E_5, E_2+E_4-E_5,E_3-E_4\\
\end{gather*}
\end{minipage} \\
 \hline
 
3 & $\IZ_2{\times}\IZ_2$ &
$P_2,~P_1$
&\begin{minipage}[c][22pt][c]{2.1in}
\begin{gather*} 
H, E_1+E_2+E_3+E_4,E_5\\
\end{gather*}
\end{minipage} \\ \hline

4 & $\IZ_2{\times}\IZ_2$ &
$P_1,~Q_2$
& \begin{minipage}[c][22pt][c]{3.2in}
\begin{gather*} 
H-{E}_5, E_1+E_2-E_5, E_3+E_4-E_5\\
\end{gather*}
\end{minipage} \\
 \hline
 
5 & $\IZ_2{\times}\IZ_2$ &
$Q_1,~P_3$
& \begin{minipage}[c][22pt][c]{3.2in}
\begin{gather*} 
H-{E}_5, E_1+E_3-E_5,E_2+E_4-E_5\\
\end{gather*}
\end{minipage} \\
 \hline
 
6 & $\IZ_2{\times}\IZ_2$ &
$Q_1,~Q_2$
& \begin{minipage}[c][22pt][c]{3.2in}
\begin{gather*} 
H-{E}_5, E_1+E_3-E_5, 
E_2+E_4-E_5\\
\end{gather*}
\end{minipage} \\
 \hline 
\end{longtable}
\end{center}
\normalsize
\newpage
There are 15 occurrences of the Hodge numbers $(19,19)$ in the CICY list. These can all be seen to correspond to the same manifold by a sequence of redundant splittings and contractions. One of these configurations is the following 
\begin{equation}\label{eq:TQSplit4}
\displaycicy{5.25in}{
X_{21}~=~~
\cicy{\IP^1 \\ \IP^1\\ \IP^1\\ \IP^1\\ \IP^1}
{ ~2 &0 ~\\
  ~0 &2 ~\\
  ~2 &0 ~\\
  ~0 &2 ~\\
  ~1 &1 ~\\}_{0}^{19,19}}{-1.0cm}{1.6in}{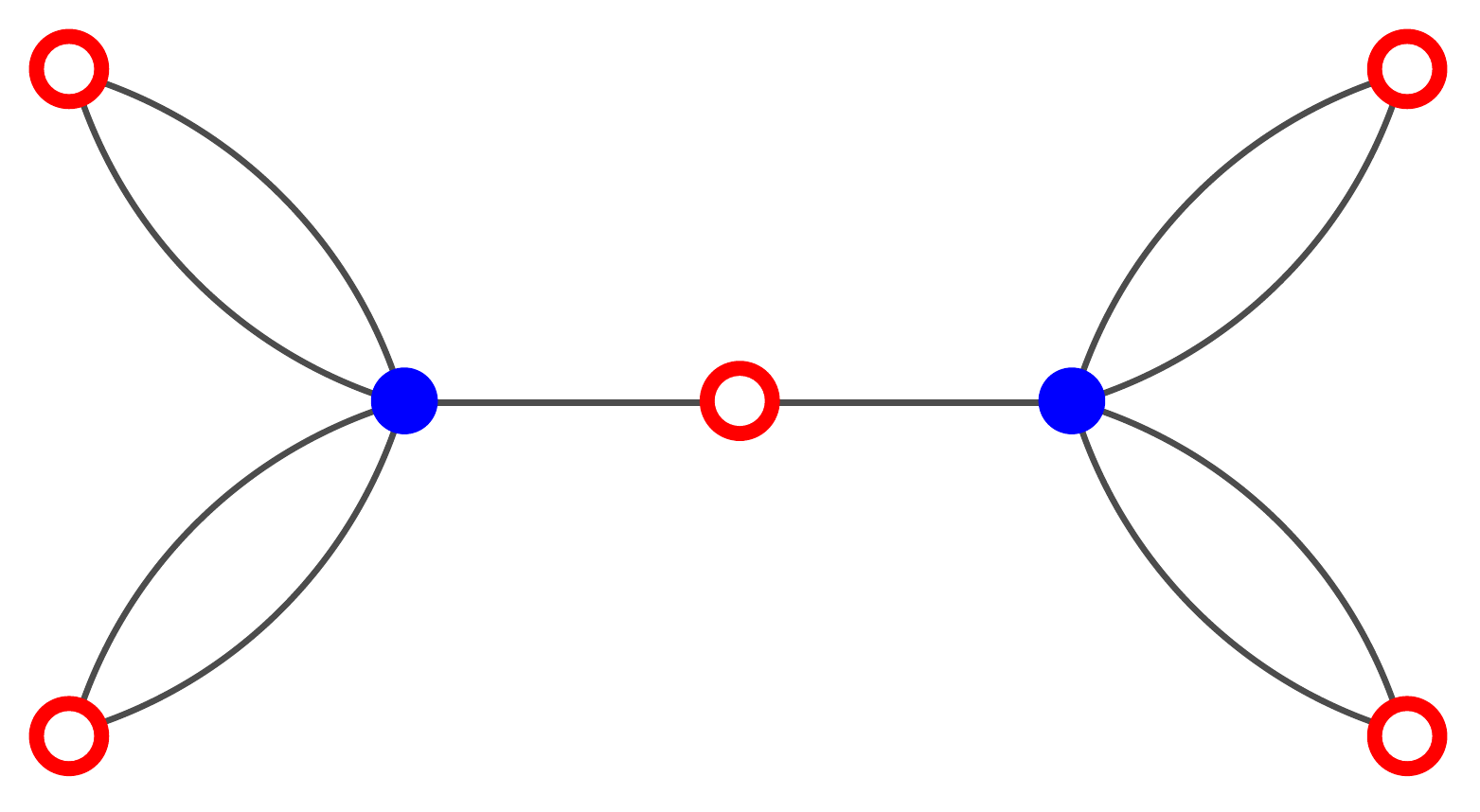}
\end{equation}
This configuration admits $53$ different free group actions by the groups
$$\IZ_2,~\IZ_4,~ \IZ_2{\times}\IZ_2,~\IZ_8,~\IZ_2{\times}\IZ_4,~ \IQ_8,~\IZ_4{\times}\IZ_4,~
\IZ_4{\rtimes}\IZ_4,~\IZ_8{\times}\IZ_2,~\IZ_8{\rtimes}\IZ_2~~\text{and}~~\IZ_2{\times} \IQ_8~.$$ 
Many of these quotients, as well as their Hodge numbers were studied in Ref.~\cite{Bouchard:2007mf}.

Although the diagram \eqref{eq:TQSplit4} is one-leg decomposable, the polynomial deformation method correctly reproduces the number of complex structure parameters for the covering manifold, and we will assume that it provides a complete parametrisation of the complex structure moduli space. With this assumption, we are able to compute the Hodge numbers for the resulting quotients, as listed in \tref{TQSplit4quotients}.

\begin{table}[H]
\vspace{12pt}
\begin{center}
\begin{tabular}{l}
\begin{tabular}{| c || c | c | c | c | c | }
\hline
\myalign{| c||}{\varstr{16pt}{10pt}$~~~~~~~  \Gamma ~~~~~~~$ } &
\myalign{m{1.31cm}|}{$~~~ \IZ_2 $} &
\myalign{m{1.31cm}|}{$~~~ \IZ_4 $} &
\myalign{m{1.31cm}|}{$~~~ \IZ_4 $} &
\myalign{m{1.9cm}|}{ $~~~\IZ_2\times\IZ_2 \ \ \ $ }  &
\myalign{m{3.05cm}|}{$~~~ \IZ_8, ~\IZ_2{\times}\IZ_4, ~\IQ_8 $} 
\\ \hline\hline
\varstr{14pt}{8pt} $h^{1,1}(X/\Gamma)$ & 11 & 5 & 6 & 7 & 3 \\
 \hline
\varstr{14pt}{8pt} $h^{2,1}(X/\Gamma)$ & 11 & 5 & 6 & 7 & 3  \\
 \hline
\varstr{14pt}{8pt} $\chi(X/\Gamma)$ & $ 0 $ & $ 0 $& $ 0 $& $ 0 $ & $ 0 $  \\
 \hline
 \end{tabular}
 \\[2cm]
\begin{tabular}{| c || c | c | c |}
\hline
\myalign{| c||}{\varstr{16pt}{10pt}$~~~~~~~  \Gamma ~~~~~~~$ } &
\myalign{m{1.9cm}|}{ $~~~\IZ_2\times\IZ_4 \ \ \ $ }  &
\begin{minipage}[c][50pt][c]{2.3in}
\begin{gather*} 
\IZ_4{\times}\IZ_4,~\IZ_4{\rtimes}\IZ_4\\
~\IZ_8{\times}\IZ_2, ~\IZ_8{\rtimes}\IZ_2,~\IZ_2{\times} \IQ_8\\
\end{gather*}
\end{minipage}
\\ \hline\hline
\varstr{14pt}{8pt} $h^{1,1}(X/\Gamma)$ & 4 & 2 \\
 \hline
\varstr{14pt}{8pt} $h^{2,1}(X/\Gamma)$ & 4 & 2 \\
 \hline
\varstr{14pt}{8pt} $\chi(X/\Gamma)$ & $ 0 $ & $ 0 $ \\
 \hline
 \end{tabular}
 \end{tabular}
\vskip 0.3cm
\capt{4.6in}{TQSplit4quotients}{Hodge numbers for the quotients of the manifold \eqref{eq:TQSplit4}.}
 \end{center}
 \vspace{-12pt}
 \end{table}

\subsection{Other splits of the tetraquadric}\label{sec:TQOtherSplits}
\vskip-10pt
\subsubsection{Three favourable splits with Hodge numbers $(5,29)$}\label{TQOtherSplitsX5,29}
The first split corresponds to the following configuration:
\begin{equation}\label{eq:TQSplit10}
\displaycicy{5.25in}{
X_{5301}~=~~
\cicy{\IP^1 \\ \IP^1\\ \IP^1\\ \IP^1\\ \IP^3}
{ ~1 &1 & 0 & 0 ~\\
  ~1 &1 & 0 & 0 ~\\
  ~0 &0 & 1 & 1~\\
  ~0 &0 & 1 & 1 ~\\
  ~1 &1 & 1 & 1~\\}_{-48}^{5,29}}{-1.0cm}{1.6in}{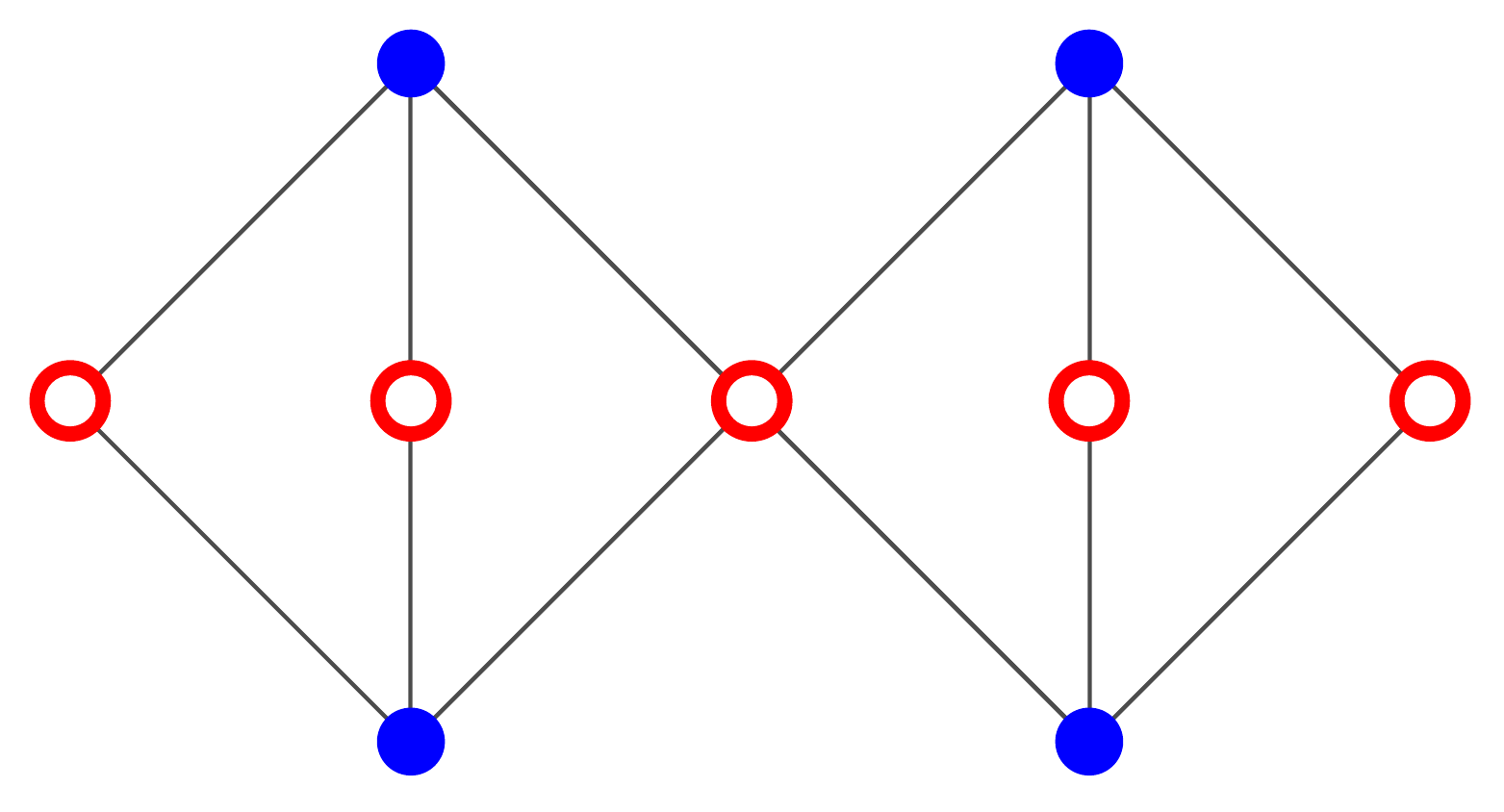}
\end{equation}

The manifold \eqref{eq:TQSplit10} admits $3$ different free actions. For the Hodge number computation for the resulting quotients we are able to use both the polynomial deformation method and the counting of the invariant K\"ahler forms. The results are listed in \tref{TQSplit10quotients}.
\begin{table}[H]
\begin{center}
\begin{tabular}{| c || c | c | c |}
\hline
\myalign{| c||}{\varstr{16pt}{10pt}$~~~~~~~  \Gamma ~~~~~~~$ } &
\myalign{m{1.31cm}|}{$~~~ \IZ_2 $} &
\myalign{m{1.31cm}|}{$~~~ \IZ_4 $} &
\myalign{m{1.9cm}|}{ $~~~\IZ_2\times\IZ_2 \ \ \ $ }  
\\ \hline\hline
\varstr{14pt}{8pt} $h^{1,1}(X/\Gamma)$ & 5 & 3 & 5 \\
 \hline
\varstr{14pt}{8pt} $h^{2,1}(X/\Gamma)$ & 17 & 9 & 11 \\
 \hline
\varstr{14pt}{8pt} $\chi(X/\Gamma)$ & $\!\!\!\!-24$ & $\!\!\!\!-12$ & $\!\!\!\!-12$  \\
 \hline
 \end{tabular}
   \vskip0.1cm
\capt{5in}{TQSplit10quotients}{Hodge numbers for the quotients of the manifolds \eqref{eq:TQSplit10} and \eqref{eq:TQSplit12}.}
 \end{center}
 \vspace{-12pt}
 \end{table}
 
The second split corresponds to the following configuration:
\begin{equation}\label{eq:TQSplit11}
\displaycicy{5.25in}{
X_{5256}~=~~
\cicy{\IP^1 \\ \IP^1\\ \IP^1\\ \IP^1\\ \IP^3}
{ ~2 &0 & 0 & 0 ~\\
  ~1 &1 & 0 & 0 ~\\
  ~0 &0 & 1 & 1~\\
  ~0 &0 & 1 & 1 ~\\
  ~1 &1 & 1 & 1~\\}_{-48}^{5,29}}{-1.18cm}{1.9in}{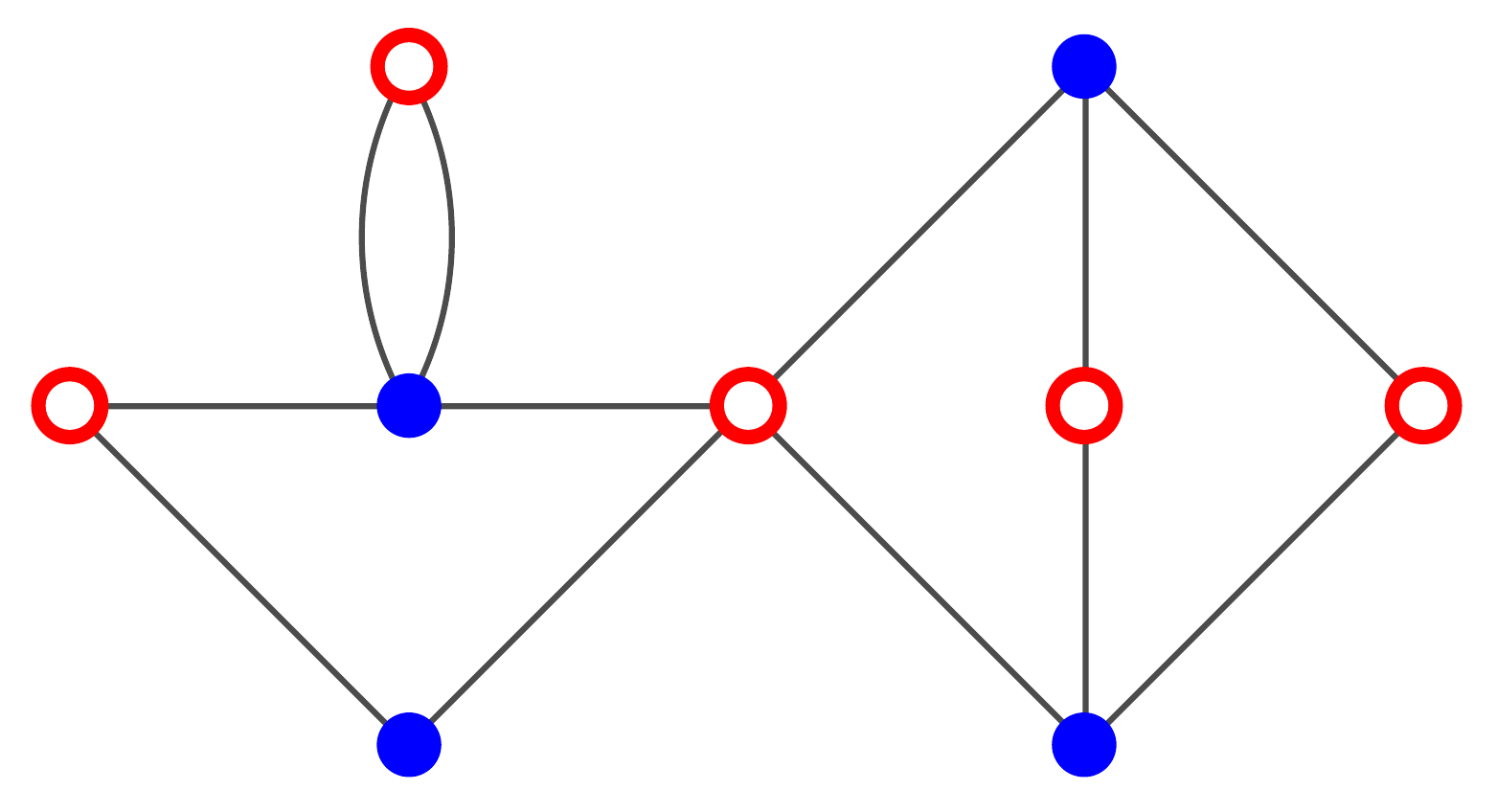}
\end{equation}
This manifold admits $6$ different free actions, resulting in two pairs of Hodge numbers for the quotients. The results are listed in \tref{TQSplit11quotients}.
\begin{table}[H]
\begin{center}
\begin{tabular}{| c || c | c |}
\hline
\myalign{| c||}{\varstr{16pt}{10pt}$~~~~~~~  \Gamma ~~~~~~~$ } &
\myalign{m{1.31cm}|}{$~~~ \IZ_2 $} &
\myalign{m{1.9cm}|}{ $~~~\IZ_2\times\IZ_2 \ \ \ $ }  
\\ \hline\hline
\varstr{14pt}{8pt} $h^{1,1}(X/\Gamma)$ & 5 & 5 \\
 \hline
\varstr{14pt}{8pt} $h^{2,1}(X/\Gamma)$ & 17 & 11 \\
 \hline
\varstr{14pt}{8pt} $\chi(X/\Gamma)$ & $\!\!\!\!-24$ & $\!\!\!\!-12$  \\
 \hline
 \end{tabular}
   \vskip 0.1cm
\capt{4.5in}{TQSplit11quotients}{Hodge numbers for the quotients of the manifold \eqref{eq:TQSplit11}.}
 \end{center}
 \end{table}

The third split corresponds to the following configuration:
\begin{equation}\label{eq:TQSplit12}
\displaycicy{5.25in}{
X_{5452}~=~~
\cicy{\IP^1 \\ \IP^1\\ \IP^1\\ \IP^1\\ \IP^3}
{ ~2 &0 & 0 & 0 ~\\
  ~1 &1 & 0 & 0 ~\\
  ~0 &0 & 2 & 0~\\
  ~0 &0 & 1 & 1 ~\\
  ~1 &1 & 1 & 1~\\}_{-48}^{5,29}}{-0.7cm}{1.6in}{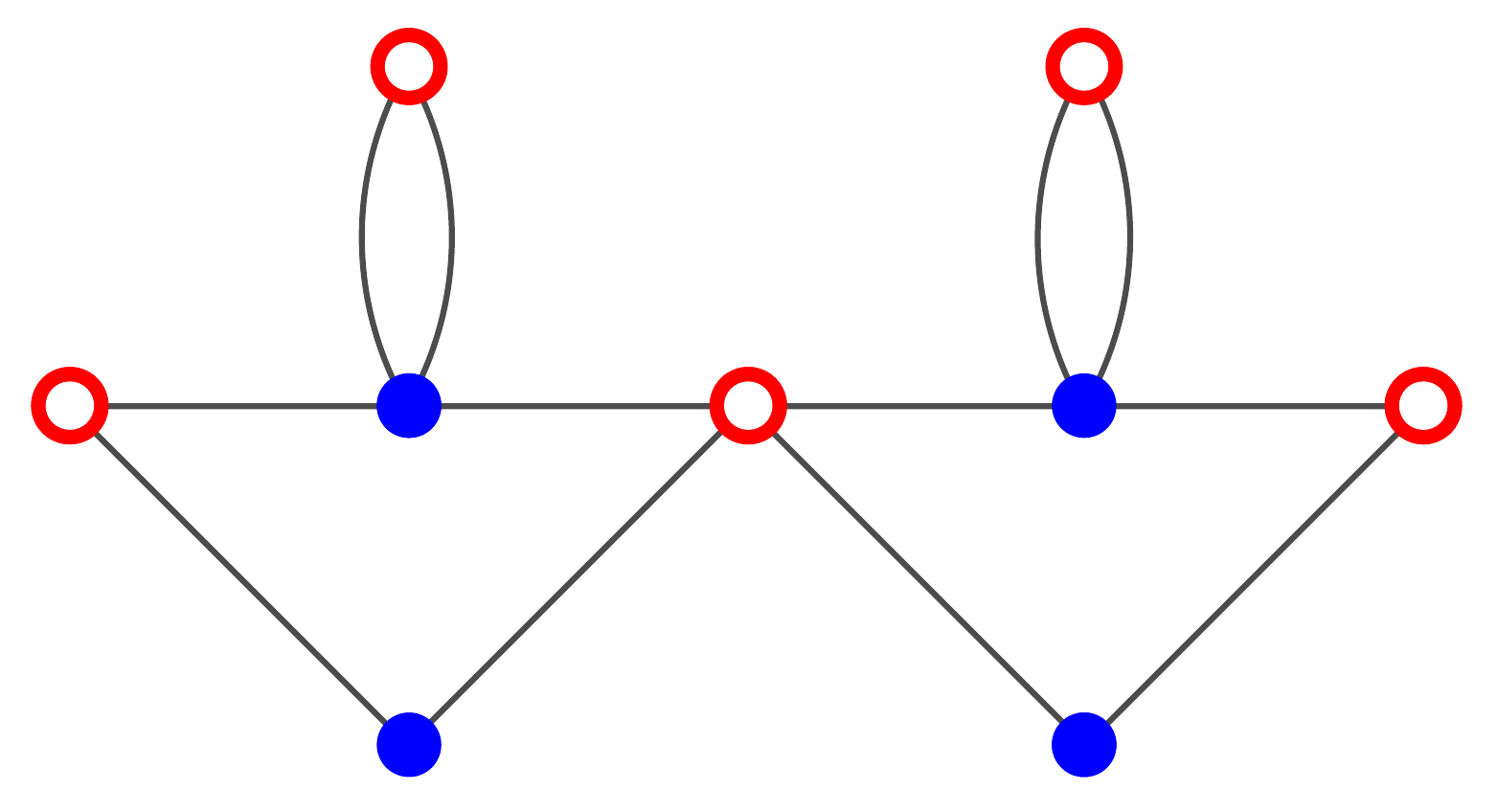}
\end{equation}
This manifold admits a total of $22$ different free group actions by the same groups as for the manifold~\eqref{eq:TQSplit10}. The Hodge numbers of the quotients are identical to those in \tref{TQSplit10quotients}.

\subsubsection{Five favourable splits with Hodge numbers $(5,37)$}\vskip-8pt
The tetraquadric also admits five different splits with Hodge numbers $(5,37)$ whose quotients are discussed in Section \ref{sec:TransposeTQSeqOfSplits}.

\subsection{Further splits}\vskip-10pt
\subsubsection{The favourable split $X^{5,45}\rightarrow X^{6,30}$}\label{TQFurtherSplits1}\vskip-8pt
The manifold $X^{6,30}$ can be obtained by splitting the first column of \eqref{eq:TQSplit2}, leading to the following configuration:
\begin{equation}\label{eq:TQSplit13}
\displaycicy{5.25in}{
X_{5302}~=~~
\cicy{\IP^1 \\ \IP^1\\ \IP^1\\ \IP^1\\ \IP^1\\ \IP^1} 	
{ ~1 &0 & 1  ~\\
  ~1 &0 & 1 ~\\
  ~1 &1 & 0~\\
  ~1 &1 & 0  ~\\
  ~0 &1 & 1 ~\\
  ~0 &1 & 1 ~\\}_{-48}^{6,30}}{-1.6cm}{1.45in}{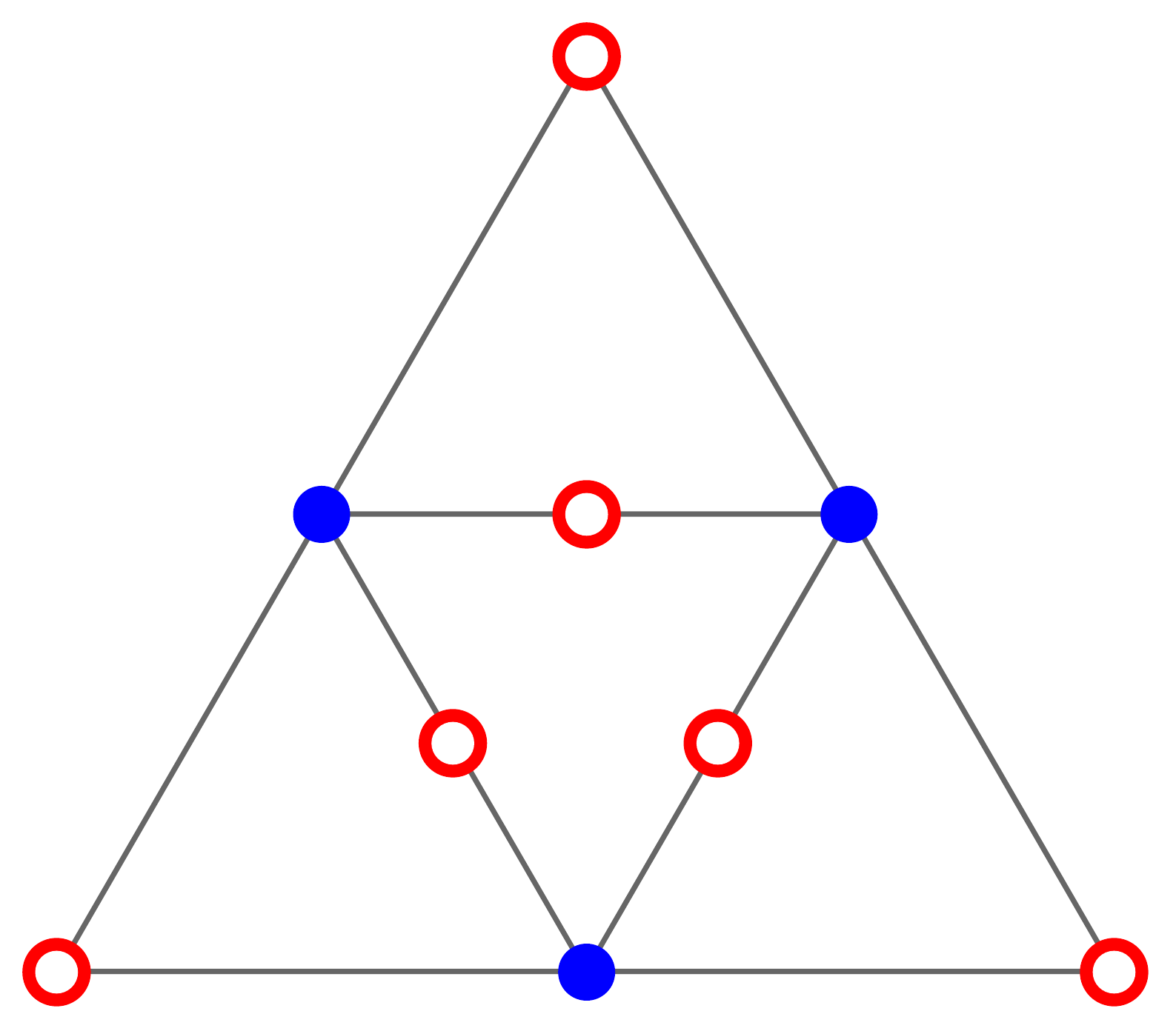}
\end{equation}
This manifold admits $20$ different free group actions. We compute the Hodge numbers for the resulting quotients using the counting of K\"ahler parameters. In this case, the polynomial deformation method does not yield the expected number of complex structure parameters for the covering manifold. We present the Hodge numbers in \tref{TQSplit13quotients}.

\begin{table}[H]
\vspace{2pt}
\begin{center}
\begin{tabular}{| c || c | c |}
\hline
\myalign{| c||}{\varstr{16pt}{10pt}$~~~~~~~  \Gamma ~~~~~~~$ } &
\myalign{m{1.31cm}|}{$~~~ \IZ_2 $} &
\myalign{m{1.9cm}|}{ $~~~\IZ_2\times\IZ_2 \ \ \ $ }  
\\ \hline\hline
\varstr{14pt}{8pt} $h^{1,1}(X/\Gamma)$ & 6 & 6 \\
 \hline
\varstr{14pt}{8pt} $h^{2,1}(X/\Gamma)$ & 18 & 12 \\
 \hline
\varstr{14pt}{8pt} $\chi(X/\Gamma)$ & $ \!\!\!\! -24 $ & $ \!\!\!\!-12$  \\
 \hline
 \end{tabular}
   \vskip 0.3cm
\capt{4.5in}{TQSplit13quotients}{Hodge numbers for the quotients of the manifold \eqref{eq:TQSplit13}.}
 \end{center}
 \vspace{2pt}
 \end{table}

\subsubsection{Splits of the manifold $X^{8,40}$: $X^{15,15}$ and $X^{9,25}$}\label{TQ_FurSpl_X8,40}\vskip-8pt
The manifold $X^{15,15}$ can be obtained by splitting the first column of \eqref{eq:TQSplit3} with a $\IP^1$, which leads to the following configuration:
\vspace{-12pt}
\begin{equation}\label{eq:TQSplit14}
\displaycicy{5.25in}{
X_{15}~=~~
\cicy{\IP^1 \\ \IP^1\\ \IP^1\\ \IP^1\\ \IP^1\\ \IP^1}
{ ~2 &0 & 0  ~\\
  ~0 &2 & 0 ~\\
  ~0 &0 & 2~\\
  ~1 &1 & 0  ~\\
  ~1 &0 & 1 ~\\
  ~0 &1 & 1 ~\\}_{0}^{15,15}}{-1.8cm}{1.5in}{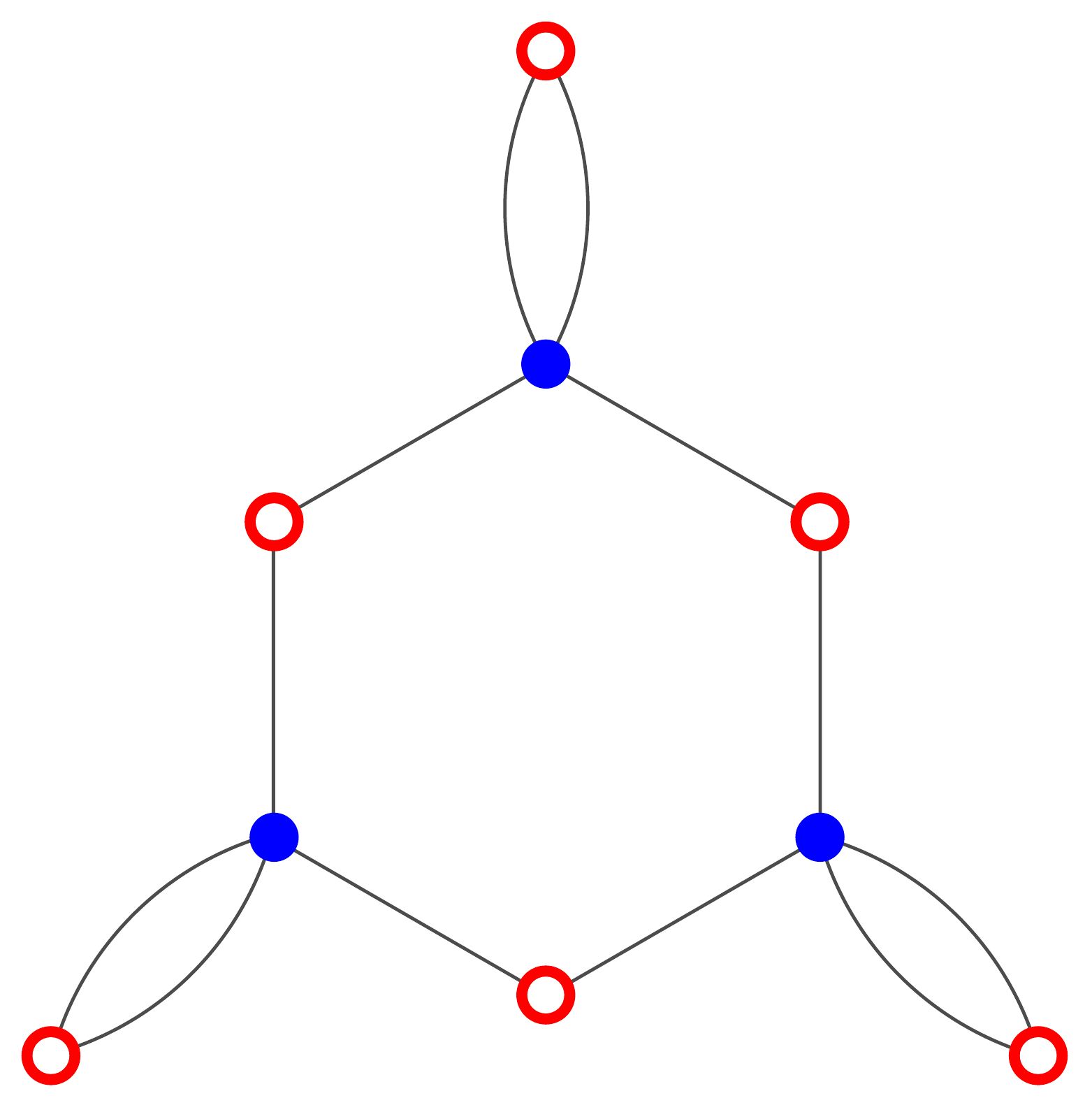}
\end{equation}
The manifold admits $20$ different free group actions by $\IZ_2$ and $\IZ_{2}{\times}\IZ_2$. 
We can compute the Hodge numbers for the corresponding quotients by using the polynomial deformation method. The polynomials defining the covering manifold contain $36$ parameters, of which $18$ account for coordinate redefinitions and $3$ for rescalings of the polynomials. 
For the $\IZ_2$-actions, the restricted polynomials contain $18$ coefficients, of which $6$ can be removed by coordinate redefinitions and $3$ by overall scalings of the polynomials. Similar considerations apply to the $\IZ_2{\times}\IZ_2$ actions. The results are summarised in \tref{TQSplit14quotients}. 
\begin{table}[H]
\vspace{8pt}
\begin{center}
\begin{tabular}{| c || c | c |}
\hline
\myalign{| c||}{\varstr{16pt}{10pt}$~~~~~~~  \Gamma ~~~~~~~$ } &
\myalign{m{1.31cm}|}{$~~~ \IZ_2 $} &
\myalign{m{1.9cm}|}{ $~~~\IZ_2\times\IZ_2 \ \ \ $ }  
\\ \hline\hline
\varstr{14pt}{8pt} $h^{1,1}(X/\Gamma)$ & 9 & 6 \\
 \hline
\varstr{14pt}{8pt} $h^{2,1}(X/\Gamma)$ & 9 & 6 \\
 \hline
\varstr{14pt}{8pt} $\chi(X/\Gamma)$ & $ 0 $ & $ 0$  \\
 \hline
 \end{tabular}
 \vskip 0.3cm
\capt{4.5in}{TQSplit14quotients}{Hodge numbers for the quotients of the manifold \eqref{eq:TQSplit14}.}
 \end{center}
 \vspace{-12pt}
 \end{table}
\goodbreak
The manifold $X^{9,25}$ can be obtained through a different splitting the first column of \eqref{eq:TQSplit3}, which corresponds to the following configuration:
\begin{equation}\label{eq:dP4_2534}
\displaycicy{5.25in}{
X_{2534}~=~~
\cicy{\IP^1 \\ \IP^1\\ \IP^1\\ \IP^1\\ \IP^1\\ \IP^1}
{ ~1 &1 & 0  ~\\
  ~1 &1 & 0 ~\\
  ~1 &1 & 0  ~\\
  ~0 &0 & 2~\\
  ~1 &0 & 1 ~\\
  ~0 &1 & 1 ~\\}_{-32}^{9,25}}{-1.1cm}{1.5in}{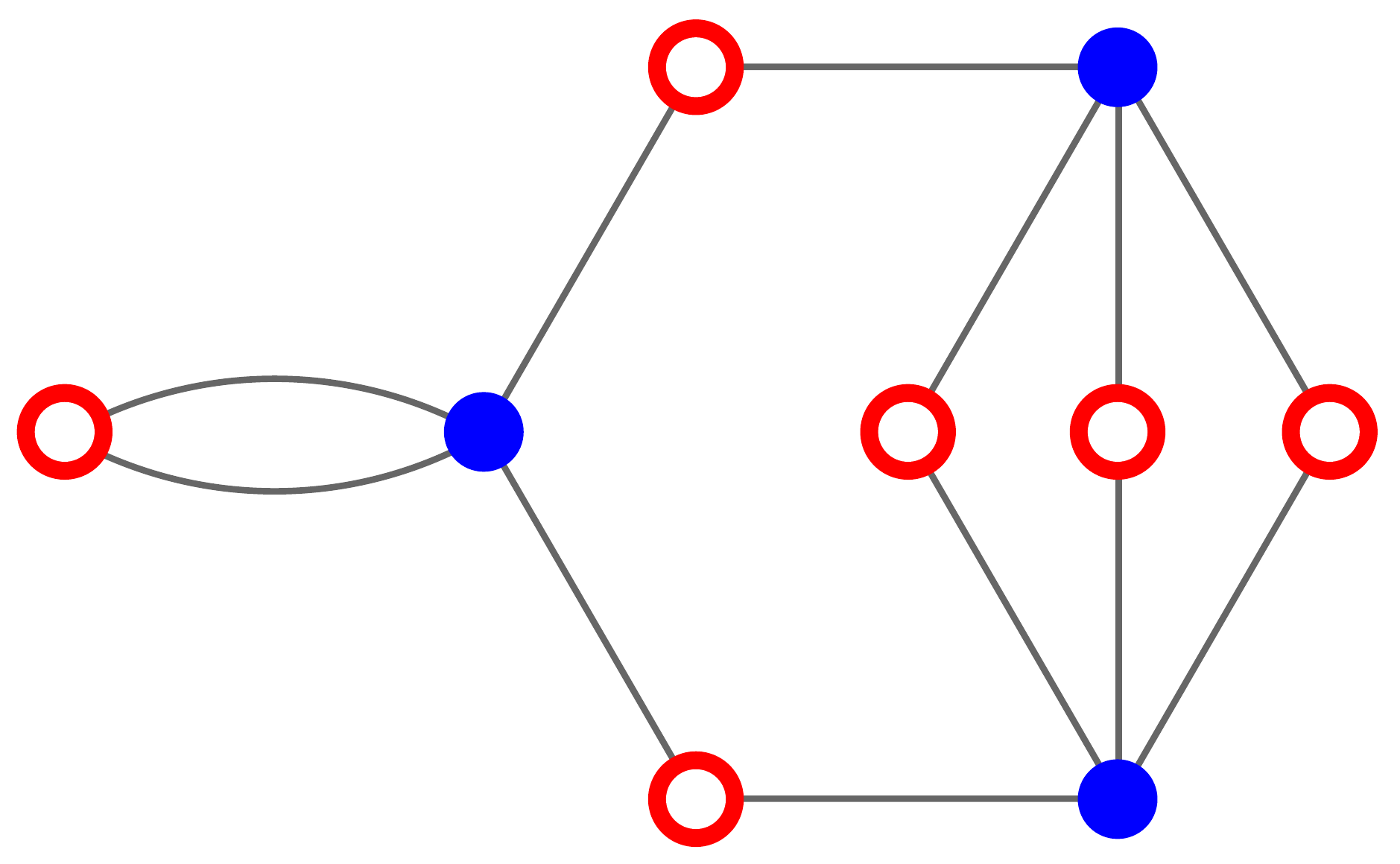}
\end{equation}
The above embedding \eqref{eq:dP4_2534} is not favourable and the polynomial deformation method does not reproduce the number of complex structure parameters. However, $X_{2534}$ is embedded in the product 
$\text{dP}_4{\times}\IP^1{\times}\IP^1{\times}\IP^1$. This will allow us to compute the $h^{1,1}$ of the quotients by computing the group action on the cohomology basis directly, using methods of section \sref{subsec:delPezzo}. The $h^{1,1}$ of this manifold is 9 since $h^{1,1}(\text{dP}_4)=6$ and the three $\IP^1$'s contribute 1 each. The distinct group actions on the coordinates are listed in \tref{dP4_2534_Symm_Action}. 

The various group actions on the cohomology basis $\{H,E_1,E_2,E_3,E_4,E_5\}$ and corresponding invariants are listed in \tref{dP4_2534_Symm_Action_Coh}. Since $X_{2534}\subset\text{dP}_4\times\IP^1\times\IP^1\times\IP^1$, and none of the symmetries mix the three additional $\IP^1$'s, the $h^{1,1}$ for the quotients is equal to three plus the number of corresponding cohomology invariants. They are listed in \tref{dP4_2534}.
\begin{table}[H]
\begin{center}
\begin{tabular}{| c || c | c |}
\hline
\myalign{| c||}{\varstr{16pt}{10pt}$~~~~~~~  \Gamma ~~~~~~~$ } &
\myalign{m{1.31cm}|}{$~~~ \IZ_2 $} &
\myalign{m{1.9cm}|}{ $~~~\IZ_2\times\IZ_2 \ \ \ $ }  
\\ \hline\hline
\varstr{14pt}{8pt} $h^{1,1}(X/\Gamma)$ & 7 & 6 \\
 \hline
\varstr{14pt}{8pt} $h^{2,1}(X/\Gamma)$ & 15 & 10 \\
 \hline
\varstr{14pt}{8pt} $\chi(X/\Gamma)$ & $\!\!\!\!-16$ & $\!\!\!\!-8$ \\
 \hline
 \end{tabular}
 \vskip 0.3cm
\capt{4.5in}{dP4_2534}{Hodge numbers for the quotients of the manifold~\eqref{eq:dP4_2534}.}
 \end{center}
\vskip-20pt
 \end{table}
 \normalsize
A still different split of the first column of \eqref{eq:TQSplit3} leads to a manifold with Hodge numbers $(12,28)$, which can be further split to yield a $(15,15)$ manifold. We discuss these in Section~\ref{sec:12_28}.
\small
\begin{center}
\begin{longtable}{|c|c|c|c|c|c|}
\captionsetup{width=0.9\textwidth}
\caption{\it Various symmetry actions on the ambient space of the manifold~\eqref{eq:dP4_2534}. The coordinate patch of the $\text{dP}_4$ is chosen to be $(1,x){\times}(1,y){\times}(1,z)$. $(p,q)$, $(r,s)$ and $(u,v)$ are taken to be coordinates of the first three $\IP^1$ spaces.} \label{dP4_2534_Symm_Action} \\

\hline \multicolumn{1}{|c|}{\str\textbf{Index}} &  \multicolumn{1}{|c|}{\str\textbf{~Group~}} & \multicolumn{1}{|c|}{$\mathbf{(\+x,\+y,\+z)}$} &  \multicolumn{1}{|c|}{$\mathbf{(\+p,\+q)}$} &  \multicolumn{1}{|c|}{$\mathbf{(\+r,\+s)}$} &  \multicolumn{1}{|c|}{$\mathbf{(\+u,\+v)}$}  \\ \hline 
\endfirsthead

\hline 
\textbf{Index} &
\textbf{~Group~} &
$\mathbf{(\+x,\+y,\+z)}$ &
$\mathbf{(\+p,\+q)}$ &
$\mathbf{(\+r,\+s)}$ &
$\mathbf{(\+u,\+v)}$ \\ \hline 
\endhead

\hline\hline \multicolumn{6}{|r|}{{\str Continued on next page}} \\ \hline
\endfoot

\endlastfoot

\hline\hline

\varstr{14pt}{8pt} 1 & $\IZ_2$ & $(-x,-y,-z)$ & $(-p,\+q)$ & $(-r,\+s)$ & $(-u,\+v)$\\
 \hline
\varstr{14pt}{8pt} 2 & $\IZ_2$ & $(\+y,\+x,-z)$ &  $(-p,\+q)$ & $(-r,\+s)$ & $(-u,\+v)$\\
 \hline
\varstr{16pt}{10pt} 3 & $\IZ_2{\times}\IZ_2$ &
  \begin{minipage}[c][40pt][c]{1.3in}
  \begin{gather*} 
 (-x,-y,-z) \\ 
 ({x^{-1}},{y^{-1}},{z^{-1}}) \\
  \end{gather*}
  \end{minipage} 
& 
 \begin{minipage}[c][40pt][c]{0.8in}
  \begin{gather*}
(\+p,-q) \\ 
(\+q,\+p)\\
  \end{gather*}
  \end{minipage} 
  & 
 \begin{minipage}[c][40pt][c]{0.8in}
  \begin{gather*}
(\+r,-s) \\ 
(\+s,\+r)\\
  \end{gather*}
  \end{minipage} 
  &
  \begin{minipage}[c][40pt][c]{0.8in}
  \begin{gather*}
(\+u,-v) \\ 
(\+v,\+u)\\
  \end{gather*}
  \end{minipage} \\
 \hline
 4 & $\IZ_2{\times}\IZ_2$ &
  \begin{minipage}[c][40pt][c]{1.3in}
  \begin{gather*} 
 (\+y,\+x,-z) \\ 
 ({-y},{-x},{z^{-1}}) \\
  \end{gather*}
  \end{minipage} 
& 
 \begin{minipage}[c][40pt][c]{0.8in}
  \begin{gather*}
(\+p,-q) \\ 
(\+q,\+p)\\
  \end{gather*}
  \end{minipage} 
  & 
 \begin{minipage}[c][40pt][c]{0.8in}
  \begin{gather*}
(\+r,-s) \\ 
(\+s,\+r)\\
  \end{gather*}
  \end{minipage} 
  &
  \begin{minipage}[c][40pt][c]{0.8in}
  \begin{gather*}
(\+u,-v) \\ 
(\+v,\+u)\\
  \end{gather*}
  \end{minipage} \\
 \hline
  5 & $\IZ_2{\times}\IZ_2$ &
  \begin{minipage}[c][40pt][c]{1.3in}
  \begin{gather*} 
 (\+y,\+x,-z) \\ 
 ({-x},{-y},{z^{-1}}) \\
  \end{gather*}
  \end{minipage} 
& 
 \begin{minipage}[c][40pt][c]{0.8in}
  \begin{gather*}
(\+p,-q) \\ 
(\+q,\+p)\\
  \end{gather*}
  \end{minipage} 
  & 
 \begin{minipage}[c][40pt][c]{0.8in}
  \begin{gather*}
(\+r,-s) \\ 
(\+s,\+r)\\
  \end{gather*}
  \end{minipage} 
  &
  \begin{minipage}[c][40pt][c]{0.8in}
  \begin{gather*}
(\+u,-v) \\ 
(\+v,\+u)\\
  \end{gather*}
  \end{minipage} \\
 \hline
 6 & $\IZ_2{\times}\IZ_2$ &
  \begin{minipage}[c][40pt][c]{1.3in}
  \begin{gather*} 
 (-x,-y,-z) \\ 
 ({\+y},{\+x},{z^{-1}}) \\
  \end{gather*}
  \end{minipage} 
& 
 \begin{minipage}[c][40pt][c]{0.8in}
  \begin{gather*}
(\+p,-q) \\ 
(\+q,\+p)\\
  \end{gather*}
  \end{minipage} 
  & 
 \begin{minipage}[c][40pt][c]{0.8in}
  \begin{gather*}
(\+r,-s) \\ 
(\+s,\+r)\\
  \end{gather*}
  \end{minipage} 
  &
  \begin{minipage}[c][40pt][c]{0.8in}
  \begin{gather*}
(\+u,-v) \\ 
(\+v,\+u)\\
  \end{gather*}
  \end{minipage} \\
 \hline
 \end{longtable}
\end{center}
\normalsize
\small
\begin{center}
\begin{longtable}{|c|c|c|c|}
\captionsetup{width=0.9\textwidth}
\caption{\it Symmetry actions on the cohomology basis and the corresponding invariants for the manifold~\eqref{eq:dP4_2534}. The matrices $P_i$ and $Q_i$ are defined in \eqref{eq:CohMatDefs2}.} \label{dP4_2534_Symm_Action_Coh} \\
\hline \multicolumn{1}{|c|}{\str\textbf{Index}}&  \multicolumn{1}{|c|}{\str\textbf{~Group~}} & \multicolumn{1}{|c|}{\str\textbf{\begin{minipage}[c][35pt][c]{0.85in}
Action on\\ 
Coh Basis
  \end{minipage}}} &  \multicolumn{1}{|c|}{\str\textbf{~~Coh Invariants~~}} \\ \hline 
\endfirsthead

\hline 
\textbf{Index} & \textbf{~Group~} & \textbf{\begin{minipage}[c][35pt][c]{0.85in}
Action on\\ 
Coh Basis
  \end{minipage}} & \textbf{~~Coh Invariants~~} \\ \hline 
\endhead

\hline\hline \multicolumn{4}{|r|}{{\str Continued on next page}} \\ \hline
\endfoot

\endlastfoot

\hline\hline
1 & $\IZ_2$ &
$P_1$
& \begin{minipage}[c][20pt][c]{2.1in}
\begin{gather*} 
H, E_1+E_2,E_3+E_4,E_5\\
\end{gather*}
\end{minipage} \\
 \hline
 
2 & $\IZ_2$ &
$Q_1$
& \begin{minipage}[c][22pt][c]{3.2in}
\begin{gather*} 
H-{E}_5, E_1+E_4-E_5, E_2+E_4-E_5,E_3-E_4\\
\end{gather*}
\end{minipage} \\
 \hline
 
3 & $\IZ_2{\times}\IZ_2$ &
$P_2$,~$P_1$
&\begin{minipage}[c][22pt][c]{2.1in}
\begin{gather*} 
H, E_1+E_2+E_3+E_4,E_5\\
\end{gather*}
\end{minipage} \\ \hline

4 & $\IZ_2{\times}\IZ_2$ &
$Q_1$,~$Q_2$
& \begin{minipage}[c][22pt][c]{2.1in}
\begin{gather*} 
H-{E}_5, E_1+E_3-E_5, E_2+E_4-E_5\\
\end{gather*}
\end{minipage} \\
 \hline
 
5 & $\IZ_2{\times}\IZ_2$ &
$Q_1$,~$P_2$
& \begin{minipage}[c][22pt][c]{2.1in}
\begin{gather*} 
H-{E}_5, E_1+E_4-E_5, E_2+E_3-E_5\\
\end{gather*}
\end{minipage} \\
 \hline
 
6 & $\IZ_2{\times}\IZ_2$ &
$P_2$,~$Q_1$
& \begin{minipage}[c][22pt][c]{2.1in}
\begin{gather*} 
H-{E}_5, E_1+E_4-E_5, E_2+E_3-E_5\\
\end{gather*}
\end{minipage} \\
 \hline 
\end{longtable}
\end{center}
%
%
\subsection{The manifold $X^{8,44}$ and its split $X^{9,25}$}\label{TQ_X8,44}\vskip-8pt
This manifold can be obtained from the tetraquadric through a sequence of two splits. The first split leads to the configuration:
\begin{equation}\label{eq:TQSplit17p}
\displaycicy{5.25in}{
X_{7709}~=~~
\cicy{\IP^1 \\ \IP^1\\ \IP^1\\ \IP^1\\ \IP^1}
{ ~2 &0  ~\\
  ~2 & 0 ~\\
  ~1 &1  ~\\
  ~1 &1 ~\\
  ~1 &1 ~\\}_{-96}^{6,54}}{-0.9cm}{1.2in}{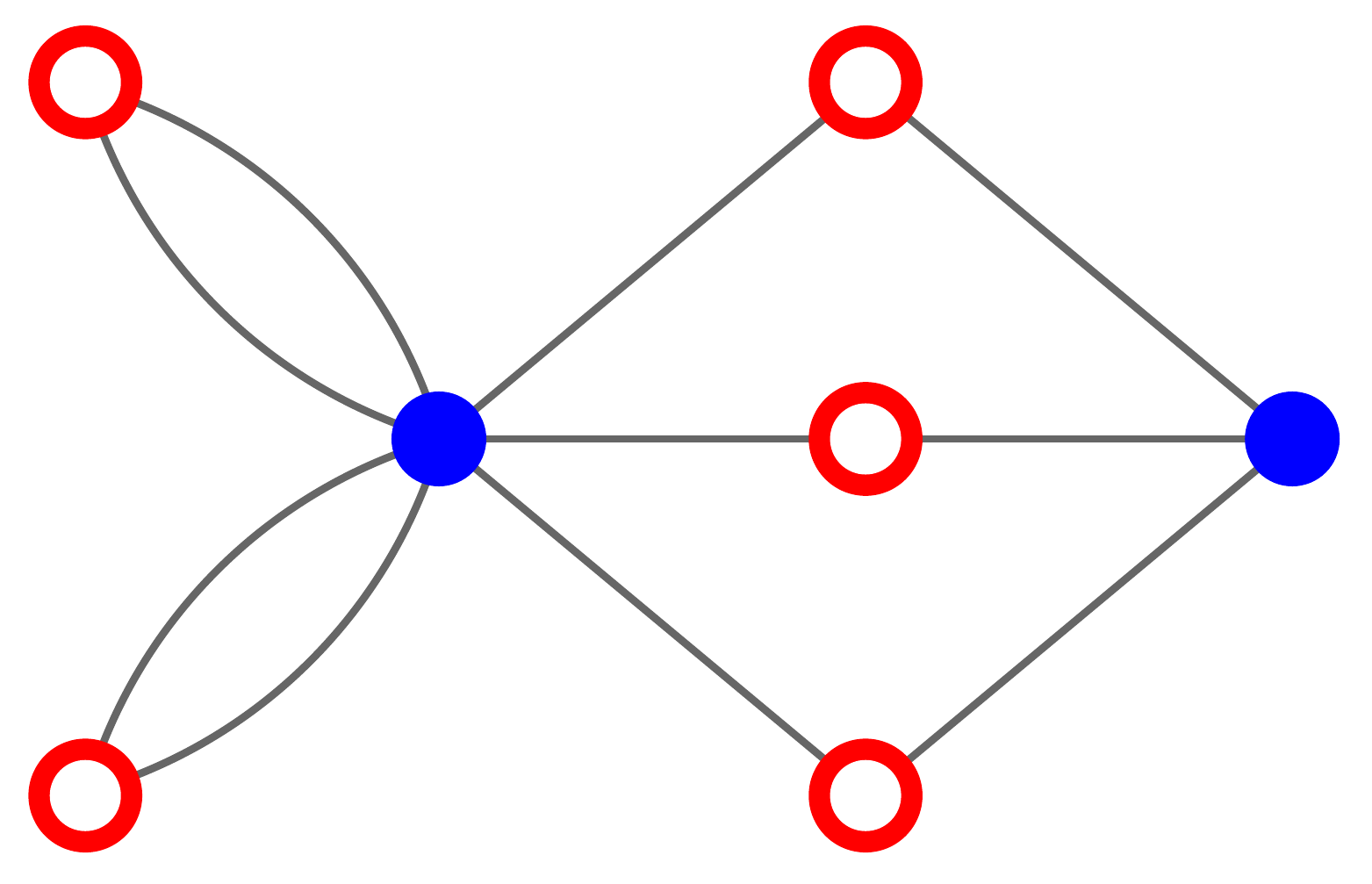}
\end{equation}
This manifold admits two free actions of $\IZ_2$. The polynomial deformation method can be applied in this case. For the covering manifold, we have a number of $80$ parameters in the defining polynomials, $15$ of which can be removed by coordinate redefinitions and $11$ by redefinitions of the polynomials. For the quotient manifold, there are $40$ coefficients in the specialised polynomials, $5$ of which can be eliminated by projective linear coordinate transformations, and $6$ by redefinitions of the polynomials. This leads to $h^{2,1}(X/\IZ_2)=29$ and hence $h^{1,1}(X/\IZ_2) = 5$. 

Splitting the first column of the above configuration matrix with a $\IP^1$ space leads us to the manifold~$X^{8,44}$ defined by the following configuration:
\begin{equation}\label{eq:TQSplit17d}
\displaycicy{5.25in}{
X_{7300}~=~~
\cicy{\IP^1 \\ \IP^1\\ \IP^1\\ \IP^1\\ \IP^1\\\IP^1}
{ ~1 &0 & 1  ~\\
  ~1 & 0&1 ~\\
  ~1 & 0&1 ~\\
  ~0&1 &1  ~\\
  ~0&1 &1 ~\\
  ~0&1 &1 ~\\}_{-72}^{8,44}}{-1.1cm}{1.7in}{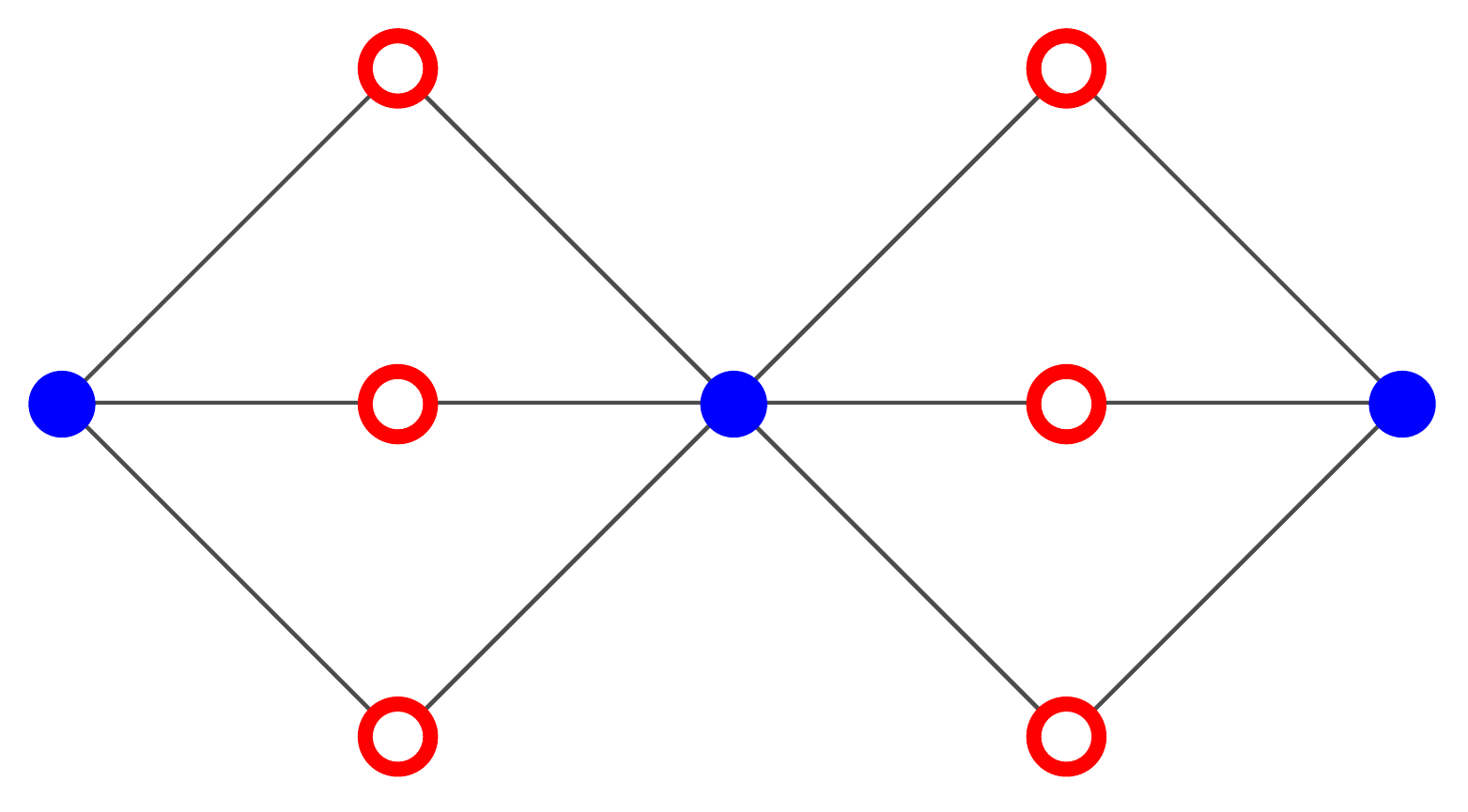}
\end{equation}
This manifold has been used in Refs.~\cite{Candelas:2008wb, Braun:2009qy} in order to construct a class of heterotic models with three generations and the MSSM spectrum. As shown in \sref{redun} the manifold $X^{8,44}$ can be represented by the following equivalent configuration: 
\begin{equation}\label{eq:TQSplit17}
\displaycicy{5.25in}{
X_{7246}~=~~
\cicy{\IP^2 \\ \IP^2\\ \IP^2\\ \IP^2}
{ ~1 &1 & 0& 0  & 1~\\
  ~1 & 1&0& 0 & 1~\\
  ~0 & 0&1& 1 & 1~\\
  ~0&0 &1& 1& 1  ~\\}_{-72}^{8,44}}{-1.1cm}{1.7in}{TQSplit17.pdf}
\end{equation}
In the embedding \eqref{eq:TQSplit17d}, the manifold admits $15$ free actions of the following groups:
\beq
\IZ_2,~ \IZ_3,~\IZ_4,~\IZ_6, ~\IZ_3\rtimes \IZ_4,~\IZ_{12},
\notag\eeq
while in the embedding~\eqref{eq:TQSplit17} it admits $26$ free actions of the same groups. Using either of the two embeddings, one can compute the Hodge numbers for the resulting quotients through the polynomial deformation method. Alternatively,  one can analyse the action of the above groups on the second cohomology of ${\rm dP}_6$, as done in~\cite{Candelas:2008wb, Braun:2009qy}. By either method, we obtain the Hodge numbers listed in \tref{TQSplit17quotients}. 
\begin{table}[!ht]
\begin{center}
\begin{tabular}{| c || c | c | c | c | c | c |}
\hline
\myalign{| c||}{\varstr{16pt}{10pt}$~~~~~~~  \Gamma ~~~~~~~$ } &
\myalign{m{1.31cm}|}{$~~~ \IZ_2 $} &
\myalign{m{1.31cm}|}{$~~~ \IZ_3 $} &
\myalign{m{1.31cm}|}{$~~~ \IZ_4 $} &
\myalign{m{1.31cm}|}{$~~~ \IZ_6 $} &
\myalign{m{1.9cm}|}{ $~~~\IZ_3\rtimes\IZ_4 \ \ \ $ }  &
\myalign{m{1.31cm}|}{$~~~ \IZ_{12} $} 
\\ \hline\hline
\varstr{14pt}{8pt} $h^{1,1}(X/\Gamma)$ & 6 & 4 & 3 & 2 & 1 & 1\\
 \hline
\varstr{14pt}{8pt} $h^{2,1}(X/\Gamma)$ & 24 & 16 & 12 & 8 & 4 & 4\\
 \hline
\varstr{14pt}{8pt} $\chi(X/\Gamma)$ & $\!\!\!\!-36$ & $\!\!\!\!-24$ & $\!\!\!\!-18$ & $\!\!\!\!-12$  & $\!\!\!\!-6$ & $\!\!\!\!-6$ \\
 \hline
 \end{tabular}
 \vskip 0.1cm
\capt{5.5in}{TQSplit17quotients}{Hodge numbers for the quotients of the manifolds \eqref{eq:TQSplit17d} and  \eqref{eq:TQSplit17}.}
 \end{center}
 \vspace*{-15pt}
 \end{table}

The last column of the manifold \eqref{eq:TQSplit17d} can be split with a $\IP^3$, giving the following configuration for the manifold $X^{9,25}$:
\begin{equation}\label{eq:TQSplit19d}
\displaycicy{5.25in}{
X_{2639}~=~~
\cicy{\IP^1 \\ \IP^1\\ \IP^1\\ \IP^1\\ \IP^1\\\IP^1\\\IP^3}
{ ~1 &0 & 1&0&0&0  ~\\
  ~1 & 0&0&1&0&0 ~\\
  ~1 & 0&0&1&0&0 ~\\
  ~0&1 &0&0&1&0  ~\\
  ~0&1 &0&0 &0&1 ~\\
  ~0&1 &0&0&0&1 ~\\
  ~0&0 &1&1&1&1 ~\\}_{-32}^{9,25}}{-1.0cm}{1.8in}{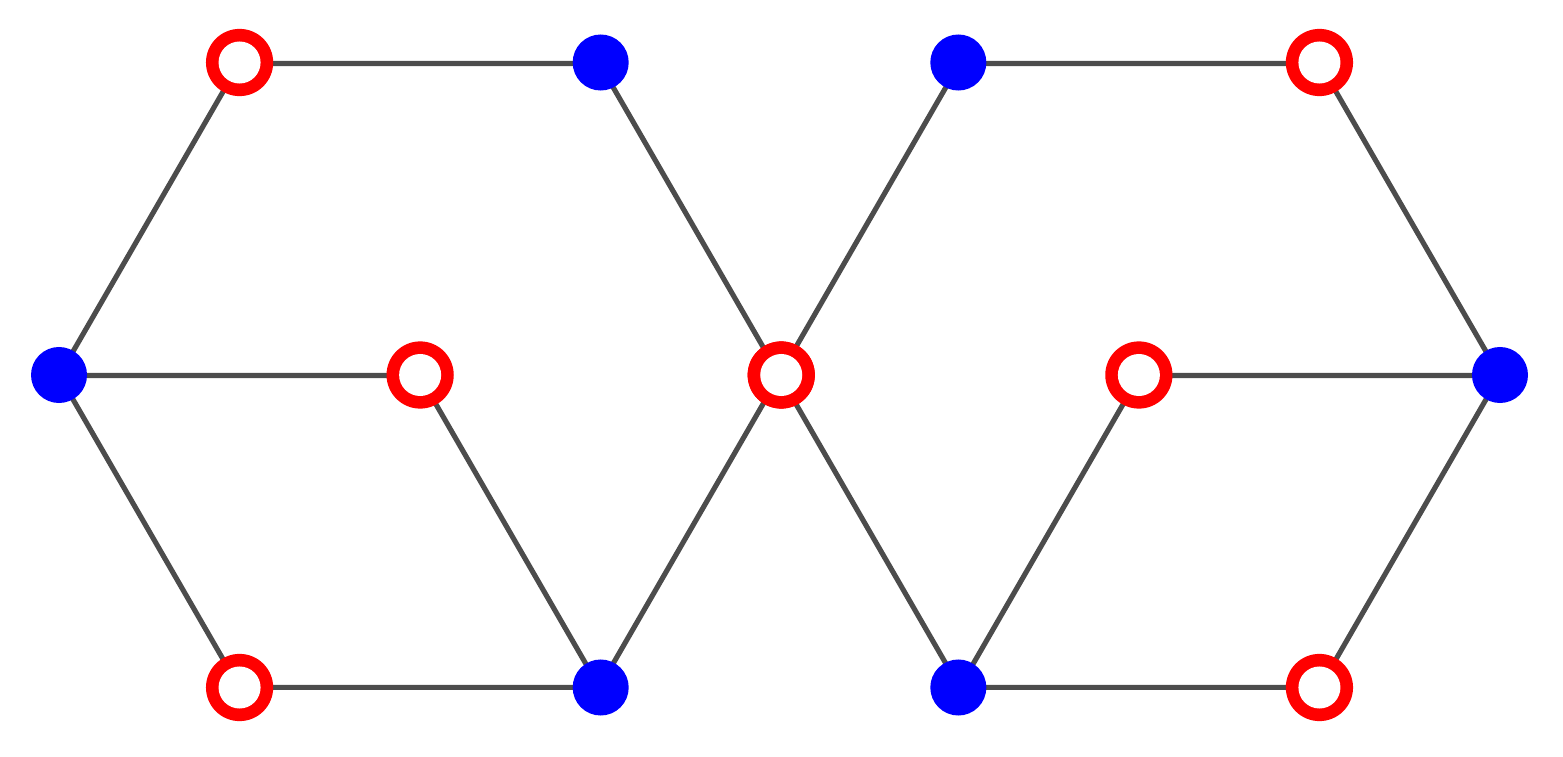}
\end{equation}
Similarly, the last column of the manifold \eqref{eq:TQSplit17} can be split with a $\IP^3$, an operation which leads to the following configuration matrix:
\begin{equation}\label{eq:TQSplit19}
\displaycicy{5.25in}{
X_{2572}~=~~
\cicy{\IP^2 \\ \IP^2\\ \IP^2\\ \IP^2\\\IP^3}
{ ~1 &1 & 0& 0  & 1& 0& 0& 0~\\
  ~1 & 1&0& 0 & 0& 1& 0& 0~\\
  ~0 & 0&1& 1 & 0& 0& 1& 0~\\
  ~0&0 &1& 1& 0& 0& 0& 1  ~\\
  ~0&0 &0& 0& 1& 1& 1& 1  ~\\}_{-32}^{9,25}}{-1.0cm}{1.8in}{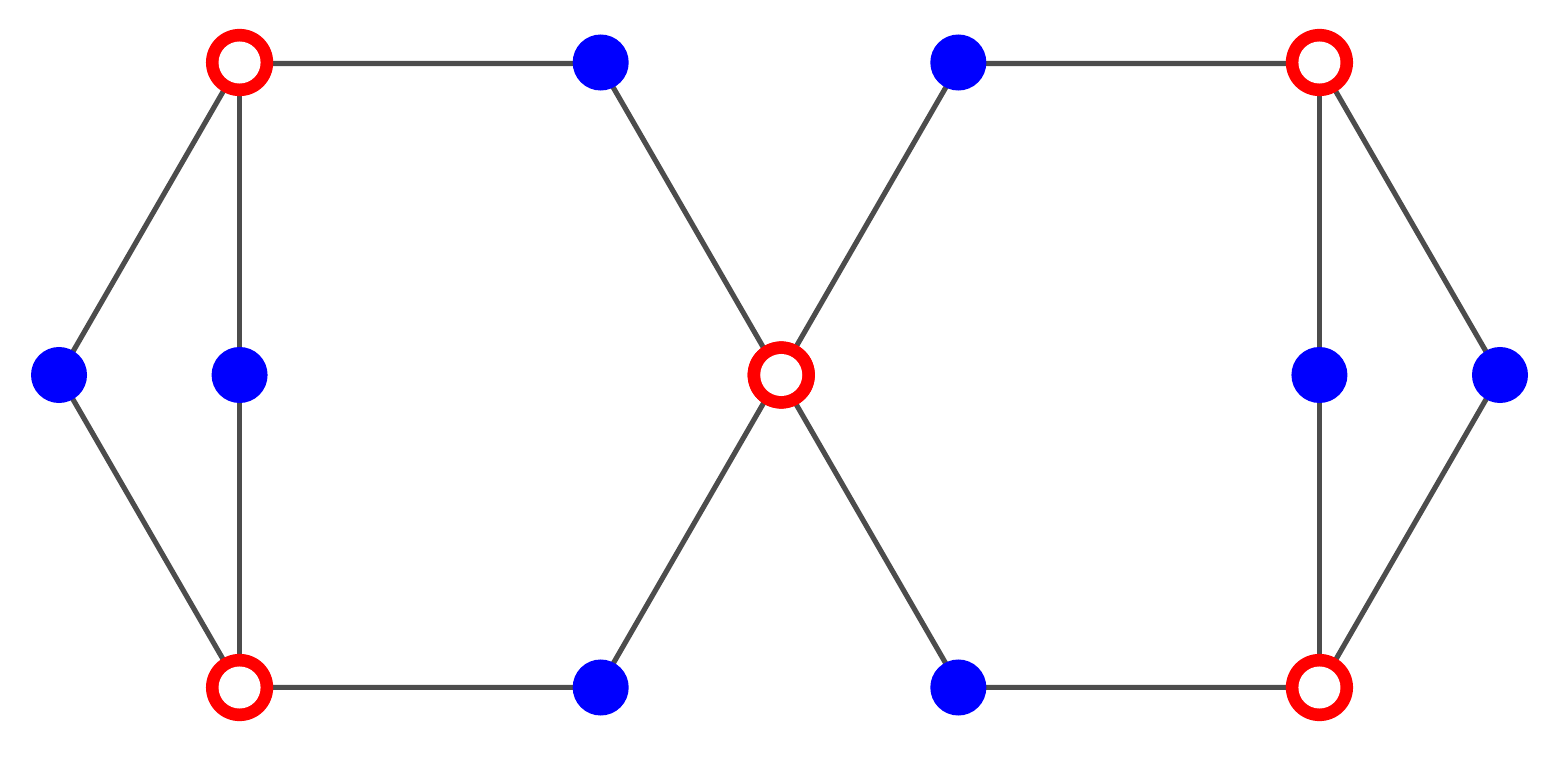}
\end{equation}
We do not have a good method for computing the Hodge numbers for the quotients of \eqref{eq:TQSplit19d} directly. However, given the identity of \eqref{eq:TQSplit17d} and \eqref{eq:TQSplit17} we will assume that \eqref{eq:TQSplit19d} and \eqref{eq:TQSplit19} are also the same and have equivalent group actions. 
The manifold \eqref{eq:TQSplit19} admits $2$ different free group actions by $\IZ_2$ and $\IZ_4$. We compute the Hodge numbers for the resulting quotients using the polynomial deformation method. 
\begin{table}[H]
\begin{center}
\begin{tabular}{| c || c | c |}
\hline
\myalign{| c||}{\varstr{16pt}{10pt}$~~~~~~~  \Gamma ~~~~~~~$ } &
\myalign{m{1.31cm}|}{$~~~ \IZ_2 $} &
\myalign{m{1.31cm}|}{$~~~ \IZ_4 $} 
\\ \hline\hline
\varstr{14pt}{8pt} $h^{1,1}(X/\Gamma)$ & 7 & 4 \\
 \hline
\varstr{14pt}{8pt} $h^{2,1}(X/\Gamma)$ & 15 & 8 \\
 \hline
\varstr{14pt}{8pt} $\chi(X/\Gamma)$ & $ \!\!\!\!-16 $ & $ \!\!\!\!- 8$  \\
 \hline
 \end{tabular}
   \vskip 0.3cm
\capt{5.5in}{TQSplit19quotients}{Hodge numbers for the quotients of the manifolds \eqref{eq:TQSplit19d} and \eqref{eq:TQSplit19}.}
 \end{center}
 \end{table}
\newpage
\section{The manifold $\IP^7[2,2,2,2]$ and its splits}\label{sec:fourquadrics}
The parent manifold in this sequence of splits is defined by the following configuration: 
\begin{equation}\label{eq:TTQ}
\displaycicy{5.25in}{
X_{7861}~=~~
\cicy{\IP^7\,\,}
{ ~2& \! 2 &\!2 &\!2~ \\}_{-128}^{\,1,\,65}}{-1.1cm}{0.95in}{TTQ.pdf}
\end{equation}
In this section we will describe smooth quotients of not only this manifold but also of manifolds that descend from this via a sequence  of conifold transitions: 
$X^{1,65}\rightarrow X^{2,58}\rightarrow X^{3,51}\rightarrow X^{4,44}\rightarrow X^{5,37}$. These manifolds appear in multiple guises so we end up considering all the configurations with the `flower'  diagrams of \fref{FlowerWeb}. 

\subsection{Transpose of the tetraquadric $X^{1,65}$ and its smooth quotients}\label{sec:TTQ}
The manifold $X^{1,65}$ admits $45$ different smooth quotients by the finite groups listed in  Table~\ref{TTQquotients}. The counting of invariant K\"ahler forms in this case is trivial, since the embedding space is a single projective space. The polynomial deformation method also yields consistent results. For this manifold the classification of free group actions was first performed by Hua {\it et al.\/} \cite{Hua:2007fq}.

\begin{table}[H]
\begin{center}
\begin{tabular}{| >{\hskip-2pt}c<{\hskip-3pt} || >{\hskip3pt}c<{\hskip-4pt} | >{\hskip0pt}c<{\hskip3pt} 
                         | >{\hskip-1pt}c<{\hskip-7pt}  | >{\hskip0pt}c<{\hskip2pt} | >{\hskip1pt}c<{\hskip-2pt}  | }
\hline
$\G$ & 
\begin{minipage}[c][65pt][c]{0.15in}
\begin{gather*}
\!\!\IZ_2\\ 
\\
\\
\end{gather*}
\end{minipage}
 & 
\begin{minipage}[c][65pt][c]{0.35in}
\begin{gather*}
\IZ_4\\ 
\! \IZ_2{\times}\IZ_2\\
\\
\end{gather*}
\end{minipage}
&
\begin{minipage}[c][65pt][c]{0.8in}
\begin{gather*}
\IZ_8\\
\!\!\IZ_2{\times}\IZ_4,\,\IQ_8\\ 
\!\!\IZ_2{\times}\IZ_2{\times}\IZ_2\\
\end{gather*}
\end{minipage}
&
\begin{minipage}[c][65pt][c]{1.25in}
\begin{gather*}
\IZ_4{\times}\IZ_4, \,\IZ_4{\rtimes}\IZ_4\\ 
\!\IZ_2{\times}\IZ_8, \IZ_4{\times}\IZ_2{\times} \IZ_2\\
\IZ_2{\times} \IQ_8 \\
\end{gather*}
\end{minipage}
& 
\begin{minipage}[c][65pt][c]{2.3in}
\begin{gather*}
(\IZ_4{\times}\IZ_2){\rtimes} \IZ_4,\,\IZ_8{\times}\IZ_4, \,\IZ_8{\rtimes}\IZ_4 \\
\!(\IZ_8{\times}\IZ_2){\rtimes}\IZ_2, \IZ_8{\rtimes}\IZ_4, \IZ_4{\times}\IZ_4{\times} \IZ_2 \\
\!\!\IZ_2{\times} (\IZ_4{\rtimes}\IZ_4), \IZ_4{\rtimes} \IQ_8,\IZ_2{\times}\IZ_2{\times} \IQ_8 \\
\end{gather*}
\end{minipage}\\ 
\hline\hline
\varstr{14pt}{8pt} $h^{1,1}(X/\Gamma)$ & 1 & 1 & 1  & 1 & 1\\
 \hline
\varstr{14pt}{8pt} $h^{2,1}(X/\Gamma)$ & 33 & 17 & 9 & 5 & 3   \\
 \hline
\varstr{14pt}{8pt} $\chi(X/\Gamma)$ & $\!\!\!\!-64$ &$\!\!\!\!-32$ & $\!\!\!\!-16$  & $\!\!\!\!-8$ &$\!\!\!\!-4$\\
 \hline
 \end{tabular}
\vskip10pt
\capt{4.5in}{TTQquotients}{Hodge numbers for the quotients of the manifold $X^{1,65}$.} 
\end{center}
\vspace{-12pt}
\end{table}
\begin{figure}[H]
\vspace{16pt}
\begin{center}
\fbox{
\begin{minipage}[c][6.3in][c]{6.5in}
\xymatrix@R-=0.27cm@C-=0.4cm{
&\hskip-16pt\raisebox{1.02cm}{$(1,65)$}\hskip-12pt
&\raisebox{0.375cm}{\includegraphics[width=.5in]{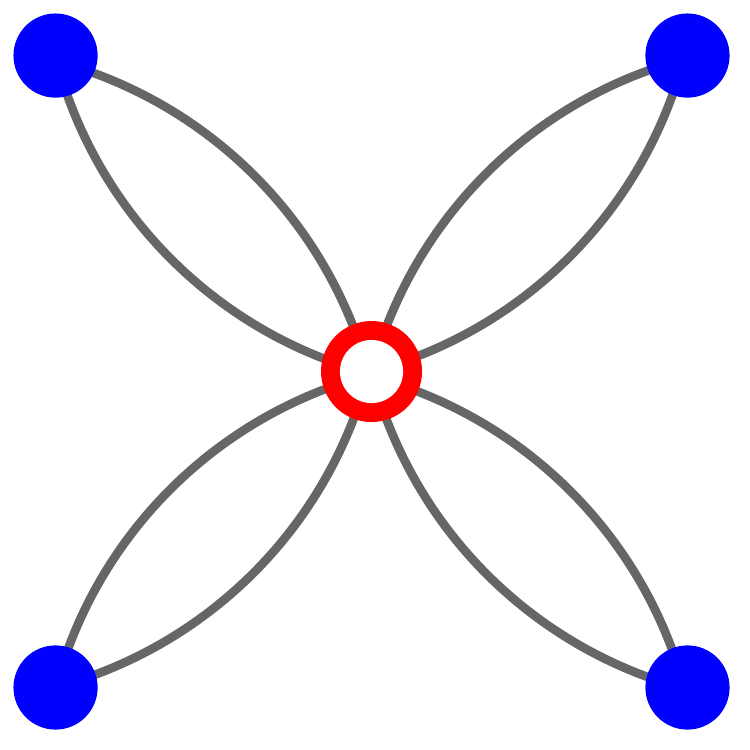}}\ar[dd] 
\\ \\
&\hskip-16pt\raisebox{1.25cm}{$(2,58)$}\hskip-12pt
&\,~~~~\raisebox{0.3cm}{\includegraphics[width=.72in]{TTQSplit1.pdf}}\ar[dd]
&\,~~~~\raisebox{0.3cm}{\includegraphics[width=.72in]{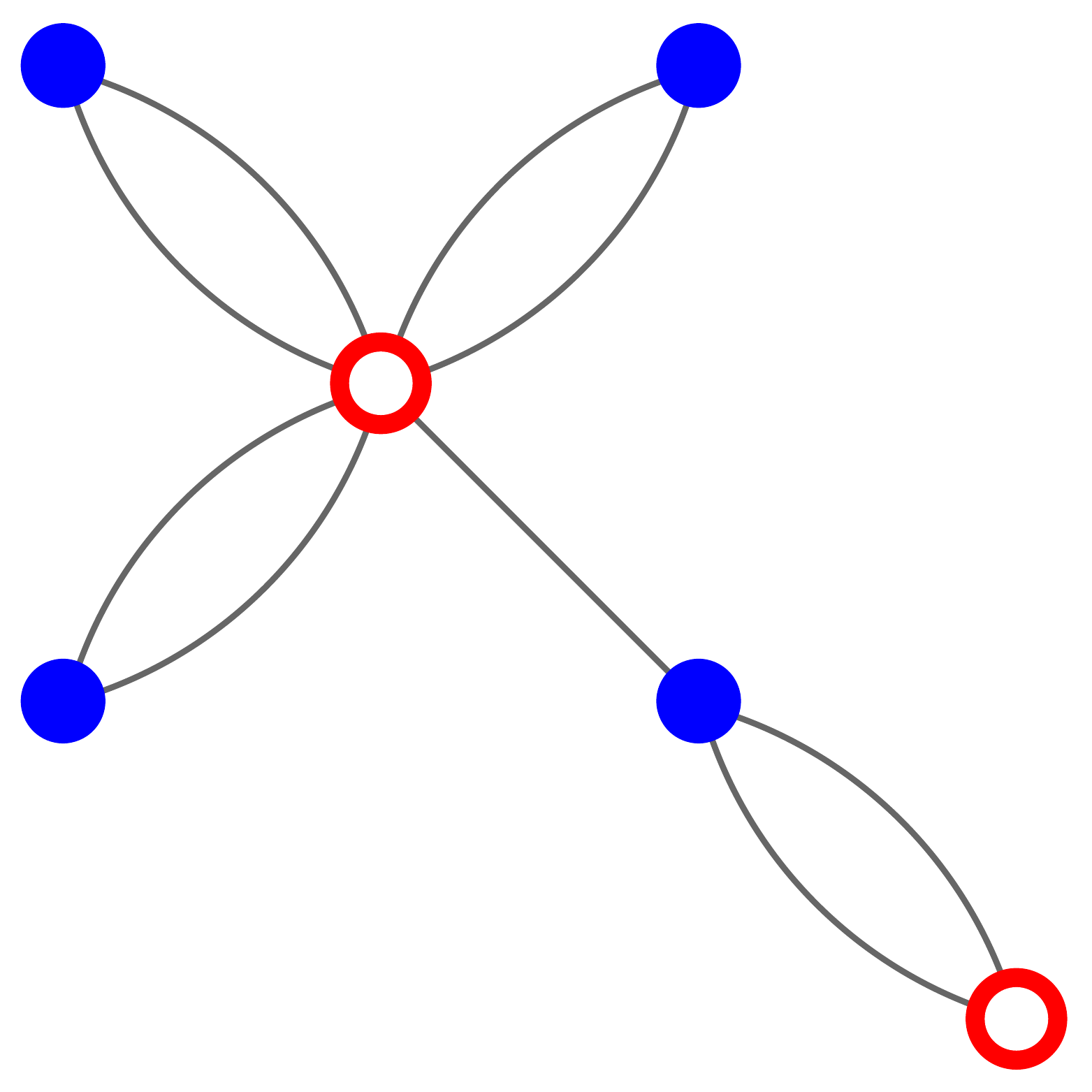}}\ar[dd] 
\\\\
&\hskip-16pt\raisebox{1.1cm}{$(3,51)$}\hskip-12pt
&\raisebox{0cm}{\includegraphics[width=.9in]{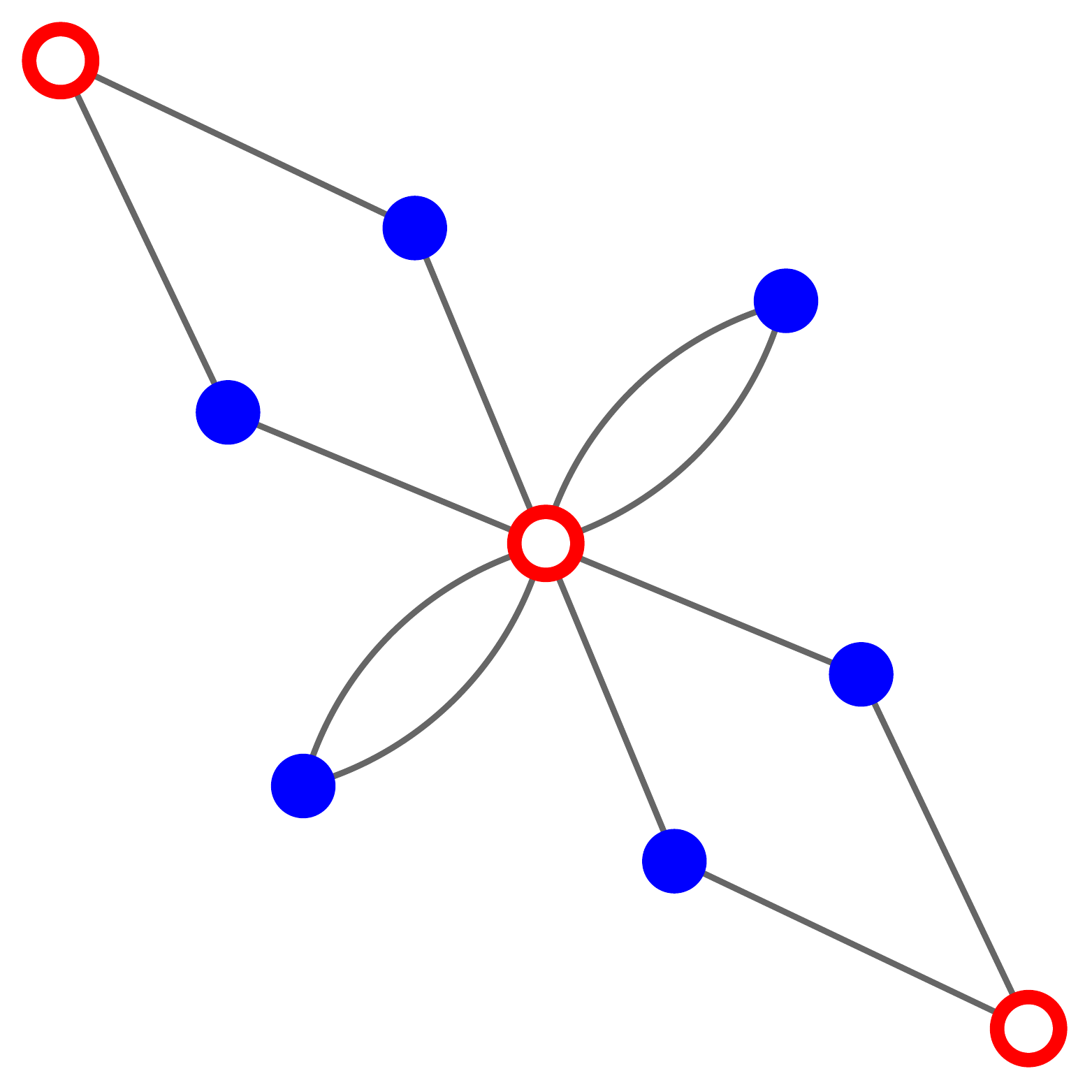}}\ar[dd]
&\raisebox{0cm}{\includegraphics[width=0.9in]{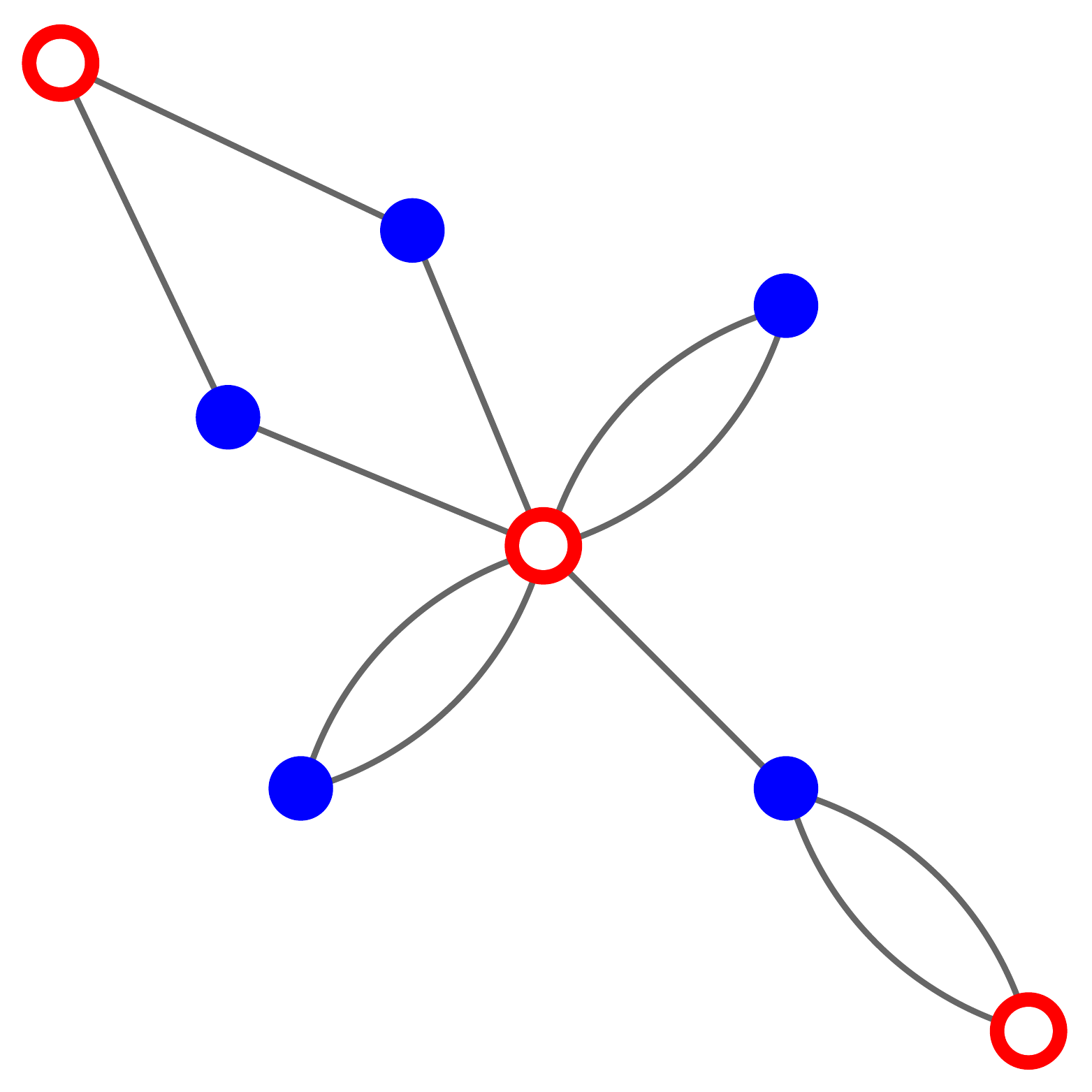}}\ar[dd]
&\raisebox{0cm}{\includegraphics[width=0.9in]{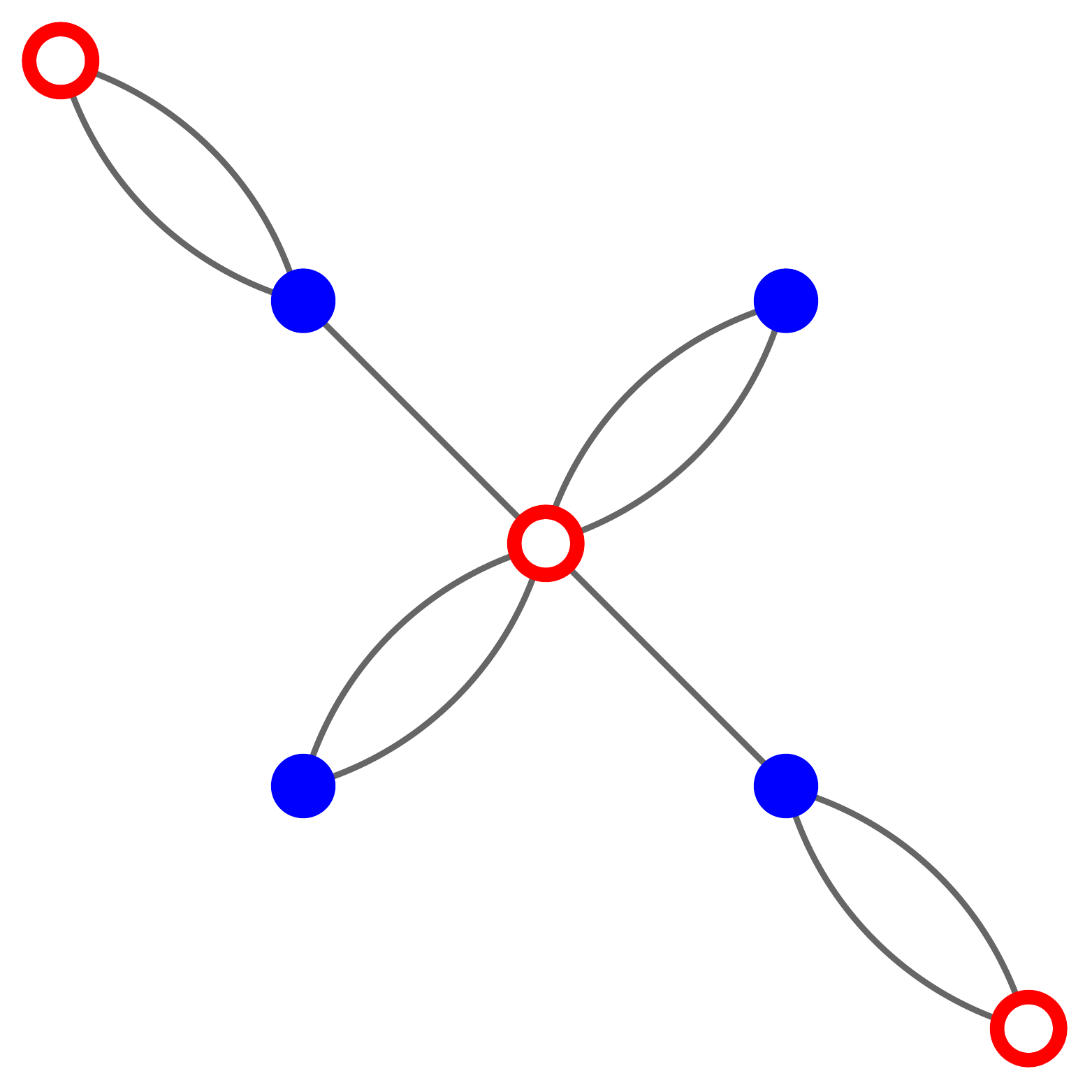}}\ar[dd]
\\\\
&\hskip-16pt\raisebox{1.1cm}{$(4,44)$}\hskip-12pt
&\raisebox{0cm}{\includegraphics[width=0.9in]{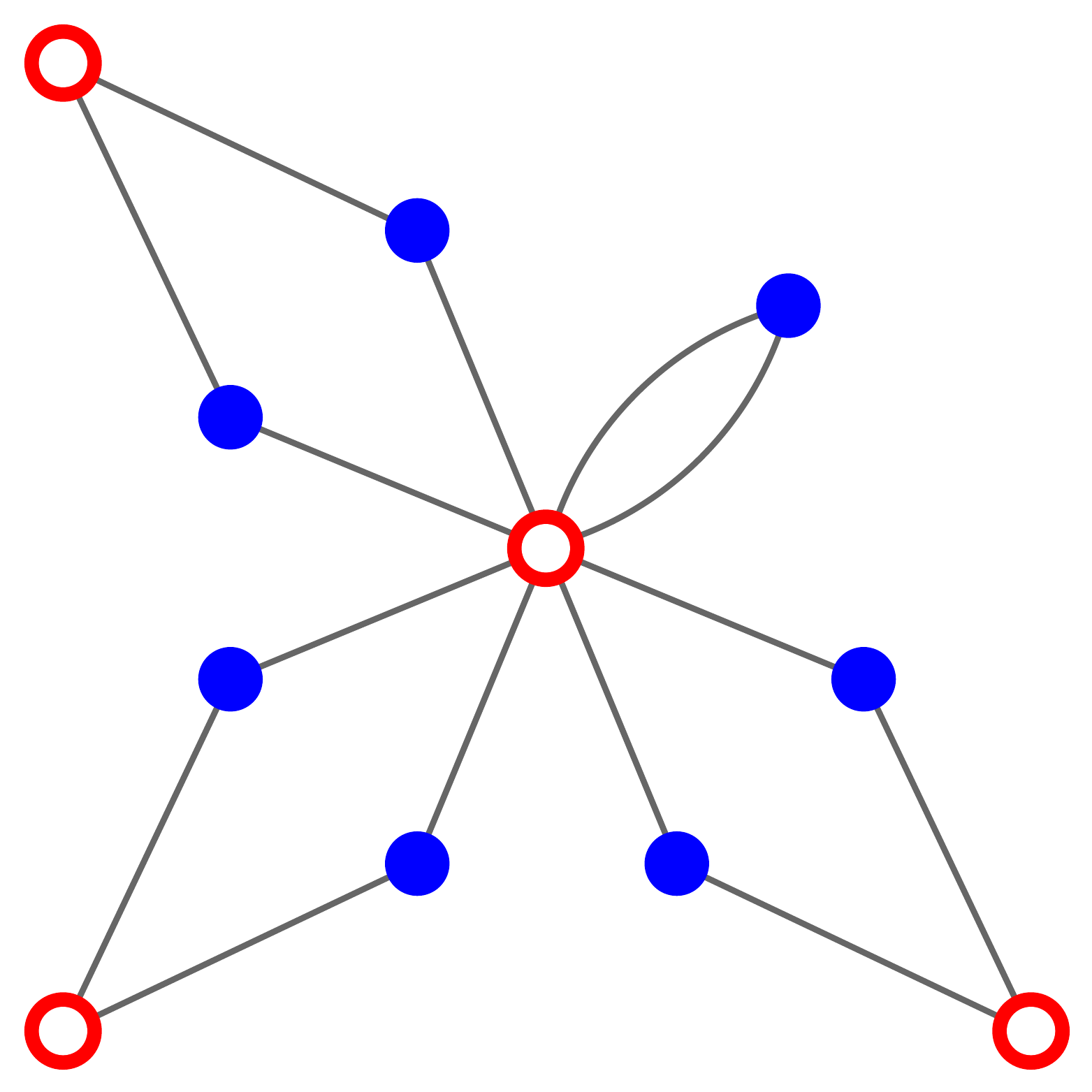}}\ar[dd]
&\raisebox{0cm}{\includegraphics[width=0.9in]{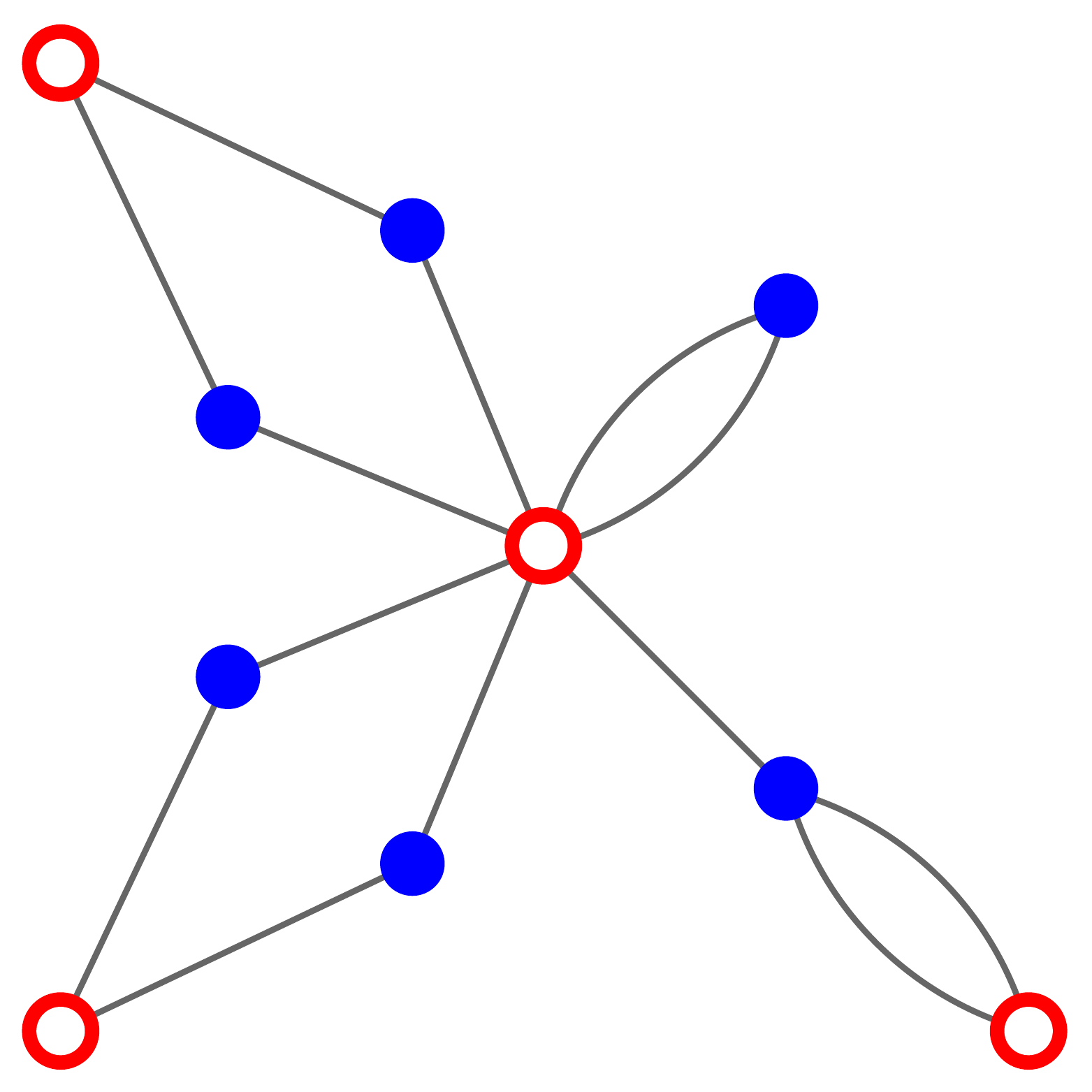}}\ar[dd]
&\raisebox{0cm}{\includegraphics[width=0.9in]{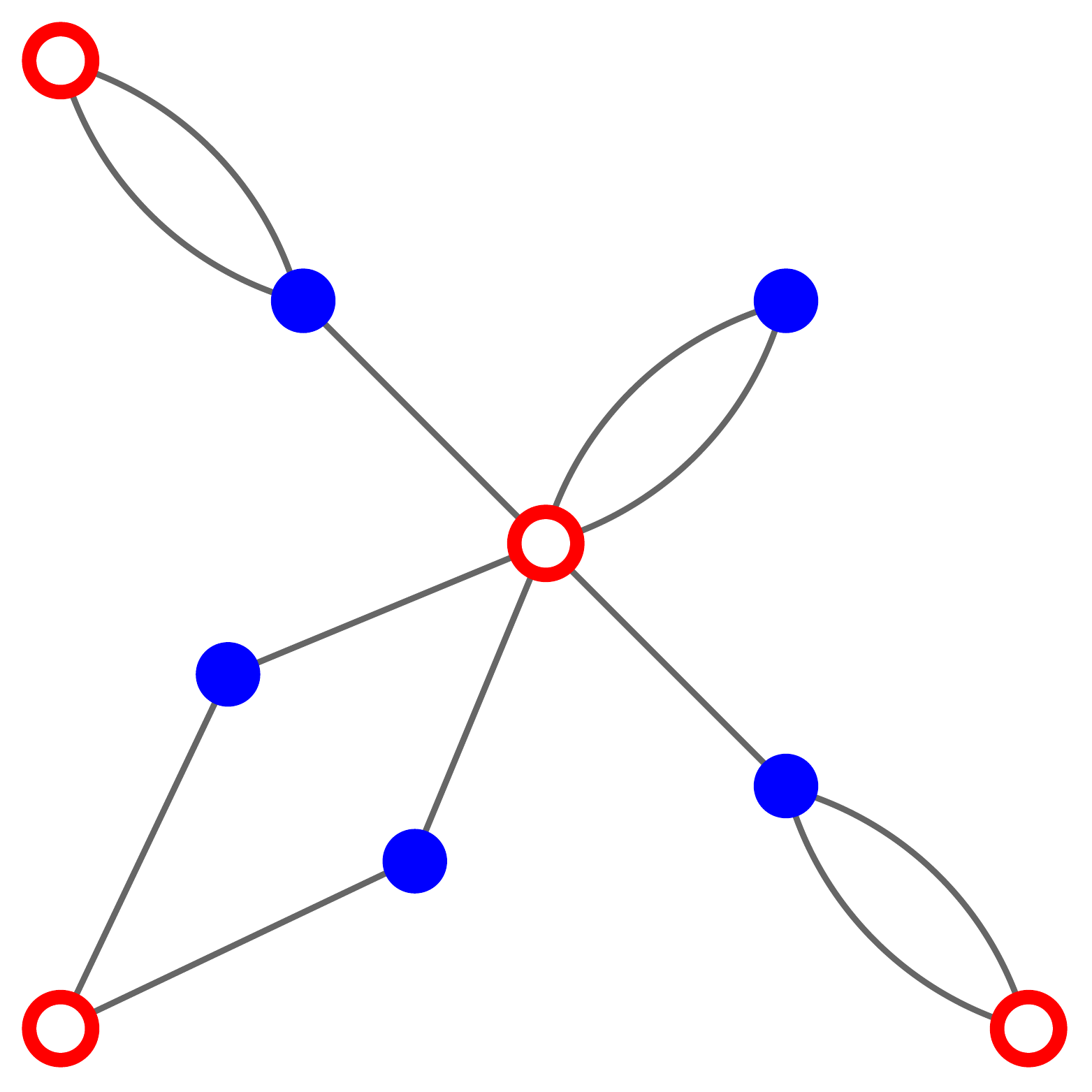}}\ar[dd]
&\raisebox{0cm}{\includegraphics[width=0.9in]{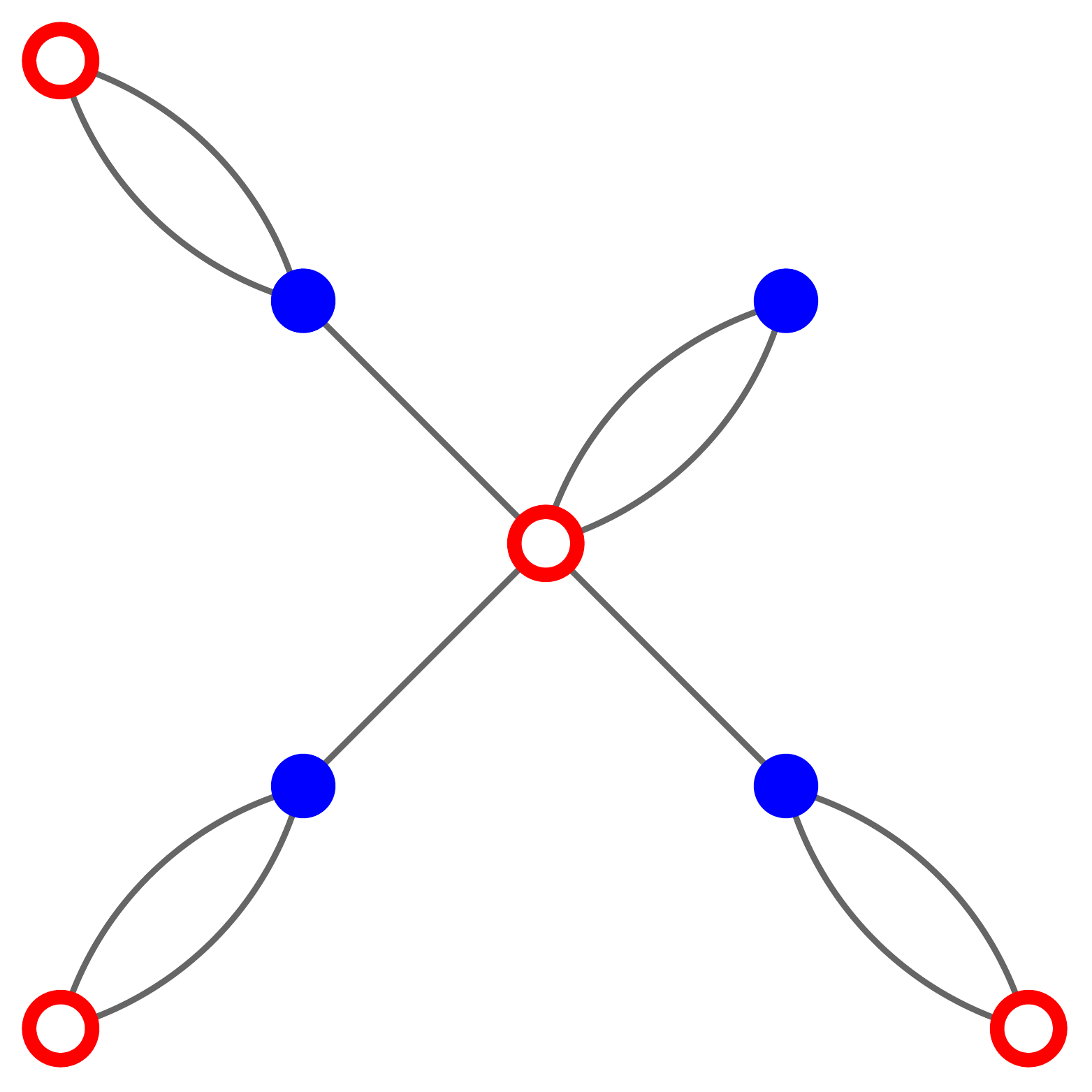}}\ar[dd]
\\\\
&\hskip-16pt\raisebox{1.05cm}{$(5,37)$}\hskip-12pt
&\raisebox{0cm}{\includegraphics[width=0.9in]{TQSplit9.pdf}}
&\raisebox{0cm}{\includegraphics[width=0.9in]{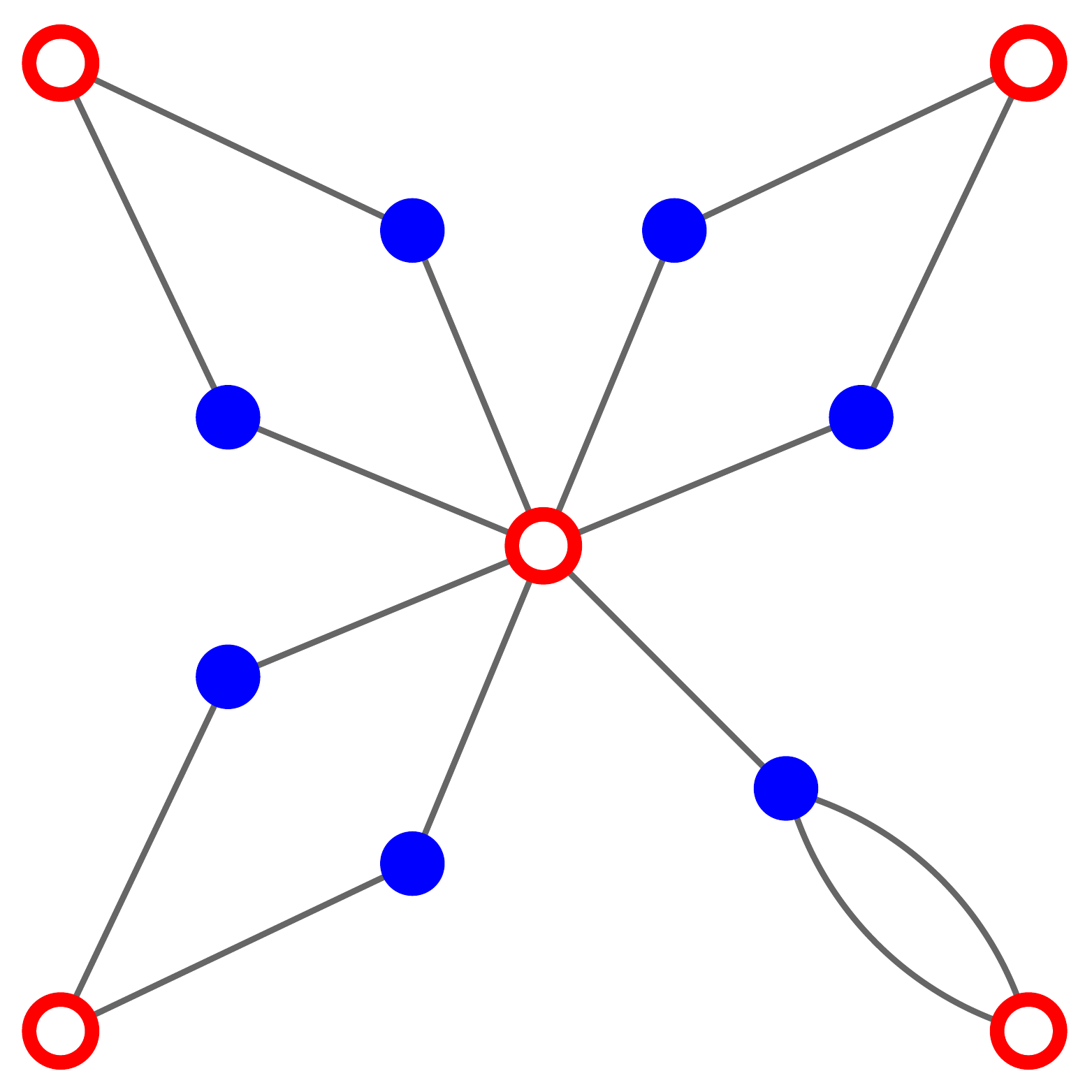}}
&\raisebox{0cm}{\includegraphics[width=0.9in]{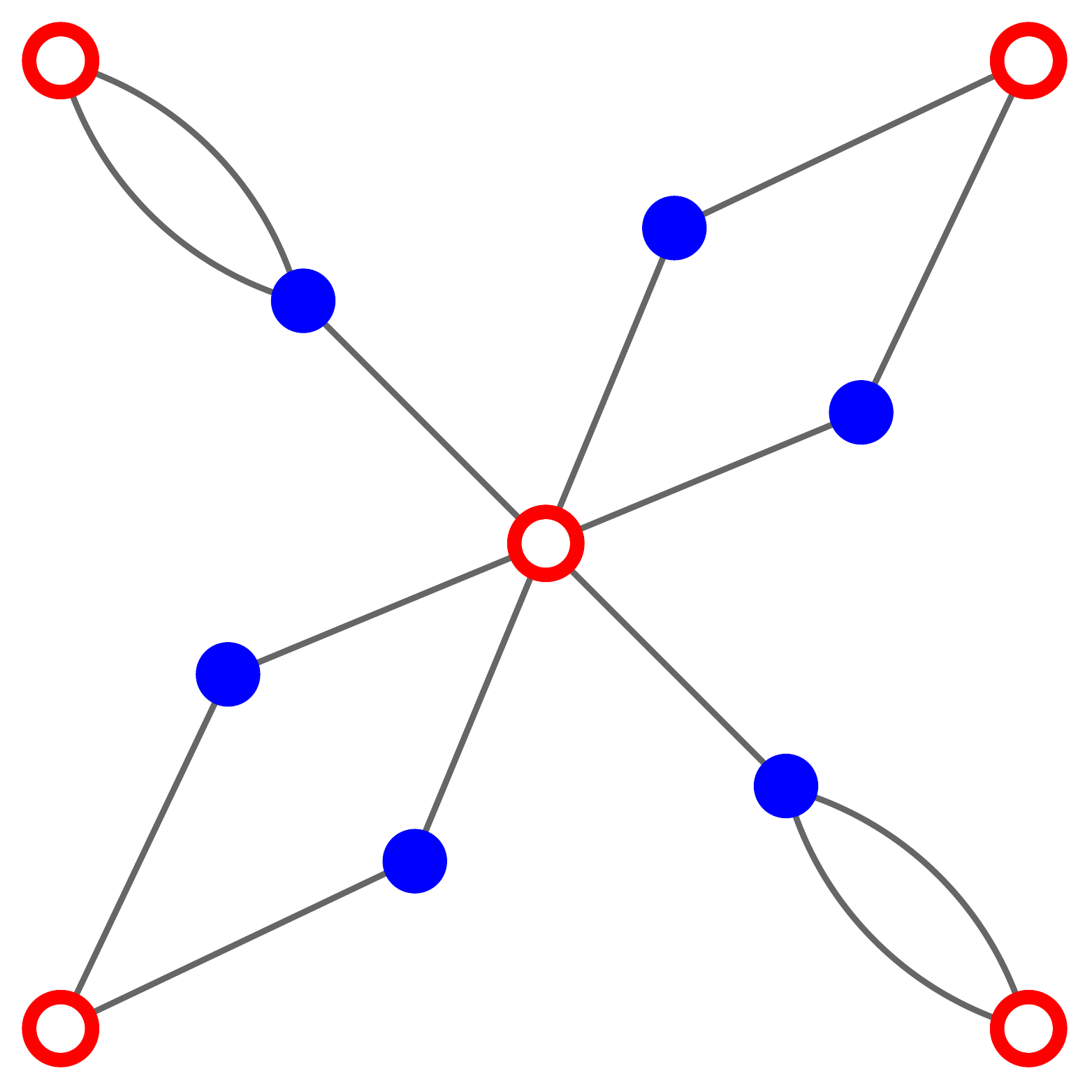}}
&\raisebox{0cm}{\includegraphics[width=0.9in]{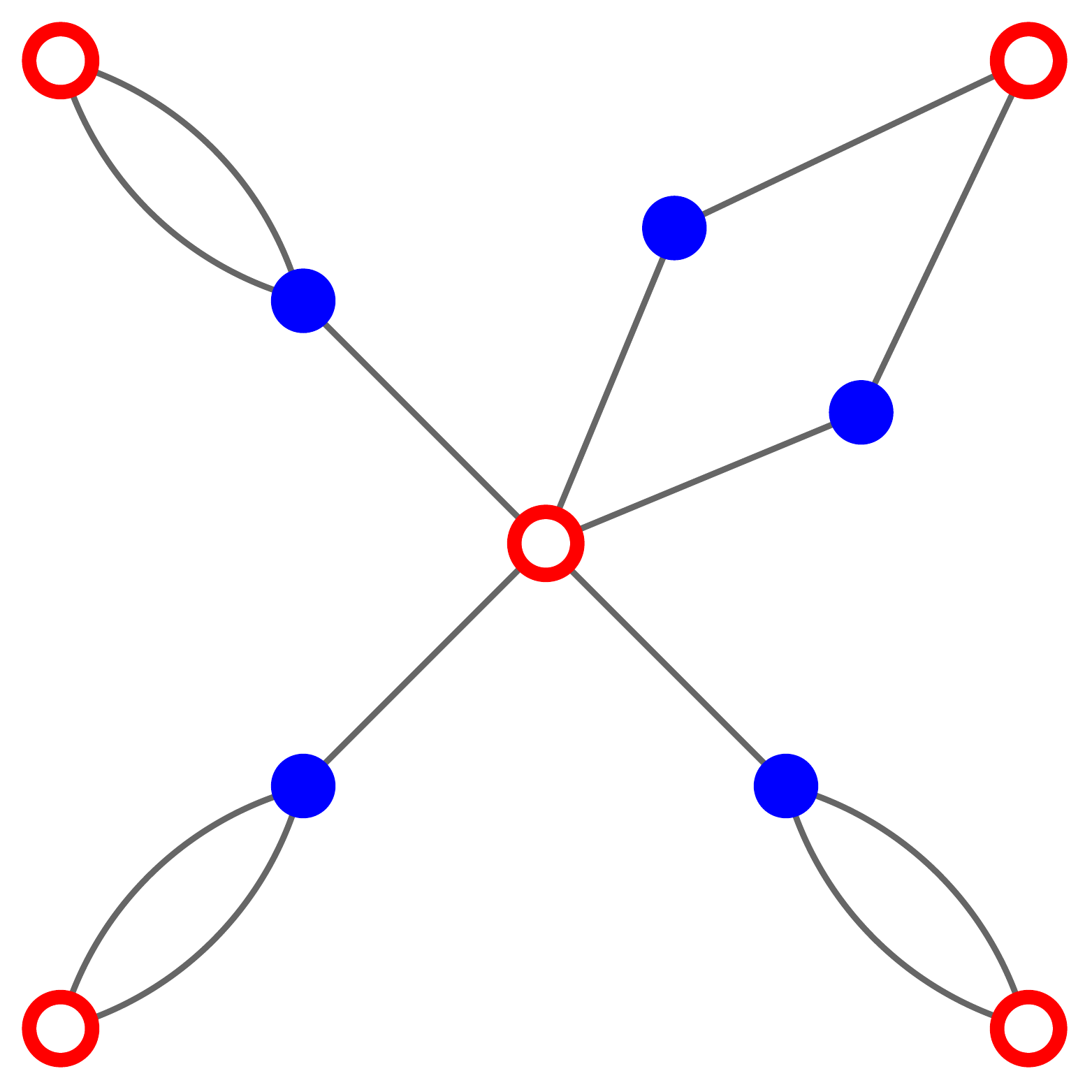}}
&\raisebox{0cm}{\includegraphics[width=0.9in]{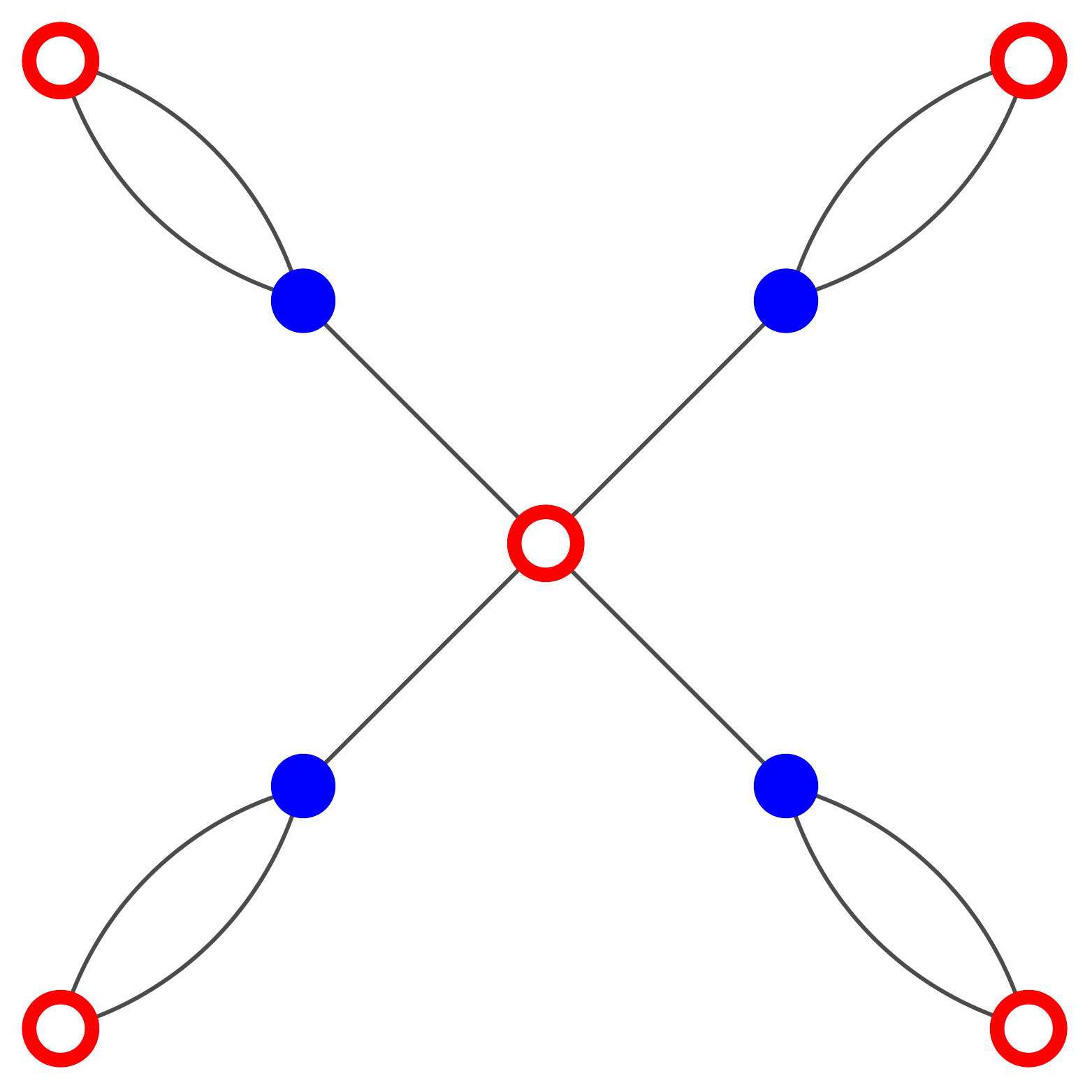}}\hskip10pt
}
\end{minipage}}
\vskip20pt
\capt{6.3in}{FlowerWeb}{\vspace{4pt}Conifold transitions depicting the splits: $X^{1,65}\rightarrow X^{2,58}\rightarrow X^{3,51}\rightarrow X^{4,44}\rightarrow X^{5,37}$. $\Delta (h^{1,1}, h^{2,1})=(1,-7)$ for each transition depicted above. The CICYs in each row are equivalent. }
\end{center}
\end{figure}
\newpage
\subsection{The sequence of splits: $X^{1,65}\rightarrow X^{2,58}\rightarrow X^{3,51}\rightarrow X^{4,44}\rightarrow X^{5,37}$}\label{sec:TransposeTQSeqOfSplits}
\vskip-7pt
\subsubsection{Two favourable splits with Hodge numbers $(2,58)$}\label{TransposeTQSeqOfSplitsX2,58}\vskip-8pt
The first split corresponds to the following configuration:
\begin{equation}\label{eq:TTQSplit1}
\displaycicy{5.25in}{\vrule height40pt depth10pt width0pt
X_{7819}~=~~
\cicy{\IP^1\\ \IP^7}
{ ~ 1 & 1 & 0 & 0 & 0~\\
~1 & 1 & 2& 2 &2 ~ \\}_{-112}^{\,2,\,58}}
{-1.3cm}{1.0in}{TTQSplit1.pdf}
\end{equation}
The manifold \eqref{eq:TTQSplit1} admits two different smooth quotients. The corresponding Hodge numbers are listed in \tref{TTQSplit1quotients}. These can be obtained by either the polynomial deformation method or counting the 
K\"ahler parameters.
\begin{table}[H]
\vspace{8pt}
\begin{center}
\begin{tabular}{| c || c | c |}
\hline
\myalign{| c||}{\varstr{16pt}{10pt}$~~~~~~~  \Gamma ~~~~~~~$ } &
\myalign{m{1.31cm}|}{$~~~ \IZ_2 $} &
\myalign{m{1.9cm}|}{ $~~~\IZ_2{\times}\IZ_2 \ \ \ $ }  
\\ \hline\hline
\varstr{14pt}{8pt} $h^{1,1}(X/\Gamma)$ & 2 & 2 \\
 \hline
\varstr{14pt}{8pt} $h^{2,1}(X/\Gamma)$ & 30 & 16 \\
 \hline
\varstr{14pt}{8pt} $\chi(X/\Gamma)$ & $ \!\!\!\!-56 $ & $ \!\!\!\!- 28$  \\
 \hline
 \end{tabular}
   \vskip 0.3cm
\capt{5in}{TTQSplit1quotients}{Hodge numbers for the quotients of the manifolds \eqref{eq:TTQSplit1} and \eqref{eq:M4}.}
 \end{center}
\vskip-20pt
 \end{table}
Another configuration that we shall show presently corresponds to the same manifold as the configuration above is 
\vspace{-5pt}
\begin{equation}\label{eq:M4}
\displaycicy{5.25in}{\vrule height10pt depth40pt width0pt
X_{7823}~=~~
\cicy{\IP^1\\\IP^6}
{ ~2 & 0& 0& 0\\
   ~1 & 2& 2& 2\\}_{-112}^{\,2,\,58}}
{-1cm}{1.0in}{M4.pdf}
\vspace{-10pt}
\end{equation}
This manifold also admits two different smooth quotients, whose Hodge numbers were computed by counting of K\"ahler parameters. The polynomial deformation method does not lead to a correct count of the complex structure parameters in this case. It is interesting to note that the Hodge numbers of the quotient manifolds are the same as in the case of the manifold~\eqref{eq:TTQSplit1}. We will show next that these manifolds are in fact the same and that this identity extends to an identity between the members of each row of \fref{TTQquotients}. Although the representations \eqref{eq:TTQSplit1} and \eqref{eq:M4} define equivalent CICYs, a symmetry linearly realised in one representation, may not be linearly realised in the other. In the discussion that follows, we will find such examples.

We wish now to show that the two configurations above, \eqref{eq:TTQSplit1} and \eqref{eq:M4} define identical manifolds. Consider first \eqref{eq:M4} and denote the first polynomial by $p$ and the remaining three by $q^i$, 
$i{=}1,2,3$. The polynomial $p$ has the form
\beq\begin{split}
p~&=~\sum_{k=1}^7( A_k\, t_1^2 + 2B_k\, t_1 t_2 + C_k\, t_2^2)\, x_k\\[3pt]
    &=~t_1 \sum_{k=1}^7 (A_k t_1  + B_k t_2)\, x_k + t_2 \sum_{k=1}^7 (B_k t_1  + C_k t_2)\, x_k~, 
\end{split}\notag\eeq
where $t$ and $x$ are coordinates on the $\IP^1$ and $\IP^6$, respectively. Define $x_8$ by the equations
\beq\begin{split}
t_1\, x_8  - \sum_{k=1}^7 (B_k t_1  + C_k t_2)\, x_k~&=~0\\[3pt]
t_2\, x_8 + \sum_{k=1}^7 (A_k t_1  + B_k t_2)\, x_k~&=~0~. 
\end{split}\notag\eeq
The quantity $x_8$ is uniquely defined since the coordinates $(t_1,t_2)$ cannot vanish simultaneously and the equations are consistent in virtue of the equation $p{\,=\,}0$. Note also that if all the $x_k$ vanish, $k{=}1,\ldots,7$, then so does $x_8$. Denoting the two equations above by $(p^1,p^2)$, we see that there is a one-to-one relation between the zero loci of the polynomials $\{ p,\,q^i\}$ and $\{p^1,\,p^2,\,q^i\}$, so taking $(x_k,\,x_8)$ as the coordinates of a $\IP^7$ we have established the equality.

The following generalisation is immediate and relates the configurations of the rows of \fref{FlowerWeb}:
\beq
\cicy{\cA_{n+3-k}\\ \IP^{1\phantom{+1}} \\ \IP^{k+1}}{0 & 0 & M\\ 1 & 1 & \bf 0\\ 1 & 1 & \bf d}~=~
\cicy{\cA_{n+3-k}\\ \IP^1 \\ \IP^{k}}{0 & M\\ 2 & \bf 0\\ 1 & \bf d}~\raisebox{-15pt}{,}
\eeq
here $\cA_{n+3-k}$ is an ambient space of the indicated dimension, $M$ is a matrix and $\bf d$ is a degree vector with $n$ components.

\subsubsection{Three favourable splits with Hodge numbers $(3,51)$}\label{TransposeTQSeqOfSplitsX3,51}
\vskip-10pt
The second split corresponds to the following configuration:
\vspace{-3pt}
\begin{equation}\label{eq:TTQSplit2}
\displaycicy{5.25in}{
X_{7745}~=~~
\cicy{\IP^1\\ \IP^1\\ \IP^7}
{ ~ 1 & 1 & 0 & 0& 0 & 0 ~\\
  ~ 0 & 0 & 1 & 1&0 & 0 ~\\
  ~ 1 & 1 & 1 & 1& 2& 2~ \\}_{-96}^{\,3,\,51}}{-1.5cm}{1.25in}{TTQSplit2v2.pdf}
\vspace{-3pt}
\end{equation}
The manifold \eqref{eq:TTQSplit2} admits five different free group actions. The Hodge numbers for the resulting quotients are listed in \tref{TTQSplit2quotients}, the computation of which are amenable to both the polynomial deformation method and the counting of K\"ahler parameters.

\begin{table}[H]
\vspace{4pt}
\begin{center}
\begin{tabular}{| c || c | c | c | c |}
\hline
\myalign{| c||}{\varstr{16pt}{10pt}$~~~~~~~  \Gamma ~~~~~~~$ } &
\myalign{m{1.31cm}|}{$~~~ \IZ_2 $} &
\myalign{m{1.31cm}|}{$~~~ \IZ_4 $} &
\myalign{m{1.9cm}|}{ $~~~\IZ_2\times\IZ_2 \ \ \ $ } &
\myalign{m{1.9cm}|}{ $~~~\IZ_2\times\IZ_4 \ \ \ $ }  
\\ \hline\hline
\varstr{14pt}{8pt} $h^{1,1}(X/\Gamma)$ & 3 & 2 & 3 & 2\\
 \hline
\varstr{14pt}{8pt} $h^{2,1}(X/\Gamma)$ & 27 & 14 & 15 & 8\\
 \hline
\varstr{14pt}{8pt} $\chi(X/\Gamma)$ & $ \!\!\!\!-48 $ & $ \!\!\!\!- 24$ & $ \!\!\!\!- 24$ & $ \!\!\!\!- 12$  \\
 \hline
 \end{tabular}
   \vskip 0.3cm
\capt{5in}{TTQSplit2quotients}{Hodge numbers for the quotients of the manifolds \eqref{eq:TTQSplit2} and \eqref{eq:M5}.}
 \end{center}
 \vspace{-25pt}
 \end{table}
There are two other favourable embeddings of the manifold described by \eqref{eq:TTQSplit2} that we discuss now. The first one corresponds to the following configuration:
\vspace{-10pt}
\begin{equation}\label{eq:M4Split1}
\displaycicy{5.25in}{\vrule height40pt depth20pt width0pt
X_{7714}~=~~
\cicy{\IP^1\\\IP^1 \\\IP^6}
{ ~0& 0& 0& 0& 2~\\
  ~0& 0& 1& 1& 0~\\
  ~2& 2& 1& 1& 1~\\}_{-96}^{\,3,\,51}}
{-1.5cm}{1.25in}{M4Split1v2.pdf}
\vspace{-3pt}
\end{equation}
This manifold admits two different smooth quotients, whose Hodge numbers are listed in \tref{M4Split1quotients}. We can make use of only the counting of K\"ahler parameters for this favourable embedding. 
\begin{table}[H]
\begin{center}
\begin{tabular}{| c || c | c |}
\hline
\myalign{| c||}{\varstr{16pt}{10pt}$~~~~~~~  \Gamma ~~~~~~~$ } &
\myalign{m{1.31cm}|}{$~~~ \IZ_2 $} &
\myalign{m{1.9cm}|}{ $~~~\IZ_2\times\IZ_2 \ \ \ $ }  
\\ \hline\hline
\varstr{14pt}{8pt} $h^{1,1}(X/\Gamma)$ & 3 & 3 \\
 \hline
\varstr{14pt}{8pt} $h^{2,1}(X/\Gamma)$ & 27 & 15 \\
 \hline
\varstr{14pt}{8pt} $\chi(X/\Gamma)$ & $ \!\!\!\!-48 $ & $ \!\!\!\! -24$  \\
 \hline
 \end{tabular}
   \vskip 0.3cm
\capt{4.5in}{M4Split1quotients}{Hodge numbers for the quotients of the manifold \eqref{eq:M4Split1}.}
 \end{center}
 \vspace{-25pt}
 \end{table}
 
The second equivalent representation of the manifold \eqref{eq:TTQSplit2} is given by the embedding:
\vspace{0pt}
\begin{equation}\label{eq:M5}
\displaycicy{5.25in}{\vrule height40pt depth20pt width0pt
X_{7735}~=~~
\cicy{\IP^1\\\IP^1\\\IP^5}
{ ~2& 0& 0& 0~ \\
  ~0& 2& 0& 0~ \\
  ~1& 1& 2& 2~ \\}_{-96}^{\,3,\,51}}
{-1.5cm}{1.25in}{M5v2.pdf}
\end{equation}
\vskip5pt
The manifold \eqref{eq:M5} admits $8$ different free group actions by $\IZ_2, \IZ_4, \IZ_2\times \IZ_2$ and $\IZ_2\times \IZ_4$. The Hodge numbers for the resulting quotients are identical to those in \tref{TTQSplit2quotients}, obtained for the manifold~\eqref{eq:TTQSplit2}. In order to compute the Hodge numbers, we count the K\"ahler parameters, since the polynomial deformation cannot be expected to give correct answers, in this case, owing to the fact that it does not count correctly the complex structure parameters of the covering space and, moreover, the diagram for the configuration is 1-leg decomposable.

\subsubsection{Four favourable splits with Hodge numbers $(4,44)$}\label{TransposeTQSeqOfSplitsX4,44}
The third split corresponds to the transition from the third to the fourth row in Figure \ref{FlowerWeb}. The first CICY in this set of four equivalent CICYs are given by the following configuration:
\begin{equation}\label{eq:TTQSplit3}
\displaycicy{5.25in}{
X_{7435}~=~~
\cicy{\IP^1\\ \IP^1\\ \IP^1\\ \IP^7}
{  ~ 1 & 1 & 0 & 0 & 0 & 0& 0 ~\\
  ~ 0 & 0 & 1 & 1 & 0 & 0& 0~\\
  ~0 & 0 & 0 & 0 & 1 & 1& 0~\\
  ~1 & 1 & 1 & 1 & 1 & 1& 2~ \\}_{-80}^{\,4,\,44}}{-1.5cm}{1.25in}{TTQSplit3.pdf}
\end{equation}
\vspace{4pt}

The four CICYs appearing in this class admit smooth quotients by $\IZ_2$ and $\IZ_2\times\IZ_2$. The Hodge numbers of the quotients are the same for each manifold and are shown in \tref{TTQSplit3quotients}. Both the polynomial deformation method and the counting of K\"ahler parameters can be employed to compute the Hodge numbers listed in \tref{TTQSplit3quotients} for the manifold \eqref{eq:TTQSplit3}.

The remaining three equivalent descriptions of \eqref{eq:TTQSplit3} are given by the  configuration matrices and diagrams below. These manifolds admit respectively 2, 4 and 19 different group actions by $\IZ_2$ and $\IZ_2\times\IZ_2$. Their Hodge numbers are listed in \tref{TTQSplit3quotients}.

 \begin{equation}\label{eq:M4Split2}
\hskip -27pt \displaycicy{5.25in}{
X_{7522}~=~~
\cicy{\IP^1\\\IP^1\\\IP^1 \\\IP^6}
{ ~0& 0& 0& 0&0 &  2~ \\
  ~0& 0& 0& 1& 1& 0\\
  ~0& 1& 1& 0& 0& 0\\
  ~2& 1& 1& 1& 1& 1~ \\}_{-80}^{\,4,\,44}}{-1.5cm}{1.25in}{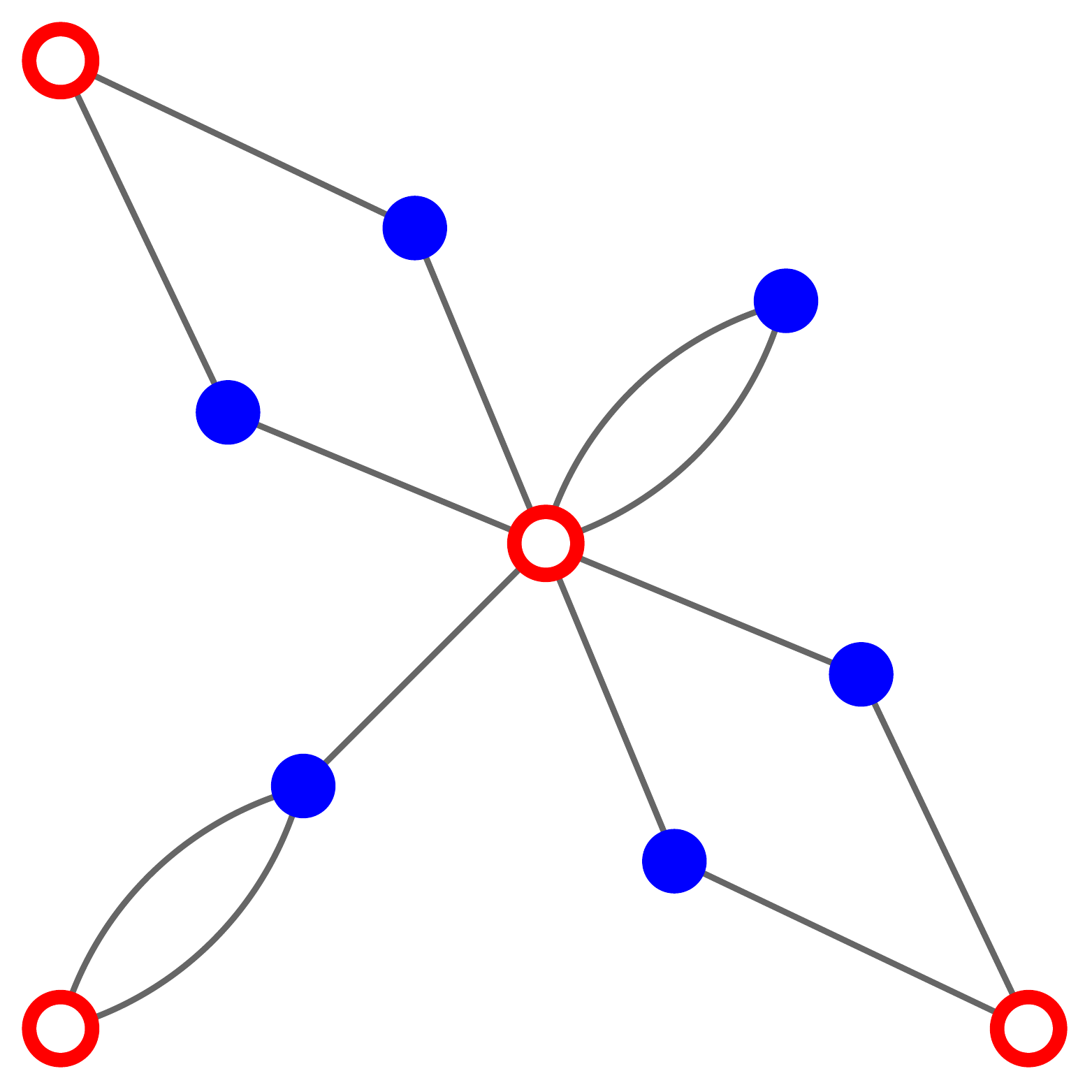}
\end{equation}

\begin{equation}\label{eq:M7}
\displaycicy{5.25in}{
X_{7462}~=~~
\cicy{\IP^1\\\IP^1\\\IP^1\\\IP^5}
{ ~2& 0& 0& 0& 0~ \\
  ~0& 2& 0& 0& 0~ \\
  ~0& 0& 1& 1& 0~ \\
  ~1& 1& 1& 1& 2~ \\}_{-80}^{\,4,\,44}}{-1.5cm}{1.25in}{M7.pdf}
\end{equation}

\begin{equation}\label{eq:M6}
\displaycicy{5.25in}{
X_{7491}~=~~
\cicy{\IP^1\\\IP^1\\\IP^1\\\IP^4}
{ ~2& 0& 0& 0~ \\
  ~0& 2& 0& 0~ \\
  ~0& 0& 2& 0~ \\
  ~1& 1& 1& 2~ \\}_{-80}^{\,4,\,44}}{-1.5cm}{1.25in}{M6.pdf}
\end{equation}

\begin{table}[H]
\vspace{10pt}
\begin{center}
\begin{tabular}{| c || c | c |}
\hline
\myalign{| c||}{\varstr{16pt}{10pt}$~~~~~~~  \Gamma ~~~~~~~$ } &
\myalign{m{1.31cm}|}{$~~~ \IZ_2 $} &
\myalign{m{1.9cm}|}{ $~~~\IZ_2\times\IZ_2 \ \ \ $ }  
\\ \hline\hline
\varstr{14pt}{8pt} $h^{1,1}(X/\Gamma)$ & 4 & 4 \\
 \hline
\varstr{14pt}{8pt} $h^{2,1}(X/\Gamma)$ & 24 & 14 \\
 \hline
\varstr{14pt}{8pt} $\chi(X/\Gamma)$ & $ \!\!\!\!-40 $ & $ \!\!\!\!- 20$  \\
 \hline
 \end{tabular}
   \vskip 0.3cm
\capt{6in}{TTQSplit3quotients}{Hodge numbers for the quotients of the manifolds \eqref{eq:TTQSplit3}, \eqref{eq:M4Split2}, \eqref{eq:M7} and \eqref{eq:M6}.}
 \end{center}
 \vspace{-15pt}
 \end{table}
\subsubsection{Five favourable splits with Hodge numbers $(5,37)$}\label{TransposeTQSeqOfSplitsX5,37}\vskip-8pt
The fourth split of the manifold $X^{1,65}$ is the manifold \eqref{eq:TQSplit9} with Hodge numbers $(5,37)$. This CICY has four other equivalent embeddings. The first CICY corresponds to the following configuration:
\begin{equation}\label{eq:TQSplit5}
\displaycicy{5.25in}{
X_{6836}~=~~
\cicy{\IP^1 \\ \IP^1\\ \IP^1\\ \IP^1\\ \IP^3}
{ ~2 &0 & 0 & 0 ~\\
  ~0 &2 & 0 & 0 ~\\
  ~0 &0 & 2 & 0~\\
  ~0 &0 & 0 & 2 ~\\
  ~1 &1 & 1 & 1~\\}_{-64}^{5,37}}{-1.5cm}{1.25in}{TQSplit5.pdf}
\end{equation}
The manifold \eqref{eq:TQSplit5} admits an impressive number of $117$ different free group actions \cite{Braun:2010vc}. However, the Hodge numbers for the quotients corresponding to a fixed group are all the same. We summarise this information in Table~\ref{TQSplit5quotients}. The polynomial deformation method does not reproduce correctly the number of complex structure parameters for the covering manifold~$X^{5,37}$, and we cannot use it for computing $h^{2,1}(X/\Gamma)$. On the other hand, the embedding~\eqref{eq:TQSplit5} is favourable; as such, we are able to count the number of linearly independent K\"ahler forms for each quotient and then infer~$h^{2,1}(X/\Gamma)$ from the Euler number. 

\begin{table}[H]
\vspace{0pt}
\begin{center}
\begin{tabular}{l}
\begin{tabular}{| c || c | c | c | c | c | c | c | c |}
\hline
\myalign{| c||}{\varstr{16pt}{10pt}$~~~~~~ \Gamma ~~~~~~$ } &
\myalign{m{1.28cm}|}{$~~~ \IZ_2 $} &
\myalign{m{1.28cm}|}{$~~~ \IZ_4 $} &
\myalign{m{1.28cm}|}{ $~\IZ_2{\times}\IZ_2 \ \ \ $ }  &
\myalign{m{1.28cm}|}{$~~~ \IZ_8 $} &
\myalign{m{1.28cm}|}{ $~\IZ_2{\times}\IZ_4 \ \ \ $ }  &
\myalign{m{1.28cm}|}{$~~~ \IQ_8 $} &
\myalign{m{1.28cm}|}{ $~\IZ_4{\rtimes}\IZ_4 \ \ \ $ }  &
\myalign{m{1.28cm}|}{ $~\IZ_2{\times}\IZ_8 \ \ \ $ } 
\\ \hline\hline
\varstr{14pt}{8pt} $h^{1,1}(X/\Gamma)$ & 5 & 3 & 5 & 2 & 3 & 2 & 2 & 2\\
 \hline
\varstr{14pt}{8pt} $h^{2,1}(X/\Gamma)$ & 21 & 11 & 13 & 6 & 7 & 6 & 4 & 4\\
 \hline
\varstr{14pt}{8pt} $\chi(X/\Gamma)$ & $\!\!\!\!-32$ & $\!\!\!\!-16$ & $\!\!\!\!-16$ & $\!\!\!\!-8$  & $\!\!\!\!-8$ & $\!\!\!\!-8$ & $\!\!\!\!-4$ & $\!\!\!\!-4$ \\
 \hline
 \end{tabular}
 \end{tabular}
 \vskip 0.3cm
\capt{5.5in}{TQSplit5quotients}{Hodge numbers for the quotients of the manifolds \eqref{eq:TQSplit5} and \eqref{eq:TQSplit9}.}
 \end{center}
 \vspace{-20pt}
 \end{table}

The second CICY in this class of $(5,37)$ manifolds corresponds to the following configuration:
\begin{equation}\label{eq:TQSplit6}
\displaycicy{5.25in}{
X_{6788}~=~~
\cicy{\IP^1 \\ \IP^1\\ \IP^1\\ \IP^1\\ \IP^4}
{ ~2 &0 & 0 & 0 & 0 ~\\
  ~0 &2 & 0 & 0 & 0~\\
  ~0 &0 & 2 & 0 & 0~\\
  ~0 &0 & 0 & 1 & 1 ~\\
  ~1 &1 & 1 & 1& 1~\\}_{-64}^{5,37}}{-1.5cm}{1.25in}{TQSplit6.pdf}
\end{equation}

The manifold \eqref{eq:TQSplit6} admits $12$ different free actions. Similar comments to the previous manifold apply here as well. The Hodge numbers for the quotients are listed in \tref{TQSplit6quotients}.
\begin{table}[H]
\vspace{12pt}
\begin{center}
\begin{tabular}{| c || c | c |}
\hline
\myalign{| c||}{\varstr{16pt}{10pt}$~~~~~~~  \Gamma ~~~~~~~$ } &
\myalign{m{1.31cm}|}{$~~~ \IZ_2 $} &
\myalign{m{1.9cm}|}{ $~~~\IZ_2\times\IZ_2 \ \ \ $ }   
\\ \hline\hline
\varstr{14pt}{8pt} $h^{1,1}(X/\Gamma)$ & 5 & 5 \\
 \hline
\varstr{14pt}{8pt} $h^{2,1}(X/\Gamma)$ & 21 & 13 \\
 \hline
\varstr{14pt}{8pt} $\chi(X/\Gamma)$ & $\!\!\!\!-32$ & $\!\!\!\!-16$ \\
 \hline
 \end{tabular}
  \vskip 0.3cm
\capt{5.5in}{TQSplit6quotients}{Hodge numbers for the quotients of the manifolds \eqref{eq:TQSplit6} and \eqref{eq:TQSplit8}.}
 \end{center}
 \vspace{-12pt}
 \end{table}

\vspace{12pt}

The third CICY in this class of $(5,37)$ manifolds corresponds to the following configuration:
\begin{equation}\label{eq:TQSplit7}
\displaycicy{5.25in}{
X_{6927}~=~~
\cicy{\IP^1 \\ \IP^1\\ \IP^1\\ \IP^1\\ \IP^5}
{ ~2 &0 & 0 & 0 & 0 & 0 ~\\
  ~0 &2 & 0 & 0 & 0 & 0~\\
  ~0 &0 & 1 & 1 & 0 & 0~\\
  ~0 &0 & 0 & 0 &1 & 1 ~\\
  ~1 &1 & 1 & 1 & 1& 1~\\}_{-64}^{5,37}}{-1.5cm}{1.25in}{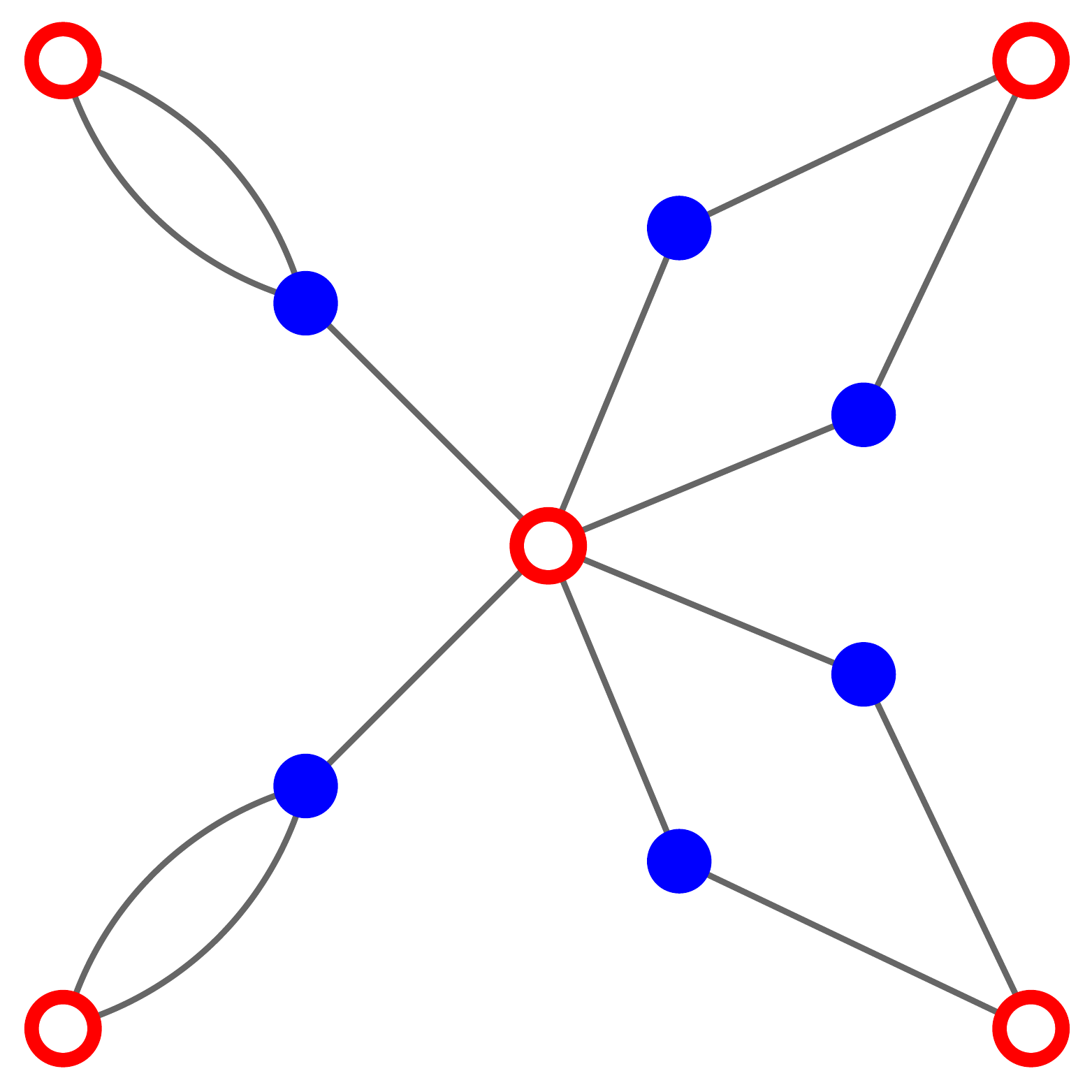}
\end{equation}

The manifold \eqref{eq:TQSplit7} admits $8$ different free actions. As before, we compute the Hodge number for the different quotients by making use of the favourable embedding, as listed in \tref{TQSplit7quotients}.

\begin{table}[!ht]
\vspace{12pt}
\begin{center}
\begin{tabular}{| c || c | c | c | c |}
\hline
\myalign{| c||}{\varstr{16pt}{10pt}$~~~~~~~  \Gamma ~~~~~~~$ } &
\myalign{m{1.31cm}|}{$~~~ \IZ_2 $} &
\myalign{m{1.31cm}|}{$~~~ \IZ_4 $} &
\myalign{m{1.9cm}|}{ $~~~\IZ_2\times\IZ_2 \ \ \ $ }  &
\myalign{m{1.9cm}|}{ $~~~\IZ_2\times\IZ_4 \ \ \ $ }  
\\ \hline\hline
\varstr{14pt}{8pt} $h^{1,1}(X/\Gamma)$ & 5 & 3 & 5 & 3 \\
 \hline
\varstr{14pt}{8pt} $h^{2,1}(X/\Gamma)$ & 21 & 11 & 13  & 7 \\
 \hline
\varstr{14pt}{8pt} $\chi(X/\Gamma)$ & $\!\!\!\!-32$ & $\!\!\!\!-16$ & $\!\!\!\!-16$ & $\!\!\!\!-8$  \\
 \hline
 \end{tabular}
  \vskip 0.3cm
\capt{4.5in}{TQSplit7quotients}{Hodge numbers for the quotients of the manifold \eqref{eq:TQSplit7}.} 
\end{center}
\vskip-20pt
 \end{table}

The fourth CICY corresponds to the following configuration:
\begin{equation}\label{eq:TQSplit8}
\displaycicy{5.25in}{
X_{6715}~=~~
\cicy{\IP^1 \\ \IP^1\\ \IP^1\\ \IP^1\\ \IP^6}
{ ~2 &0&0 & 0 & 0 & 0 & 0 ~\\
  ~0 &1&1 & 0 & 0 & 0 & 0~\\
  ~0 &0 &0& 1 & 1 & 0 & 0~\\
  ~0 &0 &0& 0 & 0 &1 & 1 ~\\
  ~1 &1 &1& 1 & 1 & 1& 1~\\}_{-64}^{5,37}}{-1.5cm}{1.25in}{TQSplit8.pdf}
\end{equation}

The manifold \eqref{eq:TQSplit8} admits only $2$ different free actions. The Hodge numbers for the resulting quotients are identical to those of the manifold~\eqref{eq:TQSplit6}, see \tref{TQSplit6quotients}.

 \vspace{8pt}
The final CICY in this class of $(5,37)$ manifolds corresponds to the following configuration:
\begin{equation}\label{eq:TQSplit9}
\displaycicy{5.25in}{
X_{6947}~=~~
\cicy{\IP^1 \\ \IP^1\\ \IP^1\\ \IP^1\\ \IP^7}
{ ~1 &1&0&0 & 0 & 0 & 0 & 0 ~\\
  ~0 &0&1&1 & 0 & 0 & 0 & 0~\\
  ~0 &0&0 &0& 1 & 1 & 0 & 0~\\
  ~0 &0&0 &0& 0 & 0 &1 & 1 ~\\
  ~1 &1&1 &1& 1 & 1 & 1& 1~\\}_{-64}^{5,37}}{-1.5cm}{1.25in}{TQSplit9.pdf}
\end{equation}

The manifold \eqref{eq:TQSplit9} admits $9$ different free actions by the same groups as for the manifold~\eqref{eq:TQSplit5}. The resulting quotients have Hodge numbers that are identical to those obtained in the previous discussion, see \tref{TQSplit5quotients}. Note, however, that unlike for the manifold~\eqref{eq:TQSplit5}, in this case the polynomial deformation method can be applied and agrees with the counting of K\"ahler parameters. 
\newpage
\section{The remaining sources, and their descendants}\label{sec:RemainingSources}
\vskip-10pt
\subsection{The manifold $X^{3,43}$ and its splits}
\vskip-10pt
\subsubsection{The manifold $X^{3,43}$ and its smooth quotients}\label{REM_X3,43_1}
A third sequence of splits starts with the manifold defined by the following configuration:
\begin{equation}\label{eq:M3}
\displaycicy{5.25in}{
X_{7484}~=~~
\cicy{\IP^1\\\IP^1 \\\IP^3}
{ ~2& 0~ \\
  ~0& 2~ \\
  ~2& 2~ \\}_{-80}^{\,3,\,43}}
{-0.10cm}{1.8in}{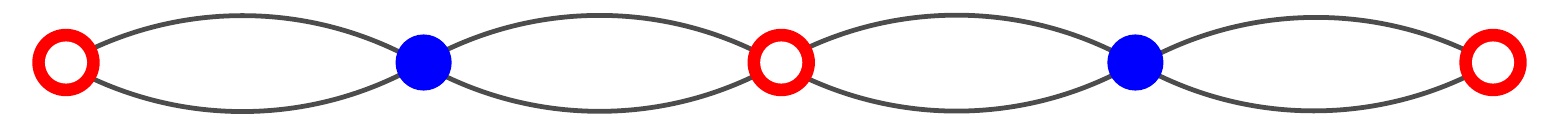}
\end{equation}
The manifold $X^{3,43}$ admits three different smooth quotients, the details of which are listed in \tref{M3quotients}. For this manifold, the polynomial deformation method does not reproduce correctly the number of complex structure parameters, and as such, it cannot be used in order to compute $h^{2,1}(X/\Gamma)$. However, we can exploit the fact that the embedding \eqref{eq:M3} is favourable and hence compute the number of K\"ahler parameters invariant under each group action. 
\begin{table}[H]
\begin{center}
\begin{tabular}{| c || c | c | c |}
\hline
\myalign{| c||}{\varstr{16pt}{10pt}$~~~~~~~  \Gamma ~~~~~~~$ } &
\myalign{m{1.31cm}|}{$~~~ \IZ_2 $} &
\myalign{m{1.31cm}|}{$~~~ \IZ_4 $} &
\myalign{m{1.9cm}|}{ $~~~\IZ_2\times\IZ_2 \ \ \ $ }  
\\ \hline\hline
\varstr{14pt}{8pt} $h^{1,1}(X/\Gamma)$ & 3 & 2 & 3 \\
 \hline
\varstr{14pt}{8pt} $h^{2,1}(X/\Gamma)$ & 23 & 12  & 13\\
 \hline
\varstr{14pt}{8pt} $\chi(X/\Gamma)$ & $ \!\!\!\!-40 $ & $ \!\!\!\!- 20$& $ \!\!\!\!- 20$  \\
 \hline
 \end{tabular}
   \vskip 0.3cm
\capt{4.5in}{M3quotients}{Hodge numbers for the quotients of the manifold \eqref{eq:M3}.}
 \end{center}
 \end{table}

\subsubsection{Two favourable splits with Hodge numbers $(4,36)$}\label{REM_X3,43_2}
The first split corresponds to the following configuration:
\begin{equation}\label{eq:M3Split1}
\displaycicy{5.25in}{
X_{6784}~=~~
\cicy{\IP^1\\\IP^1 \\\IP^1 \\\IP^3}
{ ~2& 0 & 0~ \\
  ~0& 1 & 1~ \\
  ~0& 0& 2~\\
  ~2& 1 & 1~ \\}_{-64}^{\,4,\,36}}{-1.1cm}{1.8in}{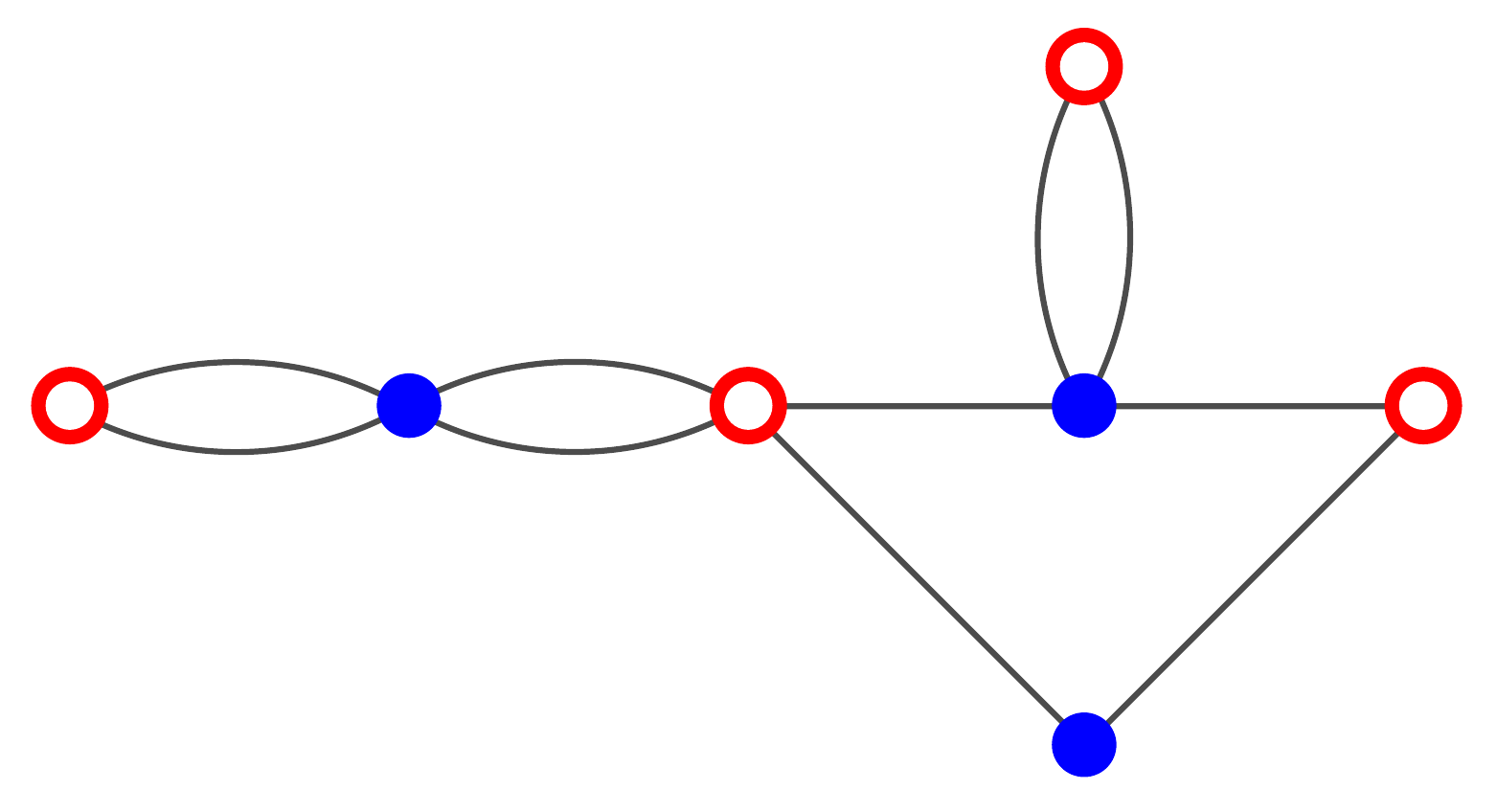}
\end{equation}
The manifold \eqref{eq:M3Split1} admits six different free group actions. Also for this case, the polynomial deformation method does not yield the correct number of complex structure parameters. We use instead the counting of K\"ahler parameters, which leads to the Hodge numbers presented in \tref{M3Split1quotients}.

The second split corresponds to the following configuration:
\begin{equation}\label{eq:M3Split2}
\displaycicy{5.25in}{
X_{6828}~=~~
\cicy{\IP^1\\\IP^1 \\\IP^1 \\\IP^3}
{ ~2& 0 & 0~ \\
  ~0& 1 & 1~ \\
  ~0& 1& 1~\\
  ~2& 1 & 1~ \\}_{-64}^{\,4,\,36}}{-1.0cm}{1.8in}{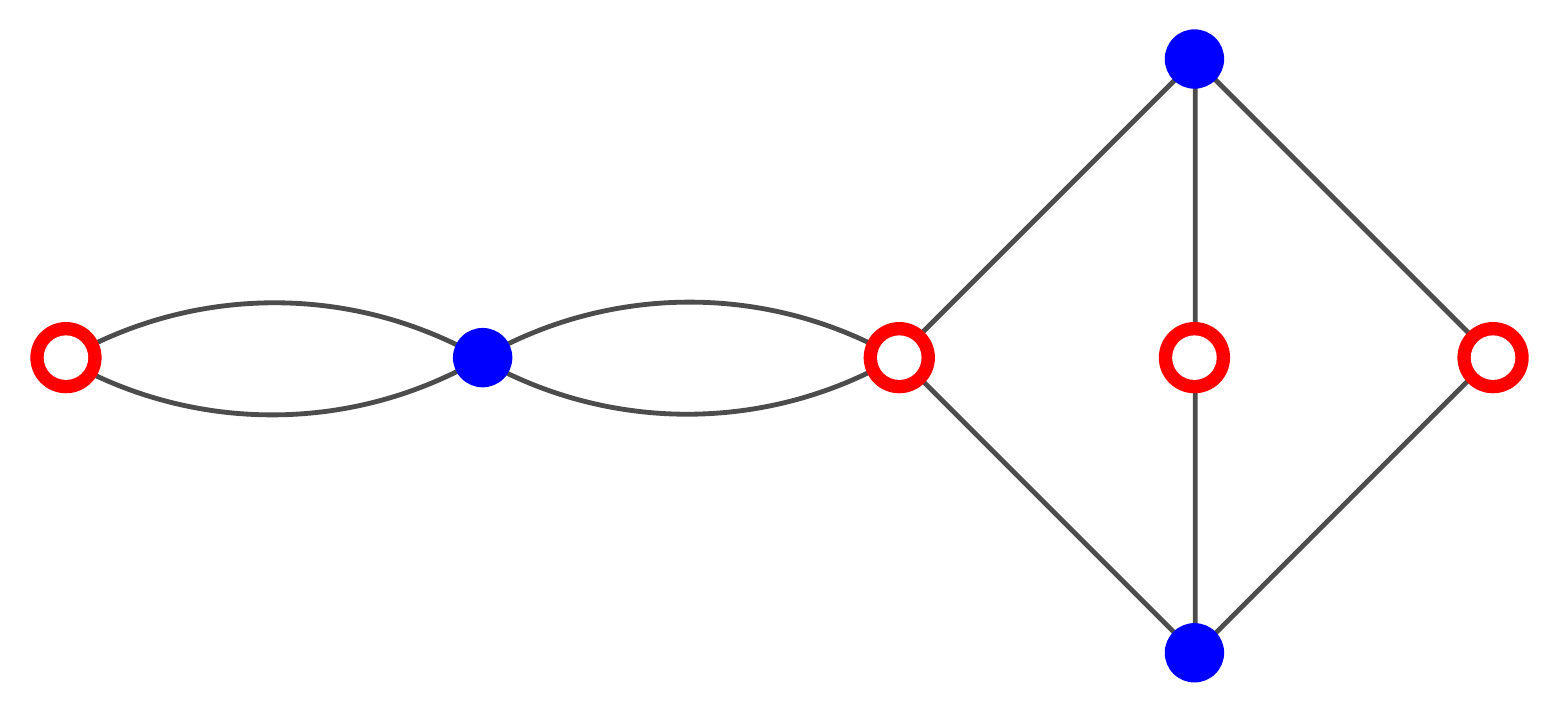}
\end{equation}

The manifold \eqref{eq:M3Split2} admits two smooth quotients, having the same Hodge numbers as the quotients of the manifold \eqref{eq:M3Split1}. 
\begin{table}[!ht]
\begin{center}
\begin{tabular}{| c || c | c |}
\hline
\myalign{| c||}{\varstr{16pt}{10pt}$~~~~~~~  \Gamma ~~~~~~~$ } &
\myalign{m{1.31cm}|}{$~~~ \IZ_2 $} &
\myalign{m{1.9cm}|}{ $~~~\IZ_2\times\IZ_2 \ \ \ $ }  
\\ \hline\hline
\varstr{14pt}{8pt} $h^{1,1}(X/\Gamma)$ & 4 & 4 \\
 \hline
\varstr{14pt}{8pt} $h^{2,1}(X/\Gamma)$ & 20 & 12 \\
 \hline
\varstr{14pt}{8pt} $\chi(X/\Gamma)$ & $ \!\!\!\!-32 $ & $ \!\!\!\!- 16$  \\
 \hline
 \end{tabular}
   \vskip 0.3cm
\capt{5in}{M3Split1quotients}{Hodge numbers for the quotients of the manifolds \eqref{eq:M3Split1} and \eqref{eq:M3Split2}.}
 \end{center}
 \vspace{-2pt}
 \end{table}

\subsubsection{Further non-favourable splits: $X^{4,36}\rightarrow X^{8,32}\rightarrow X^{9,25}\rightarrow X^{13,21}$}\label{REM_X3,43_3}
The manifold $X^{8,32}$ can be obtained by splitting the second column of the configuration matrix \eqref{eq:M3Split1}, and corresponds to the following configuration:
\begin{equation}\label{eq:M3Split3}
\displaycicy{5.25in}{
X_{5421}~=~~
\cicy{\IP^1\\\IP^1 \\\IP^1 \\\IP^1\\\IP^3}
{ ~2& 0 & 0 & 0 ~ \\
  ~0& 1 & 0 & 1~ \\
   ~0& 0& 1 & 1~\\
  ~0& 0& 0 & 2~\\
  ~2& 1 & 1 & 0~ \\}_{-48}^{\,8,\,32}}{-1.1cm}{2.25in}{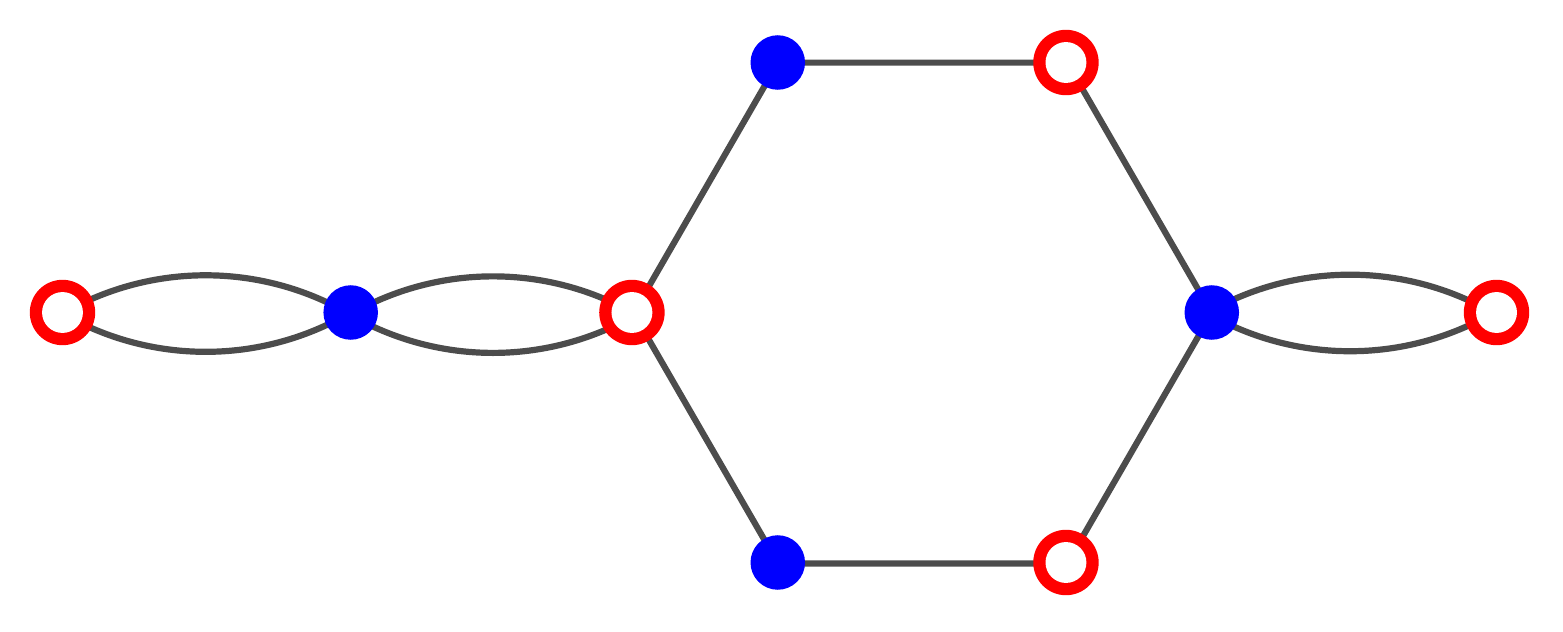}
\end{equation}
This manifold admits a total number of $27$ different free group actions, by $\IZ_2$ and $\IZ_2\times \IZ_2$. In this case, the polynomial deformation method does not reproduce correctly the number of complex structure parameters. Moreover, the embedding~\eqref{eq:M3Split3} is not favourable in the sense of Section~\ref{sec:KahlerParameters}. However, we note that $X_{5421}$ is embedded in the product $\text{dP}_4\times\IP^1\times\IP^3$. Using our knowledge of del Pezzo surfaces of degree $4$ described in \sref{subsec:delPezzo}, we compute the Hodge numbers of the quotients. In \tref{dP4_5421_Symm_Action}, we list the distinct symmetry actions on the ambient space coordinates. 
\begin{table}[H]
\vspace{12pt}
\begin{center}
\begin{tabular}{| c || c | c |}
\hline
\myalign{| c||}{\varstr{16pt}{10pt}$~~~~~~~  \Gamma ~~~~~~~$ } &
\myalign{m{1.31cm}|}{$\hfil \IZ_2 $} &
\myalign{m{1.9cm}|}{ $\hfil \IZ_2\times\IZ_2$ }
\\ \hline\hline
\varstr{14pt}{8pt} $h^{1,1}(X/\Gamma)$ & 6 & 5\\
 \hline
\varstr{14pt}{8pt} $h^{2,1}(X/\Gamma)$ & 18 & 11 \\
 \hline
\varstr{14pt}{8pt} $\chi(X/\Gamma)$ & $\!\!\!\!-24$ & $\!\!\!\!-12$\\
 \hline
 \end{tabular}
 \vskip 0.3cm
\capt{1\textwidth}{dP4_5421}{Hodge numbers for the quotients of the manifold ~\eqref{eq:M3Split3}.}
 \end{center}
 \end{table}
\small
\begin{center}
\begin{longtable}{|c|c|c|c|c|}
\captionsetup{width=0.9\textwidth}
\caption{\it Various symmetry actions on the ambient space of the manifold~\eqref{eq:M3Split3}. The coordinate patch of the $\text{dP}_4$ is chosen to be $(1,x){\times}(1,y){\times}(1,z)$. $(p,q)$ and $(a,b,c,d)$ are taken to be coordinates of the first $\IP^1$ space and the $\IP^3$ respectively.} \label{dP4_5421_Symm_Action} \\

\hline \multicolumn{1}{|c|}{\str\textbf{Index}} &  \multicolumn{1}{|c|}{\str\textbf{~Group~}} & \multicolumn{1}{|c|}{$\mathbf{(\+x,\+y,\+z)}$} &  \multicolumn{1}{|c|}{$\mathbf{(\+p,\+q)}$} &  \multicolumn{1}{|c|}{$\mathbf{(\+a,\+b,\+c,\+d)}$}  \\ \hline 
\endfirsthead

\hline 
\textbf{Index} &
\textbf{~Group~} &
$\mathbf{(\+x,\+y,\+z)}$ &
$\mathbf{(\+p,\+q)}$ &
$\mathbf{(\+a,\+b,\+c,\+d)}$ \\ \hline 
\endhead

\hline\hline \multicolumn{5}{|r|}{{\str Continued on next page}} \\ \hline
\endfoot

\endlastfoot

\hline\hline

\varstr{14pt}{8pt} 1 & $\IZ_2$ & $(-x,-y,-z)$ & $(-p,\+q)$ & $(-a,-b,\+c,\+d)$\\
 \hline
\varstr{14pt}{8pt} 2 & $\IZ_2$ & $(\+y,\+x,-z)$ &  $(-p,\+q)$ & $(-a,-b,\+c,\+d)$\\
 \hline
\varstr{16pt}{10pt} 3 & $\IZ_2{\times}\IZ_2$ &
  \begin{minipage}[c][40pt][c]{1.3in}
  \begin{gather*} 
 (-x,-y,-z) \\ 
 ({x^{-1}},{y^{-1}},{z^{-1}}) \\
  \end{gather*}
  \end{minipage} 
& 
 \begin{minipage}[c][40pt][c]{0.8in}
  \begin{gather*}
(\+p,-q) \\ 
(\+q,\+p)\\
  \end{gather*}
  \end{minipage} 
  &
   \begin{minipage}[c][40pt][c]{1.3in}
  \begin{gather*} 
  (\+a,-b,\+c,-d) \\ 
 (\+b,\+a,\+d,\+c) \\
  \end{gather*}
  \end{minipage} \\
 \hline
 4 & $\IZ_2{\times}\IZ_2$ &
  \begin{minipage}[c][40pt][c]{1.3in}
  \begin{gather*} 
 (\+y,\+x,-z) \\ 
 ({-y},{-x},{z^{-1}}) \\
  \end{gather*}
  \end{minipage} 
& 
 \begin{minipage}[c][40pt][c]{0.8in}
  \begin{gather*}
(\+p,-q) \\ 
(\+q,\+p)\\
  \end{gather*}
  \end{minipage} 
  &
   \begin{minipage}[c][40pt][c]{1.3in}
  \begin{gather*} 
  (\+a,-b,\+c,-d) \\ 
 (\+b,\+a,\+d,\+c) \\
  \end{gather*}
  \end{minipage} \\ \hline
  5 & $\IZ_2{\times}\IZ_2$ &
  \begin{minipage}[c][40pt][c]{1.3in}
  \begin{gather*} 
 (\+y,\+x,-z) \\ 
 ({-x},{-y},{z^{-1}}) \\
  \end{gather*}
  \end{minipage} 
& 
 \begin{minipage}[c][26pt][c]{0.8in}
  \begin{gather*}
(\+p,-q) \\ 
(\+q,\+p)\\
  \end{gather*}
  \end{minipage} 
  &
   \begin{minipage}[c][26pt][c]{1.3in}
  \begin{gather*} 
  (\+a,-b,\+c,-d) \\ 
 (\+b,\+a,\+d,\+c) \\
  \end{gather*}
  \end{minipage} \\ \hline
6 & $\IZ_2{\times}\IZ_2$ &
  \begin{minipage}[c][40pt][c]{1.3in}
  \begin{gather*} 
 (-x,-y,-z) \\ 
 ({\+y},{\+x},{z^{-1}}) \\
  \end{gather*}
  \end{minipage} 
& 
 \begin{minipage}[c][30pt][c]{0.8in}
  \begin{gather*}
 (\+p,-q) \\ 
(\+q,\+p)\\
 \end{gather*}
  \end{minipage} 
  &
   \begin{minipage}[c][40pt][c]{1.3in}
  \begin{gather*} 
  (\+a,-b,\+c,-d) \\ 
 (\+b,\+a,\+d,\+c) \\
  \end{gather*}
  \end{minipage} \\
\hline
\end{longtable}
\vskip-20pt
\end{center}
\normalsize
\small
\begin{center}
\begin{longtable}{|c|c|c|c|}
\captionsetup{width=0.9\textwidth}
\caption{\it Symmetry actions on the cohomology basis and the corresponding invariants for the manifold~\eqref{eq:M3Split3}. The matrices $P_i$ and $Q_i$ are defined in \eqref{eq:CohMatDefs2}.} \label{dP4_5421_Symm_Action_Coh} \\

\hline \multicolumn{1}{|c|}{\str\textbf{Index}}&  \multicolumn{1}{|c|}{\str\textbf{~~Group~~}} & \multicolumn{1}{|c|}{\str\textbf{\begin{minipage}[c][35pt][c]{0.85in}
Action on\\ 
Coh Basis
  \end{minipage}}} &  \multicolumn{1}{|c|}{\str\textbf{~~Coh Invariants~~}} \\ \hline 
\endfirsthead

\hline 
\textbf{~Index~} & \textbf{~~Group~~} & \textbf{\begin{minipage}[c][35pt][c]{0.85in}
Action on\\ 
Coh Basis
  \end{minipage}} & \textbf{~~Coh Invariants~~} \\ \hline 
\endhead

\hline\hline \multicolumn{4}{|r|}{{\str Continued on next page}} \\ \hline
\endfoot

\endlastfoot

\hline\hline
1 & $\IZ_2$ &
$P_1$
& \begin{minipage}[c][22pt][c]{2.1in}
\begin{gather*} 
H, E_1+E_2,E_3+E_4,E_5\\
\end{gather*}
\end{minipage} \\
 \hline
 
2 & $\IZ_2$ &
$Q_1$
& \begin{minipage}[c][22pt][c]{3.2in}
\begin{gather*} 
H-E_5, E_1+E_4-E_5,
E_2+E_4-E_5, E_3-E_4\\
\end{gather*}
\end{minipage} \\
 \hline
 
3 & $\IZ_2{\times}\IZ_2$ &
$P_1$,~$P_2$
&\begin{minipage}[c][22pt][c]{2.1in}
\begin{gather*} 
H, E_1+E_2+E_3+E_4,E_5\\
\end{gather*}
\end{minipage} \\ \hline

4 & $\IZ_2{\times}\IZ_2$ &
$Q_1$,~$Q_2$
& \begin{minipage}[c][22pt][c]{3.2in}
\begin{gather*} 
H-{E}_5, E_1+E_3-E_5, E_2+E_4-E_5\\
\end{gather*}
\end{minipage} \\
 \hline
 
5 & $\IZ_2{\times}\IZ_2$ &
$Q_1$,~$P_2$
& \begin{minipage}[c][22pt][c]{3.2in}
\begin{gather*} 
H-{E}_5, E_1+E_4-E_5, E_2+E_3-E_5\\
\end{gather*}
\end{minipage} \\
 \hline
 
6 & $\IZ_2{\times}\IZ_2$ &
$P_1$,~$Q_2$
& \begin{minipage}[c][22pt][c]{3.2in}
\begin{gather*} 
H-{E}_5, E_1+E_2-E_5, E_3+E_4-E_5\\
\end{gather*}
\end{minipage} \\
 \hline 
\end{longtable}
\vskip-20pt
\end{center}
\normalsize
The group action on the cohomology and the invariants are listed in \tref{dP4_5421_Symm_Action_Coh}. For the quotient manifolds, the counting of K\"ahler classes includes the two hyperplane classes corresponding to the $\IP^1$ and $\IP^3$ spaces, plus the number of cohomology invariants listed in \tref{dP4_5421_Symm_Action_Coh}. The resulting Hodge numbers are shown in \tref{dP4_5421}.
\newpage
Now we turn to the manifold $X^{9,25}$, which can be obtained by splitting the first column of the configuration matrix~\eqref{eq:M3Split3}, and corresponds to the following configuration:
\begin{equation}\label{eq:dP4_2640}
\displaycicy{5.25in}{
X_{2640}~=~~
\cicy{\IP^1\\\IP^1 \\\IP^1 \\\IP^1\\\IP^1\\\IP^3}
{ ~1 & 1 & 0& 0 &0  ~ \\
  ~1&1& 0& 0 &0 ~\\
  ~0&0& 1& 0 & 1~\\
  ~0&0& 0& 1 & 1~\\
    ~0& 0& 0 & 0 & 2~ \\
  ~1& 1 & 1 & 1 & 0~ \\}_{-32}^{\,9,\,25}}{-1.1cm}{2.25in}{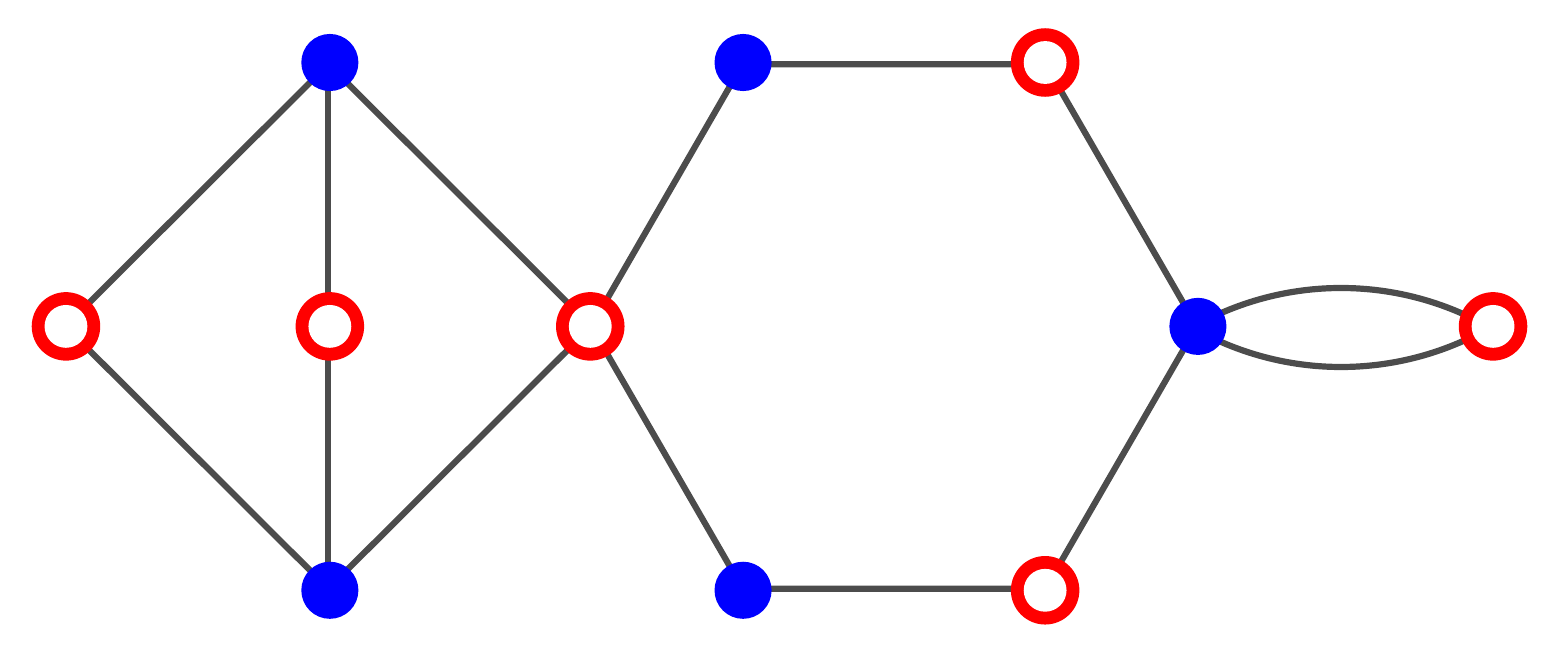}
\end{equation}
This manifold is embedded in the product $\text{dP}_4\times\IP^1\times\IP^1\times\IP^3$ and admits a total of $27$ different free group actions, by $\IZ_2$ and $\IZ_2\times \IZ_2$. Because of this embedding, $h^{1,1}=6+3=9$ for the manifold. In order to compute the Hodge numbers of the quotients, we resort to computing the group action on the cohomology basis of the $\text{dP}_4$, since this manifold is neither favourable nor amenable to polynomial deformation methods. In \tref{dP4_2640_Symm_Action}, we list the distinct symmetry actions on the ambient space coordinates. The group action on the cohomology and corresponding invariants are listed in \tref{dP4_2640_Symm_Action_Coh}. The Hodge numbers of the quotients of \eqref{eq:dP4_2640} are listed in \tref{dP4_2357}. 
\small
\begin{center}
\begin{longtable}{|c|c|c|c|c|c|}
\captionsetup{width=0.9\textwidth}
\caption{\it Various symmetry actions on the ambient space of the manifold~\eqref{eq:dP4_2640}. The coordinate patch of the $\text{dP}_4$ is chosen to be $(1,x)\times(1,y)\times(1,z)$. $(p,q)$, $(r,s)$ and $(a,b,c,d)$ are taken to be coordinates of the first two $\IP^1$ spaces and the $\IP^3$ respectively.} \label{dP4_2640_Symm_Action} \\

\hline \multicolumn{1}{|c|}{\str\textbf{Index}} &  \multicolumn{1}{|c|}{\str\textbf{~Group~}} & \multicolumn{1}{|c|}{$\mathbf{(\+x,\+y,\+z)}$} &  \multicolumn{1}{|c|}{$\mathbf{(\+p,\+q)}$} &  \multicolumn{1}{|c|}{$\mathbf{(\+r,\+s)}$} &  \multicolumn{1}{|c|}{$\mathbf{(\+a,\+b,\+c,\+d)}$}  \\ \hline 
\endfirsthead

\hline 
\textbf{Index} &
\textbf{~Group~} &
$\mathbf{(\+x,\+y,\+z)}$ &
$\mathbf{(\+p,\+q)}$ &
$\mathbf{(\+r,\+s)}$ &
$\mathbf{(\+a,\+b,\+c,\+d)}$ \\ \hline 
\endhead

\hline\hline \multicolumn{6}{|r|}{{\str Continued on next page}} \\ \hline
\endfoot

\endlastfoot

\hline\hline

\varstr{16pt}{10pt} 1 & $\IZ_2$ & $(-x,-y,-z)$ & $(-p,\+q)$ & $(-r,\+s)$ & $(-a,-b,\+c,\+d)$\\
 \hline
\varstr{16pt}{10pt} 2 & $\IZ_2$ & $(y,x,-z)$ &  $(-p,\+q)$ & $(-r,\+s)$ & $(-a,-b,\+c,\+d)$\\
 \hline
\varstr{16pt}{10pt} 3 & $\IZ_2{\times}\IZ_2$ &
  \begin{minipage}[c][44pt][c]{1.3in}
  \begin{gather*} 
 (-x,-y,-z) \\ 
 ({x^{-1}},{y^{-1}},{z^{-1}}) \\
  \end{gather*}
  \end{minipage} 
& 
 \begin{minipage}[c][30pt][c]{0.8in}
  \begin{gather*}
(\+p,-q) \\ 
(\+q,\+p)\\
  \end{gather*}
  \end{minipage} 
  & 
 \begin{minipage}[c][30pt][c]{0.8in}
  \begin{gather*}
(\+r,-s) \\ 
(\+s,\+r)\\
  \end{gather*}
  \end{minipage} 
  &
   \begin{minipage}[c][44pt][c]{1.3in}
  \begin{gather*} 
  (\+a,-b,\+c,-d) \\ 
 (\+b,\+a,\+d,\+c) \\
  \end{gather*}
  \end{minipage} \\
 \hline
 4 & $\IZ_2{\times}\IZ_2$ &
  \begin{minipage}[c][44pt][c]{1.3in}
  \begin{gather*} 
 (-x,-y,-z) \\ 
 ({\+y},{\+x},{z^{-1}}) \\
  \end{gather*}
  \end{minipage} 
& 
 \begin{minipage}[c][30pt][c]{0.8in}
  \begin{gather*}
(\+p,-q) \\ 
(\+q,\+p)\\
  \end{gather*}
  \end{minipage} 
  & 
 \begin{minipage}[c][30pt][c]{0.8in}
  \begin{gather*}
(\+r,-s) \\ 
(\+s,\+r)\\
  \end{gather*}
  \end{minipage} 
  &
   \begin{minipage}[c][44pt][c]{1.3in}
  \begin{gather*} 
  (\+a,-b,\+c,-d) \\ 
 (\+b,\+a,\+d,\+c) \\
  \end{gather*}
  \end{minipage} \\ \hline
  5 & $\IZ_2{\times}\IZ_2$ &
  \begin{minipage}[c][44pt][c]{1.3in}
  \begin{gather*} 
 (\+y,\+x,-z) \\ 
 ({-x},{-y},{z^{-1}}) \\
  \end{gather*}
  \end{minipage} 
& 
 \begin{minipage}[c][30pt][c]{0.8in}
  \begin{gather*}
(\+p,-q) \\ 
(\+q,\+p)\\
  \end{gather*}
  \end{minipage} 
& 
 \begin{minipage}[c][30pt][c]{0.8in}
  \begin{gather*}
(\+r,-s) \\ 
(\+s,\+r)\\
  \end{gather*}
  \end{minipage} 
  &
   \begin{minipage}[c][44pt][c]{1.3in}
  \begin{gather*} 
  (\+a,-b,\+c,-d) \\ 
 (\+b,\+a,\+d,\+c) \\
  \end{gather*}
  \end{minipage} \\ \hline
6 & $\IZ_2{\times}\IZ_2$ &
  \begin{minipage}[c][44pt][c]{1.3in}
  \begin{gather*} 
 (\+y,\+x,-z) \\ 
 ({-y},{-x},{z^{-1}}) \\
  \end{gather*}
  \end{minipage} 
& 
 \begin{minipage}[c][30pt][c]{0.8in}
  \begin{gather*}
 (\+p,-q) \\ 
(\+q,\+p)\\
 \end{gather*}
  \end{minipage} 
  & 
 \begin{minipage}[c][30pt][c]{0.8in}
  \begin{gather*}
 (\+r,-s) \\ 
(\+s,\+r)\\
 \end{gather*}
  \end{minipage} 
  &
   \begin{minipage}[c][44pt][c]{1.3in}
  \begin{gather*} 
  (\+a,-b,\+c,-d) \\ 
 (\+b,\+a,\+d,\+c) \\
  \end{gather*}
  \end{minipage} \\
\hline
\end{longtable}
\end{center}
\normalsize
\small
\begin{center}
\begin{longtable}{|c|c|c|c|}
\captionsetup{width=0.9\textwidth}
\caption{\it Symmetry actions on the cohomology basis and the corresponding invariants for the manifold~\eqref{eq:dP4_2640}. The matrices $P_i$ and $Q_i$ are defined in \eqref{eq:CohMatDefs2}.} \label{dP4_2640_Symm_Action_Coh} \\

\hline \multicolumn{1}{|c|}{\str\textbf{Index}}&  \multicolumn{1}{|c|}{\str\textbf{~~Group~~}} & \multicolumn{1}{|c|}{\str\textbf{\begin{minipage}[c][35pt][c]{0.85in}
Action on\\ 
Coh Basis
  \end{minipage}}} &  \multicolumn{1}{|c|}{\str\textbf{~~Coh Invariants~~}} \\ \hline 
\endfirsthead

\hline 
\textbf{~Index~} & \textbf{~~Group~~} & \textbf{\begin{minipage}[c][35pt][c]{0.85in}
Action on\\ 
Coh Basis
  \end{minipage}} & \textbf{~~Coh Invariants~~} \\ \hline 
\endhead

\hline\hline \multicolumn{4}{|r|}{{\str Continued on next page}} \\ \hline
\endfoot

\endlastfoot

\hline\hline
1 & $\IZ_2$ &
$P_1$
& \begin{minipage}[c][26pt][c]{2.1in}
\begin{gather*} 
H, E_1+E_2,E_3+E_4,E_5\\
\end{gather*}
\end{minipage} \\
 \hline
 
2 & $\IZ_2$ &
$Q_2$
& \begin{minipage}[c][26pt][c]{2.1in}
\begin{gather*} 
H-E_5, E_1+E_4,E_2+E_3,E_5\\
\end{gather*}
\end{minipage} \\
 \hline
 
3 & $\IZ_2{\times}\IZ_2$ &
$P_1$,~$P_3$
&\begin{minipage}[c][26pt][c]{2.1in}
\begin{gather*} 
H, E_1+E_2+E_3+E_4,E_5\\
\end{gather*}
\end{minipage} \\ \hline

4 & $\IZ_2{\times}\IZ_2$ &
$P_1$,~$Q_2$
& \begin{minipage}[c][26pt][c]{2.6in}
\begin{gather*} 
H-{E}_5, E_1+E_2-E_5, E_3+E_4-E_5\\
\end{gather*}
\end{minipage} \\
 \hline
 
5 & $\IZ_2{\times}\IZ_2$ &
$Q_1$,~$P_2$
& \begin{minipage}[c][26pt][c]{2.6in}
\begin{gather*} 
H-{E}_5, E_1+E_4-E_5,
E_2+E_3-E_5\\
\end{gather*}
\end{minipage} \\
 \hline
 
6 & $\IZ_2{\times}\IZ_2$ &
$Q_1$,~$Q_2$
& \begin{minipage}[c][26pt][c]{2.6in}
\begin{gather*} 
H-{E}_5, E_1+E_3-E_5, E_2+E_4-E_5\\
\end{gather*}
\end{minipage} \\
 \hline 
\end{longtable}
\vskip-30pt
\end{center}
\normalsize

Another manifold $X^{9,25}$ can be obtained through a different splitting of the first column of the configuration matrix~\eqref{eq:M3Split3}, which corresponds to the following configuration:
\begin{equation}\label{eq:M3Split5}
\displaycicy{5.25in}{
X_{2357}~=~~
\cicy{\IP^1\\\IP^1 \\\IP^1 \\\IP^1\\\IP^1\\\IP^3}
{ ~1 & 1 & 0 & 0 & 0 ~ \\
  ~0& 2& 0 & 0 & 0~ \\
  ~0&0& 1& 0 & 1~\\
  ~0&0& 0& 1 & 1~\\
  ~0&0& 0& 0 & 2~\\
  ~1& 1 & 1 & 1 & 0~ \\}_{-32}^{\,9,\,25}}{-1.1cm}{2.25in}{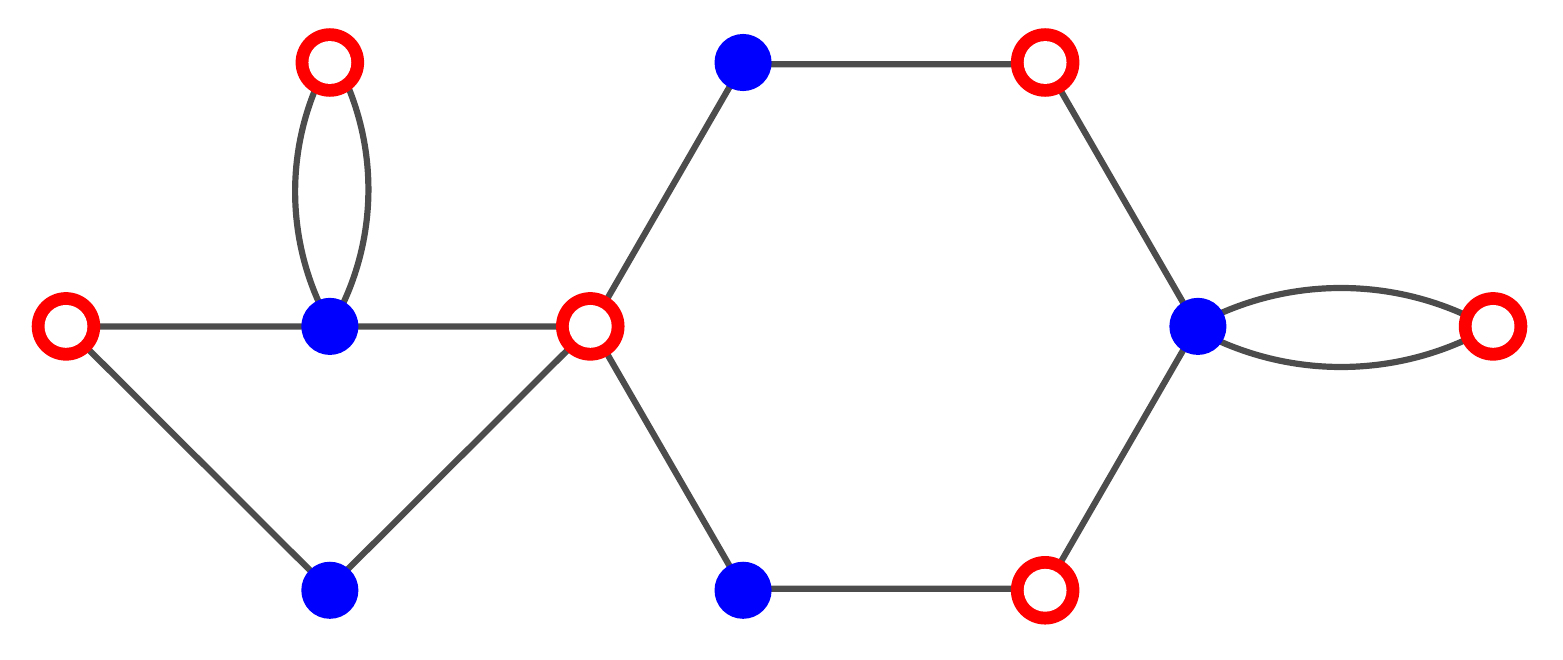}
\end{equation}
Like \eqref{eq:dP4_2640}, this manifold is embedded in the product 
$\text{dP}_4{\times}\IP^1{\times}\IP^1{\times}\IP^3$ and admits a total of $27$ different free group actions, by $\IZ_2$ and $\IZ_2\times \IZ_2$. $h^{1,1}=6+3=9$ for the manifold because of contributions from the $\text{dP}_4$ and the three other $\IP^n$'s respectively. We compute the Hodge numbers of the quotients by computing the group action on the cohomology basis of the $\text{dP}_4$, since this manifold is neither favourable nor amenable to polynomial deformation methods. In \tref{dP4_2357_Symm_Action}, we list the distinct symmetry actions on the ambient space coordinates. The group action on the cohomology and corresponding invariants are listed in \tref{dP4_2357_Symm_Action_Coh}. Finally, the Hodge numbers of the quotients of \eqref{eq:M3Split5} are listed in \tref{dP4_2357}. 
\begin{table}[H]
\begin{center}
\begin{tabular}{| c || c | c |}
\hline
\myalign{| c||}{\varstr{16pt}{10pt}$~~~~~~~  \Gamma ~~~~~~~$ } &
\myalign{m{1.31cm}|}{$\hfil \IZ_2 $} &
\myalign{m{1.9cm}|}{ $\hfil \IZ_2\times\IZ_2$ }
\\ \hline\hline
\varstr{14pt}{8pt} $h^{1,1}(X/\Gamma)$ & 7 & 6\\
 \hline
\varstr{14pt}{8pt} $h^{2,1}(X/\Gamma)$ & 15 & 10 \\
 \hline
\varstr{14pt}{8pt} $\chi(X/\Gamma)$ & $\!\!\!\!-16$ & $\!\!\!\!-8$\\
 \hline
 \end{tabular}
 \vskip 0.3cm
\capt{1\textwidth}{dP4_2357}{Hodge numbers for the quotients of the manifolds~\eqref{eq:dP4_2640} and \eqref{eq:M3Split5}.}
 \end{center}
 \vspace{-40pt}
 \end{table}
\small
\begin{center}
\begin{longtable}{|c|c|c|c|c|c|}
\captionsetup{width=0.9\textwidth}
\caption{\it Various symmetry actions on the ambient space of the manifold~\eqref{eq:M3Split5}. The coordinate patch of the $\text{dP}_4$ is chosen to be $(1,x){\times}(1,y){\times}(1,z)$. $(p,q)$, $(r,s)$ and $(a,b,c,d)$ are the coordinates of the first two $\IP^1$ spaces and the $\IP^3$ respectively.} \label{dP4_2357_Symm_Action} \\

\hline \multicolumn{1}{|c|}{\str\textbf{Index}} &  \multicolumn{1}{|c|}{\str\textbf{~Group~}} & \multicolumn{1}{|c|}{$\mathbf{(\+x,\+y,\+z)}$} &  \multicolumn{1}{|c|}{$\mathbf{(\+p,\+q)}$} &  \multicolumn{1}{|c|}{$\mathbf{(\+u,\+v)}$} &  \multicolumn{1}{|c|}{$\mathbf{(\+a,\+b,\+c,\+d)}$}  \\ \hline 
\endfirsthead

\hline 
\textbf{Index} &
\textbf{~Group~} &
$\mathbf{(\+x,\+y,\+z)}$ &
$\mathbf{(\+p,\+q)}$ &
$\mathbf{(\+r,\+s)}$ &
$\mathbf{(\+a,\+b,\+c,\+d)}$ \\ \hline 
\endhead

\hline\hline \multicolumn{6}{|r|}{{\str Continued on next page}} \\ \hline
\endfoot

\endlastfoot

\hline\hline

\varstr{14pt}{8pt} 1 & $\IZ_2$ & $(-x,-y,-z)$ & $(-p,\+q)$ & $(-r,\+s)$ & $(-a,-b,\+c,\+d)$\\
 \hline
\varstr{14pt}{8pt} 2 & $\IZ_2$ & $(\+y,\+x,-z)$ &  $(-p,\+q)$ & $(-r,\+s)$ & $(-a,-b,\+c,\+d)$\\
 \hline
\varstr{16pt}{10pt} 3 & $\IZ_2{\times}\IZ_2$ &
  \begin{minipage}[c][40pt][c]{1.3in}
  \begin{gather*} 
 (-x,-y,-z) \\ 
 ({x^{-1}},{y^{-1}},{z^{-1}}) \\
  \end{gather*}
  \end{minipage} 
& 
 \begin{minipage}[c][30pt][c]{0.8in}
  \begin{gather*}
(\+p,-q) \\ 
(\+q,\+p)\\
  \end{gather*}
  \end{minipage} 
  & 
 \begin{minipage}[c][30pt][c]{0.8in}
  \begin{gather*}
(\+r,-s) \\ 
(\+s,\+r)\\
  \end{gather*}
  \end{minipage} 
  &
   \begin{minipage}[c][40pt][c]{1.3in}
  \begin{gather*} 
  (\+a,-b,\+c,-d) \\ 
 (\+b,\+a,\+d,\+c) \\
  \end{gather*}
  \end{minipage} \\
 \hline
 4 & $\IZ_2{\times}\IZ_2$ &
  \begin{minipage}[c][40pt][c]{1.3in}
  \begin{gather*} 
 (\+y,\+x,-z) \\ 
 ({-y},{-x},{z^{-1}}) \\
  \end{gather*}
  \end{minipage} 
& 
 \begin{minipage}[c][30pt][c]{0.8in}
  \begin{gather*}
(\+p,-q) \\ 
(\+q,\+p)\\
  \end{gather*}
  \end{minipage} 
  & 
 \begin{minipage}[c][30pt][c]{0.8in}
  \begin{gather*}
(\+r,-s) \\ 
(\+s,\+r)\\
  \end{gather*}
  \end{minipage} 
  &
   \begin{minipage}[c][40pt][c]{1.3in}
  \begin{gather*} 
  (\+a,-b,\+c,-d) \\ 
 (\+b,\+a,\+d,\+c) \\
  \end{gather*}
  \end{minipage} \\ \hline
  5 & $\IZ_2{\times}\IZ_2$ &
  \begin{minipage}[c][40pt][c]{1.3in}
  \begin{gather*} 
 (-x,-y,-z) \\ 
 ({\+y},{\+x},{z^{-1}}) \\
  \end{gather*}
  \end{minipage} 
& 
 \begin{minipage}[c][30pt][c]{0.8in}
  \begin{gather*}
(\+p,-q) \\ 
(\+q,\+p)\\
  \end{gather*}
  \end{minipage} 
& 
 \begin{minipage}[c][30pt][c]{0.8in}
  \begin{gather*}
(\+r,-s) \\ 
(\+s,\+r)\\
  \end{gather*}
  \end{minipage} 
  &
   \begin{minipage}[c][40pt][c]{1.3in}
  \begin{gather*} 
  (\+a,-b,\+c,-d) \\ 
 (\+b,\+a,\+d,\+c) \\
  \end{gather*}
  \end{minipage} \\ \hline
6 & $\IZ_2{\times}\IZ_2$ &
  \begin{minipage}[c][40pt][c]{1.3in}
  \begin{gather*} 
 (\+y,\+x,-z) \\ 
 ({-x},{-y},{z^{-1}}) \\
  \end{gather*}
  \end{minipage} 
& 
 \begin{minipage}[c][30pt][c]{0.8in}
  \begin{gather*}
 (\+p,-q) \\ 
(\+q,\+p)\\
 \end{gather*}
  \end{minipage} 
  & 
 \begin{minipage}[c][30pt][c]{0.8in}
  \begin{gather*}
 (\+r,-s) \\ 
(\+s,\+r)\\
 \end{gather*}
  \end{minipage} 
  &
   \begin{minipage}[c][40pt][c]{1.3in}
  \begin{gather*} 
  (\+a,-b,\+c,-d) \\ 
 (\+b,\+a,\+d,\+c) \\
  \end{gather*}
  \end{minipage} \\
\hline
\end{longtable}
\end{center}
\normalsize
\small
\begin{center}
\begin{longtable}{|c|c|c|c|}
\captionsetup{width=0.9\textwidth}
\caption{\it Symmetry actions on the cohomology basis and the corresponding invariants for the manifold~\eqref{eq:M3Split5}. The matrices $P_i$ and $Q_i$ are defined in \eqref{eq:CohMatDefs2}.} \label{dP4_2357_Symm_Action_Coh} \\

\hline \multicolumn{1}{|c|}{\str\textbf{Index}}&  \multicolumn{1}{|c|}{\str\textbf{~~Group~~}} & \multicolumn{1}{|c|}{\str\textbf{\begin{minipage}[c][35pt][c]{0.85in}
Action on\\ 
Coh Basis
  \end{minipage}}} &  \multicolumn{1}{|c|}{\str\textbf{~~Coh Invariants~~}} \\ \hline 
\endfirsthead

\hline 
\textbf{~Index~} & \textbf{~~Group~~} & \textbf{\begin{minipage}[c][35pt][c]{0.85in}
Action on\\ 
Coh Basis
\end{minipage}} & \textbf{~~Coh Invariants~~} \\ \hline 
\endhead

\hline\hline \multicolumn{4}{|r|}{{\str Continued on next page}} \\ \hline
\endfoot

\endlastfoot

\hline\hline
1 & $\IZ_2$ &
$P_1$
& \begin{minipage}[c][22pt][c]{2.1in}
\begin{gather*} 
H, E_1+E_2,E_3+E_4,E_5\\
\end{gather*}
\end{minipage} \\
 \hline
 
2 & $\IZ_2$ &
$Q_1$
& \begin{minipage}[c][22pt][c]{2.9in}
\begin{gather*} 
H-E_5, E_1+E_4-E_5,
E_2+E_4-E_5, E_3-E_4\\
\end{gather*}
\end{minipage} \\
 \hline
 
3 & $\IZ_2{\times}\IZ_2$ &
$P_2$,~$P_1$
&\begin{minipage}[c][22pt][c]{2.1in}
\begin{gather*} 
H, E_1+E_2+E_3+E_4,E_5\\
\end{gather*}
\end{minipage} \\ \hline

4 & $\IZ_2{\times}\IZ_2$ &
$Q_1$,~$Q_2$
& \begin{minipage}[c][22pt][c]{3.2in}
\begin{gather*} 
H-{E}_5, E_1+E_3-E_5,
E_2+E_4-E_5\\
\end{gather*}
\end{minipage} \\
 \hline
 
5 & $\IZ_2{\times}\IZ_2$ &
$P_1$,~$Q_2$
& \begin{minipage}[c][22pt][c]{3.2in}
\begin{gather*} 
H-{E}_5, E_1+E_2-E_5,
E_3+E_4-E_5\\
\end{gather*}
\end{minipage} \\
 \hline
 
6 & $\IZ_2{\times}\IZ_2$ &
$Q_1$,~$P_2$
& \begin{minipage}[c][22pt][c]{3.2in}
\begin{gather*} 
H-{E}_5, E_1+E_4-E_5, 
E_2+E_3-E_5\\
\end{gather*}
\end{minipage} \\
 \hline 
\end{longtable}
\end{center}
\normalsize
\newpage
A further splitting of the second column of \eqref{eq:M3Split5} leads us to the manifold $X^{13,21}$, specified by the following configuration:
\begin{equation}\label{eq:dP4_480}
\displaycicy{5.25in}{
X_{480}~=~~
\cicy{\IP^1\\\IP^1\\\IP^1 \\\IP^1 \\\IP^1\\\IP^1\\\IP^3}
{ ~1 & 0& 1 & 0 & 0 & 0 ~ \\
  ~0 & 1& 1 & 0 & 0 & 0 ~ \\
  ~0&0& 2& 0 & 0 & 0~ \\
  ~0&0&0& 1& 0 & 1~\\
  ~0&0&0& 0& 1 & 1~\\
  ~0&0&0& 0& 0 & 2~\\
  ~1& 1 &0& 1 & 1 & 0~ \\}_{-16}^{\,13,\,21}}{-1.0cm}{2.51in}{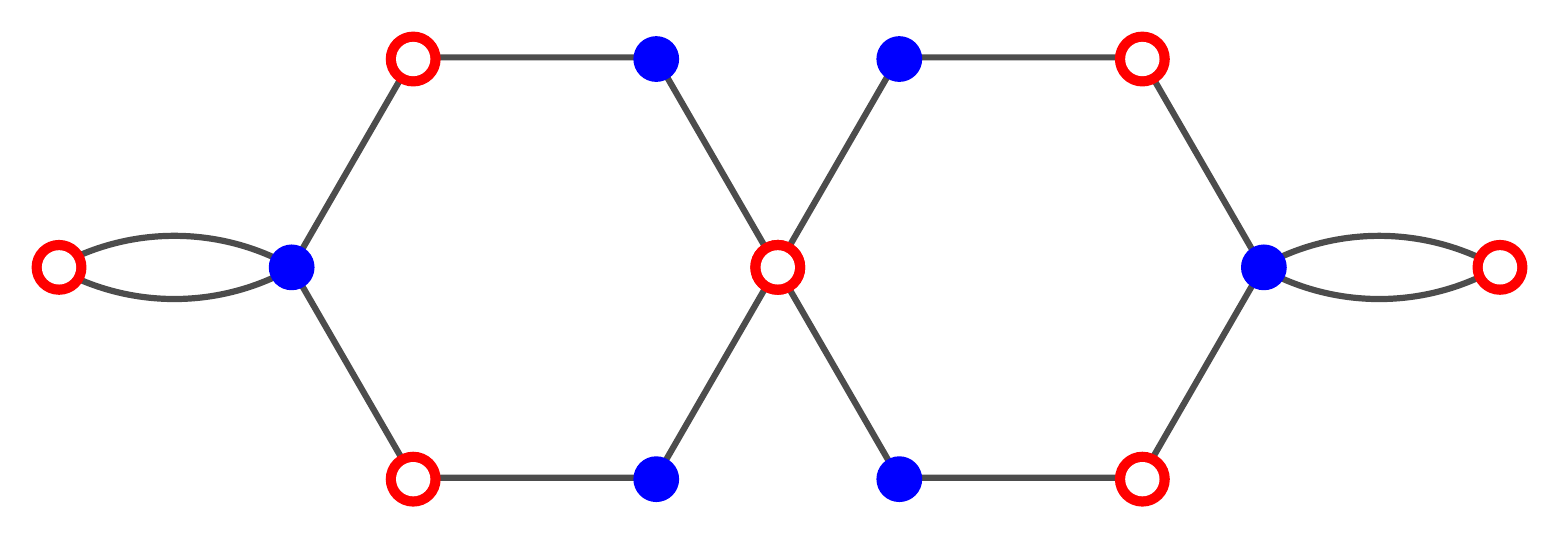}
\end{equation}
The above manifold, embedded in a product of $\text{dP}_4{\times} \text{dP}_4{\times}\IP^3$ has a large set of freely acting symmetries (394 in number). The value of $h^{1,1}$ for this manifold can be understood as the sum of $h^{1,1}$'s for the two $\text{dP}_4$'s and the $\IP^3$. Thus $h^{1,1}(X_{480})=6+6+1=13$. The symmetries although large in number, act on the coordinates in a grand total of $13$ distinct ways. The number $394$ is due to the large number of different actions on the polynomials. None of these symmetries mix the $\IP^3$ with the $\IP^1$'s, although 5 of them do interchange the two $\text{dP}_4$'s. These symmetries are listed in \tref{dP4_480_Symm_Action}. 

The action of the symmetries on the combined 32 special lines of the two $\text{dP}_4$'s are listed in \tref{dP4_480_Symm_Action_Coh}. This also contains all the cohomology invariants. Note that the $Z_4$ actions interchange the two del Pezzo surfaces. Finally, the $h^{1,1}$ of each quotient is the number of such invariants plus 1 for the $\IP^3$ space, since $X_{480}\subset\text{dP}_4\times \text{dP}_4\times\IP^3$. These values are listed in \tref{dP4_480}.
\vskip15pt
\begin{table}[H]
\begin{center}
\begin{tabular}{| c || c | c | c |}
\hline
\myalign{| c||}{\varstr{16pt}{10pt}$~~~~~~~  \Gamma ~~~~~~~$ } &
\myalign{m{1.31cm}|}{$\hfil \IZ_2 $} &
\myalign{m{1.31cm}|}{$\hfil \IZ_4 $} &
\myalign{m{1.9cm}|}{ $\hfil \IZ_2\times\IZ_2$ }
\\ \hline\hline
\varstr{14pt}{8pt} $h^{1,1}(X/\Gamma)$ & 9 & 5 & 7\\
 \hline
\varstr{14pt}{8pt} $h^{2,1}(X/\Gamma)$ & 13 & 7 & 9 \\
 \hline
\varstr{14pt}{8pt} $\chi(X/\Gamma)$ & $\!\!\!\!-8$ & $\!\!\!\!-4$ & $\!\!\!\!-4$\\
 \hline
 \end{tabular}
 \vskip 0.3cm
\capt{4.5in}{dP4_480}{Hodge numbers for the quotients of the manifold~\eqref{eq:dP4_480}.}
 \end{center}
 \vspace{-12pt}
 \end{table}
 \normalsize
\newpage
\small
\begin{center}
\begin{longtable}{|c|c|c|c|c|}
\captionsetup{width=0.9\textwidth}
\caption{\it Various symmetry actions on the ambient space of the manifold~\eqref{eq:dP4_480}. The coordinate patch of the two $\text{dP}_4$'s are chosen to be $(1,x){\times}(1,y){\times}(1,z)$ and 
$(1,\widetilde{x}){\times}(1,\widetilde{y}){\times}(1,\widetilde{z})$ respectively. $(a,b,c,d)$ is taken to be coordinates of the $\IP^3$.} 
\label{dP4_480_Symm_Action} \\

\hline \multicolumn{1}{|c|}{\str\textbf{Index}} &  \multicolumn{1}{|c|}{\str\textbf{~Group~}} & \multicolumn{1}{|c|}{$\mathbf{(\+x,\+y,\+z)}$} &  \multicolumn{1}{|c|}{$\mathbf{(\+\widetilde{x},\+\widetilde{y},\+\widetilde{z})}$} &  \multicolumn{1}{|c|}{$\mathbf{(\+a,\+b,\+c,\+d)}$}  \\ \hline 
\endfirsthead

\hline 
\textbf{Index} &
\textbf{~Group~} &
$\mathbf{(\+x,\+y,\+z)}$ &
$\mathbf{(\+\widetilde{x},\+\widetilde{y},\+\widetilde{z})}$ &
$\mathbf{(\+a,\+b,\+c,\+d)}$ \\ \hline 
\endhead

\hline\hline \multicolumn{5}{|r|}{{\str Continued on next page}} \\ \hline
\endfoot

\endlastfoot

\hline\hline

\varstr{14pt}{8pt} 1 & $\IZ_2$ & $(-x,-y,-z)$ & $(-\widetilde{x},-\widetilde{y},-\widetilde{z})$ & $(-a,-b,\+c,\+d)$\\
 \hline
\varstr{14pt}{8pt} 2 & $\IZ_2$ & $(\+y,\+x,-z)$ & $(-\widetilde{x},-\widetilde{y},-\widetilde{z})$ & $(-a,-b,\+c,\+d)$\\
 \hline
\varstr{14pt}{8pt} 3 & $\IZ_2$ & $(\+y,\+x,-z)$ & $(\+\widetilde{y},\+\widetilde{x},-\widetilde{z})$ & $(-a,-b,\+c,\+d)$\\
 \hline
\varstr{14pt}{8pt} 4 & $\IZ_4$ &$(-\widetilde{y},-\widetilde{x},-\widetilde{z})$ & $(\+y,\+x,\+z)$ & $(\+a,-b,\+ \ii c,-\ii d)$\\
 \hline
\varstr{14pt}{8pt}  5 & $\IZ_4$ & $(\+\widetilde{x},\+\widetilde{y},-\widetilde{z})$ & $(\+y,\+x,\+z)$ & $(\+a,-b,\+ \ii c,-\ii d)$\\
 \hline
6 & $\IZ_2{\times}\IZ_2$ &
  \begin{minipage}[c][40pt][c]{1.3in}
  \begin{gather*} 
 (-x,-y,-z) \\ 
 ({x^{-1}},{y^{-1}},{z^{-1}}) \\
  \end{gather*}
  \end{minipage} 
& 
 \begin{minipage}[c][40pt][c]{1.3in}
  \begin{gather*}
  (-\widetilde{x},-\widetilde{y},-\widetilde{z}) \\ 
  ({\widetilde{x}^{-1}},{\widetilde{y}^{-1}},{\widetilde{z}^{-1}})\\
  \end{gather*}
  \end{minipage} 
  &
   \begin{minipage}[c][40pt][c]{1.3in}
  \begin{gather*} 
  (\+a,-b,\+c,-d) \\ 
 (\+b,\+a,\+d,\+c) \\
  \end{gather*}
  \end{minipage} 
\\
 \hline
 \varstr{16pt}{10pt} 7 & $\IZ_2{\times}\IZ_2$ &
  \begin{minipage}[c][40pt][c]{1.3in}
  \begin{gather*} 
 (\+y,\+x,-z) \\ 
 (-y,-x,{z^{-1}}) \\
  \end{gather*}
  \end{minipage} 
& 
 \begin{minipage}[c][40pt][c]{1.3in}
  \begin{gather*}
  (-\widetilde{x},-\widetilde{y},-\widetilde{z}) \\ 
  ({\widetilde{x}^{-1}},{\widetilde{y}^{-1}},{\widetilde{z}^{-1}})\\
  \end{gather*}
  \end{minipage} 
  &
   \begin{minipage}[c][40pt][c]{1.3in}
  \begin{gather*} 
  (\+a,-b,\+c,-d) \\ 
 (\+b,\+a,\+d,\+c) \\
  \end{gather*}
  \end{minipage} 
\\
 \hline
\varstr{16pt}{10pt} 8 & $\IZ_2{\times}\IZ_2$ &
  \begin{minipage}[c][40pt][c]{1.3in}
  \begin{gather*} 
 (-x,-y,-z) \\ 
 (\+y,\+x,{z^{-1}}) \\
  \end{gather*}
  \end{minipage} 
& 
 \begin{minipage}[c][40pt][c]{1.3in}
  \begin{gather*}
  (-\widetilde{x},-\widetilde{y},-\widetilde{z}) \\ 
  ({\widetilde{x}^{-1}},{\widetilde{y}^{-1}},{\widetilde{z}^{-1}})\\
  \end{gather*}
  \end{minipage} 
  &
   \begin{minipage}[c][40pt][c]{1.3in}
  \begin{gather*} 
  (\+a,-b,\+c,-d) \\ 
 (\+b,\+a,\+d,\+c) \\
  \end{gather*}
  \end{minipage} 
\\
 \hline
\varstr{16pt}{10pt} 9 & $\IZ_2{\times}\IZ_2$ &
  \begin{minipage}[c][40pt][c]{1.3in}
  \begin{gather*} 
 (\+y,\+x,-z) \\ 
 (-{x},-{y},{z}^{-1}) \\
  \end{gather*}
  \end{minipage} 
& 
 \begin{minipage}[c][40pt][c]{1.3in}
  \begin{gather*}
  (-\widetilde{x},-\widetilde{y},-\widetilde{z}) \\ 
  ({\widetilde{x}^{-1}},{\widetilde{y}^{-1}},{\widetilde{z}^{-1}})\\
  \end{gather*}
  \end{minipage} 
  &
   \begin{minipage}[c][40pt][c]{1.3in}
  \begin{gather*} 
  (\+a,-b,\+c,-d) \\ 
 (\+b,\+a,\+d,\+c) \\
  \end{gather*}
  \end{minipage} 
\\
 \hline
\varstr{16pt}{10pt} 10 & $\IZ_2{\times}\IZ_2$ &
  \begin{minipage}[c][40pt][c]{1.3in}
  \begin{gather*} 
 (\+y,\+x,-z) \\ 
 (-{x},-{y},{z}^{-1}) \\
  \end{gather*}
  \end{minipage} 
& 
 \begin{minipage}[c][40pt][c]{1.3in}
  \begin{gather*}
  (\+\widetilde{y},\+\widetilde{x},-\widetilde{z}) \\ 
  ({-\widetilde{x}},{-\widetilde{y}},{\widetilde{z}^{-1}})\\
  \end{gather*}
  \end{minipage} 
  &
   \begin{minipage}[c][40pt][c]{1.3in}
  \begin{gather*} 
  (\+a,-b,\+c,-d) \\ 
 (\+b,\+a,\+d,\+c) \\
  \end{gather*}
  \end{minipage} 
\\
 \hline
\varstr{16pt}{10pt} 11 & $\IZ_2{\times}\IZ_2$ &
  \begin{minipage}[c][40pt][c]{1.3in}
  \begin{gather*} 
 (-x,-y,-z) \\ 
 (\+y,\+x,{z}^{-1}) \\
  \end{gather*}
  \end{minipage} 
& 
 \begin{minipage}[c][40pt][c]{1.3in}
  \begin{gather*}
  (-\widetilde{x},-\widetilde{y},-\widetilde{z}) \\ 
  ({\+\widetilde{y}},{\+\widetilde{x}},{\widetilde{z}^{-1}})\\
  \end{gather*}
  \end{minipage} 
  &
   \begin{minipage}[c][40pt][c]{1.3in}
  \begin{gather*} 
  (\+a,-b,\+c,-d) \\ 
 (\+b,\+a,\+d,\+c) \\
  \end{gather*}
  \end{minipage} 
\\
 \hline
\varstr{16pt}{10pt} 12 & $\IZ_2{\times}\IZ_2$ &
  \begin{minipage}[c][44pt][c]{1.3in}
  \begin{gather*} 
 (\+y,\+x,-z) \\ 
 (-y,-x,z^{-1}) \\ 
  \end{gather*}
  \end{minipage} 
& 
 \begin{minipage}[c][40pt][c]{1.3in}
  \begin{gather*}
  ({\+\widetilde{y}},{\+\widetilde{x}},{-\widetilde{z}})\\
  ({-\widetilde{y}},{-\widetilde{x}},{\widetilde{z}^{-1}})\\
  \end{gather*}
  \end{minipage} 
  &
   \begin{minipage}[c][40pt][c]{1.3in}
  \begin{gather*} 
  (\+a,-b,\+c,-d) \\ 
 (\+b,\+a,\+d,\+c) \\
  \end{gather*}
  \end{minipage} 
\\
 \hline
\varstr{16pt}{10pt} 13 & $\IZ_2{\times}\IZ_2$ &
  \begin{minipage}[c][40pt][c]{1.3in}
  \begin{gather*} 
  (\+y,\+x,-z) \\ 
 (-y,-x,z^{-1}) \\ 
  \end{gather*}
  \end{minipage} 
& 
 \begin{minipage}[c][40pt][c]{1.3in}
  \begin{gather*}
  ({\+\widetilde{y}},{\+\widetilde{x}},{-\widetilde{z}})\\
  ({-\widetilde{x}},{-\widetilde{y}},{\widetilde{z}^{-1}})\\
  \end{gather*}
  \end{minipage} 
  &
   \begin{minipage}[c][40pt][c]{1.3in}
  \begin{gather*} 
  (\+a,-b,\+c,-d) \\ 
 (\+b,\+a,\+d,\+c) \\
  \end{gather*}
  \end{minipage} \\
\hline
\end{longtable}
\end{center}
\normalsize
\small
\begin{center}
\begin{longtable}{|c|c|c|c|}
\captionsetup{width=0.9\textwidth}
\caption{\it Symmetry actions on the cohomology basis and the corresponding invariants for the manifold~\eqref{eq:dP4_480}. The matrices $P_i$ and $Q_i$ are defined in \eqref{eq:CohMatDefs2}.} \label{dP4_480_Symm_Action_Coh} \\

\hline \multicolumn{1}{|c|}{\str\textbf{Index}}&  \multicolumn{1}{|c|}{\str\textbf{~Group~}} & \multicolumn{1}{|c|}{\str\textbf{Action on Coh Basis}} &  \multicolumn{1}{|c|}{\str\textbf{~~Coh Invariants~~}} \\ \hline 
\endfirsthead

\hline 
\textbf{Index} & \textbf{~Group~} & \textbf{Action on Coh Basis} & \textbf{~~Coh Invariants~~} \\ \hline 
\endhead

\hline\hline \multicolumn{4}{|r|}{{\str Continued on next page}} \\ \hline
\endfoot

\endlastfoot

\hline\hline
1 & $\IZ_2$ &
$\left[\begin{array}{cc}
P_2 & 0 \\
0 & P_2
\end{array}\right]$
& \begin{minipage}[c][38pt][c]{2.8in}
\begin{gather*} 
H, E_1+E_4,E_2+E_3,E_5,\\
\widetilde{H}, \widetilde{E}_1+\widetilde{E}_4, \widetilde{E}_2+\widetilde{E}_3,\widetilde{E}_5 \\
\end{gather*}
\end{minipage} \\
 \hline
 
2 & $\IZ_2$ &
$\left[\begin{array}{cc}
Q_1 & 0 \\
0 & P_2
\end{array}\right]$
& \begin{minipage}[c][38pt][c]{3in}
\begin{gather*} 
H-E_5, E_1+E_4-E_5, E_2+E_4-E_5,\\
 E_3-E_4,~\widetilde{H}, \widetilde{E}_1+\widetilde{E}_4,\widetilde{E}_2+\widetilde{E}_3,\widetilde{E}_5 \\
\end{gather*}
\end{minipage} \\
 \hline
 
3 & $\IZ_2$ &
$\left[\begin{array}{cc}
Q_1 & 0 \\
0 & Q_1
\end{array}\right]$
&\begin{minipage}[c][38pt][c]{3.1in}
\begin{gather*} 
H-E_5, E_1+E_4-E_5, E_2+E_4-E_5, E_3-E_4,\\
\widetilde{H}-\widetilde{E}_5, \widetilde{E}_1+\widetilde{E}_4-\widetilde{E}_5, \widetilde{E}_2+\widetilde{E}_4-\widetilde{E}_5,\widetilde{E}_3-\widetilde{E}_4 \\
\end{gather*}
\end{minipage} \\ \hline

4 & $\IZ_4$ &
$\left[\begin{array}{cc}
0 & Q_1 \\
Q_4 & 0
\end{array}\right]$
& \begin{minipage}[c][67pt][c]{2.8in}
\begin{gather*} 
H+3\widetilde{H}-\widetilde{E}_1-\widetilde{E}_2-\widetilde{E}_3-\widetilde{E}_4-2\widetilde{E}_5, \\[-3pt]
E_1+E_2+2\widetilde{H}-\widetilde{E}_1-\widetilde{E}_2-2\widetilde{E}_5,\\[-3pt]
E_3+E_4+2\widetilde{H}-\widetilde{E}_3-\widetilde{E}_4-2\widetilde{E}_5,\\[-3pt]
E_5+2\widetilde{H}-\widetilde{E}_1-\widetilde{E}_2-\widetilde{E}_3-\widetilde{E}_4-\widetilde{E}_5 \\[-3pt]
\end{gather*}
\end{minipage} \\
 \hline
 
5 & $\IZ_4$ &
$\left[\begin{array}{cc}
0 & P_1 \\
Q_4 & 0
\end{array}\right]$
& \begin{minipage}[c][51pt][c]{2.8in}
\begin{gather*} 
H-E_5+\widetilde{H}-\widetilde{E}_5,
E_3-E_4-(\widetilde{E}_3-\widetilde{E}_4), \\[-3pt]
E_1+E_4-E_5+\widetilde{E}_2+\widetilde{E}_3-\widetilde{E}_5,\\[-3pt]
E_2+E_4-E_5+\widetilde{E}_1+\widetilde{E}_3-\widetilde{E}_5 \\[-3pt]
\end{gather*}
\end{minipage} \\
 \hline
 
6 & $\IZ_2{\times}\IZ_2$ &
$\left[\begin{array}{cc}
P_1 & 0 \\
0 & P_1
\end{array}\right]$, $\left[\begin{array}{cc}
P_2 & 0 \\
0 & P_2
\end{array}\right]$
& \begin{minipage}[c][38pt][c]{1.8in}
\begin{gather*} 
H, E_1+E_2+E_3+E_4,E_5, \\
~\widetilde{H}, \widetilde{E}_1+\widetilde{E}_2+\widetilde{E}_3+\widetilde{E}_4,\widetilde{E}_5 \\
\end{gather*}
\end{minipage} \\
 \hline
 
7 & $\IZ_2{\times}\IZ_2$ &
$\left[\begin{array}{cc}
Q_1 & 0 \\
0 & P_2
\end{array}\right]$, $\left[\begin{array}{cc}
Q_2 & 0 \\
0 & P_1
\end{array}\right]$& \begin{minipage}[c][38pt][c]{2.8in}
\begin{gather*} 
H-E_5, E_1+E_3-E_5, E_2+E_4-E_5,\\
 \widetilde{H}, \widetilde{E}_1+\widetilde{E}_2+\widetilde{E}_3+\widetilde{E}_4,\widetilde{E}_5 \\
\end{gather*}
\end{minipage} \\
 \hline
8 & $\IZ_2{\times}\IZ_2$ &
$\left[\begin{array}{cc}
P_1 & 0 \\
0 & P_2
\end{array}\right]$, $\left[\begin{array}{cc}
Q_2 & 0 \\
0 & P_1
\end{array}\right]$
& \begin{minipage}[c][38pt][c]{2.8in}
\begin{gather*} 
H-E_5, E_1+E_2-E_5, E_3+E_4-E_5,\\
 \widetilde{H}, \widetilde{E}_1+\widetilde{E}_2+\widetilde{E}_3+\widetilde{E}_4,\widetilde{E}_5 \\
\end{gather*}
\end{minipage} \\
 \hline
 9 & $\IZ_2{\times}\IZ_2$ &
$\left[\begin{array}{cc}
Q_1 & 0 \\
0 & P_2
\end{array}\right]$, $\left[\begin{array}{cc}
P_2 & 0 \\
0 & P_1
\end{array}\right]$
& \begin{minipage}[c][38pt][c]{2.8in}
\begin{gather*} 
H-E_5, E_1+E_4-E_5, E_2+E_3-E_5,\\
 \widetilde{H}, \widetilde{E}_1+\widetilde{E}_2+\widetilde{E}_3+\widetilde{E}_4,\widetilde{E}_5 \\
\end{gather*}
\end{minipage} \\
 \hline
10 & $\IZ_2{\times}\IZ_2$ &
$\left[\begin{array}{cc}
Q_1 & 0 \\
0 & Q_1
\end{array}\right]$, $\left[\begin{array}{cc}
P_2 & 0 \\
0 & P_2
\end{array}\right]$
& \begin{minipage}[c][38pt][c]{2.8in}
\begin{gather*} 
H-E_5, E_1+E_4-E_5, E_2+E_3-E_5,\\
~ \widetilde{H}-\widetilde{E}_5, \widetilde{E}_1+\widetilde{E}_4-\widetilde{E}_5,\widetilde{E}_2+\widetilde{E}_3-\widetilde{E}_5 \\\end{gather*}
\end{minipage} \\
 \hline
 11 & $\IZ_2{\times}\IZ_2$ &
$\left[\begin{array}{cc}
P_1 & 0 \\
0 & P_2
\end{array}\right]$, $\left[\begin{array}{cc}
Q_2 & 0 \\
0 & Q_1
\end{array}\right]$
& \begin{minipage}[c][38pt][c]{2.8in}
\begin{gather*} 
H-E_5, E_1+E_2-E_5, E_3+E_4-E_5,\\
~ \widetilde{H}-\widetilde{E}_5, \widetilde{E}_1+\widetilde{E}_4-\widetilde{E}_5,\widetilde{E}_2+\widetilde{E}_3-\widetilde{E}_5 \\\end{gather*}
\end{minipage} \\
 \hline
12 & $\IZ_2{\times}\IZ_2$ &
$\left[\begin{array}{cc}
Q_1 & 0 \\
0 & Q_2
\end{array}\right]$, $\left[\begin{array}{cc}
Q_2 & 0 \\
0 & Q_1
\end{array}\right]$
& \begin{minipage}[c][38pt][c]{2.8in}
\begin{gather*} 
H-E_5, E_1+E_3-E_5, E_2+E_4-E_5,\\
~ \widetilde{H}-\widetilde{E}_5, \widetilde{E}_1+\widetilde{E}_3-\widetilde{E}_5,\widetilde{E}_2+\widetilde{E}_4-\widetilde{E}_5 \\\end{gather*}
\end{minipage} \\
 \hline
13 & $\IZ_2{\times}\IZ_2$ &
$\left[\begin{array}{cc}
Q_1 & 0 \\
0 & Q_1
\end{array}\right]$, $\left[\begin{array}{cc}
Q_2 & 0 \\
0 & Q_2
\end{array}\right]$
& \begin{minipage}[c][38pt][c]{2.8in}
\begin{gather*} 
H-E_5, E_1+E_3-E_5, E_2+E_4-E_5,\\
~ \widetilde{H}-\widetilde{E}_5, \widetilde{E}_1+\widetilde{E}_4-\widetilde{E}_5,\widetilde{E}_2+\widetilde{E}_3-\widetilde{E}_5 \\\end{gather*}
\end{minipage} \\
 \hline 
\end{longtable}
\vspace{-30pt}
\end{center}

\subsection{The manifold $X^{12,28}$ and its splits}\label{sec:12_28}
In our sub-class of CICYs admitting freely acting symmetries that are either $\IZ_2$ or of order divisible by $4$, there are three manifolds with Hodge numbers $(12,28)$. All three of them can be argued to be embedded in a product of two del Pezzo surfaces of degree 4. 

\subsubsection{$X^{12,28}$ and its split $X^{15,15}$}\label{REM_X12,28_1}\vskip-8pt
The first of the $(12,28)$ manifolds can be obtained by splitting the first column of the $X^{8,40}$ manifold \eqref{eq:TQSplit3}. It is defined by the following configuration:
\begin{equation}\label{eq:TQSplit16}
\displaycicy{5.25in}{
X_{2568}~=~~
\cicy{\IP^1 \\ \IP^1\\ \IP^1\\ \IP^1\\ \IP^1\\ \IP^1}
{ ~1 &1 & 0  ~\\
  ~1 &1 & 0  ~\\
  ~0 &2 & 0 ~\\
  ~1 &0 & 1 ~\\
  ~1 &0 & 1 ~\\
    ~0 &0 & 2~\\}_{-32}^{12,28}}{-1.0cm}{2.5in}{TQSplit16.pdf}
\end{equation}

This manifold as well as its quotients have been discussed in detail in Section~\ref{sec:X2568}. %
A splitting of the first column of \eqref{eq:TQSplit16}, leads to the manifold $X^{15,15}$, defined by the following configuration:
\begin{equation}\label{eq:TQSplit18}
\displaycicy{5.25in}{
X_{22}~=~~
\cicy{\IP^1 \\ \IP^1 \\ \IP^1\\ \IP^1\\ \IP^1\\ \IP^1\\ \IP^1}
{ ~1 &1 & 0& 0  ~\\
  ~1 &0 & 1& 0  ~\\
  ~0 &1 &1 & 0  ~\\
  ~0 &0 &2 & 0 ~\\
  ~1 &0 &0 & 1 ~\\
  ~0 &1 &0 & 1 ~\\
  ~0 &0 &0 & 2~\\}_{0}^{15,15}}{-1.0cm}{2.4in}{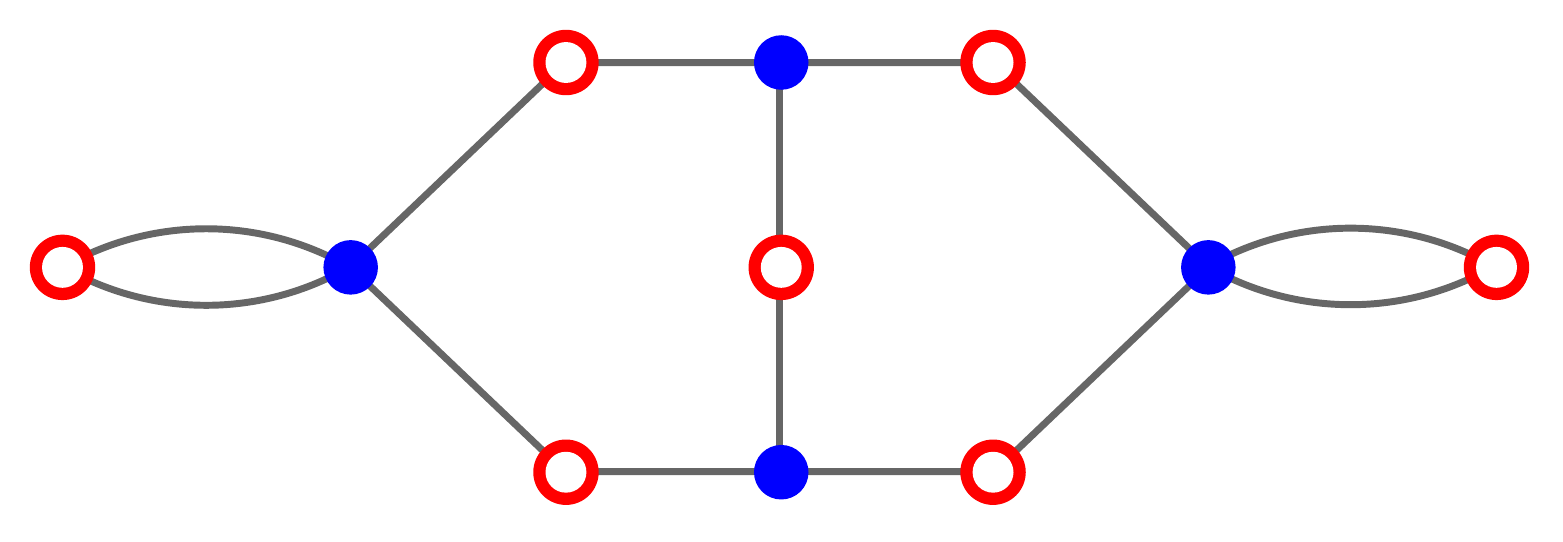}
\end{equation}
The manifold \eqref{eq:TQSplit18} admits $15$ different free group actions by $\IZ_2$ and $\IZ_2\times \IZ_2$. We compute the Hodge numbers for the resulting quotients using the polynomial deformation method. The results are listed in \tref{TQSplit18quotients}.

\begin{table}[!ht]
\vspace{20pt}
\begin{center}
\begin{tabular}{| c || c | c | c |}
\hline
\myalign{| c||}{\varstr{16pt}{10pt}$~~~~~~~  \Gamma ~~~~~~~$ } &
\myalign{m{1.31cm}|}{$~~~ \IZ_2 $} &
\myalign{m{1.31cm}|}{$~~~ \IZ_2 $} &
\myalign{m{1.9cm}|}{ $~~~\IZ_2\times\IZ_2 \ \ \ $ }  
\\ \hline\hline
\varstr{14pt}{8pt} $h^{1,1}(X/\Gamma)$ & 9 & 10& 7 \\
 \hline
\varstr{14pt}{8pt} $h^{2,1}(X/\Gamma)$ & 9 & 10& 7 \\
 \hline
\varstr{14pt}{8pt} $\chi(X/\Gamma)$ & $ 0 $ & $ 0$& $ 0 $   \\
 \hline
 \end{tabular}
   \vskip 0.3cm
\capt{4.5in}{TQSplit18quotients}{Hodge numbers for the quotients of the manifold \eqref{eq:TQSplit18}.}
 \end{center}
 \end{table}
\newpage
\subsubsection{$X^{12,28}$ and its split $X^{19,19}$} \label{REM_X12,28_2}\vskip-8pt
The second $(12,28)$ manifold corresponds to the following configuration:
\vskip-12pt
\begin{equation}\label{eq:M9}
\displaycicy{5.25in}{
X_{2564}~=~~
\cicy{\IP^4\\\IP^4}
{  ~2&2&1& 0& 0~ \\
  ~0&0&1& 2& 2~ \\}_{-32}^{12,\,28}}{-1.0cm}{0.9in}{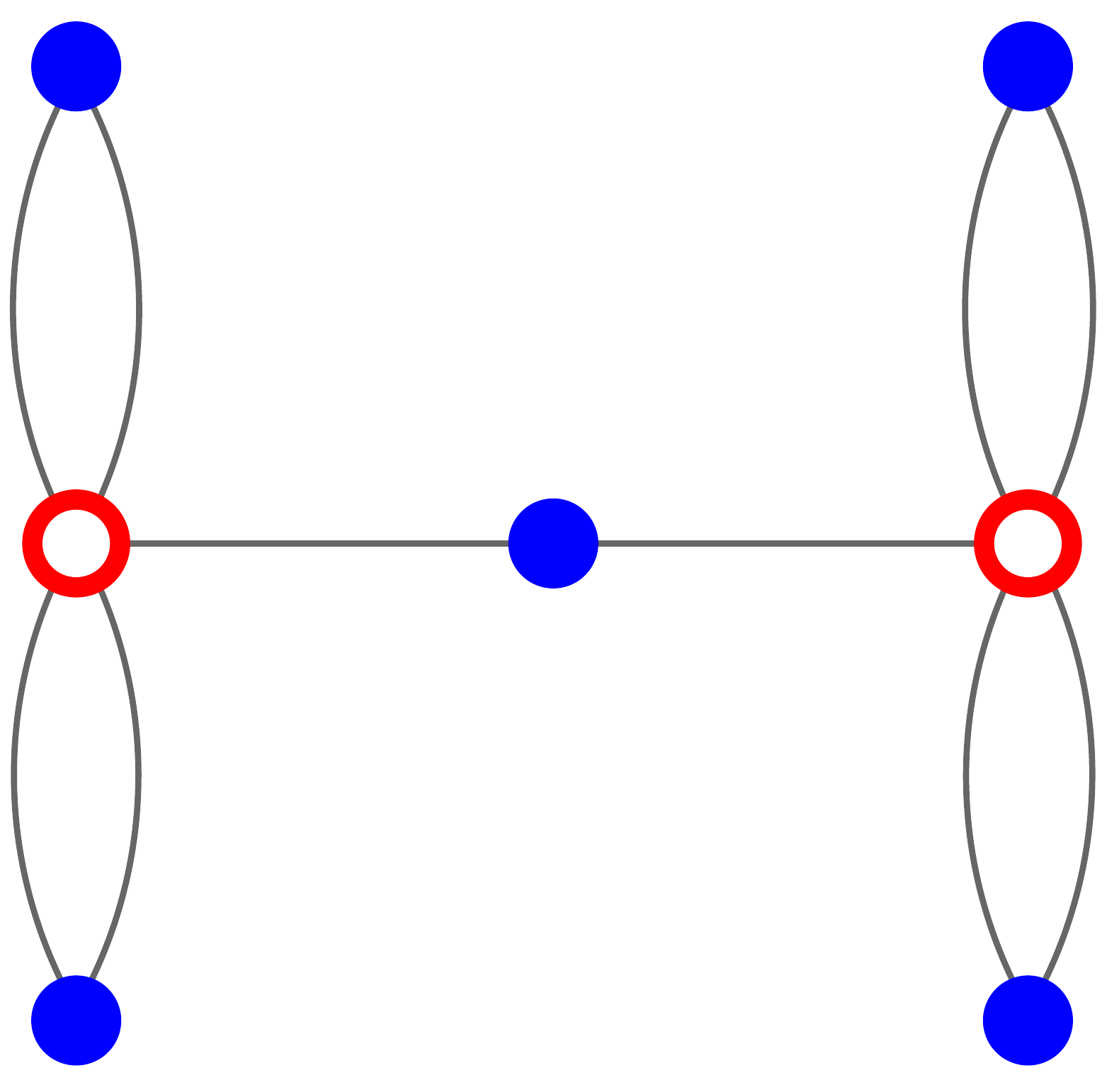}
\end{equation}

\vskip-5pt
The manifold \eqref{eq:M9} admits $9$ different free group actions by $\IZ_2, \IZ_4, \IZ_2\times \IZ_2, \IZ_8, \IZ_2\times \IZ_4$ and~$\IQ_8$. The polynomial deformation method correctly reproduces the number of complex structure parameters for the covering manifold, and, although the diagram is one-leg decomposable, we will assume that it provides a complete parametrisation of the complex structure moduli space. With this assumption, we are able to compute the Hodge numbers listed in \tref{M9quotients}. 
\vskip3pt
\begin{table}[H]
\begin{center}
\begin{tabular}{| c || c | c | c | c | c | c | c |}
\hline
\myalign{| c||}{\varstr{16pt}{10pt}$~~~~~~~  \Gamma ~~~~~~~$ } &
\myalign{m{1.31cm}|}{$~~~ \IZ_2 $} &
\myalign{m{1.31cm}|}{$~~~ \IZ_4 $} &
\myalign{m{1.9cm}|}{ $~~~\IZ_2\times\IZ_2 \ \ \ $ }  &
\myalign{m{1.31cm}|}{$~~~ \IZ_8 $} &
\myalign{m{1.9cm}|}{ $~~~\IZ_2\times\IZ_4 \ \ \ $ }  &
\myalign{m{1.31cm}|}{$~~~ \IQ_8 $} 
\\ \hline\hline
\varstr{14pt}{8pt} $h^{1,1}(X/\Gamma)$ & 8 & 4 & 6 & 2 & 3 & 2\\
 \hline
\varstr{14pt}{8pt} $h^{2,1}(X/\Gamma)$ & 16 & 8 & 10 & 4 & 5 & 4 \\
 \hline
\varstr{14pt}{8pt} $\chi(X/\Gamma)$ & $ \!\!\!\!-16 $ &  $ \!\!\!\!-8 $& $ \!\!\!\! -8 $ & $ \!\!\!\! -4 $& $ \!\!\!\! -4 $& $ \!\!\!\! -4 $  \\
 \hline
 \end{tabular}
   \vskip 0.3cm
\capt{4.5in}{M9quotients}{Hodge numbers for the quotients of the manifold \eqref{eq:M9}.}
 \end{center}
 \vspace{-33pt}
 \end{table}

The configuration matrix \eqref{eq:M9} can be split with a $\IP^1$ to give the manifold $X^{19,19}$: 
This manifold corresponds to the following configuration:
\vskip-10pt
\begin{equation}\label{eq:M9Split1}
\displaycicy{5.25in}{
X_{19}~=~~
\cicy{\IP^1\\\IP^4\\\IP^4}
{ ~1&1&0&0& 0& 0~ \\ 
  ~1&0&2&2& 0& 0~ \\
  ~0&1&0&0& 2& 2~ \\}_{0}^{19,\,19}}{-1.1cm}{1.8in}{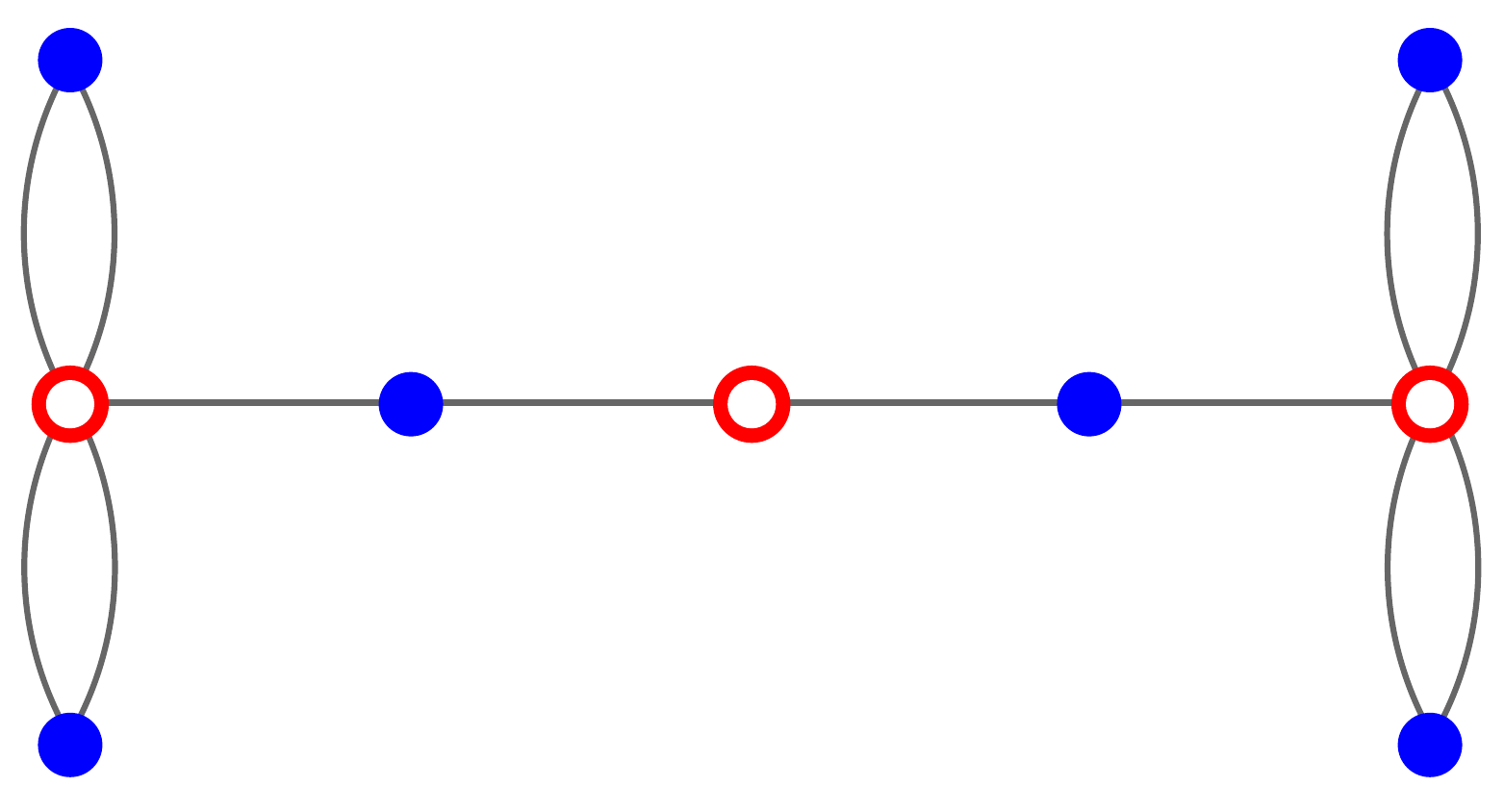}
\end{equation}
\vskip-5pt
The manifold \eqref{eq:M9Split1} admits $29$ different free group actions by the same groups as the previous manifold. We compute the Hodge numbers for the quotient manifolds using the same assumption as above. The results are listed in \tref{M9Split1quotients}. 
\vskip5pt
\begin{table}[H]
\begin{center}
\begin{tabular}{| c || c | c | c | c | c | c | c |}
\hline
\myalign{| c||}{\varstr{16pt}{10pt}$~~~~~~~  \Gamma ~~~~~~~$ } &
\myalign{m{1.31cm}|}{$~~~ \IZ_2 $} &
\myalign{m{1.31cm}|}{$~~~ \IZ_4 $} &
\myalign{m{1.31cm}|}{$~~~ \IZ_4 $} &
\myalign{m{1.9cm}|}{ $~~~\IZ_2\times\IZ_2 \ \ \ $ }  &
\myalign{m{1.31cm}|}{$~~~ \IZ_8 $} &
\myalign{m{1.9cm}|}{ $~~~\IZ_2\times\IZ_4 \ \ \ $ }  &
\myalign{m{1.31cm}|}{$~~~ \IQ_8 $} 
\\ \hline\hline
\varstr{14pt}{8pt} $h^{1,1}(X/\Gamma)$ & 11 & 5 & 6 & 7 & 3 & 4 & 3\\
 \hline
\varstr{14pt}{8pt} $h^{2,1}(X/\Gamma)$ & 11 & 5 & 6 & 7 & 3 & 4 & 3 \\
 \hline
\varstr{14pt}{8pt} $\chi(X/\Gamma)$ & $ 0 $ & $ 0 $& $ 0 $& $ 0 $ & $ 0 $& $ 0 $& $ 0 $  \\
 \hline
 \end{tabular}
   \vskip 0.3cm
\capt{4.5in}{M9Split1quotients}{Hodge numbers for the quotients of the manifold \eqref{eq:M9Split1}.}
 \end{center}
 \end{table}
\newpage
\subsubsection{Another $X^{12,28}$}\label{sec:12_28_2566}\vskip-8pt
An $X^{12,28}$ manifold can be obtained through a splitting of the first column of a $(8,40)$ manifold~\eqref{eq:M8}. This corresponds to the following configuration:
\begin{equation}\label{eq:M8Split2}
\displaycicy{5.25in}{
X_{2566}~=~~
\cicy{\IP^1\\\IP^1\\\IP^1\\\IP^4}
{ ~2&0& 0& 0~ \\
  ~1&1& 0& 0~ \\
  ~1&1& 0& 0~ \\
  ~0&1& 2& 2~ \\}_{-32}^{12,\,28}}{-1.1cm}{1.7in}{M8Split2.pdf}
\end{equation}
This manifold, embedded in a product of two $\text{dP}_4$'s, admits $10$ different free group actions by $\IZ_2$ and $\IZ_2\times \IZ_2$. The polynomial deformation method does not correctly reproduce the number of complex structure parameters in this case. We have listed the Hodge numbers of the quotient manifolds in \tref{dP4_2566}.

\subsection{Other non-favourable manifolds}
\vskip-10pt
\subsubsection{The manifold $X^{8,40}$ and its split $X^{19,19}$}\label{REM_OTHER_X8,40}\vskip-8pt
The manifold $X^{8,40}$ is defined by the following configuration:
\begin{equation}\label{eq:M8}
\displaycicy{5.25in}{
X_{6826}~=~~
\cicy{\IP^1\\\IP^1\\\IP^4}
{ ~2& 0& 0~ \\
  ~2& 0& 0~ \\
  ~1& 2& 2~ \\}_{-64}^{\,8,\,40}}{-1.05cm}{0.5in}{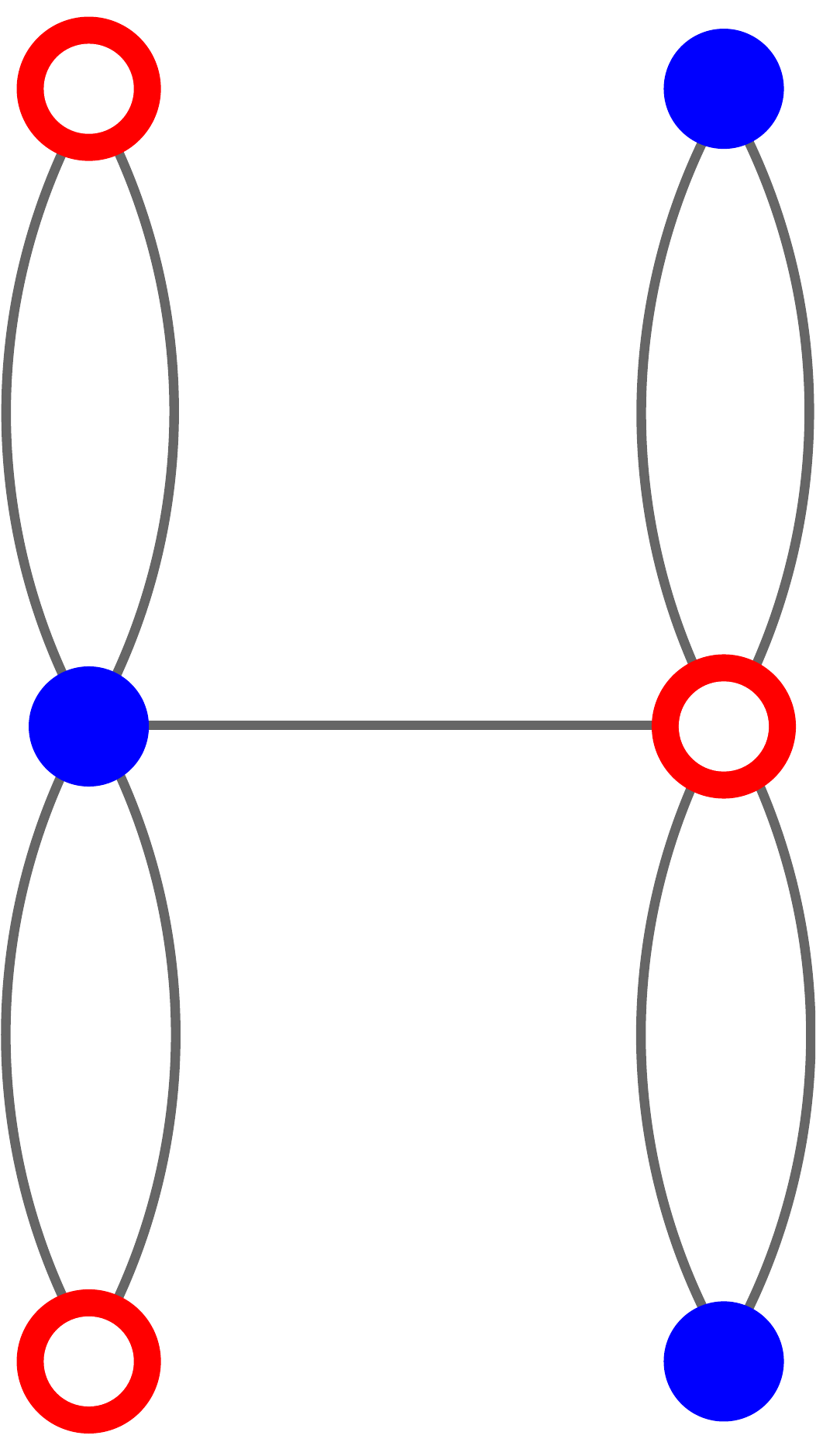}
\end{equation}
The manifold \eqref{eq:M8} admits $3$ different free group actions by $\IZ_2, \IZ_4$ and $\IZ_2\times \IZ_2$. Although the diagram is one-leg decomposable, the polynomial deformation method correctly reproduces the number of complex structure parameters for the covering manifold. Assuming that the method provides a complete parametrisation of the complex structure moduli space, we compute in this way the Hodge numbers $h^{2,1}(X/\Gamma)$. 

\begin{table}[H]
\vspace{2pt}
\begin{center}
\begin{tabular}{| c || c | c | c |}
\hline
\myalign{| c||}{\varstr{16pt}{10pt}$~~~~~~~  \Gamma ~~~~~~~$ } &
\myalign{m{1.31cm}|}{$~~~ \IZ_2 $} &
\myalign{m{1.31cm}|}{$~~~ \IZ_4 $} &
\myalign{m{1.9cm}|}{ $~~~\IZ_2\times\IZ_2 \ \ \ $ }  
\\ \hline\hline
\varstr{14pt}{8pt} $h^{1,1}(X/\Gamma)$ & 6 & 3 & 5\\
 \hline
\varstr{14pt}{8pt} $h^{2,1}(X/\Gamma)$ & 22 & 11 & 13\\
 \hline
\varstr{14pt}{8pt} $\chi(X/\Gamma)$ & $ \!\!\!\!-32 $ & $ \!\!\!\! - 16$& $ \!\!\!\! -16$  \\
 \hline
 \end{tabular}
   \vskip 0.3cm
\capt{4.5in}{M8quotients}{Hodge numbers for the quotients of the manifold \eqref{eq:M8}.}
 \end{center}
 \vspace{2pt}
 \end{table}
\newpage
The manifold $X^{19,19}$ can be obtained by splitting the first column of \eqref{eq:M8}, leading to the following favourable embedding:
\begin{equation}\label{eq:M8Split1}
\displaycicy{5.25in}{
X_{20}~=~~
\cicy{\IP^1\\\IP^1\\\IP^1\\\IP^4}
{ ~2&0& 0& 0~ \\
  ~2&0& 0& 0~ \\
  ~1&1& 0& 0~ \\
  ~0&1& 2& 2~ \\}_{0}^{19,\,19}}{-1.27cm}{1.5in}{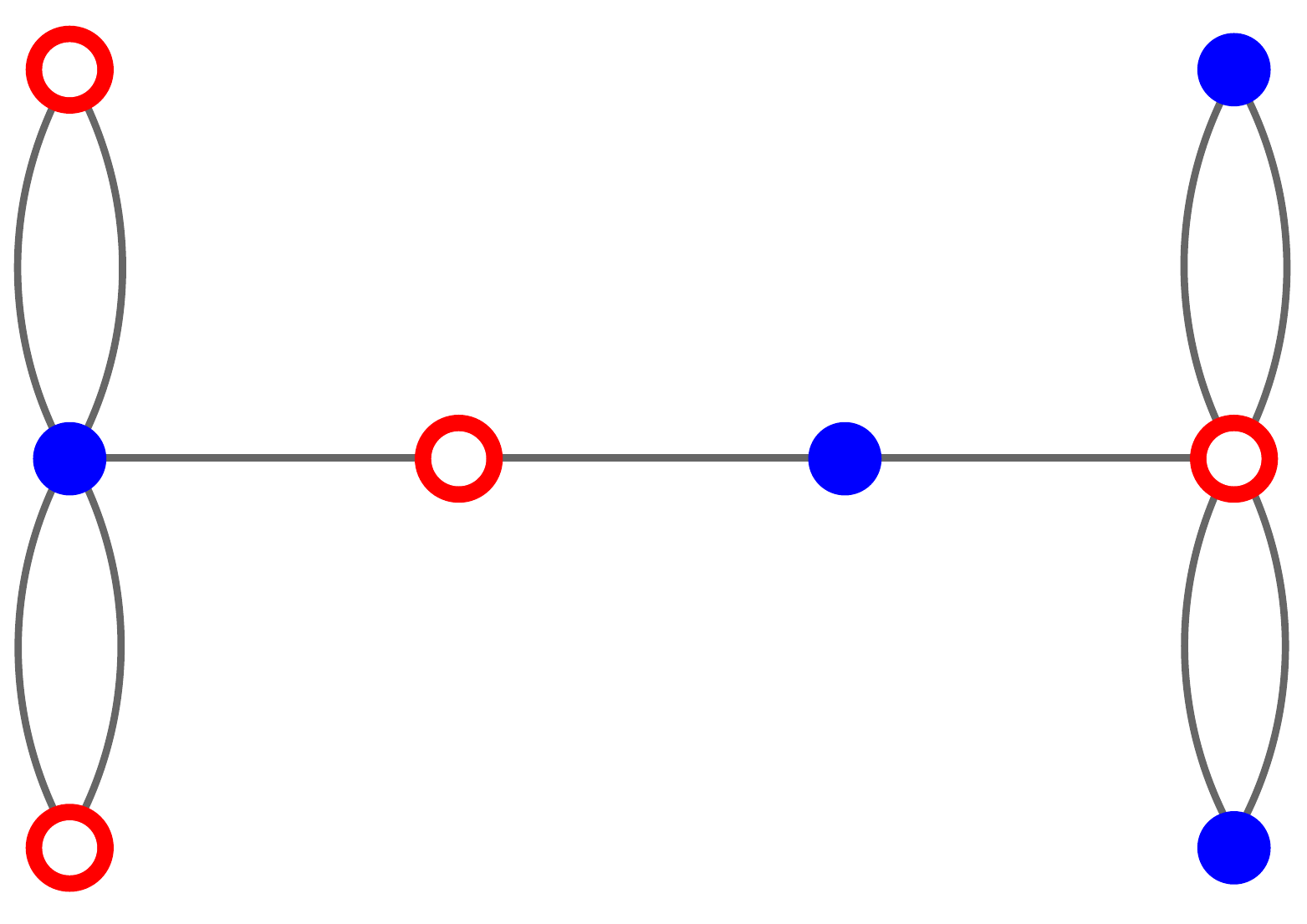}
\end{equation}
The manifold \eqref{eq:M8Split1} admits $14$ free group actions by $\IZ_2, \IZ_4$ and $\IZ_2\times \IZ_2$. For the computation of the Hodge numbers for the quotient manifolds, the same comments as for the previous manifold apply. We present the results in \tref{M8Split1quotients}. 
\begin{table}[H]
\vspace{2pt}
\begin{center}
\begin{tabular}{| c || c | c | c |}
\hline
\myalign{| c||}{\varstr{16pt}{10pt}$~~~~~~~  \Gamma ~~~~~~~$ } &
\myalign{m{1.31cm}|}{$~~~ \IZ_2 $} &
\myalign{m{1.31cm}|}{$~~~ \IZ_4 $} &
\myalign{m{1.9cm}|}{ $~~~\IZ_2\times\IZ_2 \ \ \ $ }  
\\ \hline\hline
\varstr{14pt}{8pt} $h^{1,1}(X/\Gamma)$ & 11 & 5 & 7\\
 \hline
\varstr{14pt}{8pt} $h^{2,1}(X/\Gamma)$ & 11 & 5 & 7\\
 \hline
\varstr{14pt}{8pt} $\chi(X/\Gamma)$ & $ 0 $ & $ 0 $& $ 0 $  \\
 \hline
 \end{tabular}
   \vskip 0.3cm
\capt{4.5in}{M8Split1quotients}{\it\small Hodge numbers for the quotients of the manifold \eqref{eq:M8Split1}.}
 \end{center}
 \vskip-20pt
 \end{table}

A different splitting of the first column of~\eqref{eq:M8} yields the $X^{12,28}$ that was discussed in \ref{sec:12_28_2566}.
\subsubsection{The manifold $X^{19,19}$ once again} \label{REM_OTHER_X19,19}\vskip-8pt
There is another configuration matrix which leads to the manifold $X^{19,19}$:
\begin{equation}\label{eq:M10}
\displaycicy{5.5in}{
X_{30}~=~~
\cicy{\IP^1\\\IP^1\\\IP^1\\\IP^2\\\IP^2}
{ ~2&0&0&0~ \\ 
  ~0&2&0&0~ \\
  ~0&0&1&1~ \\
  ~2&0&1&0~ \\
  ~0&2&0&1~ \\}_{0}^{19,\,19}}{-0.2cm}{3.4in}{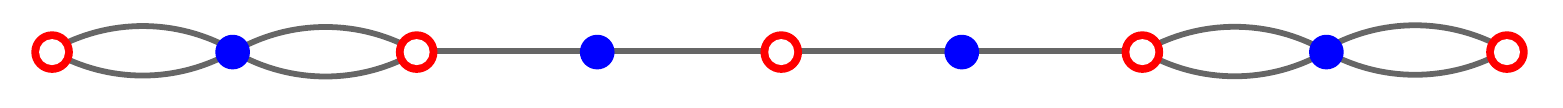}
\end{equation}

In this embedding, $X^{19,19}$ admits only $4$ different, linearly represented, free group actions, by $\IZ_2$ and $\IZ_4$. The Hodge numbers for the $\IZ_2$--quotients are $(11,11)$, while for the $\IZ_4$--quotient $(6,6)$. We have obtained these through the application of the polynomial deformation method. 
\newpage
\section{Conclusions}
Since we are, in a sense, summarising a series of papers, it seems worthwhile to take stock of the manifolds that have been found. The following has some overlap with \cite{Candelas:2008wb} and \cite{Davies:2011fr}, which give more detail and contain references to the original literature. 

A graphical depiction of our present knowledge of the existence of \cys of small height ($=h^{1,1}+ h^{1,2}$) is provided by~\fref{TipHodgePlot}. For the purposes of the present discussion we follow~\cite{Davies:2011fr} in making the subjective choice of considering Hodge numbers to be small if $h^{1,1}{+} h^{1,2}{\,\leq\,}24$. Most of the manifolds of the plot have been found as quotients of CICYs. Although many of these manifolds are interesting in themselves, it is also evident that the resolution of singular limits and singular quotients of these yields other important manifolds. Examples of this will be indicated below.

Any list of interesting manifolds must include Yau's manifold 
\beqnn
\cicy{\IP^3\\\IP^3}{1&3&0\\ 1&0&3}^{6,9}\quotient{\IZ_3}
\eeqnn
that led to the first of the three generation models and which inspired the CICY class \cite{Greene:1986bm, Greene:1986jb}. 
The three generation UPenn models~\cite{Donagi:2000zf, Donagi:2000zs, Donagi:2004ub, Braun:2005nv, Bouchard:2005ag} are based on a $\IZ_3{\times}\IZ_3$ quotient of the split bicubic.
\beq
\cicy{\IP^1\\ \IP^2 \\ \IP^2}{1&1\\ 3&0\\ 0&3\\}^{19,19}
\notag\eeq
The split bicubic itself
appears in many guises in the CICY list. As remarked previously, all 15 occurences of the Hodge numbers $(19,19)$ can be seen to correspond to the same manifold by a series of ineffective splittings and contractions. By considering these different representations and  drawing also on \cite{Bouchard:2007mf} we have the following quotients, all of which, of course, have $\chi{\,=\,}0$.
\vskip5pt
\begin{table}[H]
\begin{center}
\begin{tabular}{|c|| *{10}{@{}c@{}|}}
\hline
\varstr{16pt}{10pt}$\Gamma$  &
\myalign{m{1.0cm}|}{\centering$\IZ_2 $} &
\myalign{m{1.0cm}|}{\centering$\IZ_3 $} &
\myalign{m{1.0cm}|}{\centering$\IZ_4 $} &
\myalign{m{1.0cm}|}{\centering$\IZ_4 $} &
\myalign{m{1.0cm}|}{\centering$\!\IZ_2{\times}\IZ_2$} &
\myalign{m{1.0cm}|}{\centering$\IZ_5 $} &
\myalign{m{1.0cm}|}{\centering$\IZ_6 $} &
\myalign{m{1.0cm}|}{\centering$\IZ_8 $} &
\myalign{m{1.0cm}|}{\centering$\IQ_8 $} & 
\myalign{m{1.0cm}|}{\centering$\!\IZ_4{\times}\IZ_2$} 
\\ \hline\hline
\varstr{14pt}{8pt} $\!\!(h^{1,1}, h^{2,1})\!\!$ & \!(11,11)\! & (7,\,7) & (5,\,5) & (6,\,6) & (7,\,7) & (3,\,3) & (3,\,3) &(3,\,3) & (3,\,3) & (4,\,4)\\
 \hline
\varstr{14pt}{8pt} Ref. & \!\!\cite{Bouchard:2007mf,Candelas:2008wb}\!\! & \!\!\cite{Bouchard:2007mf,Candelas:2008wb}\!\! & \cite{Bouchard:2007mf} &  & \cite{Bouchard:2007mf}  & \cite{Bouchard:2007mf}  &  \!\!\cite{Bouchard:2007mf,Candelas:2008wb}\!\! && \cite{Candelas:2008wb} &\\
 \hline
 \end{tabular}
 \vskip 12pt
 
\begin{tabular}{|c|| *{10}{@{}c@{}|}}
\hline
\varstr{16pt}{10pt}$\Gamma$  &
\myalign{m{1.0cm}|}{\centering$\!\IZ_4{\times}\IZ_2 $} &
\myalign{m{1.0cm}|}{\centering$\!\IZ_3{\times}\IZ_3 $} &
\myalign{m{1.0cm}|}{\centering$\IZ_{12} $}            &
\myalign{m{1.0cm}|}{\centering$\text{Dic}_3 $}         &
\myalign{m{1.0cm}|}{\centering$\!\IZ_4{\times}\IZ_4 $} &
\myalign{m{1.0cm}|}{\centering$\!\IZ_4{\rtimes}\IZ_4 $} &
\myalign{m{1.0cm}|}{\centering$\!\IZ_8{\times}\IZ_2 $} &
\myalign{m{1.0cm}|}{\centering$\!\IZ_8{\rtimes}\IZ_2 $} &
\myalign{m{1.0cm}|}{\centering$\hskip-3pt\IQ_8{\times}\IZ_2$} &
\myalign{m{1.0cm}|}{\centering${}$}

\\ \hline\hline
\varstr{14pt}{8pt} $\!\!(h^{1,1}, h^{2,1})\!\!$  & (3,\,3) & (3,\,3) & (2,\,2) & (2,\,2)
 & (2,\,2) & (2,\,2) & (2,\,2) & (2,\,2) & (2,\,2) &\\
 \hline
\varstr{14pt}{8pt} Ref. &\cite{Bouchard:2007mf} &\cite{Bouchard:2007mf}  &\cite{Braun:2009qy}
&\cite{Braun:2009qy}&&&&&&\\
 \hline
 \end{tabular}
  \vskip 0.3cm
\capt{6.0in}{19,19Quotients}{Hodge numbers for the quotients of the manifold $X^{19,19}$. Whenever a reference is not given, the corresponding quotients have not, to our knowledge, been previously discussed.}
 \end{center}
\vskip-20pt
 \end{table}
The manifold with Hodge numbers $(15,15)$, which we have met here as the configuration \eqref{eq:TQSplit14}, also occurs 15 times in the CICY list in different guises. There are also extended representations that are not in the list, but which might manifest additional symmetries, including as a $15{\times}18$ matrix (see \cite[Table 15]{Candelas:2008wb}), which is the maximum size for a  CICY matrix. One way to think of this manifold is as a codimension 3 submanifold of 
$\text{dP}_6{\times}\text{dP}_6{\times}\text{dP}_6$. This manifold is special because it can be highly symmetric and, like the split bicubic, it has many quotients. A somewhat more detailed discussion of the manifold may be found in \cite{Candelas:2008wb}. We give here the quotients that we know.
\vskip5pt
\begin{table}[H]
\begin{center}
\begin{tabular}{| c || c | c | c | c | c | c |}
\hline
\myalign{| c||}{\varstr{16pt}{10pt}$~~~~~~~  \Gamma ~~~~~~~$ } &
\myalign{m{1.28cm}|}{$~~~ \IZ_2 $} &
\myalign{m{1.28cm}|}{$~~~ \IZ_2 $} &
\myalign{m{1.28cm}|}{$~~~ \IZ_3 $} &
\myalign{m{1.28cm}|}{$~\IZ_2{\times}\IZ_2$} &
\myalign{m{1.28cm}|}{$~\IZ_2{\times}\IZ_2$} &
\myalign{m{1.28cm}|}{$~\IZ_3{\times}\IZ_2$} 
\\ \hline\hline
\varstr{14pt}{8pt} $(h^{1,1},\,h^{2,1})$ & (10,\,10) & (9,\,9) & (7,\,7) & (7,\,7) & (6,\,6) & (3,\,3)  \\
 \hline
\varstr{14pt}{8pt} Ref. & \cite{Candelas:2008wb} & & \cite{Candelas:2008wb} &  &  & \cite{Candelas:2008wb}    \\
 \hline
 \end{tabular}
   \vskip 0.3cm
\capt{5.8in}{15,15Quotients}{Hodge numbers for the quotients of the manifold $X^{15,15}$. 
}
 \end{center}
 \vskip-20pt
 \end{table}
It is interesting that there is a conifold transition~\cite{Candelas:2007ac} from the Yau manifold to $X^{19,19}/\IZ_3$, a quotient with Hodge numbers $(7,7)$. Particularly intriguing is the role of  
$X^{19,19}/\IZ_3$ as one of the two superimposed sinks in the $\IZ_3$ web of CICYs, see \fref{fig:Z3Web}  which we reproduce in a slightly updated version from \cite{Candelas:2010ve}. This is a single web with two endpoints. These are the $\IZ_3$ quotients of $X^{19,19}$ and of $X^{15,15}$, both of which have Hodge numbers $(7,7)$.
\begin{figure}[!t]
\begin{center}
\includegraphics[width=6.4in]{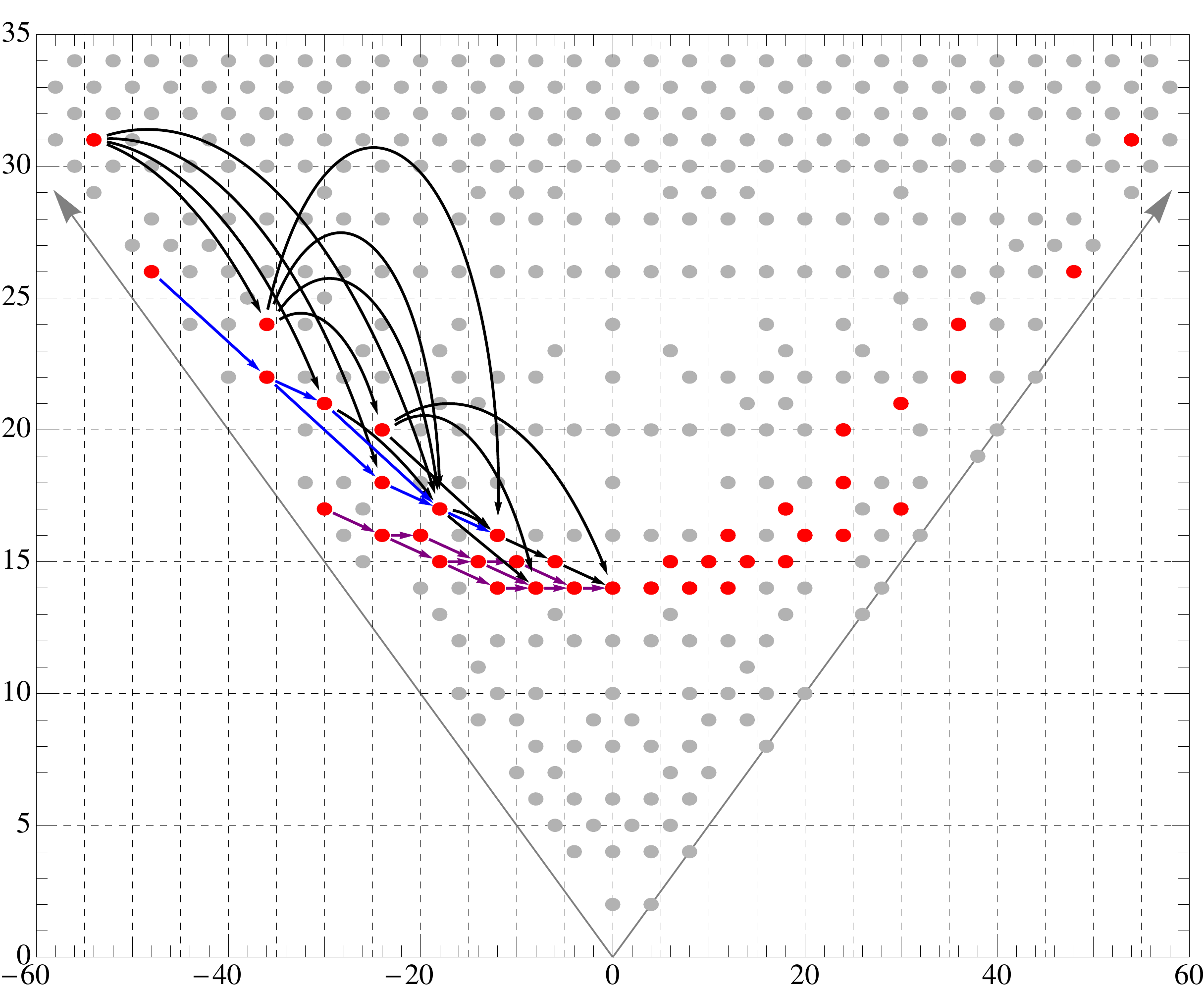}
\vskip3pt
\capt{6.0in}{fig:Z3Web}{The web of $\IZ_3$ quotients of CICY manifolds and their mirrors \cite{Candelas:2010ve}. The red points indicate $\IZ_3$ quotients. The arrows correspond to conifold transitions, and the different colours indicate three different webs of $\IZ_3$ quotients. The manifold $X^{19,19}/\IZ_3$ with Hodge numbers $(7,7)$ is the endpoint of two such sequences.}
\end{center}
\end{figure}

Notable also are the manifolds of \sref{TQ_X8,44}, with Hodge numbers (8,44), that admit smooth quotients by $\IZ_{12}$ and $\text{Dic}_3$ which also lead directly (i.e.~via the standard embedding or deformations thereof) to three generation models. It is interesting to note that these manifolds also have conifold transitions to  quotients of the split bicubic. For the $\IZ_3$ quotient we have a transition~\cite{Candelas:2008wb} to the sink just mentioned
\beq
\Big(X^{8,44}/\IZ_3\Big)^{4,16} \to~~ \Big(X^{19,19}/\IZ_3\Big)^{7,7}~.
\notag\eeq
For quotients by groups $G=\IZ_{12}$ or $G=\text{Dic}_3$ we have conifold transitions to quotients of the split bicubic that have Hodge numbers $(2,2)$.
\beq
\Big( X^{8,44}/G \Big)^{1,4} ~\to~~~\Big(X^{19,19}/G\Big)^{2,2}~.
\notag\eeq

We have hitherto concentrated on conifold transitions that preserve the fundamental group of the manifold. Davies~\cite{Davies:2009ub, Davies:2011is, Davies:2013pna} has exploited hyperconifold transitions, in which a quotient of a node is blown up into a divisor. These transitions are more drastic in that they do not preserve the fundamental group. In 
\cite{Davies:2011fr} Davies constructs two explicit examples that relate to our results. One is a hyperconifold transition from $X^{8,44}/\text{Dic}_3$ that yields a manifold with Hodge numbers $(2,3)$ and fundamental group $S_3$. This is the origin of the gray points at height 5 in \fref{TipHodgePlot}. The second example relates to a $\IZ_{10}$ quotient of the $\IP^4$-split of the quintic. A hyperconifold transition creates, in this case, a manifold with Hodge numbers $(2,5)$  and with fundamental group $\IZ_5$. This manifold has $\chi=-6$ but does not appear promising for the purposes of model building.

It was realised very early \cite{Strominger:1985it} that the $\IP^7[2,2,2,2]$ family admits smooth quotients by freely acting groups of order 32 and has manifolds with 64 nodes that admit free quotients by groups of order~64. The nodes are identified under the group action so the quotients have a single node. The smooth quotients by groups of order 32 have Hodge numbers $(1,3)$, so are very near the tip of the distribution. The nodal quotient by a group of order 64 can be resolved to give one of the remarkable Gross-Popescu manifolds with Hodge numbers~$(2,2)$. The Gross-Popescu manifolds are fibered by abelian varieties and occupy some of the sites with $\chi=0$ in \fref{TipHodgePlot}. For a slightly more detailed discussion of the Gross-Popescu manifolds see \cite{Candelas:2008wb}, which makes reference to the original papers. The tetraquadric, together with its quotients by groups of order 16, which all have Hodge numbers $(1,5)$, also deserves special mention. 

Some of the CICY's can be represented in different ways. The Yau manifold can be thought of as a quotient of a hypersurface in $\text{dP}_3{\times}\text{dP}_3$ and the three-generation manifolds with Hodge numbers $(1,4)$ can be thought of as quotients of hypersurfaces in $\text{dP}_6{\times}\text{dP}_6$. More generally many of the special CICY's can be thought of as hypersurfaces in an embedding space $\ccS{\times}\ccS'$ where $\ccS$ and $\ccS'$ are del Pezzo surfaces. The importance of this class of \cys was recognised by Bini and Favale~\cite{bini2012}.
Since $\IP^2$ and $\IP^1{\times}\IP^1$ are also del Pezzo surfaces this class contains also the configuration that follows, as well as the tetraquadric.  
\beq
X_{7884}~=~~\cicy{\IP^2\\ \IP^2}{3\\3}
\notag\eeq

Another of these spaces is the manifold
\beq
X_{2564}~=~~
\cicy{\IP^4\\\IP^4}
{ ~2&2&1& 0& 0~ \\
  ~0&0&1& 2& 2~ \\}_{-32}^{12,\,28}~,
\notag\eeq
being a hypersurface in $\text{dP}_4{\times}\text{dP}_4$, is, in many ways, an analogue of the covering space of Yau's manifold and is notable since it admits $\IQ_8$ as a freely acting symmetry.  

Returning to the embedding space $\text{dP}_6{\times}\text{dP}_6$, we note that this embedding space is a toric variety and has moreover a high degree of symmetry since each $\text{dP}_6$ has 6 exceptional lines that form a hexagon, so each $\text{dP}_6$ has $\text{Dih}_6$, the group of the hexagon, as a symmetry group. The product 
$\text{dP}_6{\times}\text{dP}_6$ has then a symmetry group 
$(\text{Dih}_6{\times}\text{Dih}_6){\,\ltimes\,}\IZ_2$
and leads to the unexplored topic of symmetric reflexive polyhedra. 

Volker Braun's manifolds, with Hodge numbers $(2,2)$ are not the quotient of a CICY, but were found by an extension of the techniques used to analyse the $(1,4)$ three-generation manifold. These manifolds have Hodge numbers $(1,1)$ and are clearly very remarkable \cite{Braun:2011hd}. They are free quotients, by groups of order 24, of manifolds specified by a reflexive polyhedron, where the polyhedron in question is the 24-cell, a four dimensional regular and self-dual polyhedron with 24 vertices and 24~faces.

The role of transposition of configurations remains an ill understood though intriguing phenomenon. It is known, and easy to see, that the transpose of a configuration matrix is again a configuration matrix of a \cy threefold. The intriguing fact is that the transposes of interesting configurations are themselves interesting. The reader will already have noted that the covering space for the Yau manifold corresponds to the transpose of the split bicubic. The transpose of $X_{2564}$, 
above,~is 
\beq
X_{21}~=~~\cicy{\IP^1\\ \IP^1\\  \IP^1\\ \IP^1\\ \IP^1}{1&1\\ 2&0\\ 2&0\\ 0&2\\ 0&2}^{19,19}
\notag\eeq
which is another of the avatars of the split bicubic and which admits freely acting groups of order 16 including 
$\IZ_2{\times}\IQ_8$.

Consider the split
\beq
\IP^4[5]^{1,101}_{\raisebox{-5pt}{$\scriptstyle\chi=-200$}}~\to~~\cicy{\IP^4\\ \IP^4}{1&1&1&1&1\\ 1&1&1&1&1}^{2,52}_{\chi=-100}~,
\notag\eeq
which has not been considered in the present paper owing to the fact that the freely acting symmetries have $\IZ_5$
as a subgroup rather than $\IZ_4$ or $\IZ_2{\times}\IZ_2$. There are quintics that have $\IZ_5{\times}\IZ_5$ as a freely acting symmetry and members of the family, shown on the right that have freely acting symmetry 
$\IZ_5{\times}\IZ_2$. The quintic is notable, in part for the group $\IZ_5{\times}\IZ_5$ which is large. The largest free groups that act on smooth CICYs have order 32 and these act on just one space, $\IP^7[2,2,2,2]$. Now the configuration on the right of the above split is a resolution of a quintic with 50 nodes. We may go to a singular limit in which the configuration on the right has a a freely acting symmetry $\IZ_5{\times}\IZ_5$ and has 50 nodes. The singular variety can be resolved to give the Gross Popescu manifold $\text{GP}^{4,4}$, which as the notation suggests, has Hodge numbers $(4,4)$ and vanishing Euler number. The manifold $\text{GP}^{4,4}$ inherits the freely acting symmetry 
$\IZ_5{\times}\IZ_5$. The point being made here is that if the quintic and the manifold $\text{GP}^{4,4}$ are notable,
then so is the half way house corresponding to the right hand side of the above split.
The transpose of this manifold is the configuration
\beq
\hskip20pt X_{7447}~=~
\smallcicy{\IP^1\\ \IP^1\\ \IP^1\\ \IP^1\\ \IP^1\\}{1&1\\ 1&1\\ 1&1\\ 1&1\\ 1&1\\}^{5,45}_{\chi=-80}\hskip50pt
\notag\eeq
which is notable because it admits the large group $\IZ_5{\times}\IZ_2{\times}\IZ_2$ and so leads to a manifold with the small Hodge numbers $(1,3)$.

The tetraquadric and its transpose, the configuration $\IP^7[2,2,2,2]$, have free quotients of order up to 16 and 32 respectively. By allowing nodal limits of $\IP^7[2,2,2,2]$ it is possible to find freely acting symmetries of order 64. These nodal varieties can be resolved to give Gross-Popescu manifolds $\text{GP}^{2,2}$ that inherit the symmetry.

One last example: consider the transpose of one of the avatars of the manifold $X^{8,44}$. 
\beq
X_{7240}~=~~
\smallcicy{\IP^2\\ \IP^2\\ \IP^5}{1&1&1&0&0&0\\ 0&0&0&1&1&1\\ 1&1&1&1&1&1}^{3,39}_{\chi=-72}~.
\notag\eeq
Here one can find smooth manifolds that admit a freely acting $\IZ_3{\times}\IZ_3$ symmetry. However one can also find singular varieties with 36 nodes that admit a free action by $\IZ_6{\times}\IZ_6$. These nodes may be resolved to give the Gross Popescu manifold $\text{GP}^{6,6}$, which inherits the free action by $\IZ_6{\times}\IZ_6$.

The real test for these manifolds is whether they admit interesting holomorphic vector bundles. This question has been partly addressed in a recent ongoing program (see Refs.~\cite{Anderson:2007nc, Anderson:2009mh, Anderson:2011ns, Anderson:2012yf, Anderson:2013xka}) aiming to construct in a systematic manner large classes of holomorphic and poly-stable vector bundles, realised as monad bundles or sums of line bundles over CICY quotients, in a search for phenomenologically viable models. So far, these searches were limited to favourable CICY quotients with a small number ($<7$) of K\"ahler parameters, but have already revealed hundreds of models with the correct gauge group, an exact MSSM spectrum and one or several pairs of Higgs doublets. Interestingly, the study undertaken in Ref.~\cite{Anderson:2014hia} reinforced the role played by manifolds with non-trivial fundamental group, and hence by CICY quotients, revealing a significant conflict between direct symmetry breaking approaches in heterotic compactifications and a realistic particle spectrum. 
A~parallel search for interesting vector bundles over the $16$ manifolds from the Kreuzer-Skarke list exhibiting a non-trivial fundamental group was undertaken in Refs.~\cite{He:2009wi, He:2011rs, He:2013ofa}. The full extension of this work to hypersurfaces in toric varieties would require a systematic study of discrete quotients thereof, analogous to Braun's classification of CICY quotients. 
\section*{Acknowledgments}
The authors would like to thank A.~Braun, A.~Lukas and Y.~Tschinkel for interesting discussions and valuable comments, and R.~Davies for comments on a draft of this paper. The work of PC is supported by EPSRC grant BKRWDM00. PC also wishes to thank the Yau Center for Mathematical Sciences of Tsinghua University for hospitality during the closing phases of this work and for the opportunity to teach a course on this material.  AC~is partially supported by the COST Short Term Scientific Mission MP1210-28996 and would like to thank the Theoretical Physics Department at Oxford University for hospitality during part of the preparation of this paper. CM would like to thank the Rhodes Trust, Exeter College Oxford, and the Yusuf and Farida Hamied Foundation for support during this work.
\newpage
\begin{appendix}
\section{Other $\IZ_2$-quotients}\label{app:Z2quotients}
In this appendix we list the Hodge numbers for the $\IZ_2-$quotients of manifolds that have not been discussed in the previous sections. According to~\cite{Braun:2010vc}, the CICY list contains $166$ manifolds that admit smooth quotients by $\IZ_2$. Out of these, $46$ have been presented in the main part of the text. From the remaining $121$, $36$ correspond to favourable embeddings, and hence one can compute the Hodge numbers for the resulting quotients by counting K\"ahler parameters. 
\subsection{Favourable embeddings}
For each of the $36$ manifolds mentioned above, we list below the position in the CICY list, the configuration matrix and the Hodge numbers for the $\IZ_2-$quotients. 

\begin{center}
\begin{longtable}{|c|>{\quad}c<{\quad}|c|}

\hline \multicolumn{1}{|c|}{\str\textbf{~~CICY \#~~}}&  \multicolumn{1}{|c|}{\textbf{~~~~~Configuration Matrix~~~~~}} & \multicolumn{1}{|c|}{$ ~~~( h^{1,1},\, h^{2,1})~~~$} \\ \hline 
\endfirsthead

\hline 
\str\textbf{CICY \#} &
\textbf{Configuration Matrix} &
$ ( h^{1,1},\, h^{2,1})$ \\ \hline 
\endhead

\hline\hline \multicolumn{3}{|r|}{{\str Continued on next page}} \\ \hline
\endfoot

\hline\hline\multicolumn{3}{|c|}{\str}\\ \hline
\endlastfoot

\hline\hline

$4109$ 
& \simpentry{\cicy{\IP^1\\\IP^1\\\IP^1\\\IP^1\\\IP^2\\\IP^2}
{ 1 & 0 & 1 & 0 & 0 \\
 1 & 0 & 1 & 0 & 0 \\
 0 & 1 & 0 & 0 & 1 \\
 0 & 1 & 0 & 0 & 1 \\
 1 & 0 & 0 & 1 & 1 \\
 0 & 1 & 1 & 1 & 0 \\}_{-64}^{\,8,\,40}
 }
  & $(5,15)$  \\
\hline

$5273$
& \simpentry{\cicy{\IP^1\\\IP^1\\\IP^1\\\IP^1\\\IP^2\\\IP^2}
{ 1 & 1 & 0 & 0 & 0 \\
 0 & 0 & 0 & 0 & 2 \\
 0 & 0 & 1 & 1 & 0 \\
 0 & 0 & 1 & 1 & 0 \\
 1 & 0 & 1 & 0 & 1 \\
 0 & 1 & 0 & 1 & 1 \\}_{-48}^{\,6,\,30}}
   
 & $(5,17)$ \\ 
\hline

$5425$
&\simpentry{\cicy{\IP^1\\\IP^1\\\IP^1\\\IP^1\\\IP^3\\\IP^3}
{ 1 & 1 & 0 & 0 & 0 & 0 & 0 \\
 0 & 0 & 1 & 1 & 0 & 0 & 0 \\
 0 & 0 & 0 & 0 & 1 & 1 & 0 \\
 0 & 0 & 0 & 0 & 1 & 1 & 0 \\
 1 & 0 & 1 & 0 & 1 & 0 & 1 \\
 0 & 1 & 0 & 1 & 0 & 1 & 1 \\}^{6,30}_{-48}}
& $(5,17)$ \\ \hline

$5958$& 
\simpentry{\cicy{\IP^1\\\IP^1\\\IP^1\\\IP^1\\\IP^4\\\IP^4}
{ 1 & 1 & 0 & 0 & 0 & 0 & 0 & 0 & 0 \\
 0 & 0 & 1 & 1 & 0 & 0 & 0 & 0 & 0 \\
 0 & 0 & 0 & 0 & 1 & 1 & 0 & 0 & 0 \\
 0 & 0 & 0 & 0 & 0 & 0 & 1 & 1 & 0 \\
 1 & 0 & 1 & 0 & 1 & 0 & 1 & 0 & 1 \\
 0 & 1 & 0 & 1 & 0 & 1 & 0 & 1 & 1 \\}^{6,32}_{-52}}& $(5,18)$ \\ \hline

$6204$& 
\simpentry{\cicy{\IP^1\\\IP^1\\\IP^1\\\IP^3\\\IP^3}
{ 1 & 1 & 0 & 0 & 0 & 0 \\
 0 & 0 & 1 & 1 & 0 & 0 \\
 0 & 0 & 1 & 1 & 0 & 0 \\
 1 & 0 & 1 & 0 & 1 & 1 \\
 0 & 1 & 0 & 1 & 1 & 1 \\}^{5,33}_{-56}}& $(4,18)$ \\ \hline

$6225$& 
\simpentry{\cicy{\IP^1\\\IP^1\\\IP^1\\\IP^2\\\IP^2}
{
 0 & 0 & 0 & 2 \\
 0 & 1 & 1 & 0 \\
 0 & 1 & 1 & 0 \\
 1 & 1 & 0 & 1 \\
 1 & 0 & 1 & 1 \\
}^{5,33}_{-56}}& $(4,18)$ \\ \hline

 $6724$& \simpentry{\cicy{\IP^1\\\IP^1\\\IP^1\\\IP^4\\\IP^4}
{
 1 & 1 & 0 & 0 & 0 & 0 & 0 & 0 \\
 0 & 0 & 1 & 1 & 0 & 0 & 0 & 0 \\
 0 & 0 & 0 & 0 & 1 & 1 & 0 & 0 \\
 1 & 0 & 1 & 0 & 1 & 0 & 1 & 1 \\
 0 & 1 & 0 & 1 & 0 & 1 & 1 & 1 \\
}^{5,37}_{-64}}& $(4,20)$ \\ \hline

 $6732$& \simpentry{\cicy{\IP^1\\\IP^1\\\IP^1\\\IP^1\\\IP^5}
{
 1 & 1 & 0 & 0 & 0 & 0 \\
 0 & 0 & 1 & 1 & 0 & 0 \\
 0 & 0 & 0 & 0 & 1 & 1 \\
 0 & 0 & 0 & 0 & 2 & 0 \\
 1 & 1 & 1 & 1 & 1 & 1 \\
}^{5,37}_{-64}}& $(5,21)$ \\ \hline

$6738$& \simpentry{\cicy{\IP^1\\\IP^1\\\IP^1\\\IP^1\\\IP^3\\\IP^3}
{
 1 & 1 & 0 & 0 & 0 & 0 & 0 \\
 0 & 0 & 1 & 1 & 0 & 0 & 0 \\
 0 & 0 & 0 & 0 & 1 & 1 & 0 \\
 0 & 0 & 0 & 0 & 0 & 0 & 2 \\
 1 & 0 & 1 & 0 & 1 & 0 & 1 \\
 0 & 1 & 0 & 1 & 0 & 1 & 1 \\
}^{6,38}_{-64}}& $(5,21)$ \\ \hline

 $6770$& \simpentry{\cicy{\IP^1\\\IP^1\\\IP^1\\\IP^1\\\IP^1}{
 1 & 1 \\
 1 & 1 \\
 1 & 1 \\
 2 & 0 \\
 0 & 2 \\}^{5,37}_{-64}}& $(5,21)$ \\ \hline

 $6777$& \simpentry{\cicy{\IP^1\\\IP^1\\\IP^1\\\IP^1\\\IP^3}
{
 1 & 1 & 0 & 0 \\
 0 & 0 & 0 & 2 \\
 0 & 0 & 2 & 0 \\
 2 & 0 & 0 & 0 \\
 1 & 1 & 1 & 1 \\
}^{5,37}_{-64}}& $(5,21)$ \\ \hline

 $6802$& \simpentry{\cicy{\IP^1\\\IP^1\\\IP^1\\\IP^1\\\IP^4}
{
 1 & 1 & 0 & 0 & 0 \\
 0 & 0 & 2 & 0 & 0 \\
 0 & 0 & 0 & 1 & 1 \\
 0 & 0 & 0 & 1 & 1 \\
 1 & 1 & 1 & 1 & 1 \\
}^{5,37}_{-64}}& $(5,21)$ \\ \hline

 $6804$& \simpentry{\cicy{\IP^1\\\IP^1\\\IP^1\\\IP^2\\\IP^2}
{
 1 & 1 & 0 & 0 \\
 0 & 0 & 2 & 0 \\
 0 & 0 & 0 & 2 \\
 1 & 0 & 1 & 1 \\
 0 & 1 & 1 & 1 \\}^{5,37}_{-64}}& $(4,20)$ \\ \hline

$6831$& \simpentry{\cicy{\IP^1\\\IP^1\\\IP^3\\\IP^3}
{
 1 & 1 & 0 & 0 & 0 \\
 1 & 1 & 0 & 0 & 0 \\
 1 & 0 & 1 & 1 & 1 \\
 0 & 1 & 1 & 1 & 1 \\
}^{4,36}_{-64}}& $(3,19)$ \\ \hline

 $6834$& \simpentry{\cicy{\IP^1\\\IP^1\\\IP^1\\\IP^1\\\IP^3}
{
 2 & 0 & 0 & 0 \\
 0 & 0 & 0 & 2 \\
 0 & 1 & 1 & 0 \\
 0 & 1 & 1 & 0 \\
 1 & 1 & 1 & 1 \\
}^{5,37}_{-64}}& $(5,21)$ \\ \hline

 $6890$& \simpentry{\cicy{\IP^1\\\IP^1\\\IP^1\\\IP^1\\\IP^4}
{
 1 & 1 & 0 & 0 & 0 \\
 0 & 0 & 1 & 1 & 0 \\
 0 & 0 & 0 & 0 & 2 \\
 0 & 0 & 2 & 0 & 0 \\
 1 & 1 & 1 & 1 & 1 \\}^{5,37}_{-64}}& $(5,21)$ \\ \hline
 
 $6896$& \simpentry{\cicy{\IP^1\\\IP^1\\\IP^1\\\IP^1\\\IP^5}{
 1 & 1 & 0 & 0 & 0 & 0 \\
 0 & 0 & 1 & 1 & 0 & 0 \\
 0 & 0 & 0 & 0 & 1 & 1 \\
 0 & 0 & 0 & 0 & 1 & 1 \\
 1 & 1 & 1 & 1 & 1 & 1 \\}^{5,37}_{-64}}& $(5,21)$ \\ \hline

 $7204$& \simpentry{\cicy{\IP^1\\\IP^1\\\IP^1\\\IP^4}{1 & 1 & 0 & 0 \\
 0 & 0 & 2 & 0 \\
 2 & 0 & 0 & 0 \\
 1 & 1 & 1 & 2 \\}^{4,40}_{-72}}& $(4,22)$ \\ \hline

 $7218$& \simpentry{\cicy{\IP^1\\\IP^1\\\IP^1\\\IP^5}
{ 1 & 1 & 0 & 0 & 0 \\
 0 & 0 & 1 & 1 & 0 \\
 0 & 0 & 1 & 1 & 0 \\
 1 & 1 & 1 & 1 & 2 \\}^{4,40}_{-72}}& $(4,22)$ \\ \hline

 $7241$& \simpentry{\cicy{\IP^1\\\IP^1\\\IP^1\\\IP^4}{2 & 0 & 0 & 0 \\
 0 & 1 & 1 & 0 \\
 0 & 1 & 1 & 0 \\
 1 & 1 & 1 & 2 \\}^{4,40}_{-72}}& $(4,22)$ \\ \hline

 $7245$& \simpentry{\cicy{\IP^1\\\IP^1\\\IP^2\\\IP^2}{ 2 & 0 & 0 \\
 0 & 0 & 2 \\
 1 & 1 & 1 \\
 1 & 1 & 1 \\}^{4,40}_{-72}}& $(3,21)$ \\ \hline

 $7270$& \simpentry{\cicy{\IP^1\\\IP^1\\\IP^1\\\IP^5}{ 1 & 1 & 0 & 0 & 0 \\
 0 & 0 & 1 & 1 & 0 \\
 0 & 0 & 2 & 0 & 0 \\
 1 & 1 & 1 & 1 & 2 \\}^{4,40}_{-72}}& $(4,22)$ \\ \hline

 $7279$& \simpentry{\cicy{\IP^1\\\IP^1\\\IP^1\\\IP^3\\\IP^3}{ 1 & 1 & 0 & 0 & 0 & 0 \\
 0 & 0 & 1 & 1 & 0 & 0 \\
 0 & 0 & 0 & 0 & 2 & 0 \\
 1 & 0 & 1 & 0 & 1 & 1 \\
 0 & 1 & 0 & 1 & 1 & 1 \\}^{5,41}_{-72}}& $(4,22)$ \\ \hline

 $7403$& \simpentry{\cicy{\IP^1\\\IP^1\\\IP^4\\\IP^4}{ 1 & 1 & 0 & 0 & 0 & 0 & 0 \\
 0 & 0 & 1 & 1 & 0 & 0 & 0 \\
 1 & 0 & 1 & 0 & 1 & 1 & 1 \\
 0 & 1 & 0 & 1 & 1 & 1 & 1 \\}^{4,42}_{-76}}& $(3,22)$ \\ \hline

 $7450$& \simpentry{\cicy{\IP^1\\\IP^1\\\IP^5}{ 0 & 0 & 1 & 1 \\
 0 & 0 & 0 & 2 \\
 2 & 2 & 1 & 1 \\}^{3,43}_{-80}}& $(3,23)$ \\ \hline

 $7468$& \simpentry{\cicy{\IP^1\\\IP^1\\\IP^3\\\IP^3}{ 1 & 1 & 0 & 0 & 0 \\
 0 & 0 & 2 & 0 & 0 \\
 1 & 0 & 1 & 1 & 1 \\
 0 & 1 & 1 & 1 & 1 \\}^{4,44}_{-80}}& $(3,23)$ \\ \hline

 $7481$& \simpentry{\cicy{\IP^1\\\IP^1\\\IP^5}{ 0 & 0 & 1 & 1 \\
 0 & 0 & 1 & 1 \\
 2 & 2 & 1 & 1 \\}^{3,43}_{-80}}& $(3,23)$ \\ \hline

 $7636$& \simpentry{\cicy{\IP^1\\\IP^4\\\IP^4}{ 1 & 1 & 0 & 0 & 0 & 0 \\
 1 & 0 & 1 & 1 & 1 & 1 \\
 0 & 1 & 1 & 1 & 1 & 1 \\}^{3,47}_{-88}}& $(2,24)$ \\ \hline

 $7647$& \simpentry{\cicy{\IP^1\\\IP^3\\\IP^3}{
 0 & 0 & 0 & 2 \\
 1 & 1 & 1 & 1 \\
 1 & 1 & 1 & 1 \\}^{3,47}_{-88}}& $(2,24)$ \\ \hline

 $7719$& \simpentry{\cicy{\IP^1\\\IP^1\\\IP^1\\\IP^4}{ 1 & 1 & 0 & 0 \\
 0 & 0 & 0 & 2 \\
 0 & 0 & 2 & 0 \\
 1 & 1 & 2 & 1 \\}^{4,52}_{-96}}& $(4,28)$ \\ \hline

 $7736$& \simpentry{\cicy{\IP^1\\\IP^1\\\IP^1\\\IP^3}{0 & 0 & 2 \\
 2 & 0 & 0 \\
 0 & 2 & 0 \\
 1 & 2 & 1 \\}^{4,52}_{-96}}& $(4,28)$ \\ \hline

 $7742$& \simpentry{\cicy{\IP^1\\\IP^1\\\IP^1\\\IP^5}{1 & 1 & 0 & 0 & 0 \\
 0 & 0 & 1 & 1 & 0 \\
 0 & 0 & 0 & 0 & 2 \\
 1 & 1 & 1 & 1 & 2 \\}^{4,52}_{-96}}& $(4,28)$ \\ \hline

 $7761$& \simpentry{\cicy{\IP^4\\\IP^4}{
 1 & 1 & 1 & 1 & 1 \\
 1 & 1 & 1 & 1 & 1 \\}^{2,52}_{-100}}& $(1,26)$ \\ \hline

 $7788$& \simpentry{\cicy{\IP^1\\\IP^1\\\IP^5}{1 & 1 & 0 & 0 \\
 0 & 0 & 2 & 0 \\
 1 & 1 & 2 & 2 \\}^{3,55}_{-104}}& $(3,29)$ \\ \hline

 $7792$& \simpentry{\cicy{\IP^1\\\IP^1\\\IP^4}{0 & 2 & 0 \\
 2 & 0 & 0 \\
 2 & 1 & 2 \\}^{3,55}_{-104}}& $(3,29)$ \\ \hline

 $7822$& \simpentry{\cicy{\IP^1\\\IP^5}{0 & 0 & 2 \\
 2 & 2 & 2 \\}^{2,58}_{-112}}& $(2,30)$ \\ \hline

\end{longtable}
\end{center}

\newpage
\subsection{Hodge numbers obtained by counting complex structure parameters}
Apart from the manifolds discussed in the main body of the text, the CICY list contains $17$ manifolds that admit smooth $\IZ_2-$quotients, and for which the polynomial deformation method is applicable.

\begin{center}
\begin{longtable}{|c|>{\quad}c<{\quad}|c|}

\hline \multicolumn{1}{|c|}{\str\textbf{~~CICY \#~~}}&  \multicolumn{1}{|c|}{\textbf{~~~~~Configuration Matrix~~~~~}} & \multicolumn{1}{|c|}{$ ~~~( h^{1,1},\, h^{2,1})~~~$} \\ \hline 
\endfirsthead

\hline 
\str\textbf{CICY \#} &
\textbf{Configuration Matrix} &
$ ( h^{1,1},\, h^{2,1})$ \\ \hline 
\endhead

\hline\hline \multicolumn{3}{|r|}{{\str Continued on next page}} \\ \hline
\endfoot

\hline\hline\multicolumn{3}{|c|}{\str}\\ \hline
\endlastfoot

\hline\hline

$4$& \simpentry{\cicy{\IP^1\\\IP^1\\\IP^1\\\IP^1\\\IP^1\\\IP^1\\\IP^1\\\IP^1}{1 & 1 & 0 & 0 & 0 \\
 0 & 0 & 1 & 0 & 1 \\
 1 & 0 & 0 & 1 & 0 \\
 0 & 1 & 0 & 0 & 1 \\
 0 & 1 & 1 & 0 & 0 \\
 0 & 0 & 1 & 1 & 0 \\
 0 & 0 & 0 & 1 & 1 \\
 2 & 0 & 0 & 0 & 0 \\}^{15,15}_{0}}& $(9,9)$ \\ \hline

$5$& \simpentry{\cicy{\IP^1\\\IP^1\\\IP^1\\\IP^1\\\IP^1\\\IP^1\\\IP^1\\\IP^1\\\IP^1}{1 & 1 & 0 & 0 & 0 & 0 \\
 0 & 0 & 1 & 0 & 0 & 1 \\
 0 & 0 & 0 & 1 & 1 & 0 \\
 1 & 0 & 0 & 0 & 0 & 1 \\
 0 & 1 & 1 & 0 & 0 & 0 \\
 0 & 0 & 1 & 0 & 1 & 0 \\
 0 & 0 & 0 & 1 & 0 & 1 \\
 1 & 0 & 0 & 1 & 0 & 0 \\
 0 & 1 & 0 & 0 & 1 & 0 \\}^{15,15}_{0}}& $(9,9)$ \\ \hline

$6$& \simpentry{\cicy{\IP^1\\\IP^1\\\IP^1\\\IP^1\\\IP^1\\\IP^1\\\IP^1\\\IP^1\\\IP^1}{1 & 1 & 0 & 0 & 0 & 0 \\
 0 & 0 & 1 & 0 & 0 & 1 \\
 0 & 0 & 0 & 1 & 1 & 0 \\
 1 & 0 & 0 & 0 & 0 & 1 \\
 0 & 1 & 1 & 0 & 0 & 0 \\
 0 & 0 & 1 & 0 & 1 & 0 \\
 0 & 0 & 0 & 1 & 0 & 1 \\
 0 & 1 & 0 & 1 & 0 & 0 \\
 1 & 0 & 0 & 0 & 1 & 0 \\}^{15,15}_{0}}& $(9,9)$ \\ \hline

$90$& \simpentry{\cicy{\IP^1\\\IP^1\\\IP^1\\\IP^1\\\IP^2\\\IP^2\\\IP^2\\\IP^2}{1 & 1 & 0 & 0 & 0 & 0 & 0 & 0 & 0 \\
 0 & 0 & 1 & 0 & 1 & 0 & 0 & 0 & 0 \\
 0 & 0 & 0 & 1 & 1 & 0 & 0 & 0 & 0 \\
 0 & 0 & 0 & 0 & 2 & 0 & 0 & 0 & 0 \\
 0 & 0 & 0 & 0 & 0 & 1 & 1 & 0 & 1 \\
 0 & 0 & 0 & 0 & 0 & 1 & 0 & 1 & 1 \\
 1 & 0 & 0 & 1 & 0 & 0 & 1 & 0 & 0 \\
 0 & 1 & 1 & 0 & 0 & 0 & 0 & 1 & 0 \\}^{13,17}_{-8}}& $(9,11)$ \\ \hline

$261$& \simpentry{\cicy{\IP^1\\\IP^1\\\IP^1\\\IP^1\\\IP^1\\\IP^1\\\IP^3\\\IP^3}{1 & 1 & 0 & 0 & 0 & 0 & 0 & 0 & 0 \\
 0 & 0 & 1 & 1 & 0 & 0 & 0 & 0 & 0 \\
 0 & 0 & 0 & 0 & 0 & 1 & 1 & 0 & 0 \\
 0 & 0 & 0 & 0 & 1 & 0 & 0 & 1 & 0 \\
 0 & 0 & 0 & 0 & 1 & 0 & 0 & 0 & 1 \\
 0 & 0 & 0 & 0 & 2 & 0 & 0 & 0 & 0 \\
 1 & 0 & 1 & 0 & 0 & 1 & 0 & 0 & 1 \\
 0 & 1 & 0 & 1 & 0 & 0 & 1 & 1 & 0 \\}^{11,19}_{-16}}& $(8,12)$ \\ \hline

$343$& \simpentry{\cicy{\IP^1\\\IP^1\\\IP^1\\\IP^1\\\IP^1\\\IP^1\\\IP^2\\\IP^2}{1 & 1 & 0 & 0 & 0 & 0 & 0 \\
 0 & 0 & 1 & 0 & 1 & 0 & 0 \\
 0 & 0 & 0 & 1 & 1 & 0 & 0 \\
 0 & 0 & 0 & 0 & 2 & 0 & 0 \\
 0 & 0 & 0 & 0 & 0 & 1 & 1 \\
 0 & 0 & 0 & 0 & 0 & 1 & 1 \\
 1 & 0 & 0 & 1 & 0 & 0 & 1 \\
 0 & 1 & 1 & 0 & 0 & 1 & 0 \\}^{11,19}_{-16}}& $(8,12)$ \\ \hline

$376$& \simpentry{\cicy{\IP^1\\\IP^1\\\IP^1\\\IP^1\\\IP^2\\\IP^2}{1 & 1 & 0 & 0 & 0 \\
 0 & 0 & 0 & 0 & 2 \\
 1 & 0 & 0 & 0 & 1 \\
 0 & 1 & 0 & 0 & 1 \\
 0 & 1 & 1 & 1 & 0 \\
 1 & 0 & 1 & 1 & 0 \\}^{11,19}_{-16}}& $(8,12)$ \\ \hline

$379$& \simpentry{\cicy{\IP^1\\\IP^2\\\IP^2\\\IP^2\\\IP^2\\\IP^2\\\IP^2}{1 & 1 & 0 & 0 & 0 & 0 & 0 & 0 & 0 & 0 \\
 0 & 0 & 1 & 1 & 1 & 0 & 0 & 0 & 0 & 0 \\
 0 & 0 & 1 & 1 & 0 & 1 & 0 & 0 & 0 & 0 \\
 1 & 0 & 0 & 0 & 1 & 0 & 1 & 0 & 0 & 0 \\
 0 & 1 & 0 & 0 & 0 & 1 & 0 & 1 & 0 & 0 \\
 0 & 0 & 0 & 0 & 0 & 0 & 1 & 0 & 1 & 1 \\
 0 & 0 & 0 & 0 & 0 & 0 & 0 & 1 & 1 & 1 \\}^{11,19}_{-16}}& $(8,12)$ \\ \hline

 $1262$& \simpentry{\cicy{\IP^1\\\IP^1\\\IP^1\\\IP^2\\\IP^2\\\IP^2\\\IP^2}{1 & 1 & 0 & 0 & 0 & 0 & 0 & 0 \\
 0 & 0 & 0 & 0 & 0 & 1 & 1 & 0 \\
 0 & 0 & 0 & 0 & 0 & 1 & 1 & 0 \\
 0 & 0 & 1 & 1 & 0 & 0 & 0 & 1 \\
 0 & 0 & 1 & 0 & 1 & 0 & 0 & 1 \\
 1 & 0 & 0 & 1 & 0 & 1 & 0 & 0 \\
 0 & 1 & 0 & 0 & 1 & 0 & 1 & 0 \\}^{9,21}_{-24}}& $(7,13)$ \\ \hline

 $1701$& \simpentry{\cicy{\IP^1\\\IP^1\\\IP^1\\\IP^2\\\IP^2\\\IP^3\\\IP^3}{1 & 1 & 0 & 0 & 0 & 0 & 0 & 0 & 0 & 0 \\
 0 & 0 & 1 & 1 & 0 & 0 & 0 & 0 & 0 & 0 \\
 0 & 0 & 0 & 0 & 1 & 1 & 0 & 0 & 0 & 0 \\
 0 & 0 & 0 & 0 & 0 & 0 & 1 & 1 & 1 & 0 \\
 0 & 0 & 0 & 0 & 0 & 0 & 1 & 1 & 0 & 1 \\
 1 & 0 & 1 & 0 & 1 & 0 & 0 & 0 & 1 & 0 \\
 0 & 1 & 0 & 1 & 0 & 1 & 0 & 0 & 0 & 1 \\}^{9,23}_{-28}}& $(7,14)$ \\ \hline

 $2544$& \simpentry{\cicy{\IP^1\\\IP^1\\\IP^1\\\IP^1\\\IP^1\\\IP^2\\\IP^2}{1 & 1 & 0 & 0 & 0 & 0 \\
 0 & 0 & 1 & 0 & 1 & 0 \\
 0 & 0 & 1 & 0 & 1 & 0 \\
 0 & 0 & 0 & 1 & 0 & 1 \\
 0 & 0 & 0 & 1 & 0 & 1 \\
 1 & 0 & 1 & 0 & 0 & 1 \\
 0 & 1 & 0 & 1 & 1 & 0 \\}^{7,23}_{-32}}& $(6,14)$ \\ \hline

$3381$& \simpentry{\cicy{\IP^1\\\IP^2\\\IP^2\\\IP^2\\\IP^2}{1 & 1 & 0 & 0 & 0 & 0 \\
 1 & 0 & 1 & 1 & 0 & 0 \\
 0 & 1 & 1 & 1 & 0 & 0 \\
 0 & 1 & 0 & 0 & 1 & 1 \\
 1 & 0 & 0 & 0 & 1 & 1 \\}^{9,27}_{-36}}& $(7,16)$ \\ \hline

 $3929$& \simpentry{\cicy{\IP^1\\\IP^1\\\IP^1\\\IP^1\\\IP^1\\\IP^3\\\IP^3}{1 & 1 & 0 & 0 & 0 & 0 & 0 & 0 \\
 0 & 0 & 1 & 1 & 0 & 0 & 0 & 0 \\
 0 & 0 & 0 & 0 & 1 & 1 & 0 & 0 \\
 0 & 0 & 0 & 0 & 0 & 0 & 1 & 1 \\
 0 & 0 & 0 & 0 & 0 & 0 & 1 & 1 \\
 1 & 0 & 1 & 0 & 1 & 0 & 1 & 0 \\
 0 & 1 & 0 & 1 & 0 & 1 & 0 & 1 \\}^{7,27}_{-40}}& $(6,16)$ \\ \hline

$4108$& \simpentry{\cicy{\IP^1\\\IP^1\\\IP^2\\\IP^2\\\IP^3}{0 & 0 & 0 & 1 & 1 & 0 \\
 0 & 0 & 0 & 1 & 1 & 0 \\
 1 & 1 & 0 & 0 & 0 & 1 \\
 1 & 0 & 1 & 0 & 0 & 1 \\
 0 & 1 & 1 & 1 & 1 & 0 \\}^{7,27}_{-40}}& $(6,16)$ \\ \hline

$4335$& \simpentry{\cicy{\IP^1\\\IP^1\\\IP^1\\\IP^1\\\IP^1\\\IP^4\\\IP^4}{1 & 1 & 0 & 0 & 0 & 0 & 0 & 0 & 0 & 0 \\
 0 & 0 & 1 & 1 & 0 & 0 & 0 & 0 & 0 & 0 \\
 0 & 0 & 0 & 0 & 1 & 1 & 0 & 0 & 0 & 0 \\
 0 & 0 & 0 & 0 & 0 & 0 & 1 & 1 & 0 & 0 \\
 0 & 0 & 0 & 0 & 0 & 0 & 0 & 0 & 1 & 1 \\
 1 & 0 & 1 & 0 & 1 & 0 & 1 & 0 & 1 & 0 \\
 0 & 1 & 0 & 1 & 0 & 1 & 0 & 1 & 0 & 1 \\}^{7,27}_{-40}}& $(6,16)$ \\ \hline

$5423$& \simpentry{\cicy{\IP^1\\\IP^1\\\IP^2\\\IP^2\\\IP^5}{1 & 1 & 0 & 0 & 0 & 0 & 0 & 0 \\
 0 & 0 & 1 & 1 & 0 & 0 & 0 & 0 \\
 0 & 0 & 0 & 0 & 1 & 1 & 1 & 0 \\
 0 & 0 & 0 & 0 & 1 & 1 & 0 & 1 \\
 1 & 1 & 1 & 1 & 0 & 0 & 1 & 1 \\}^{7,31}_{-48}}& $(6,18)$ \\ \hline

$6173$& \simpentry{\cicy{\IP^1\\\IP^1\\\IP^1\\\IP^2\\\IP^2}{1 & 1 & 0 & 0 \\
 1 & 1 & 0 & 0 \\
 1 & 1 & 0 & 0 \\
 1 & 0 & 1 & 1 \\
 0 & 1 & 1 & 1 \\}^{7,35}_{-56}}& $(6,20)$ \\ \hline
\end{longtable}
\end{center}
\end{appendix}
\newpage
\renewcommand{\baselinestretch}{0.9}\normalsize
\bibliographystyle{utcaps}
\bibliography{bibfile}

\providecommand{\href}[2]{#2}\begingroup\raggedright\begin{thebibliography}{10}

\bibitem{Candelas:1985en}
P.~Candelas, G.~T. Horowitz, A.~Strominger, and E.~Witten, ``{Vacuum
  Configurations for Superstrings},'' {\em Nucl.Phys.} {\bf B258} (1985)
46--74.

\bibitem{Braun:2005ux}
V.~Braun, Y.-H. He, B.~A. Ovrut, and T.~Pantev, ``{A Heterotic standard
  model},'' {\em Phys.Lett.} {\bf B618} (2005) 252--258,
\href{http://arXiv.org/abs/hep-th/0501070}{{\tt hep-th/0501070}}.

\bibitem{Braun:2005bw}
V.~Braun, Y.-H. He, B.~A. Ovrut, and T.~Pantev, ``{A Standard model from the
  E(8) x E(8) heterotic superstring},'' {\em JHEP} {\bf 0506} (2005) 039,
\href{http://arXiv.org/abs/hep-th/0502155}{{\tt hep-th/0502155}}.

\bibitem{Anderson:2009mh}
L.~B. Anderson, J.~Gray, Y.-H. He, and A.~Lukas, ``{Exploring Positive Monad
  Bundles And a New Heterotic Standard Model},'' {\em JHEP} {\bf 1002} (2010)
  054,
\href{http://arXiv.org/abs/0911.1569}{{\tt 0911.1569}}.

\bibitem{Bouchard:2005ag}
V.~Bouchard and R.~Donagi, ``{An SU(5) heterotic standard model},'' {\em
  Phys.Lett.} {\bf B633} (2006) 783--791,
\href{http://arXiv.org/abs/hep-th/0512149}{{\tt hep-th/0512149}}.

\bibitem{Braun:2011ni}
V.~Braun, P.~Candelas, R.~Davies, and R.~Donagi, ``{The MSSM Spectrum from
  (0,2)-Deformations of the Heterotic Standard Embedding},'' {\em JHEP} {\bf
  1205} (2012) 127,
\href{http://arXiv.org/abs/1112.1097}{{\tt 1112.1097}}.

\bibitem{Anderson:2011ns}
L.~B. Anderson, J.~Gray, A.~Lukas, and E.~Palti, ``{Two Hundred Heterotic
  Standard Models on Smooth Calabi-Yau Threefolds},'' {\em Phys.Rev.} {\bf D84}
  (2011) 106005,
\href{http://arXiv.org/abs/1106.4804}{{\tt 1106.4804}}.

\bibitem{Anderson:2012yf}
L.~B. Anderson, J.~Gray, A.~Lukas, and E.~Palti, ``{Heterotic Line Bundle
  Standard Models},'' {\em JHEP} {\bf 1206} (2012) 113,
\href{http://arXiv.org/abs/1202.1757}{{\tt 1202.1757}}.

\bibitem{Anderson:2013xka}
L.~B. Anderson, A.~Constantin, J.~Gray, A.~Lukas, and E.~Palti, ``{A
  Comprehensive Scan for Heterotic SU(5) GUT models},'' {\em JHEP} {\bf 01}
  (2014) 047,
\href{http://arXiv.org/abs/1307.4787}{{\tt 1307.4787}}.

\bibitem{Gross:2001as}
M.~Gross and S.~Popescu, ``Calabi--Yau threefolds and moduli of abelian
  surfaces I,'' {\em Compositio Mathematica} {\bf 127} (2001), no.~2, 169--228.

\bibitem{Candelas:2008wb}
P.~Candelas and R.~Davies, ``{New Calabi-Yau Manifolds with Small Hodge
  Numbers},'' {\em Fortsch.Phys.} {\bf 58} (2010) 383--466,
\href{http://arXiv.org/abs/0809.4681}{{\tt 0809.4681}}.

\bibitem{Braun:2010vc}
V.~Braun, ``{On Free Quotients of Complete Intersection Calabi-Yau
  Manifolds},'' {\em JHEP} {\bf 1104} (2011) 005,
\href{http://arXiv.org/abs/1003.3235}{{\tt 1003.3235}}.

\bibitem{Candelas:1987kf}
P.~Candelas, A.~Dale, C.~Lutken, and R.~Schimmrigk, ``{Complete Intersection
  Calabi-Yau Manifolds},'' {\em Nucl.Phys.} {\bf B298} (1988)
493.

\bibitem{Candelas:2010ve}
P.~Candelas and A.~Constantin, ``{Completing the Web of $Z_3$ - Quotients of
  Complete Intersection Calabi-Yau Manifolds},'' {\em Fortsch.Phys.} {\bf 60}
  (2012) 345--369,
\href{http://arXiv.org/abs/1010.1878}{{\tt 1010.1878}}.

\bibitem{cicylist2}
CICY list, compiled by Andre Lukas. Includes Hodge numbers and freely-acting
  discrete symmetries. Data available online at
  http://www-thphys.physics.ox.ac.uk/projects/CalabiYau/cicylist/index.html.

\bibitem{Hubsch:1992nu}
T.~Hubsch, {\em {Calabi-Yau manifolds: A Bestiary for physicists}}.
\newblock World Scientific, Singapore,
1994.
\newblock

\bibitem{He:1990pg}
A.-m. He and P.~Candelas, ``{On the Number of Complete Intersection
  {Calabi-Yau} Manifolds},'' {\em Commun.Math.Phys.} {\bf 135} (1990)
193--200.

\bibitem{Yau:1986gu}
S.-T. Yau, ``{Compact Three-Dimensional K\"ahler Manifolds with Zero Ricci
  Curvature},'' in {\em {In *Argonne/chicago 1985, Proceedings, Anomalies,
  Geometry, Topology*, 395-406}}.
\newblock
1986.
\newblock

\bibitem{Green:1987cr}
P.~S. Green, T.~Hubsch, and C.~A. Lutken, ``{All Hodge Numbers of All Complete
  Intersection Calabi-Yau Manifolds},'' {\em Class. Quant. Grav.} {\bf 6}
  (1989)
105--124.

\bibitem{Rodland:1998pm}
E.~A. Rodland, ``The Pfaffian Calabi--Yau, its mirror, and their link to the
  Grassmannian G(2, 7),'' {\em Compositio Mathematica} {\bf 122} (2000),
  no.~02, 135--149.

\bibitem{Kreuzer:2000xy}
M.~Kreuzer and H.~Skarke, ``{Complete classification of reflexive polyhedra in
  four-dimensions},'' {\em Adv.Theor.Math.Phys.} {\bf 4} (2002) 1209--1230,
\href{http://arXiv.org/abs/hep-th/0002240}{{\tt hep-th/0002240}}.

\bibitem{Kreuzer:2003xx}
M.~Kreuzer, E.~Riegler, and D.~Sahakyan, ``{Toric complete intersections and
  weighted projective space},'' {\em J.~Geom.~Phys.} {\bf 46} (2003) 159--173,
  \href{http://arXiv.org/abs/0103214}{{\tt 0103214}}.

\bibitem{Klemm:2004km}
A.~Klemm, M.~Kreuzer, E.~Riegler, and E.~Scheidegger, ``{Topological string
  amplitudes, complete intersection Calabi-Yau spaces and threshold
  corrections},'' {\em JHEP} {\bf 05} (2005) 023,
\href{http://arXiv.org/abs/hep-th/0410018}{{\tt hep-th/0410018}}.

\bibitem{Tonoli:2004ps}
F.~Tonoli, ``{Construction of Calabi-Yau 3-folds in P6},'' {\em J. Algebraic
  Geometry} {\bf 13} (2004) 209--232.

\bibitem{Kreuzer:2007ni}
M.~Kreuzer and B.~Nill, ``{Classification of toric Fano 5-folds},'' {\em
  Adv.~Geom.} {\bf 9} (2009) 85--97, \href{http://arXiv.org/abs/0702890}{{\tt
  0702890}}.

\bibitem{Candelas:2007ac}
P.~Candelas, X.~de~la Ossa, Y.-H. He, and B.~Szendroi, ``{Triadophilia: A
  Special Corner in the Landscape},'' {\em Adv. Theor. Math. Phys.} {\bf 12}
  (2008) 429--473,
\href{http://arXiv.org/abs/0706.3134}{{\tt 0706.3134}}.

\bibitem{Hua:2007fq}
Z.~Hua {\em et al.}, ``Classification of free actions on complete intersections
  of four quadrics,'' {\em Advances in Theoretical and Mathematical Physics}
  {\bf 15} (2011), no.~4, 973--990, \href{http://arXiv.org/abs/0707.4339}{{\tt
  0707.4339}}.

\bibitem{Kapustka:2007pc}
G.~Kapustka, ``Primitive contractions of Calabi--Yau threefolds II,'' {\em
  Journal of the London Mathematical Society} {\bf 79} (2009), no.~1, 259--271.

\bibitem{Bouchard:2007mf}
V.~Bouchard and R.~Donagi, ``{On a class of non-simply connected Calabi-Yau
  threefolds},'' {\em Commun.Num.Theor.Phys.} {\bf 2} (2008) 1--61,
\href{http://arXiv.org/abs/0704.3096}{{\tt 0704.3096}}.

\bibitem{Batyrev:2008rp}
V.~Batyrev and M.~Kreuzer, ``{Constructing new Calabi-Yau 3-folds and their
  mirrors via conifold transitions},'' {\em Adv. Theor. Math. Phys.} {\bf 14}
  (2010) 879--898,
\href{http://arXiv.org/abs/0802.3376}{{\tt 0802.3376}}.

\bibitem{Braun:2009qy}
V.~Braun, P.~Candelas, and R.~Davies, ``{A Three-Generation Calabi-Yau Manifold
  with Small Hodge Numbers},'' {\em Fortsch.Phys.} {\bf 58} (2010) 467--502,
\href{http://arXiv.org/abs/0910.5464}{{\tt 0910.5464}}.

\bibitem{Kapustka:2010pr}
G.~Kapustka, ``Projections of del Pezzo surfaces and Calabi--Yau threefolds,''
  {\em Advances in Geometry} {\bf 15} (2015), no.~2, 143--158.

\bibitem{Garbagnati:2010um}
A.~Garbagnati, ``{New families of Calabi-Yau 3-folds without maximal unipotent
  monodromy},'' \href{http://arXiv.org/abs/1005.0094}{{\tt 1005.0094}}.

\bibitem{Stapledon:2010mo}
A.~Stapledon, ``{New Mirror Pairs of Calabi-Yau Orbifolds},''
  \href{http://arXiv.org/abs/1011.5006}{{\tt 1011.5006}}.

\bibitem{Davies:2011fr}
R.~Davies, ``{The Expanding Zoo of Calabi-Yau Threefolds},'' {\em Adv. High
  Energy Phys.} {\bf 2011} (2011) 901898,
\href{http://arXiv.org/abs/1103.3156}{{\tt 1103.3156}}.

\bibitem{Davies:2011is}
R.~Davies, ``{Hyperconifold Transitions, Mirror Symmetry, and String Theory},''
  {\em Nucl. Phys.} {\bf B850} (2011) 214--231,
\href{http://arXiv.org/abs/1102.1428}{{\tt 1102.1428}}.

\bibitem{Braun:2011hd}
V.~Braun, ``{The 24-Cell and Calabi-Yau Threefolds with Hodge Numbers (1,1)},''
  {\em JHEP} {\bf 1205} (2012) 101,
\href{http://arXiv.org/abs/1102.4880}{{\tt 1102.4880}}.

\bibitem{Freitag:2011st}
E.~Freitag and R.~Salvati~Manni, ``{On Siegel threefolds with a projective
  Calabi-Yau model},'' \href{http://arXiv.org/abs/1103.2040}{{\tt 1103.2040}}.

\bibitem{Filippini:2011rf}
S.~A. Filippini and A.~Garbagnati, ``{A Rigid Calabi-Yau 3-fold},'' {\em Adv.
  Theor. Math. Phys.} {\bf 15} (2011) 1745--1788,
\href{http://arXiv.org/abs/1102.1854}{{\tt 1102.1854}}.

\bibitem{Borisov:2012xx}
L.~A. Borisov and H.~J. Nuer, ``{On (2,4) complete intersection threefolds that
  contain an Enriques surface},'' \href{http://arXiv.org/abs/1210.1903}{{\tt
  1210.1903}}.

\bibitem{bini2012}
G.~Bini and F.~F. Favale, ``Groups acting freely on Calabi-Yau threefolds
  embedded in a product of del Pezzo surfaces,'' {\em Adv. Theor. Math. Phys.}
  {\bf 16} (06, 2012) 887--993.

\bibitem{Anderson:2015iia}
L.~B. Anderson, F.~Apruzzi, X.~Gao, J.~Gray, and S.-J. Lee, ``{A New
  Construction of Calabi-Yau Manifolds: Generalized CICYs},''
\href{http://arXiv.org/abs/1507.03235}{{\tt 1507.03235}}.

\bibitem{Witten:1985xc}
E.~Witten, ``{Symmetry Breaking Patterns in Superstring Models},'' {\em
  Nucl.Phys.} {\bf B258} (1985)
75.

\bibitem{Green:1987rw}
P.~Green and T.~Hubsch, ``{Polynomial Deformations and Cohomology of Calabi-yau
  Manifolds},'' {\em Commun.Math.Phys.} {\bf 113} (1987)
505.

\bibitem{Anderson:2013qca}
L.~B. Anderson, J.~Gray, A.~Lukas, and B.~Ovrut, ``Vacuum varieties,
  holomorphic bundles and complex structure stabilization in heterotic
  theories,'' {\em Journal of High Energy Physics} {\bf 2013} (2013), no.~7,
  1--47.

\bibitem{zbMATH06053083}
I.~V. {Dolgachev}, {\em {Classical algebraic geometry. A modern view.}}
\newblock Cambridge: Cambridge University Press, 2012.

\bibitem{reid1972complete}
M.~A. Reid, {\em The complete intersection of two or more quadrics}.
\newblock PhD thesis, University of Cambridge, 1972.

\bibitem{Buchbinder:2013dna}
E.~I. Buchbinder, A.~Constantin, and A.~Lukas, ``{The Moduli Space of Heterotic
  Line Bundle Models: a Case Study for the Tetra-Quadric},'' {\em JHEP} {\bf
  1403} (2014) 025,
\href{http://arXiv.org/abs/1311.1941}{{\tt 1311.1941}}.

\bibitem{Buchbinder:2014qda}
E.~I. Buchbinder, A.~Constantin, and A.~Lukas, ``{A heterotic standard model
  with $B - L$ symmetry and a stable proton},'' {\em JHEP} {\bf 1406} (2014)
  100,
\href{http://arXiv.org/abs/1404.2767}{{\tt 1404.2767}}.

\bibitem{Buchbinder:2014sya}
E.~I. Buchbinder, A.~Constantin, and A.~Lukas, ``{Non-generic Couplings in
  Supersymmetric Standard Models},'' {\em Phys. Lett.} {\bf B748} (2015)
  251--254,
\href{http://arXiv.org/abs/1409.2412}{{\tt 1409.2412}}.

\bibitem{Constantin:2015bea}
A.~Constantin, A.~Lukas, and C.~Mishra, ``{The Family Problem: Hints from
  Heterotic Line Bundle Models},''
\href{http://arXiv.org/abs/1509.02729}{{\tt 1509.02729}}.

\bibitem{Buchbinder:2014qca}
E.~I. Buchbinder, A.~Constantin, and A.~Lukas, ``{Heterotic QCD axion},'' {\em
  Phys.Rev.} {\bf D91} (2015), no.~4, 046010,
\href{http://arXiv.org/abs/1412.8696}{{\tt 1412.8696}}.

\bibitem{Greene:1986bm}
B.~R. Greene, K.~H. Kirklin, P.~J. Miron, and G.~G. Ross, ``{A Three Generation
  Superstring Model. 1. Compactification and Discrete Symmetries},'' {\em
  Nucl.Phys.} {\bf B278} (1986)
667.

\bibitem{Greene:1986jb}
B.~R. Greene, K.~H. Kirklin, P.~J. Miron, and G.~G. Ross, ``{A Three Generation
  Superstring Model. 2. Symmetry Breaking and the Low-Energy Theory},'' {\em
  Nucl.Phys.} {\bf B292} (1987)
606.

\bibitem{Donagi:2000zf}
R.~Donagi, B.~A. Ovrut, T.~Pantev, and D.~Waldram, ``{Standard model bundles on
  nonsimply connected Calabi-Yau threefolds},'' {\em JHEP} {\bf 0108} (2001)
  053,
\href{http://arXiv.org/abs/hep-th/0008008}{{\tt hep-th/0008008}}.

\bibitem{Donagi:2000zs}
R.~Donagi, B.~A. Ovrut, T.~Pantev, and D.~Waldram, ``{Standard model
  bundles},'' {\em Adv.Theor.Math.Phys.} {\bf 5} (2002) 563--615,
\href{http://arXiv.org/abs/math/0008010}{{\tt math/0008010}}.

\bibitem{Donagi:2004ub}
R.~Donagi, Y.-H. He, B.~A. Ovrut, and R.~Reinbacher, ``{The Spectra of
  heterotic standard model vacua},'' {\em JHEP} {\bf 0506} (2005) 070,
\href{http://arXiv.org/abs/hep-th/0411156}{{\tt hep-th/0411156}}.

\bibitem{Braun:2005nv}
V.~Braun, Y.-H. He, B.~A. Ovrut, and T.~Pantev, ``{The Exact MSSM spectrum from
  string theory},'' {\em JHEP} {\bf 0605} (2006) 043,
\href{http://arXiv.org/abs/hep-th/0512177}{{\tt hep-th/0512177}}.

\bibitem{Davies:2009ub}
R.~Davies, ``{Quotients of the conifold in compact Calabi-Yau threefolds, and
  new topological transitions},'' {\em Adv.Theor.Math.Phys.} {\bf 14} (2010)
  965--990,
\href{http://arXiv.org/abs/0911.0708}{{\tt 0911.0708}}.

\bibitem{Davies:2013pna}
R.~Davies, ``{Classification and Properties of Hyperconifold Singularities and
  Transitions},''
\href{http://arXiv.org/abs/1309.6778}{{\tt 1309.6778}}.

\bibitem{Strominger:1985it}
A.~Strominger and E.~Witten, ``{New Manifolds for Superstring
  Compactification},'' {\em Commun.Math.Phys.} {\bf 101} (1985)
341.

\bibitem{Anderson:2007nc}
L.~B. Anderson, Y.-H. He, and A.~Lukas, ``{Heterotic Compactification, An
  Algorithmic Approach},'' {\em JHEP} {\bf 0707} (2007) 049,
\href{http://arXiv.org/abs/hep-th/0702210}{{\tt hep-th/0702210}}.

\bibitem{Anderson:2014hia}
L.~B. Anderson, A.~Constantin, S.-J. Lee, and A.~Lukas, ``{Hypercharge Flux in
  Heterotic Compactifications},''
\href{http://arXiv.org/abs/1411.0034}{{\tt 1411.0034}}.

\bibitem{He:2009wi}
Y.-H. He, S.-J. Lee, and A.~Lukas, ``{Heterotic Models from Vector Bundles on
  Toric Calabi-Yau Manifolds},'' {\em JHEP} {\bf 1005} (2010) 071,
\href{http://arXiv.org/abs/0911.0865}{{\tt 0911.0865}}.

\bibitem{He:2011rs}
Y.-H. He, M.~Kreuzer, S.-J. Lee, and A.~Lukas, ``{Heterotic Bundles on
  Calabi-Yau Manifolds with Small Picard Number},'' {\em JHEP} {\bf 1112}
  (2011) 039,
\href{http://arXiv.org/abs/1108.1031}{{\tt 1108.1031}}.

\bibitem{He:2013ofa}
Y.-H. He, S.-J. Lee, A.~Lukas, and C.~Sun, ``{Heterotic Model Building: 16
  Special Manifolds},''
\href{http://arXiv.org/abs/1309.0223}{{\tt 1309.0223}}.

\end{thebibliography}\endgroup
\end{document}